\documentstyle[pictex,12pt]{report}
\newcommand{\nc}{\newcommand}
\nc{\bi}{\bibitem}
\nc{\ben}{\begin{equation}}
\nc{\een}{\end{equation}}
\nc{\bea}{\begin{eqnarray}}
\nc{\eea}{\end{eqnarray}}
\nc{\bdm}{\begin{displaymath}}
\nc{\edm}{\end{displaymath}}
\nc{\NP}[1]{Nucl.\ Phys.\ {\bf #1}}
\nc{\PL}[1]{Phys.\ Lett.\ {\bf #1}}
\nc{\CMP}[1]{Commun.\ Math.\ Phys.\ {\bf #1}}
\nc{\PR}[1]{Phys.\ Rev.\ {\bf #1}}
\nc{\PRL}[1]{Phys.\ Rev.\ Lett.\ {\bf #1}}
\nc{\PTP}[1]{Prog.\ Theor.\ Phys.\ {\bf #1}}
\nc{\PTPS}[1]{Prog.\ Theor.\ Phys.\ Suppl.\ {\bf #1}}
\nc{\MPL}[1]{Mod.\ Phys.\ Lett.\ {\bf #1}}
\nc{\IJMP}[1]{Int.\ Jour.\ Mod.\ Phys.\ {\bf #1}}
\nc{\IM}[1]{Invent.\ Math.\ {\bf #1}}
\nc{\SJNP}[1]{Sov. J. Nucl. Phys.\ {\bf #1}}
\nc{\spa}{\hspace{1 cm},\hspace{1 cm}}
\nc{\C}{\mbox{\hspace{1.24mm}\rule{0.2mm}{2.5mm}\hspace{-2.7mm} C}}
\newcommand{\R}{\mbox{\hspace{.04mm}\rule{0.2mm}{2.8mm}\hspace{-1.5mm} R}}
\newcommand{\Z}{\mbox{$Z\hspace{-2mm}Z$}}
\nc{\binomial}[2]{\left (\begin{array}{c} {#1}\\ {#2} \end{array}
\right )}
\nc{\al}{\alpha}
\nc{\g}{\gamma}
\nc{\eps}{\epsilon}
\nc{\G}{\Gamma}
\nc{\D}{\Delta}
\nc{\var}{\varphi}
\nc{\ba}{\beta_\al}
\nc{\bb}{\beta_\beta}
\nc{\ga}{\g^\al}
\nc{\gb}{\g^\beta}
\nc{\kvt}{\sqrt{t}}
\nc{\hn}{h^\nu}
\nc{\kn}{k^\nu}
\nc{\dab}{{\delta_\al}^\beta}
\nc{\pa}{\partial}
\nc{\nn}{\nonumber \\ }
\nc{\hf}{\frac{1}{2}}         
\nc{\paj}{P_{-\al}^j}
\nc{\vmab}{V_{-\al}^\beta}
\nc{\vab}{V_\al^\beta}
\nc{\vib}{V_i^\beta}
\nc{\db}{\pa_\beta}
\nc{\dtb}{\delta_\theta^\beta}
\nc{\fabc}{{f_{ab}}^c}
\nc{\rton}{R^2_{(n)}}
\nc{\rjn}{R^j_{(n)}}
\nc{\br}{\langle}
\nc{\kt}{\rangle}
\nc{\bra}[1]{\langle {#1}|}
\nc{\ket}[1]{|{#1}\rangle}
\nc{\vac}{\ket{0}}
\nc{\zb}{\bar{z}}
\nc{\dz}{\frac{dz}{2\pi i}}
\nc{\oz}{\oint_0\frac{dz}{2\pi i}}
\nc{\ow}{\oint_0\frac{dw}{2\pi i}}
\nc{\wz}{\oint_w\frac{dz}{2\pi i}}
\nc{\dtp}[1]{\frac{d{#1}}{2\pi i}}
\begin{document}
\begin{titlepage}
\vspace{8cm}
\begin{center}
{\Huge\bf Applications of Free Fields}\\[.7 cm]
{\Huge\bf in 2D Current Algebra}\\[5.9 cm]

\font\thinlinefont=cmr5
\begingroup\makeatletter\ifx\SetFigFont\undefined
% extract first six characters in \fmtname
\def\x#1#2#3#4#5#6#7\relax{\def\x{#1#2#3#4#5#6}}%
\expandafter\x\fmtname xxxxxx\relax \def\y{splain}%
\ifx\x\y   % LaTeX or SliTeX?
\gdef\SetFigFont#1#2#3{%
  \ifnum #1<17\tiny\else \ifnum #1<20\small\else
  \ifnum #1<24\normalsize\else \ifnum #1<29\large\else
  \ifnum #1<34\Large\else \ifnum #1<41\LARGE\else
     \huge\fi\fi\fi\fi\fi\fi
  \csname #3\endcsname}%
\else
\gdef\SetFigFont#1#2#3{\begingroup
  \count@#1\relax \ifnum 25<\count@\count@25\fi
  \def\x{\endgroup\@setsize\SetFigFont{#2pt}}%
  \expandafter\x
    \csname \romannumeral\the\count@ pt\expandafter\endcsname
    \csname @\romannumeral\the\count@ pt\endcsname
  \csname #3\endcsname}%
\fi
\fi\endgroup
\mbox{\beginpicture
\setcoordinatesystem units <1.00000cm,1.00000cm>
\unitlength=1.00000cm
\linethickness=1pt
\setplotsymbol ({\makebox(0,0)[l]{\tencirc\symbol{'160}}})
\setshadesymbol ({\thinlinefont .})
\setlinear
%
% Fig POLYLINE object
%
\linethickness= 0.500pt
\setplotsymbol ({\thinlinefont .})
\putrule from  2.540 22.860 to  3.810 22.860
%
% Fig POLYLINE object
%
\linethickness= 0.500pt
\setplotsymbol ({\thinlinefont .})
\plot  3.810 22.860  5.080 24.130 /
%
% Fig POLYLINE object
%
\linethickness= 0.500pt
\setplotsymbol ({\thinlinefont .})
\plot  2.540 22.860  1.270 21.590 /
%
% Fig POLYLINE object
%
\linethickness= 0.500pt
\setplotsymbol ({\thinlinefont .})
\plot  3.810 22.860  5.080 21.590 /
%
% Fig POLYLINE object
%
\linethickness= 0.500pt
\setplotsymbol ({\thinlinefont .})
\plot 10.160 24.750 11.430 23.480 /
%
% Fig POLYLINE object
%
\linethickness= 0.500pt
\setplotsymbol ({\thinlinefont .})
\plot 11.430 23.480 12.700 24.750 /
%
% Fig POLYLINE object
%
\linethickness= 0.500pt
\setplotsymbol ({\thinlinefont .})
\putrule from 11.430 23.480 to 11.430 22.210
%
% Fig POLYLINE object
%
\linethickness= 0.500pt
\setplotsymbol ({\thinlinefont .})
\plot 11.430 22.210 10.160 20.940 /
%
% Fig POLYLINE object
%
\linethickness= 0.500pt
\setplotsymbol ({\thinlinefont .})
\plot 11.430 22.210 12.700 20.940 /
%
% Fig POLYLINE object
%
\linethickness= 0.500pt
\setplotsymbol ({\thinlinefont .})
\plot  1.270 24.130  2.540 22.860 /
%
% Fig TEXT object
%
\put{=} [lB] at  7.000 22.770
%
% Fig TEXT object
%
\put{$\sum$} [lB] at 9.500 22.770
%
% Fig TEXT object
%
\put{$\sum$} [lB] at  0.500 22.770
\linethickness=0pt
\putrectangle corners at  0.635 25.546 and 12.859 20.739
\endpicture}

\vspace{5 cm}
{\Large\bf J{\o}rgen Rasmussen} \\[2cm]
Thesis submitted for the degree of Doctor of Philosophy\\
in the Niels Bohr Institute at the University of Copenhagen,\\
{\bf October 1996}
\end{center}
\end{titlepage}
\newpage
\mbox{}
\newpage
\pagenumbering{roman}
\tableofcontents
\newpage
\pagenumbering{arabic}
\chapter{Introduction}

In recent years great efforts have been spent in the quest for a theory of
{\em quantum gravity}, which probably must be a unifying theory of all known 
interactions.
Based on the principles of quantum mechanics and local gauge invariance,
the standard model is a widely recognised and experimentally tested 
quantum field theory of electro-weak and strong interactions.
Einstein's theory of general relativity which is based on general
covariance and the equivalence principle, is likewise an accepted and tested
classical field theory of gravity. Though of dominant importance
for large-scale phenomena due to it's long-range attractive
property, gravity is a weak force and negligible in particle scattering
experiments done in order to test the standard model. However, in
sufficiently strong gravitational fields as occurring near black hole
singularities  and in
current descriptions of the early universe, gravity must be taken into account
in a revision of the standard model. Attempts to formulate gravity as a
quantum field theory have failed because of the non-renormalisability of
gravity, which stems from the fact that the coupling constant (Newton's
constant) has dimension as the Planck length squared in units
where $\hbar=c=1$, leading to a breakdown of standard perturbative approach.

A prime candidate for a theory of quantum gravity is (super-)string theory,
or for brevity {\em strings}. The terminology originates from the basic
description of elementary particles as excitations of one dimensional
objects, strings, of the size of the Planck length. There is definitely a need
for strings to be tiny since there is no experimental evidence for
stringy extendedness. An intriguing observation is that the graviton, the
elementary quantum of the gravitational field, is inherent in string theory
as one of the excitations. Furthermore, the stringy nature introduces
a natural cut-off rendering problems with short-distance divergences
tractable. Among the references to strings are
\cite{Pol81, FMS, GSW, GM93, AADZ}. Very recent developments on
dualities in string theory and $p$-branes (higher dimensional objects 
than strings) \cite{Polchin, Vafa, Schwarz} tend to indicate that strings
might not be favoured after all and that there exists a more fundamental
paradigm yet to be revealed. 

Strings are not the subject of the present thesis but serve as a motivation
for studying conformal field theory (CFT). A propagating string sweeps
out a two dimensional surface (the world sheet) and it is thus a two 
dimensional quantum field theory that describes the string coordinates as
functions of the world sheet coordinates. Consistency conditions demand it
to be a CFT. Studying conformally
invariant {\em quantum} field theory within string theory may be seen as a
study of {\em classical} solutions to string theory, with
interactions according to tree diagrams. Perturbative quantum corrections
are dealt with by allowing the world sheet to have increasingly 
complicated topology as characterised by it's genus (number of handles)
which is then playing the number of loops in ordinary field theory
Feynman diagrams. 

The other main physical interest in CFT   
is due to it's use in descriptions of critical phenomena in
statistical mechanics. Two dimensional statistical systems enjoy
scale (and indeed conformal \cite{Pol70, Car89}) invariance at second order 
phase transitions due to the divergence of correlation lengths. 

Besides yet other applications of CFT's in physics and in mathematics, 
their relevance may be argued for by somewhat more philosophical comments.
A CFT may be defined without referring to any specific classical theory from 
the outset and is thus free of more or less well-defined
quantisation procedures and subsequent renormalisation programmes.
Most of the ingredients in CFT inspired by similar properties of usual
quantum field theory may be seen as defining properties of CFT and are
independent of regularisations and renormalisations. Such considerations
naturally lead one to axiomatic approaches to CFT 
about which we shall have no more to say here.     

Consistency in the standard bosonic string theory demands the dimension of
the target space to take on the critical value 26. However, at the cost
of introducing a new field the dimensionality is no longer fixed
and one speaks of non-critical strings. It is the two dimensional world
sheet {\em metric} 
that becomes a dynamical quantum field and in a certain gauge the
surviving freedom is assigned the so-called Liouville field.  
Within the framework of non-critical strings, much progress has
been made in describing the coupling of minimal conformal matter to $2D$ 
gravity. Extending the Virasoro algebra, which is inherent in every CFT, 
leads to more complicated and perhaps even more realistic string theories. 
The best known ones are superstrings based on supersymmetric extensions. 
However, also bosonic extensions exist and are based on $W$-algebras. A broad 
class of these may be constructed by Hamiltonian (Drinfeld-Sokolov) reduction
of affine Lie algebras. The Virasoro algebra itself is obtained by
Hamiltonian reduction of affine $SL(2)$ algebra and in order to produce
the minimal CFT one needs to consider so-called {\em admissible} 
representations of the affine $SL(2)$ current algebra. 
Such representations will play a significant role in this thesis and are 
generalisations of the well-known integrable (unitary) representations. The
former allow the spins $j$ of the fields to be fractional, while in integrable 
representations $2j$ is integer.

$W$-strings provide us with means of defining strings consistently in
various backgrounds, but they do not admit an immediately
transparent way to compute correlation functions. However, H.-L. Hu and
M. Yu \cite{HY} and Aharony, Ganor, Sonnenschein, Yankielowicz and Sochen
\cite{AGSYS} have outlined an alternative approach based on (topological) 
$G/G$ WZNW models. In the non-critical
string theory where the (extended) conformal matter is coupled to $2D$ gravity,
one performs essentially two reductions in order to eliminate degrees of
freedom. The matter and gravity parts are
obtained using Hamiltonian reduction of WZNW models and subsequently
further reduced by imposing a BRST cohomology condition to remove unwanted 
surviving degrees of freedom. In the $G/G$ approach the elimination is 
effectuated in only one step as a BRST condition. A comparison of these two
procedures was carried out
in \cite{HY, AGSYS} in the case of $G=SL(2,\R)$. The equivalence 
of the $SL(2,\R)/SL(2,\R)$ model and the usual non-critical bosonic
string, where minimal conformal matter is coupled to Liouville gravity,
was established at the level of isomorphic field and state contents. 
Later this equivalence for $G=SL(N,\R)$ and $W_N$-strings has been 
addressed in \cite{ASY, Sad}. The ultimate test in the case of $SL(2,\R)$
is to calculate the correlation functions and compare with known results
from e.g. matrix models. For the description of coupling minimal matter
to $2D$ gravity, the powerful methods of matrix models in the discrete 
approach are still superior to the prospects of the $G/G$ approach. However,
more general string theories do not seem tractable in terms of matrix models. 
For higher groups and supergroups the $G/G$ approach
might ultimately provide us with a way of formulating and understanding
non-critical string theory, in a way for which computation of correlation
functions is feasible. 

This brings us to the subject of the present thesis, namely computation
of Greens functions in CFT based on affine current algebra. 
We shall use free field realizations which often facilitate computations.
The first obstacle is thus to work out the free field realizations of the 
affine current algebra, the primary fields and the screening currents. The 
latter constitute essential technical ingredients in the constructions.
In this thesis we shall present explicit free field realizations of 
simple affine current algebras, thus completing the long-standing search for a 
generalisation of the Wakimoto construction of $SL(2)$ current
algebra to any simple Lie group. In the case of the screening 
currents and the primary fields we still only have partial results in the
most general case, but are currently witnessing further progress \cite{PRY3}. 

In the case of $SL(2)$ such realizations have been known for some time. 
The free field realization
of the current algebra is due to Wakimoto \cite{Wak}, the completion of
the set of screening currents is due to Bershadsky and Ooguri \cite{BO},
while the construction of the primary fields in the framework of depending
on an extra isotopic coordinate, is due to Furlan, Ganchev,
Paunov and Petkova \cite{FGPP}. $N$-point functions of two dimensional
WZNW theories based on affine $SL(2)$ current algebra have been studied
much already. They are typically constructed, either by applying the free
field realization of Wakimoto or by solving the Knizhnik-Zamolodchikov 
equations.
The results of these various pieces of works are quite complete as far
as unitary, integrable representations are concerned, but are incomplete
in the case of admissible representations, which as already mentioned
are relevant in the $SL(2)/SL(2)$ approach to non-critical bosonic strings. 
The origin of this incompleteness may be traced 
to the need in the free field approach of ghost fields raised to fractional 
powers. In this thesis we overcome this difficulty by showing how the 
techniques of fractional calculus naturally provide a way to handle such items.
A major part of the thesis is then concerning applications of these
techniques in the free field realization, working out general $N$-point
functions, 4-point Greens functions and
operator algebra coefficients in CFT based on affine $SL(2)$ current algebra
in the case of admissible representations. In terms of free field realizations
we have thereby solved completely CFT (on the sphere)
based on affine $SL(2)$ current algebra for admissible representations.
\\[.8 cm]
This thesis is organised as follows. The chapters 2 to 5 all start with
overviews. In Chapter 2 we discuss some
concepts in CFT. Notably, we consider free fields and
introduce the aforementioned techniques for handling rational powers of
ghost fields. Since a major part of the thesis is concerned with a 
generalisation of the Dotsenko-Fateev construction \cite{DF} from pure CFT to 
CFT based on affine $SL(2)$ current algebra, we briefly review elements of
that construction. In that connection we present a proof 
at the level of correlators of a hidden duality in those constructions.
The underlying integral identity will be needed and generalised in Chapter 5.

In Chapter 3 we present explicit free field realizations of 
simple affine current algebras. The method employed is to first work out 
differential operator realizations of the simple Lie algebras using Gauss 
decompositions. Secondly one quantises the realization by translating
into free fields and subsequently adding appropriate anomalous terms to the 
lowering operators. These terms take care of multiple contractions.
Furthermore, free field realizations of
screening currents of both kinds and of primary fields are discussed.
Many new results of general character are presented. For $SL(3)$ the
results are complete. All of these free field realizations are needed for 
future generalisations to higher groups, of the work on $SL(2)$ presented in 
chapters 4 and 5 in this thesis. In connection with the construction
of the screening currents we digress on a quantum group structure.

In Chapter 4 we develop integral representations for chiral 
$N$-point blocks in CFT based on affine $SL(2)$ current algebra in the case of
admissible representations. We show at the level of correlators that they 
indeed satisfy the Knizhnik-Zamolodchikov equations and that they are
projectively invariant. Considering the 3-point functions and utilising a
freedom (leading to the notion of over-screening) in the choice of numbers 
of integrations in the realization, we re-derive the fusion rules for
admissible representations within our framework.
The proposal in \cite{FGPP} for how correlators in $SL(2)$ current algebra
for admissible representations reduce to correlators in conformal minimal
models is discussed and a simple proof is presented.
The reduction is then verified explicitly for the blocks just found.
We further discuss the relation to more standard formulations of
Hamiltonian reduction.

In Chapter 5 we confine ourselves to the study of 4-point functions and
perform the generalisation mentioned above of the famous work \cite{DF} by 
Dotsenko and Fateev on minimal CFT. 
A different integral realization of the 4-point functions by Andreev \cite{An}
is discussed. We present a proof at the level of highly non-trivial
integral manipulations generalising the similar proof in Chapter 2 for
minimal models, of the equivalence of the two representations. 
For both representations, the integration contours are found and shown
to produce the correct singular behaviour for certain blocks. In his work
Andreev does not address choosing contours. Due to simpler contours,
we use Andreev's representation to work out the crossing matrix, connecting
different bases of blocks, in
order to determine the monodromy invariant 4-point Greens functions.
We then use the monodromy coefficients to obtain the operator algebra
coefficients for theories based on admissible representations. Again the
notion of over-screening is utilised, and in such a way that we may 
normalise the operator algebra coefficients unambiguously, contrary to
the results in \cite{An}.

Finally, Chapter 6 contains some concluding remarks and a brief 
discussion of present and future works.\\[.8 cm]
Chapter 2 contains results published in \cite{PRY1} and \cite{PRY2}.\\
Chapter 3 is based on \cite{PRY3}. The subject is currently under further 
investigation.\\
Chapters 4 and 5 comprise an exposition of \cite{PRY1, PRY1a, PRY2}.
\newpage
\begin{flushleft}
{\bf Acknowledgement} \\[.6cm]
\end{flushleft} 
The author has the greatest admiration for his supervisor Jens Lyng Petersen,
both as a teacher and as a person. His tremendous efforts are sincerely
appreciated.\\
It is with great pleasure that the author also thanks
Yu Ming, without whom this work could never have been undertaken. 
Throughout the collaboration Yu Ming has been an indispensable source
of inspiration.\\
The author is also indebted to Anna Tollst\'en for discussions at the
early stages of this work.\\
Furthermore, he wants to thank fellow students and the staff and guests
at The Niels Bohr Institute for providing a pleasant and inspiring
environment.\\
Finally, he is grateful for the very kind hospitality he met at Institute
of Theoretical Physics, Academia Sinica in Beijing, where parts of the work
were carried out.

\chapter{Conformal Field Theory}

The list of references to original works and reviews on conformal field
theory (CFT) seems
inexhaustible. First of all there is the seminal work by Belavin, Polyakov
and Zamolodchikov \cite{BPZ} while the following list reflects some personal
preferences \cite{Pet1, Gin, ID, AGS, Pet2, Ket, Sch} and \cite{ISZ} is a 
reprint collection. These will be used without further notice and may be 
consulted for details.

We shall introduce techniques of fractional calculus 
\cite{Ross, McR, SKM, Kir} 
in order to handle ghost fields raised to non-integer powers \cite{PRY1}.
Similar techniques were recently discussed by Andreev \cite{An}. Also 
fractional (indeed complex) powers of algebra generators
have been successfully employed in \cite{MFF, AY, FIM, AF}.
However, the application was to non-integral powers of algebra generators, 
whereas we shall consider non-integer powers of free fields where the 
question of Wick contractions brings in new features.

In connection with a brief review of some aspects in minimal models, we
shall present at the level of integral manipulations
a proof of a hidden duality in those models.

\section{Some Basic Concepts}

The content of this section serves to fix notation and present some
background material.

\subsection{Conformal Transformations}

The conformal group is defined as the subgroup of the general coordinate
transformations (group of diffeomorphisms) that leaves the metric invariant
up to a space-time dependent scale factor, such that it acts on the metric as
a Weyl transformation
\bea
  x^{\mu}&\rightarrow&{x'}^{\mu}(x)\nn
  g_{\mu\nu}(x)&\rightarrow&{g'}_{\mu\nu}(x')=\Omega(x)g_{\mu\nu}(x)
  \spa \Omega(x)=e^{\omega(x)}
\eea
It is not difficult to verify that
in a flat $d$-dimensional space-time of signature $(p,q)$ the generators form
an algebra isomorphic to $so(p+1,q+1)$. In two dimensions the situation is
radically different and in complex coordinates $z=x^1+ix^2$ and $\bar{z}=x^1
-ix^2$ ($g_{ij}=\delta_{ij}$) the local conformal transformations can be 
identified with locally analytic coordinate transformations
\ben
  z\rightarrow f(z) \spa \bar{z}\rightarrow \bar{f}(\bar{z}) 
\een
where $\pa=\pa_z=\hf(\pa_1-i\pa_2)$ 
and $\bar{\pa}=\pa_{\bar{z}}=\hf(\pa_1+i\pa_2)$.
Being careful with the Jacobian one finds $d^2z=2dx^1dx^2$ while the 
components of the metric 
are $g_{z\bar{z}}=g_{\bar{z}z}=\hf$ and $g_{zz}=g_{\bar{z}\bar{z}}=0$.
Eventually one is interested in computing correlation functions which are
functions of a finite number of points. Therefore, it is indeed
natural to extend
the class of transformations considered to include 'truly local' or
infinitesimal conformal transformations. These may be defined as 
analytic transformations admitting the infinitesimal coordinate change 
$\epsilon(z)$ to be holomorphic in the interior of some (not
necessarily connected) domain $D$ and identically zero in the exterior. In the 
boundary region this is not a usual conformal mapping. The emphasis here is on
locality.

The infinitely many generators of the two dimensional conformal transformations
are described using the basis (labelled by $n\in\Z$)
\bea
  z\rightarrow z'+\eps_n(z)&\hspace{.5 cm},\hspace{.5 cm}&\zb 
    \rightarrow\zb'+\bar{\eps}_n(\zb)\nn
  \eps_n(z)=-z^{n+1}&\hspace{.5 cm},\hspace{.5 cm}&\bar{\eps}_n(\zb)=-\zb^{n+1}
\eea
and are given by
\ben
  l_n=-z^{n+1}\pa_z\hspace{.5 cm},\hspace{.5 cm}\bar{l}_n=-\zb^{n+1}\pa_{\zb}
\een
The local conformal algebra is 
achieved by working out the commutation relations
\ben
  \left[ l_m,l_n\right]=(m-n)l_{m+n}\ \ ,\ \ \left[ \bar{l}_m,\bar{l}_n\right]
  =(m-n)\bar{l}_{m+n}\ \ ,\ \
   \left[ l_m,\bar{l}_n\right]=0
\een
The last relation indicates the splitting of the full algebra into a direct
sum of two isomorphic subalgebras (each of which is isomorphic to the
Witt algebra), which justifies the common use of $z$
and $\zb$ as independent coordinates. In the bigger complexified space
($\C\rightarrow\C^2$) on which the full algebra works, one may then choose
various reality conditions. Conventionally one imposes the 'physical'
condition $\zb=z^{\ast}$ (here $\ast$ means complex conjugation) and recovers
the Euclidean plane. It is customary and convenient to make use of the 
splitting of the algebra by ignoring the anti-holomorphic part and only in
the end to reconstruct the full theory simply by adding terms with bars
when appropriate. This will be demonstrated in connection with Greens 
functions.

Only a finite subalgebra of the local algebra is well-defined globally
on the Riemann sphere ($S^2=\C\bigcup\infty$).
Non-singularity of the vector field 
$v(z)=\sum_n a_nz^{n+1}\pa_z$ as $z\rightarrow0$ resp. as $z\rightarrow\infty$
allows $a_n\neq0$ only for $n\leq1$ resp. $n\geq-1$ (seen by using the
transformation $z=-1/w$). The global conformal group on $S^2$ 
is therefore generated by $\{l_{-1},l_0,l_1\}\bigcup
\{\bar{l}_{-1},\bar{l}_0,\bar{l}_1\}$. The finite transformations 
(projective conformal transformations) form the complex M\"{o}bius group
\ben
 z\rightarrow\frac{az+b}{cz+d}\hspace{1 cm}a,b,c,d\in\C\ \ ,\ \ ad-bc=1
\een
and is isomorphic to $SL(2,\C)/\Z_2$ where the quotient by $\Z_2$ is due to 
invariance under $(a,b,c,d)\rightarrow (-a,-b,-c,-d)$. It is well-known that 
these are the only analytic one-to-one maps of $S^2$. 

It is the existence of an infinite dimensional symmetry algebra that makes
the {\it two} dimensional case unique by implying severe restrictions on
the conformally invariant field theory. In the remaining part of the thesis
considerations are confined to two dimensional CFT.   

\subsection{Fields and Operator Product Expansions}

Consider the response of the correlator or expectation value
\ben
 \br F[\phi]\rangle=\int D\phi e^{-S[\phi]}F[\phi]
\een
to a transformation
\ben
 \phi'(x)=\phi(x)+\delta_{\epsilon}\phi(x)
\een
$\phi$ is collectively denoting the fields in the system. One has
\bea
 0&=&\int D\phi'e^{-S[\phi']}F\left[\phi'\right]-\int D\phi e^{-S[\phi]}F
  \left[\phi\right]\nn
 &=&\int\delta_{\epsilon}\left(D\phi e^{-S[\phi]}\right)F\left[\phi\right]+\int
  D\phi e^{-S[\phi]}\delta_{\epsilon}F\left[\phi\right]\nn
 &=&-\frac{1}{2\pi}\int D\phi e^{-S[\phi]}\int d^2xT^{\mu\nu}(x)\pa_{\mu}
  \epsilon_{\nu}(x)F\left[\phi\right]+\br\delta_{\epsilon}F\left[\phi
  \right]\rangle
\eea
This is the definition of the energy-momentum tensor $T_{\mu\nu}$
suggested here. It is not useful to define the energy-momentum tensor in the 
standard way as the variation of $S$ wrt a change in the metric
$T_{\mu\nu}\sim\frac{\delta S}{\delta g^{\mu\nu}}$
since for a generally covariant theory  this defines a {\it tensor} object. As
will become clear $T$ is often not a true tensor in CFT.
Invariance under scaling transformations immediately shows that $T$ is
traceless, $\br {T^{\mu}}_{\mu}F[\phi]\rangle=0$, while invariance under
rotations makes $T$ symmetric, $\br T_{\mu\nu}F[\phi]\rangle=\br T_{\nu\mu}
F[\phi]\rangle$. Invariance under translations gives a conserved (divergence 
free) energy-momentum tensor, $\br\pa_{\mu}T^{\mu\nu}F[\phi]\rangle=0$, since 
one may integrate by parts
due to the vanishing outside $D$. In complex notation this means
(introducing $T\equiv T_{zz},\bar{T}\equiv T_{\zb\zb}$)
\ben
 \br T_{z\zb}F\left[\phi\right]\rangle=\br T_{\zb z}F\left[\phi\right]\rangle
  =\br\bar{\pa} 
  TF\left[\phi\right]\rangle=\br\pa\bar{T}F\left[\phi\right]\rangle=0
\een
Using these properties one has
\bea
 \br\delta_{\epsilon}F\left[\phi\right]\rangle&=&\frac{1}{4\pi}\int_Dd^2z\br(
  T^{\zb\zb}\bar{\pa}
  \epsilon_{\zb}+T^{zz}\pa\epsilon_z)F\left[\phi\right]\rangle\nn
 &=&\frac{1}{2\pi}\int_Dd^2z\br(\bar{\pa}(T_{zz}\epsilon^z)+\pa(T_{\zb\zb}
  \epsilon^{\zb}))F\left[\phi\right]\rangle\nn
 &=&\oint_{\pa D}\left(\frac{dz}{2\pi i}\br\epsilon(z)T(z)F\left[\phi\right]
  \rangle
  -\frac{d\zb}{2\pi i}\br\bar{\epsilon}(\zb)\bar{T}(\zb)F\left[\phi\right]
 \rangle
  \right)
\label{Ward1}
\eea
where the second equality is due to ($\bar{\epsilon}$) $\epsilon$ being 
(anti-)holomorphic
except in an infinitesimal region. The last equality uses Stoke's theorem
\ben
 \frac{1}{2\pi}\int_Dd^2z\bar{\pa} f(z,\zb)=\oint_{\pa D}\frac{dz}{2\pi i}f(
  z,\zb)\spa
 \frac{1}{2\pi}\int_Dd^2z\pa f(z,\zb)=-\oint_{\pa D}\frac{d\zb}{2\pi i}f(
  z,\zb)
\label{Stoke}
\een
(\ref{Ward1}) is the conformal Ward identity and is a relation between
correlation functions. It is customary to write such relations as
operator equations
\ben
 \delta_{\epsilon}F[\phi]=\oint_C\frac{dz}{2\pi i}\epsilon(z)T(z)F[\phi]-
  \oint_C\frac{d\zb}{2\pi i}\bar{\epsilon}(\zb)\bar{T}(\zb)F[\phi]
\label{Ward2}
\een
and likewise ${T^{\mu}}_{\mu}=0, \bar{\pa} T=0,T_{z\zb}=0$ etc. 
remembering that they are valid within correlation functions. (\ref{Ward2})
is also denoted the conformal Ward identity. In the sequel the anti-holomorphic
sector will often be neglected.

In quantum field theory time ordering in operator products is imposed.
After the convenient conformal map 
\ben
  z=e^{\zeta}=e^{\tau+i\sigma}
\label{cylplan}
\een
from the cylinder (regularised
spacetime, $(\tau,\sigma)$ are the Euclidean time and compactified space 
coordinates respectively) into the complex plane, time ordering becomes a 
radial ordering. In CFT operator products are always radial ordered
(here only bosonic fields are encountered)
\ben
 R(A(z)B(w))=\left\{ \begin{array}{l}
   A(z)B(w)\spa|w|<|z|\\
   B(w)A(z)\spa|z|<|w| \end{array}\right.
\een
though usually $R$ is left understood. One speaks of a radial time
$\tau\sim\ln|z|$.

The generator of infinitesimal conformal transformations (\ref{Ward2}) is
\ben
 L_{\epsilon}=\oint_C\left(\dz\epsilon(z)T(z)-\frac{d\zb}{2\pi i}\bar{
  \epsilon}({\zb})\bar{T}(\zb)\right)
\een
Among the fields in the theory those that transform\footnote{The sign in the
anti-holomorphic sector is accounted for by $\oint_{0}\frac{d\zb}{2\pi i}
\zb^{-1}=-1$} as
\ben
 \delta_{\epsilon}\phi(z,\zb)=\left((h\pa\epsilon(z)+\epsilon(z)\pa)+(\bar{h}
  \bar{\pa}\bar{\epsilon}(\zb)+\bar{\epsilon}(\zb)\bar{\pa})\right)
  \phi(z,\zb)
\label{prim}
\een
showing that $\phi(z,\zb)(dz)^\D(d\zb)^{\bar{\D}}$ 
is conserved, are called primary
or conformal tensor fields. $(\D,\bar{\D})$ are the weights of $\phi$ and are 
independent. $\D+\bar{\D}$ determines the behaviour under scalings and 
is called the dimension, while $\D-\bar{\D}$ determines the behaviour under 
rotations and is called the spin. For chiral fields $\phi(z,\zb)=\phi(z)$ 
one has $\bar{\D}=0$. The radial ordering gives the following 
description of the generator $L_{\epsilon}$
\ben
 \left[L_{\epsilon},\phi(w)\right]=\left(
  \oint_{|w|<|z|}-\oint_{|z|<|w|}\right)
  \dz\epsilon(z)T(z)\phi(w)
  =\wz\epsilon(z)T(z)\phi(w)
\een
using a contour deformation, where the last integration is around $w$.
Cauchy's integral formula gives an alternative way of classifying primary
fields
\ben
 T(z)\phi(w)=\frac{\D(\phi)}{(z-w)^2}\phi(w)+\frac{1}{z-w}\pa\phi(w)+nst
\label{OPEprim}
\een
This is an operator product expansion (OPE) where it is assumed that the lhs is
analytic in a neighbourhood around $w$. $nst$ (non-singular terms) is usually
not written.
The algebra of the fields under operator products (the operator product 
algebra) is associative. The Ward identity for an
$n$-point function of primary fields may now be written
\ben
 \br T(z)\phi_1(z_1)...\phi_n(z_n)\rangle=\sum_{i=1}^n\left(\frac{h_i}{
  (z-z_i)^2}+\frac{1}{z-z_i}\pa_{z_i}\right)\br\phi_1(z_1)...\phi_n(z_n)
  \rangle
\label{Wardprim}
\een
Since $\delta_\epsilon
\br\prod_i\phi_i\rangle=0$ for a global conformal transformation
where $\epsilon$ is a second degree polynomial, Cauchy's integral formula on 
the integrated version (\ref{Ward1}) of (\ref{Wardprim}) shows that
\ben
 \br T(z)\phi_1(z_1)...\phi_n(z_n)\rangle_{z\rightarrow\infty}\sim z^{-4}
  f(z_1,...,z_n)
\label{Tscale}
\een
for some function $f$.

Assume here that the operator product $A(z)B(w)$ can be expanded as
\ben
 A(z)B(w)=\sum_{n=-n_0}^\infty(z-w)^n(AB)_{n}(w)
\een
A normal ordering is a prescription for handling the singular limit
$z\rightarrow w$ for $n_0>0$. There are many possibilities. One is to use a
point-splitting regularisation \cite{BBSS}: Denote the normal ordered product
$:A(w)B(w):$ and define it by
\bea
 :A(w)B(w):&=&\oint_w\dz \frac{A(z)B(w)}{z-w}\nn
 &=&(AB)_0(w)
\label{(AB)}
\eea
This is neither commutative nor associative, e.g. one finds
\ben
 (AB)_0(z)-(BA)_0(z)=\sum_{n=1}^{n_0}\frac{(-1)^{n+1}}{n!}\pa^n(AB)_{(-n)}(z)
\een
It satisfies Wick's rule
\ben
 A\underbrace{(z)(B}C)(w)=\oint_w\frac{dx}{2\pi i}\frac{1}{x-w}\left(
  A\underbrace{(z)B}(x)C(w)+B(x)A\underbrace{
  (z)C}(w)\right)
\een
where the under-brace indicates that one is only considering the contraction
or singular part of that OPE.

\subsection{Virasoro Algebra}
It is a standard exercise in CFT to show that a 2-point function of
quasi-primary fields is fixed up to a normalisation constant $C_{12}$ by the
global invariances
\ben
 \br A_1(z_1)A_2(z_2)\rangle=\frac{C_{12}}{z_{12}^{2h_1}}\delta_{h_1,h_2}
  \hspace{1 cm}, \ \ z_{12}=z_1-z_2
\een
A quasi-primary field is primary under projective (global) M\"{o}bius 
transformations. Using (\ref{Tscale}) one expects
\ben
 \br T(z)T(w)\rangle=\frac{c/2}{(z-w)^4}
\een
where the constant $\hf$ is chosen as to produce $c=1$ for free bosons
(see below). $c$ is called the central charge and plays a prominent
role in CFT. Thus $T$ has $\D=2$ and the Ward identity gives
\ben
 \br\delta_\eps T\rangle=\frac{1}{12}c\pa^3\epsilon
\een
For $c\neq0$ $T$ is not primary. The minimal deviation is in the postulate
\ben
 \delta_\eps T=(2\pa\epsilon+\epsilon\pa)T+\frac{1}{12}c\pa^3\epsilon
\label{deltaT}
\een
or equivalently in the defining OPE
\ben
 T(z)T(w)=\frac{c/2}{(z-w)^4}+\frac{2}{(z-w)^2}T(w)+\frac{1}{z-w}\pa T(w)
\label{TT}
\een
The finite version of (\ref{deltaT}) is given the ansatz under $z
\rightarrow w(z)$
\ben
 T(z)=\left(\frac{dw}{dz}\right)^2T(w)+\frac{c}{12}\{w,z\}
\label{Tfinite}
\een
where $\{w,z\}$ is to be determined. Two successive transformations
$z\rightarrow u(z)\rightarrow w(u(z))$ result in
\ben
 \{w,z\}=\left(\frac{du}{dz}\right)^2\{w,u\}+\{u,z\}
\een
It may be argued that the solution to this equation consistent
with (\ref{deltaT}) is
\ben
 \{w,z\}=\frac{\pa^3w(z)\pa w(z)-\frac{3}{2}\left(\pa^2w(z)\right)^2}{(\pa
  w(z))^2}
\een
which is called the Schwartzian derivative. It is easily seen to vanish for a
M\"{o}bius transformation verifying that $T$ is quasi-primary.

The commutator of two generators is found to be
\ben
 [L_{\epsilon_1},L_{\epsilon_2}]=L_{\pa\epsilon_1\epsilon_2-\epsilon_1\pa
  \epsilon_2}+\frac{c}{12}\oz\epsilon_2(z)\pa^3\epsilon_1(z)
\een
Choosing the basis $\epsilon_n(z)=z^{n+1}$ and performing a Laurant or mode
expansion
\ben
 T(z)=\sum_{n\in\Z}L_nz^{-n-2}\hspace{1 cm},\hspace{1 cm}L_n=\oz T(z)z^{n+1}
\label{LaurantT}
\een
the commutator becomes the conventional form of the Virasoro algebra
\ben
 [L_n,L_m]=(n-m)L_{n+m}+\frac{c}{12}n(n^2-1)\delta_{n+m,0}
\label{Vir}
\een
In terms of the modes the condition (\ref{prim}) reads
\ben
 [L_n,\phi(z)]=z^n(\D(n+1)+z\pa)\phi(z)
\label{prim2}
\een

Another standard exercise in CFT is to show that the Virasoro algebra ($Vir$)
is a one-dimensional central extension of the Witt algebra. In the physics 
literature $c$ is usually treated as a c-number (as in this thesis)
though for $Vir$ to be an algebra $c\in Vir$. The treatment is justified
by the fact that in any representation $c$ has constant eigenvalue.
The central charge also appears in other respects. 
In a free bosonic scalar field theory 
$c$ counts the degrees of freedom. In a CFT defined on a
curved background the tracelessness of $T$ is broken by an anomaly proportional
to $c$ and the curvature scalar \cite{Fri}. 
On the cylinder one has using (\ref{Tfinite}) on (\ref{cylplan})
\ben
 T_{cyl}(z)=z^2T(z)-\frac{c}{24}
\een
so $c$ enters in the eigenvalues of the Hamiltonian on the cylinder,
$H=L_0-\frac{c}{24}+\bar{L}_0-\frac{\bar{c}}{24}$ (in general the
central charge $\bar{c}$ of the anti-holomorphic sector is not
necessarily equal to $c$). This
is sometimes referred to as understanding the central charge as a Casimir
effect, due to the finite geometry of the cylinder.

\subsection{Descendant Fields}

Descendant fields wrt $\phi$ are defined as finite linear combinations of 
fields of the form
\ben
 [...[L_{\epsilon_2},[L_{\epsilon_1},\phi]]...]
\een
and comprise the infinite dimensional conformal family $[\phi]$. Choosing
the above conventional basis a descendant field may be written ($\vec{n}=
(n_1,...,n_m)$)
\bea
 \phi^{(-\vec{n})}(z)&=&\hat{L}_{-n_1}(z)...\hat{L}_{-n_m}(z)\phi(z)\nn
  &=&\oint_{C_1}\frac{dz_1}{2\pi i}\frac{T(z_1)}{(z_1-z)^{n_1-1}}...
  \oint_{C_m}\frac{dz_m}{2\pi i}\frac{T(z_m)}{(z_m-z)^{n_m-1}}\phi(z)
\eea
where $C_j$ surrounds $z_{j+1},...,z_m,z$ and $\hat{L}_n(w)$ appears in a
formal expansion around $w$
\ben
 T(z)=\sum_{n\in\Z}\frac{\hat{L}_n(w)}{(z-w)^{n+2}}\hspace{1 cm},\hspace{1 cm}
 \hat{L}_n(w)=\wz T(z)(z-w)^{n+1}
\een
Obviously one has
\ben
 T(z)\phi(w)=\sum_{n\geq0}(z-w)^{n-2}\phi^{(-n)}(w)
\een
and by evaluating $\delta_\eps
(T(z)\phi(w))$ one finds by straightforward calculations
\bea
 \delta_\eps
  \phi^{(-n)}(w)&=&\left((\D+n)\pa\epsilon(w)+\epsilon(w)\pa\right)\phi^{
  (-n)}(w)\nn
 &+&\sum_{i=1}^n\frac{n+i}{(i+1)!}\pa^{i+1}\epsilon(w)\phi^{(i-n)}(w)
  +\frac{c}{12(n-2)!}\pa^{n+1}\epsilon(w)\phi(w)
\eea
and
\bea
 T(z)\phi^{(-n)}(w)&=&\frac{c}{12}n(n^2-1)(z-w)^{-n-2}\phi(w)\nn
 &+&\sum_{i=1}^n(n+i)(z-w)^{-i-2}\phi^{(i-n)}(w)
  +\sum_{i\geq0}(z-w)^{i-2}\phi^{(-i,-n)}(w)
\eea
$T$ itself is a descendant of the identity $I$, $T(z)=I^{(-2)}(z)$. 

\subsection{States}

The vacuum state $|0\rangle$ may be defined by the condition that it
respects the maximum number of symmetries. This means that it is annihilated
by the maximum number of conserved charges (the $L_{n}$'s in pure CFT, see
later for more general situations). Due to
the central charge there are only two possibilities, $L_n\vac=0$ for either
$n\geq-1$ or $n\leq1$. Since the Hamiltonian is bounded (and Hermitian, 
allowing it to be diagonalised) from below one has
\ben
 L_n\vac=0\spa n\geq-1
\label{vacuum}
\een
and similarly in the anti-holomorphic sector.
Primary states are defined as
\ben
 |\D,\bar{\D}\rangle=\lim_{z,\zb\rightarrow0}\phi(z,\zb)\vac
\een
and using (\ref{prim2}), seen to respect
\ben
 L_0\ket{\D}=\D\ket{\D} \spa L_n|\D\rangle=0\spa n\geq1
\een
This defines a highest weight vector (state) of weight $\D$. A highest 
weight module or Verma module (see \cite{Kacbook, MP} for mathematical
accounts) is characterised by the central charge $c$ and the weight $\D$
of the highest weight vector, and consists of states created from $\ket{\D}$
by acting thereupon by linear combinations of lowering operators $L_{-n}$, 
$n\geq0$. The subspace at level $N\geq0$ spanned by vectors 
$L_{-n_1}...L_{-n_m}\ket{\D}$ for which $n_1+...+n_m=N$, has dimension
$p(N)$, where $p(N)$ is the partition of $N$ (the number of possible
ways $N$ can be written as a sum of positive integers). $p(N)$ appears
in the generating function
\ben
 \prod_{n\geq1}(1-x^n)^{-1}=\sum_{n\geq0}p(n)x^n
\een  

Again using (\ref{prim2}) one has the correspondence $I\sim\vac$.
The act of taking the adjoint of a Hermitian operator in Euclidean space-time
corresponds to time reversal. This translates into the CFT notion of adjoint
being
\ben
 \phi(z,\zb)^\dagger=\frac{1}{\zb^{2\D}}\frac{1}{z^{2\bar{\D}}}\phi
  \left(\frac{1}{\zb},\frac{1}{z}\right)
\label{Adjointphi}
\een
and one defines ($\br0|=\vac^\dagger$)
\ben
 \br \D,\bar{\D}|=\lim_{z,\zb\rightarrow\infty}\br0|\phi(z,\zb)z^{2\D}\zb^{
  2\bar{\D}}
\label{infstate}
\een
For $T$ the notion of adjoint 
leads to the following hermiticity properties
\ben
 L_n^\dagger=L_{-n}
\een

\subsection{Singular States and the Kac Determinant}
A singular vector in a highest weight module is a highest weight vector
itself. The existence of a singular vector at a level higher than zero
renders the generically irreducible highest weight 
module reducible. Explicit forms for singular vectors in Verma modules for
the Virasoro algebra are considered in \cite{BDiFIZ}. See \cite{FeFu90} for
a general discussion of representations of the Virasoro algebra.
One may obtain knowledge of Virasoro singular vectors by 
considering the Kac determinants \cite{Kac}. 
They appear as determinants of the $p(N)\times p(N)$ dimensional matrices
\ben
 M_{{\bf k}{\bf k'}}^{(N)}=\bra{\D}\left(\prod_{n\geq1}(L_{-n})^{k_n}
  \right)^\dagger\prod_{m\geq1}(L_{-m})^{k_m'}\ket{\D}
\een
where $\sum nk_n=\sum mk_m'=N$, $k_n\geq0$. The products are ordered,
$\prod(L_{-n})^{k_n}=...(L_{-2})^{k_2}(L_{-1})^{k_1}$, such that we only
consider a basis of states at level $N$. Such a matrix is obviously Hermitian
and has zero determinant if there exists an eigenvector ${\bf v}^t=(v_1,...,
v_{p(N)})$ satisfying ${\bf v}^tM{\bf v}=0$, indicating that
$||\sum v_i\ket{\D^{(N)}_i}||=0$ where $\{\ket{\D^{(N)}_i}\}$ 
is the basis chosen. Since determinants are essentially basis independent 
they uniquely reveal the existence of zero norm vectors.
Such vectors are called null-vectors and are easily seen to be orthogonal to
all states in the Verma module. Their number at a given level $N$ is equal
to the number of zero eigenvalues of the Kac determinant $K_N$.
Null-states appear in submodules generated by singular states different 
from the highest weight state $\ket{\D}$.
Kac found the general expression for the determinant $K_N$ \cite{Kac}
\ben
 K_N=c_N\prod_{n,m>0,nm\leq N}(\D-\D_{n,m})^{p(N-nm)}
\een
with positive constants $c_N$. In \cite{FeFu} the first proof appeared while
\cite{Tho, R-C} contain alternative proofs.
The central charge and the conformal weights (the Kac table) are given by
\bea
 \D_{n,m}&=&-\frac{\al_0^2}{2}+\frac{1}{8}(n\al_++m\al_-)^2\spa n,m\geq1\nn
 \al_\pm&=&\al_0\pm\sqrt{\al_0^2+2}=\frac{\sqrt{1-c}\pm\sqrt{25-c}}{\sqrt{12}}
  \nn
 c=1-12\al_0^2\ \ \ &,&\ \ \  2\al_0=\al_++\al_-\spa \al_+\al_-=-2
\label{Dnm}
\eea

For $c,\D>1,0$ the Kac determinant has no zeros. A detailed study \cite{FQS}
of $K_N$ reveals that unitarity (only positive norm states) imposes severe
restrictions. By looking for zeros for $K_N$ in the plane $c,\D$, separating
regions of positive and negative norm states, one finds that in order to 
have unitarity for $c<1$, the central charge is given by
\ben
 c=1-\frac{6}{n(n+1)}\spa n=3,4,5,...
\label{cunitary}
\een
with the associated parametrisations of $\D$ (\ref{Dnm}). These are only 
necessary conditions for unitarity ($c<1$) while the proof of existence
is by coset constructions, see below.
Minimal models, to be discussed in a subsequent section, are obtained
by imposing less severe restrictions.

\subsection{Conformal Blocks and Greens Functions}
The projective invariance, generated by $\{L_{-1},L_0,L_1\}$, puts
restrictions on the $N$-point functions, leaving them essentially only
dependent on the anharmonic ratios $z_{ijkl}=(z_i-z_j)(z_k-z_l)/((z_j-z_k)
(z_l-z_i))$ of which only $N-3$ are independent. The 3-point Greens
functions are thereby fixed
\bea
 &&\langle\phi_3(z_3,\zb_3)\phi_2(z_2,\zb_2)\phi_1(z_1,\zb_1)\rangle\nn
  &=&C_{123}
  (z_2-z_1)^{-\D_2-\D_1+\D_3}(z_3-z_2)^{-\D_3-\D_2+\D_1}(z_3-z_1)^{-\D_3
  -\D_1+\D_2}\nn
 &\cdot&(\zb_2-\zb_1)^{-\bar{\D}_2-\bar{\D}_1+\bar{\D}_3}
  (\zb_3-\zb_2)^{-\bar{\D}_3-\bar{\D}_2+\bar{\D}_1}(\zb_3-\zb_1)^{-\bar{\D}_3
  -\bar{\D}_1+\bar{\D}_2}
\label{3pointgen}
\eea
See Chapter 5 for a discussion on the structure constants $C_{123}$.

Due to the decomposition $Vir\oplus \bar{Vir}$, a general 4-point Greens
function (in the 'standard projective limit') 
\ben
 G_{ijkl}(z,\zb)=\langle\phi_l(\infty,\infty)\phi_k(1,1)\phi_j(z,\zb)
  \phi_i(0,0)\rangle
\een
may be expanded in conformal blocks $F_{ijkl}^{(m)}$ (or chiral blocks
for bigger symmetry groups than $Vir$, see later)
\ben
 G_{ijkl}(z,\zb)=\sum_m{C_{ij}}^mC_{mkl}F_{ijkl}^{(m)}(z)\bar{F}_{ijkl}^{(m)}
  (\zb)
\een
If the fusion rule coefficients \cite{Ver, DV, MS, CPR, AGS} are larger than 
one, then the sum over $m$ must be supplemented by a degeneracy index labelling
the distinct ways of coupling to the field $\phi_m$. However, our prime 
interest lies in the study of CFT based on $SL(2)$ current algebra 
where such subtleties are absent, so we will ignore them here.
The blocks behave like and are normalised according to
\ben
 \lim_{z\rightarrow0}F_{ijkl}^{(m)}(z)=z^{\D_m-\D_i-\D_j}(1+{\cal O}(z))
\een
and are generically multi-valued functions. Yet, the physical Greens
functions are supposed to be single-valued functions.
The power-like singular behaviours of the blocks of a given 4-point Greens
function are all distinct and determine the non-trivial monodromies for
these blocks. Graphically one associates to the blocks, or to the Greens
functions, Feynman skeleton diagrams for a $\var^3$
type quantum field theory with the fields $\phi$ being the external legs,
while the internal lines describe the exchanged fields, see Fig. 1 in
Chapter 5. The associativity of the operator product algebra leads to some 
duality relations (crossing symmetry or bootstrap equations) expressing the
Greens function in different expansions
\bea
 \sum_m{C_{ij}}^mC_{mkl}F_{ijkl}^{(m)}(z)\bar{F}_{ijkl}^{(m)}
  (\zb)&=&\sum_m{C_{jk}}^mC_{mil}F_{jkil}^{(m)}(1-z)\bar{F}_{jkil}^{(m)}
  (1-\zb)\nn
 &=&z^{-2\D_j}\zb^{-2\bar{\D}_j}
  \sum_m{C_{jl}}^mC_{mik}F_{ikjl}^{(m)}(1/z)\bar{F}_{ikjl}^{(m)}(1/\zb)
\label{dualF}
\eea
We shall refer to these channels as s-, t- and u-channels as is standard
Mandelstam terminology in relativistic kinematics describing particle 
scattering. Similarly one may define conformal
blocks for higher $N$-point Greens functions. Greens functions are always
supposed to be single-valued. When the summations (\ref{dualF}) truncate
(which is the case in rational CFT, see later), the duality means that the 
blocks transform into finite linear combinations of themselves under analytic 
continuation. Being multi-valued functions the blocks are indeed supposed to be
defined first in one domain (for a given order of the arguments)
and then analytically continued to other domains. Since the analytic 
continuation depends on the path chosen, it is the braid group and not the
permutation group that acts on the blocks.
In this respect a quantum group structure is revealed \cite{MS, AGS1, MR, AGS}.

The decoupling of a null field (the descendant field corresponding to the null
vector in the Verma module headed by the corresponding primary field)
gives rise to certain differential equations satisfied by the correlation 
functions and the conformal blocks. Utilising this is sometimes a 
convenient way of determining the blocks.

The Ward identities for insertion of symmetry generators (here only $T$)
makes it possible to obtain correlation functions involving descendant
fields from those of primary fields. 

\section{Free Fields}

\subsection{A Digression on Lie Algebras}

Let {\bf g} be a simple Lie algebra of dim {\bf g} = $d$ and rank {\bf g} = 
$r$.
{\bf h} is a Cartan subalgebra of {\bf g}. The set of (positive) roots
is denoted ($\Delta_+$) $\Delta$ and we write $\al>\beta$ if for two roots
$\al-\beta\in\Delta_+$, and analogously $\al>0$ means $\al\in\Delta_+$.
The simple roots are $\{\al_i\}_{i=1,...,r}$. $\theta$ is the highest root
and $\hn$ is the dual Coxeter number.
The only non-vanishing elements of the Cartan-Killing form are
\ben
 \kappa_{\al,-\al}=\kappa(e_\al,f_\al)=\frac{2}{\al^2}\spa\kappa_{ij}=\kappa
  (h_i,h_j)=G_{ij}
\een
The Cartan matrix is $A_{ij}=\al_i^\nu\cdot\al_j=(\al_i^\nu,\al_j)=G_{ij}
\al_j^2/2$, while Freudenthal-de Vries strange formula states that
\ben
 \rho^2=\frac{d}{24}\hn\theta^2
\label{FdeV}
\een 
Using the triangular decomposition 
\ben
 \mbox{{\bf g}}=\mbox{{\bf g}}_-\oplus\mbox{{\bf h}}\oplus\mbox{{\bf g}}_+
\een
we will denote the raising and lowering operators $e_\al\in$ {\bf g}$_+$ and
$f_\al\in$ {\bf g}$_-$ respectively with $\al\in\Delta_+$, and Cartan operators
$h_i\in$ {\bf h}. We use the common notation $j_a$ for all
the algebra elements. 
For simple roots we sometimes write $e_i=e_{\al_i}, f_i=f_{\al_i}$ and
collectively denote the 3$r$ generators $e_i,h_i,f_i$ Chevalley generators.
Their
commutator relations are
\bea
 \left[h_i,h_j\right]=0&&\left[e_i,f_j\right]=\delta_{ij}h_j\nn
 \left[h_i,e_j\right]=A_{ij}e_j&&\left[h_i,f_j\right]=-A_{ij}f_j
\eea
In addition one has the Serre relations
\bea
 (\mbox{ad}_{e_i})^{1-A_{ij}}e_j&=&0\spa i\neq j\nn
 (\mbox{ad}_{f_i})^{1-A_{ij}}f_j&=&0\spa i\neq j
\eea
The Weyl vector is defined as $\rho=\hf\sum_{\al>0}\al$ and satisfies
$\rho\cdot\al_i^\nu=1$. We use the convention ${f_{-\al,-\beta}}^{-\g}=
-{f_{\al\beta}}^\g$. Furthermore
\bea
 \left[e_\al,f_\al\right]&=&h_\al=G^{ij}(\al_i^\nu,\al^\nu)h_j\nn
 h_{\al+\beta}&=&
  \frac{1}{(\al+\beta)^2}\left(\al^2h_\al+\beta^2h_\beta\right)\nn
 \left[h_i,e_\al\right]&=&(\al_i^\nu,\al)e_\al\nn
 \left[h_i,f_\al\right]&=&-(\al_i^\nu,\al)f_\al
\eea

\subsection{Bosonic Scalar Field}

Let us consider a set of free bosonic scalar fields $X_i(z,\zb)$ labelled
by Cartan indices $i=1,...,r$
and described by the action
\ben
 S=-\frac{1}{4\pi}\int d^2z\pa X(z,\zb)\cdot\bar{\pa}X(z,\zb)
\een
The metric in $X$-space is the Cartan part $G_{ij}$ of the Cartan-Killing
form. In this notation the fields are purely imaginary.
The classical equations of motion are
\ben
 \pa\bar{\pa}X_i(z,\zb)=0
\een
from which it follows that $\pa X_i$ $(\bar{\pa}X_i)$ is (anti-)analytic and
is therefore written $\pa X_i(z)$ $(\bar{\pa}X_i(\zb))$. 
The general solution splits the scalar field into a sum of holomorphic
and anti-holomorphic pieces 
\ben
 X_i(z,\zb)=\var_i(z)+\bar{\var}_i(\zb)
\een
The property of 
path integrals that the integral of a total derivative vanishes, gives
\bea
 0&=&\int DX\frac{\delta}{\delta X^i(z,\zb)}\left(e^{-S}X_j
   (w,\bar{w})\right)\nn
 &=&\langle G_{ij}\delta^2(z-w,\zb-\bar{w})\rangle
  -\frac{1}{2\pi}\pa_z\pa_{\zb}
   \langle X_i(z,\zb)X_j(w,\bar{w})\rangle
\eea
Using this together with
\ben
 \pa_z\bar{\pa}_{\zb}\ln|z-w|^2=2\pi\delta^2(z-w,\zb-\bar{w})
\label{papaln}
\een
following from Stoke's theorem (\ref{Stoke}), we find the 2-point function
\ben
 \langle X_i(z,\zb)X_j(w,\bar{w})\rangle=G_{ij}\ln|z-w|^2
\een
and the OPE
\ben
 X_i(z,\zb)X_j(w,\bar{w})=G_{ij}\ln|z-w|^2+:X_i(z,\zb)X_j(w,\bar{w}):
\een
or equivalently
\bea
 \var_i(z)\var_j(w)&=&G_{ij}\ln(z-w)+:\var_i(z)\var_j(w):\nn
 \var_i(z)\bar{\var}_j(\bar{w})&=&:\var_i(z)\bar{\var}_j(\bar{w}):\nn
 \bar{\var}_i(\zb)\bar{\var}_j(\bar{w})&=&G_{ij}\ln(\zb-\bar{w})+
  :\bar{\var}_i(\zb)\bar{\var}_j(\bar{w}):
\eea
whereby we have introduced the normal ordering of free scalar fields
by subtracting the singular parts.
The 2-point function demonstrates that the free fields do not have
well-defined scaling dimensions, hence they do not rigorously exist
as quantum fields \cite{Col}. Nevertheless, we shall only need to consider
constructions such as $\pa X_i$ of dimension 1 and vertex operators to be
defined in (\ref{vertex}).
According to (\ref{papaln}) the normal ordered product satisfies the 
classical equations of motion and may be Taylor expanded. The first term
in the expansion precisely corresponds to the definition of composite
operators in (\ref{(AB)}). We have found that inside the normal ordering
the scalar fields act as classical fields. 
The corresponding energy-momentum tensor is
\ben
 T=\hf:\pa X\cdot\pa X:=\hf:\pa\var\cdot\pa\var:
\een
and has central charge $c=r$. This result is the origin of the conventional
factor of 1/2 in front of $c$ in (\ref{TT}). 
The notion of normal ordering is easily generalised along the lines
of
\ben
 X_i(z,\zb):\prod_{j=1}^nX_{i_j}(z_j,\zb_j):=\sum_{j=1}^nG_{ii_j}\ln|z-z_j|^2:
  \prod_{l\neq j}X_{i_l}(z_l,\zb_l):+:X_i(z,\zb)\prod_{j=1}^nX_{i_j}
  (z_j,\zb_j):
\een
and one has that 
the OPE of two normal ordered operators can be written as the formal 
expression
\ben
 :G[X]::F[X]:=\exp\left(\int d^2zd^2wG_{ij}\ln|z-w|^2\frac{\delta_G}{\delta 
  X_i(z,\zb)}
  \frac{\delta_F}{\delta X_j(w,\bar{w})}\right):G[X]F[X]:
\label{GF}
\een

The outset for a much richer structure is the introduction of a background
charge $2a_0$ for the fields $X_i$ \cite{FeFu, DF, FMS}. 
It amounts to adding to the action a term
\ben
 S_{2a_0}=\frac{-a_0}{8\pi}\int d^2z\sqrt{g}R\rho\cdot X
\een
coupling the fields $X_i$ to the curvature of the underlying surface.
In what follows we will mostly be interested in the Riemann sphere $S^2$ 
in which case it is possible to localise the curvature at the point of
infinity (the north pole). Hence, everywhere except in the vicinity
of the north pole we may use a flat metric and still consider the fields
to be free. The equations of motion become
\ben
 \pa\bar{\pa}X_i=\frac{1}{4}a_0(\al_i^\nu,\rho)\sqrt{g}R
\een
The corresponding energy-momentum tensor is (found by using the conformal
gauge or iso-thermal coordinates $g_{\mu\nu}(z)=e^{\sigma(z)}\delta_{\mu\nu}$)
\ben
 T_\var=\hf:\pa\var\cdot\pa\var:-a_0\rho\cdot\pa^2\var
\een
which has central charge
\ben
 c_\var=r-12a_0^2\rho^2=r-\frac{\theta^2}{2}a_0^2\hn d
\een
where we have used (\ref{FdeV}).
In the sequel we will concentrate on the holomorphic part $\var$ 
and put $a_0=1/\kvt$, see (\ref{sl2conv1}),(\ref{sl2conv2}).

The vertex operators later to be used in building primary fields are
\ben
 V_\lambda(z)=:e^{\frac{1}{\kvt}\lambda\cdot\var(z)}:
\label{vertex}
\een
and are seen to have conformal weights
\ben
 \Delta(V_\lambda)=\frac{1}{2t}(\lambda,\lambda+2\rho)
\een
Furthermore
\ben
 V_\lambda(z)V_{\lambda'}(w) =(z-w)^{\frac{1}{t}\lambda\cdot\lambda'}
  :e^{\frac{1}{\kvt}(\lambda\cdot\var(z)+\lambda'\cdot\var(w))}:
\een 

\subsection{Ghost Fields}

Ghost fields appear in gauge theory in the Faddeev-Popov procedure of gauge
fixing and in string theory. Usually one
denotes a fermionic ghost system $b,c$ arising when gauge fixing a bosonic
symmetry, and a bosonic ghost system $\beta,\g$ arising when gauge fixing
a fermionic symmetry. They are described by first order actions (characterised
by having first order kinetic terms as for the free Majorana fermion) and are 
conveniently treated simultaneously by introducing the parameter $\eta$,
$\eta=-1$ for bosonic statistics and $\eta=+1$ for fermionic statistics.
Hence, we will now let $b,c$ denote any ghost pair disregarding their
statistics and write
\ben
 S_\eta=\frac{\eta}{2\pi}\int d^2z\left(b\bar{\pa}c+\bar{b}\pa\bar{c}\right)
\een
The rationale for discussing ghost fields in this thesis is not a wish to
treat gauge fixing procedures, rather they appear
as building blocks in Wakimoto free field realizations, see Chapter 3.
The equations of motion ensure that the fields are (anti-)holomorphic.
This action is conformally invariant if $b$ and $c$ have dimensions
$(\lambda,0)$ and $(1-\lambda,0)$ respectively, in which case the 
energy-momentum tensor becomes (focusing on the holomorphic part)
\ben
 T=\eta\left((1-\lambda):\pa bc:-\lambda:b\pa c:\right)
\een
As for the free scalar field the 2-point function is easy to determine,
this time using $0=\int\frac{\delta}{\delta b}\left(e^{-S}b\right)$, and 
one arrives at the OPE\footnote{The somewhat unconventional factor $\eta$ 
in the action was chosen to produce $\beta(z)\g(w)=\frac{+1}{z-w}$.}
\bea
 c(z)b(w)&=&\frac{\eta}{z-w}+:c(z)b(w):\nn
 b(z)c(w)&=&\frac{1}{z-w}+:b(z)c(w):
\eea
The conformal dimensions are readily verified and the central charge is 
easily found to be
\ben
 c=-2\eta(6\lambda^2-6\lambda+1)=\eta(1-3Q^2)
\een
where $Q=\eta(2\lambda-1)$.

The standard mode expansions read
\ben
 b(z)=\sum_{n}b_nz^{-n-\lambda}\spa 
  c(z)=\sum_{n}c_nz^{-n-(1-\lambda)}
\een
and give rise to the (anti-)commutation relation
\ben
 [b_m,c_n]_\eta=\delta_{m+n,0}
\een
where $[\cdot,\cdot]_\eta$ is a (anti-)commutator for $\eta=-1$ $(\eta=+1)$.
The ranges for parameters $n$ depend on the choice of sector 
(Neveu-Schwarz versus Ramond). 
A zero-mode is one for which $\bar{\pa}$ vanishes throughout the Riemann
surface, here the sphere $S^2$. Using that for the primary field $b$,
$b(z)\left( dz\right)^\lambda$ is conserved under the conformal map 
$z\rightarrow1/z$, one finds the zero-modes of $b$ as the ones multiplying
functions of $z$ well-defined both in the neighbourhood of 0 and in the
neighbourhood of $\infty$. This analysis gives an index theorem 
(Riemann-Roch theorem) for ghost fields on the sphere. In general \cite{FMS}
\ben
 \sharp\left(\mbox{zero-modes of $c$}\right)-\sharp
  \left(\mbox{zero-modes of $b$}
  \right)=\hf\eta Q\chi
\een
where $\chi=2(1-g)$ is the Euler characteristic of the Riemann surface and
$g$ is the genus.

The ghost number operator is
\ben
 N_{gh}=\oint_0\dz j(z)\spa j(z)=\eta:b(z)c(z):
\een
and satisfies $[N_{gh},b(w)]=b(w)$ and $[N_{gh},c(w)]=-c(w)$. The ghost
number current is non-primary (or anomalous) for $Q\neq0$
\ben
 T(z)j(w)=\frac{Q}{(z-w)^3}+\frac{j(w)}{(z-w)^2}+\frac{\pa j(w)}{z-w}
\een
The hermiticity properties are
\bea
 b_n^\dagger&=&\eta b_{-n}\nn
 c_n^\dagger&=&c_{-n}\nn
 j_n^\dagger&=&-j_{-n}-Q\delta_{n,0}
\eea
In Chapter 4 we elaborate on the case of bosonic ghost fields $\beta,\g$
of dimension (1,0).

In the applications considered in this thesis we shall only need the bosonic
ghost fields.
It is clear that the following rules apply in computations of OPE's
when either $\beta$ or $\gamma$ fields appear with integral powers  
\bea
 \beta(z)^n F(\gamma(w))&=&
  :(\beta(z)+\frac{1}{z-w}\pa_{\gamma(w)})^n F(\gamma(w)):\nn
 \gamma(z)^n F(\beta(w))&=&
  :(\gamma(z)-\frac{1}{z-w}\pa_{\beta(w)})^n F(\beta(w)):
\label{bgcon}
\eea
It turns out (see Chapter 4)
that we need to be able to treat ghost fields raised to fractional powers.
The proposal in \cite{PRY1} to deal with those cases consists in generalising
(\ref{bgcon}) as
\ben
 G(\beta(z)) F(\gamma(w))
  =:G(\beta(z)+\frac{1}{z-w}\pa_{\gamma(w)})F(\gamma(w)):
\label{funcbgcon}
\een
where the asymptotic expansions for $G(\beta(z)+\frac{1}{z-w}\pa_{\gamma(w)})$
and $F(\gamma(w))$ would depend on their monodromy conditions in the 
$z$ and $w$ variables respectively.
An example relevant in later chapters is
\bea
 \beta(z)^{-t} F(\gamma(w))&=&:(\beta(z)+\frac{1}{z-w}\pa_{\gamma(w)})^{-t} 
  F(\gamma(w)):\nn   
 &=&\sum_{n=0}^\infty\binomial{-t}{n}
  :\beta^n(z)(z-w)^{t+n}\pa_{\gamma(w)}^{-t-n}F(\gamma(w)):
\label{frbgcon}
\eea
where $t$ is fractional for admissible representations (to be introduced in
Chapter 4) 
and we then see the need for fractional calculus \cite{Ross, McR, SKM, Kir}.
As an example of how the technique works we provide in a subsequent 
section an explicit non-trivial proof (\ref{Ia,b}) that
\ben
 (\beta^a(z)\gamma^a(w))(\beta^b(z)\gamma^b(w))=\beta^{a+b}(z)\gamma^{a+b}(w)
\een
Additionally, the explicit verification that the $N$-point functions
in $SL(2)$ current algebra based on this proposal
satisfy the Knizhnik-Zamolodchikov equations (see Chapter 4),
may be viewed as a check that fractional calculus
does indeed provide us with the requisite properties for Wick contractions as
defined in (\ref{frbgcon}).

An alternative to the proposal (\ref{funcbgcon}) is to write 
(\ref{bgcon}) as in (\ref{GF})
\ben
 G(\beta(z))F(\g(w))=\exp\left(\frac{1}{z-w}\pa_{\beta(z)}\pa_{\g(w)}
  \right):G(\beta(z))F(\g(w)):
\label{alterGF}
\een
and then generalise it by allowing asymptotic expansions (see below)
of the exponential as in
\ben
 e^z=\pa^a e^z=\sum_{n\in\Z}\frac{1}{\G(n-a+1)}z^{n-a}\spa a\in\C
\label{exp}
\een
To reproduce (\ref{frbgcon}) we put $a=t$. The gamma-function $\G(a+1)$ is 
the standard analytic continuation of 'factorial': $\G(n+1)=n!$, $n$ 
non-negative integer. 
 
For every positive root $\al>0$ we now introduce a pair of free bosonic ghost
fields ($\ba,\ga$) of conformal weights (1,0) satisfying the OPE
\ben
 \ba(z)\gb(w)=\frac{\dab}{z-w}
\een
The corresponding energy-momentum tensor is
\ben
 T_{\beta\g}=:\ba\pa\ga:
\label{Tbg}
\een
with central charge
\ben
 c_{\beta\g}=d-r
\een
We will understand 'properly' repeated root indices as in (\ref{Tbg})
to be summed over the positive roots.

\subsection{Fractional Calculus and Expansions}

Here we briefly introduce fractional calculus \cite{Ross, McR, SKM, Kir}.
For an analytic function $f(z)$ the fractional derivative 
(or integration) $\pa^af(z)$
is defined for complex number $a$. It satisfies the following axioms:
\begin{enumerate}
\item
If $f(z)$ is an analytic function, the fractional derivative 
$\pa^af(z)$ is an analytic function of $z$ and $a$.
\item For $a$ integer, the result must agree with ordinary differentiation
($a$ positive) or integration ($a$ negative). By default the 
integration constants are put to zero, so that the function together with
a maximum number of derivatives vanish at some point, like $z=0$.
\item $a=0$ is the identity.
\item Fractional differentiation is linear.
\item For fractional integration, $\Re a>0,\Re b>0$
$$\pa^{-a}\pa^{-b}f(z)=\pa^{-(a+b)}f(z)$$
\end{enumerate}
For $\Re a>0$, fractional integration $\pa^{-a}$ may be represented by the
Riemann-Liouville operator
\ben
\pa^{-a}f(z)=\frac{1}{\Gamma(a)}\int_0^z(z-t)^{a-1}f(t)dt
\een
By induction in the integer $n$ it is shown that for $a=n$ this coincides
with ordinary $n$-fold integration given by
\ben
\pa^{-n}f(z)=\int_0^zdz_1\int_0^{z_1}dz_2...\int_0^{dz_{n-1}}dz_nf(z_n)
\een
The third axiom follows from a Taylor expansion of $f(t)$.
The semi-group property of axiom 5 follows from
\bea
 \pa^{-a}\pa^{-b}f(z)&=&\frac{1}{\G(a)\G(b)}\int_0^zdz_1\int_{z_1}^zdz_2
  (z-z_2)^{a-1}(z_2-z_1)^{b-1}f(z_1)\nn
 &=&\pa^{-(a+b)}f(z)
\eea
where we have used the substitution $u=\frac{z_2-z_1}{x-z_1}$ and the Euler
integral
\ben
 \int_0^1u^{b-1}(1-u)^{a-1}du=\frac{\G(a)\G(b)}{\G(a+b)}
\label{Euler}
\een
Having defined fractional differentiation for arbitrary negative powers, we
may obtain fractional derivatives of arbitrary positive powers from these
by acting with an integer number of ordinary differentiations. Thus, for 
$a>0$ we write $a=n-a'$ where $n$ is a positive integer and $0\leq a'<1$,
and define
\ben
 \pa^af(z)=\pa^n\left(\frac{1}{\G(a')}\int_0^z(z-t)^{a'-1}f(t)dt\right)
\een
For every pair of complex numbers $a,b$ we have $\pa^a\pa^b=\pa^{a+b}$.
We now consider the particular case $f(z)=z^b$ and find
\ben
\pa^a z^b=\frac{\Gamma(b+1)}{\Gamma(b-a+1)}z^{b-a}
\label{fracpol}
\een
This holds for both fractional integration and fractional differentiation.
Notice that $\pa^a 1\neq 0$ for $a$ not a positive integer.

Let us conclude this section by commenting on various expansions as the
one employed in (\ref{exp}).
Since we shall need contractions with the operator (see chapters 4 and 5)
\ben
 (1+\gamma(z)x)^{2j}
\een
we shall find it convenient to represent the expansion of the associated
analytic function
\ben
 f(z)=(1+z)^{2j}
\een
in several different ways. Indeed, from
\ben
 (1+z)^{2j}=z^{-a}{[}(1+z)^{2j}z^{a}{]}
\label{1plusz2j}
\een
we may imagine that the last bracket is expanded in integer powers of $z$
in a way convergent on the unit circle.
Although ${[}(1+z)^{2j}z^{a}{]}$ has a discontinuity on the circle, it
vanishes when we take it to be for negative $z$, corresponding to choosing
the branch cuts along the negative real axis. This
renders many equivalent representations for the function
\ben
 (1+z)^{2j}=(1+z)^{2j}_{(-a)}=\sum_{n\in\Z}\binomial{2j}{n-a}z^{n-a}
\een
which are all equivalent in the sense of analytic function theory but which 
correspond to expansions with different monodromies for the individual 
terms.

When deciding on what expansion to adopt for the operator
\ben
 (1+\gamma(z)x)^{2j}_{(a)}
  =\sum_{n\in \Z}\binomial{2j}{n+a}(\gamma(z)x)^{n+a}
\een
we use the criterion, that {\em after} all Wick contractions have been 
performed, powers of $\beta$ and $\gamma$ inside normal ordering signs are
non-negative {\em integers}. 
Only then, these terms have an obvious interpretation 
when sandwiched between states. In other terms, the existence of external 
states and other primary fields in the correlator decides what monodromies to 
choose for individual terms in expansions. All of that will be illustrated 
further in chapters 4 and 5.

Next, consider the fractional derivative of the exponential function
\ben
 D^a\exp (z)=\sum_{n\in\Z}\frac{1}{\Gamma(n-a+1)}z^{n-a}, \ \ a\in\C
\label{Daexp}
\een
The right hand side is covariant under further differentiation 
corresponding to the fact that it represents the original exponential 
function. The representation is a peculiar realization of the exponential 
function, which converges asymptotically for $|z|\rightarrow \infty$.
Again the representation may be better understood by writing
\ben
 \exp(z)=z^{-a}{[}\exp(z)z^a{]}
\een
and then introducing for the last bracket a Fourier expansion with integer 
powers of $z$ on a circle in the complex $z$ plane. On the circle
\ben
 {[}\exp(z)z^a{]}
\een
has a discontinuity which we may take to be for negative $z$. The Fourier
coefficients depend on the radius of the circle. However, for large $|z|$ 
where the discontinuity becomes vanishingly small, they converge to
the expansion coefficients in (\ref{Daexp}). Thus we take
\ben
 D^a\exp(z)=\exp(z)
\een
for any $a$. However, we shall find it convenient to use the fractional 
derivative to represent a generating functional for the integrals
\ben
 \lim_{R\rightarrow\infty}\oint_{RS^1}\dtp{u}\frac{e^u}{u^{a+n+1}}=
 \frac{1}{\Gamma(a+n+1)}
\een
Different $a$'s give rise to representations or expansions of the
exponential function in which individual terms have different non-trivial 
monodromies. We shall use $D^a\exp(z)$ simply as a reminder of the 
appropriate expansion needed.
Let us notice the following identity for fractional derivatives of exponentials
\ben
 D_z^a\exp(z)D_w^b\exp(w)=D^{a+b}\exp(z+w)
\label{DDfrac}
\een

\subsection{Consistency Check}

In this section we illustrate the non-trivial nature of the workings of the
Wick contractions proposed in (\ref{funcbgcon}), by considering the evaluation
\cite{PRY1} of 
\ben
 I_{a,b}(z,w)= (\beta^a(z)\gamma^a(w))(\beta^b(z)\gamma^b(w))
\label{Ia,b}
\een
First we notice the OPE
\ben
 \beta^a(z)\gamma^a(w)=\frac{\Gamma(a+1)}{(z-w)^a}\sum_n\binomial{a}{n}
  \frac{1}{n!}(z-w)^n:(\beta(z)\gamma(w))^n:
\label{baga}
\een
following from (\ref{funcbgcon}) and (\ref{fracpol}) (or alternatively
from (\ref{alterGF}-\ref{exp}) and (\ref{fracpol})). 
Now, (\ref{Ia,b}) may be evaluated in two ways: either (i) by first using 
(\ref{baga})
for both parentheses to reduce both of them to normal ordered products of
integer powers of $\beta$ and $\gamma$, and then subsequently carrying out all
remaining contractions, or else (ii) by simply using (\ref{baga}) with $a+b$
replacing $a$. Obviously these two ways should lead to the same result
for consistency. This requirement is part of the associativity properties
for operators, and we consider them to be justified by the fact that the
$N$-point functions of Chapter 4 satisfy the Knizhnik-Zamolodchikov equations. 
Here we demonstrate that the above condition
gives rise to non-trivial identities for which we indicate an independent 
elementary proof.

The general contractions between integer powers of $\beta$ and $\gamma$ are 
carried out using (\ref{alterGF})
\bea
 \beta^n(z)\gamma^m(w)&=&\exp\left(\frac{1}{z-w}\pa_{\beta(z)}\pa_{\g(w)}
  \right):\beta^n(z)\g^m(w):\nn
 &=&\sum_\ell \frac{1}{(z-w)^{\ell}\ell!}\frac{\Gamma(n+1)}{\Gamma(n-\ell +1)}
  \frac{\Gamma(m+1)}{\Gamma(m-\ell +1)}:\beta^{n-\ell}(z)\gamma^{m-\ell}(w):
\eea
Then one obtains after a few steps
\bea
 I_{a,b}(z,w)&=&\frac{\Gamma(a+1)\Gamma(b+1)}{(z-w)^{a+b}}
  \sum_{m,n,k,\ell}\binomial{a}{m}\binomial{b}{n}\binomial{m}{k}
  \binomial{n}{\ell}\nn
 &\cdot&\frac{(z-w)^{m+n-k-\ell}}{(m-\ell)!(n-k)!}
  :(\beta(z)\g(w))^{m+n-k-\ell}:
\eea
On the other hand the second way of evaluation simply gives the result
\ben
 I_{a,b}(z,w)=\frac{\Gamma(a+b+1)}{(z-w)^{a+b}}\sum_N\binomial{a+b}{N}
  (z-w)^N\frac{:(\beta(z)\gamma(w))^N:}{N!}
\een
Defining the generating functions
\bea
 F_a(x)&=&\sum_N\binomial{a}{N}\frac{x^N}{N!}\nn
 G_{ab}(x)&=&\sum_{m,n,k,\ell}\binomial{a}{m}\binomial{b}{n}\binomial{m}{k}
  \binomial{n}{\ell}\frac{x^{m+n-k-\ell}}{(m-\ell)!(n-k)!}
\eea
we see that the consistency condition may be expressed as
\ben
 G_{ab}(x)=\binomial{a+b}{b}F_{a+b}(x)
\label{cc}
\een
We now briefly indicate how this identity may be proven. First we notice
(see below) that we may write
\ben
 \binomial{a+b}{a}=\binomial{a+b}{b}=\oint\dtp{t}\frac{(1+t)^{a+b}}{t^{a+1}}
\label{binab}
\een
where the contour may be taken as the unit circle, passing through the
branch point at $t=-1$ of the integrand (for suitable values of the exponents).
Then we write
\bea
 G_{ab}(x)&=&\sum_{p,q,k,\ell}\binomial{a}{q+\ell}\binomial{b}{p+k}
  \binomial{q+\ell}{k}\binomial{p+k}{\ell}\frac{x^{p+q}}{p!q!}\nn
 &=&\sum_{p,q,k,\ell}\prod_{i=1}^4\oint_{{\cal C}_0}\dtp{t_i}\frac{1}{t_i}
  \frac{(1+t_1)^a}{t_1^{q+\ell}}\frac{(1+t_2)^b}{t_2^{p+k}}
  \frac{(1+t_3)^{q+\ell}}{t_3^k}
  \frac{(1+t_4)^{p+k}}{t_4^\ell}\frac{x^{p+q}}{p!q!}
\eea
The identity is now obtained by successively doing 1) the sum over $\ell$,
2) the integral over $t_4$, 3) the sum over $k$, 4) the integral over $t_3$,
5) the integral over $t_1$ and $t_2$ in any order, and 6) the sum over $q$ 
with $p+q=N$. In each case one picks up 
residues after suitable deformations of the contours.

Let us illustrate techniques of contour deformations by proving (\ref{binab}).
We consider
\ben
 \oint_{S^1}\frac{dt}{2\pi i}(1+t)^at^{b-1}=-\oint_{S^1}\frac{dt}{2\pi i}
  (1-t)^a(-t)^{b-1}
\een
On the rhs we take the branch cuts to be along the positive real axis. We
may then deform the contour to be one starting in 1 (where the integrand
vanishes) and surrounding 0 in a way such that it 'runs' first just above
and then just below the interval from 0 to 1. Compared to the negative
real axis on which the integrand is positive, we pick up phases when
being just above/below the positive real axis. Since the contour is always
to the left of 1 we only need to consider the power of $-t$. We find
\bea
 \oint_{S^1}\frac{dt}{2\pi i}(1+t)^at^{b-1}&=&-\frac{1}{2\pi i}\left(
  \int_1^0(1-t)^at^{b-1}e^{-i\pi(b-1)}dt+\int_0^1(1-t)^at^{b-1}e^{i\pi(b-1)}
  dt\right)\nn
 &=&\binomial{a}{-b}
\label{famous}
\eea
where we have used the Euler integral (\ref{Euler}) and the gamma-function
identity
\ben
 \Gamma(b)\G(1-b)=\frac{\pi}{s(b)}
\een
with
\ben
 s(b)=\sin(\pi b)
\een
Similarly, one may show that
\ben
 \oint_{S^1}\frac{dt}{2\pi i}(1-t)^at^{b-1}=e^{i\pi b}\binomial{a}{-b}
\label{famous2}
\een
Now, to make the lhs well-defined ($S^1$ a closed contour) we must choose
the cuts as before. Letting the function on the upper sheet be real near the 
positive real axis between 0 and 1, we find
\bea
 \oint_{S^1}\frac{dt}{2\pi i}(1-t)^at^{b-1}&=&\frac{1-e^{2\pi i(b-1)}}{2\pi i}
  \int_1^0(1-t)^at^{b-1}dt\nn
 &=&e^{i\pi b}\binomial{a}{-b}
\eea
These formulas and the techniques of contour manipulations will be used
repeatedly in the following chapters.

\section{Minimal Models}
Referring to the notation in (\ref{Dnm}) it can be shown \cite{BPZ} that for
$\al_-/\al_+$ being rational
\bea
 \al_-/\al_+&=&-p/q\spa c=1-\frac{6(p-q)^2}{pq}\nn
 \D_{n,m}&=&\frac{(nq-mp)^2-(p-q)^2}{4pq}
\label{minmod}
\eea
where $p$ and $q$ are co-prime, the operator algebra becomes finite 
dimensional. In general, a CFT with a finite number of primary fields 
(wrt the chiral algebra, see below) is
called a rational CFT (RCFT). The nomenclature is due to the fact 
that the weights and the central charge can then be shown to
be rational \cite{Ver, Vaf, AM}. 
The CFT's based on (\ref{minmod}) are called minimal models
and have multiplicities one in their spectra of primary fields.
The reflection symmetry $\D_{n,m}=\D_{p-n,q-m}$ leads to the fundamental
region
\ben
 0<n<p\spa 0<m<q\spa p<q
\een
For $q=p+1$ one recovers the unitary subset (\ref{cunitary}).

In Chapter 4 we discuss Hamiltonian reduction of $SL(2)$ current algebra
based on admissible representation to minimal models. There we put $p/q=t$
whereby $\al_+=\sqrt{2/t}$ and $\al_-=-\sqrt{2t}$, while the background
charge becomes $-2\al_0=-\sqrt{2/t}+\sqrt{2t}$.

\subsection{Dotsenko-Fateev Construction}
Here we discuss a realization of the minimal models due to Dotsenko and
Fateev \cite{DF} following ideas of Feigin and Fuks \cite{FeFu84}. 
It is based on a single bosonic scalar field $\var$ coupled to a background 
charge $-2\al_0$, as described in a preceding section
\bea
 \var(z)\var(w)&=&\ln(z-w)\nn
 T(z)&=&\hf:\pa\var(z)\pa\var(z):+\al_0\pa^2\var(z)\nn
 c&=&1-12\al_0^2\nn
 V_p(z)&=&:e^{p\var(z)}:\nn
 \D_p&=&\hf p(p-2\al_0)
\eea
$V_p$ is a free field realization of a primary field and $\D(V_p)=\D_p=
\D_{2\al_0-p}$. We shall be interested in conformal blocks
\ben
 \langle\prod_{i=1}^{\cal N}V_{p_i}(z_i)\rangle
\label{naiveN}
\een
subject to the charge conservation $\sum_{i=1}^{\cal N}p_i=2\al_0$.
Charges are defined wrt the (anomalous) current $j=\pa\var$
\ben
 [Q,V_p(w)]=pV_p(w)\spa Q=\oint\dz j(z)
\een
Let us define screening currents as primary fields (vertex operators)
of dimension 1. We easily find the two solutions 
\ben
 S_\pm(z)=:V_{\al_\pm}(z):\spa\al_\pm=\al_0\pm\sqrt{\al_0^2+2}
\een
These currents are useful since their conformal properties ensure that the
corresponding screening charges (integrated currents) can be inserted
into correlators without violating the conformal structure. They only
affect the boundary configuration or total charge, and may be inserted 
into (\ref{naiveN}) in order that the charge conservation be met.

The mode expansions read
\bea
 \var(z)&=&q_\var+a_0\ln z+\sum_{n\in\Z}\frac{a_n}{-n}z^{-n}\nn
 L_n&=&\hf\sum_{m\in\Z}:a_{n-m}a_m:-\al_0(n+1)a_n\nn
 \left[a_n,a_m\right]&=&n\delta_{n+m,0}\spa \left[a_0,q_\var\right]=1\nn
 :e^{-p\var(z)}:&=&\exp\left(-p\sum_{n>0}\frac{a_{-n}}{n}z^n\right)
  e^{-pq_\var}
  z^{-pa_0}\exp\left(-p\sum_{n>0}\frac{a_n}{-n}z^{-n}\right)
\eea
and the projective ($SL(2)$) invariant ket-vacuum $\ket{0}=\ket{sl_2}$
satisfies
\bea
 L_n\ket{0}&=&0\spa n\geq-1\nn
 a_n\ket{0}&=&0\spa n\geq0
\eea
The $SL(2)$ invariant bra-vacuum $\bra{sl_2}$ satisfies
\bea
 \bra{sl_2}L_n&=&0\spa n\leq1\nn
 \bra{sl_2}a_n&=&0\spa n\leq-1
\eea
From considering $\bra{sl_2}L_0$ one finds the hermiticity property
\ben
 a_n^\dagger=a_{-n}-2\al_0\delta_{n,0}
\een
Let us define an $SL(2)$ non-invariant dual bra-vacuum
\ben
 \bra{0}=\lim_{z\rightarrow\infty}\bra{sl_2}:e^{2\al_0\var(z)}:=
  \bra{sl_2}e^{2\al_0 q_\var}
\een
which has the advantage that it satisfies
\ben
 \bra{0}a_0=0\spa \bra{0}0\rangle=1
\een
It corresponds to inserting at infinity the vertex operator 
$:e^{2\al_0\var(z)}:$ of conformal dimension zero. The ket-states are 
defined by 
\ben
 \ket{p}=\lim_{z\rightarrow0}:e^{p\var(z)}:\ket{0}
\een
while dual (bra-)states are defined by
\ben
 \bra{p}=\lim_{z\rightarrow\infty}\bra{sl_2}:e^{(2\al_0-p)\var(z)}:z^{2\D_p}
  =\bra{0}e^{-pq_\var}
\label{dualp}
\een
Introduce intertwining fields \cite{F, BF,BMP}
\bea
 \left[\phi_{p_2}(z,x)\right]_{p_1}^{p_3}&=&\int\prod_{k=1}^n
  dv_k\prod_{i=1}^mdw_i\phi_{p_2}(z,x)P(v_1,...,v_n;w_1,...,w_m)\nn
 P(v_1,...,v_n;w_1,...,w_m)&=&\prod_{k=1}^nS_+(v_k)\prod_{i=1}^mS_-(w_i)
 =\prod_{k=1}^n:e^{\al_+\var(v_k)}:\prod_{i=1}^m:e^{\al_-\var(w_i)}:
\label{intertwmin}
\eea
with charge conservation $p_1+p_2-p_3+n\al_++m\al_-=0$.
Now, in the ${\cal N}$-point conformal block
\ben
 W_{\cal N}=\bra{p_{\cal N}}{[}\phi_{p_{{\cal N}-1}}(z_{{\cal N}-1})
  {]}^{p_{\cal N}}_{\kappa_{{\cal N}-2}}
  ...{[}\phi_{p_{j}}(z_{j}){]}^{\kappa_{j}}_{\kappa_{j-1}}...
  {[}\phi_{p_{2}}(z_2){]}^{\kappa_2}_{p_1}\ket{p_1}
\een
we have the overall charge conservation
\ben
 p_1+...+p_{{\cal N}-1}-p_{\cal N}+N\al_++M\al_-=0
\een
where $N,M$ are the total numbers of screening charges inserted. The 
notation follows from the pictorial version
\ben
\begin{picture}(300,70)
 \put(0,0){$p_{\cal N}$}  \put(10,10){\line(1,0){50}}  
 \put(55,0){$\kappa_{{\cal N}-2}$}
 \put(35,10){\line(0,1){40}} \put(35,55){$p_{{\cal N}-1}$} 
 \put(110,10){\line(1,0){50}}
 \put(110,0){$\kappa_j$}   \put(155,0){$\kappa_{j-1}$}  \put(135,55){$p_j$}
 \put(135,10){\line(0,1){40}} \put(210,10){\line(1,0){75}} \put(280,0){$p_1$}
 \put(235,10){\line(0,1){40}} \put(260,10){\line(0,1){40}}
 \put(235,55){$p_3$}\put(260,55){$p_2$} \put(245,0){$\kappa_2$}
 \put(210,0){$\kappa_3$}
 \put(180,9){$\cdots$} \put(80,9){$\cdots$}
\end{picture}
\een
where
\bea
 p_1+p_2-\kappa_2+n_2\al_++m_2 \al_-&=&0\nn
 \kappa_2+p_3-\kappa_3+n_3\al_++m_3 \al_-&=&0\nn
 &\vdots&\nn
 \kappa_{j-1}+p_j-\kappa_j+n_j\al_++m_j \al_-&=&0\nn
 &\vdots&\nn
 \kappa_{{\cal N}-2}+p_{{\cal N}-1}-p_{\cal N}+n_{{\cal N}-1}\al_+
  +m_{{\cal N}-1}\al_-&=&0\nn
 n_2+...+n_{{\cal N}-1}&=&N\nn
 m_2+...+m_{{\cal N}-1}&=&M
\eea
with $n_j,m_j$ non-negative integers.
The charge conservation ensures that formal powers of $\infty$ arising from
the contractions involving $:e^{(2\al_0-p_{\cal N})\var(\infty)}:$, is 
cancelled by $\infty^{{2\D}_{p_{\cal N}}}$ from (\ref{dualp}). 
Thus we may concentrate on the remaining contractions and we find
\bea
 W_{\cal N}&=&\prod_{1\leq j<j'<{\cal N}}(z_j-z_{j'})^{p_jp_{j'}}
  \int\prod_{k=1}^Ndv_k\prod_{i=1}^Mdw_i
  \prod_{j,k}(z_j-v_k)^{p_j\al_+}\prod_{j,i}(z_j-w_i)^{p_j\al_-}\nn
 &\cdot&\prod_{k<k'}(v_k-v_{k'})^{\al_+^2}\prod_{i<i'}(w_i-w_{i'})^{\al_-^2}
  \prod_{k,i}(v_k-w_i)^{-2}
\label{WNminmod}
\eea

For a 4-point function with all $p_i=p$, the charge conservation reads
$p=-(N\al_++M\al_-)/2$ and leads to a discretisation or quantisation
of the common conformal weights
\ben
 \D_p=\frac{1}{8}\left((N+1)\al_++(M+1)\al_-\right)^2-\hf\al_0^2
\een
which re-produces the Kac table (\ref{Dnm}) since here $N,M\geq0$.
An alternative derivation of this quantisation also based on free fields is
discussed in \cite{KM, FZ87}. In Chapter 4 we will use the notation
\ben
 \al_{n,m}=p_{n,m}=\frac{1-n}{2}\al_++\frac{1-m}{2}\al_-
\een

Here we will not go into a discussion of choosing contours, finding the fusion
rules, the monodromy invariant Greens functions or the operator product
algebra coefficients, but refer to the original works by Dotsenko and
Fateev \cite{DF, DF85} and by Felder \cite{F}. However, in chapters 4 and 5
we will consider such issues in the case of a CFT based on $SL(2)$ 
current algebra. In that connection we shall comment on the fusion rules for
minimal models.

\subsection{A Hidden Duality}
For 4-point functions in the limit 
$(z_1,z_2,z_3,z_4)\rightarrow(0,z,1,\infty)$, (\ref{WNminmod}) reduces to
\bea
 W_4&=&z^{p_1p_2}(1-z)^{p_1p_3}
  \int\prod_{k=1}^Ndv_k\prod_{i=1}^Mdw_i
  \prod_{k=1}^Nv_k^{p_1\al_+}(z-v_k)^{p_2\al_+}(1-v_k)^{p_3\al_+}
  \prod_{k<k'}(v_k-v_{k'})^{\al_+^2}\nn
 &\cdot&\prod_{i=1}^Mw_i^{p_1\al_-}(z-w_i)^{p_2\al_-}(1-w_i)^{p_3\al_-}
  \prod_{i<i'}(w_i-w_{i'})^{\al_-^2}
  \prod_{k,i}(v_k-w_i)^{-2}
\label{WNminmod4}
\eea
For minimal models where the momenta or charges are quantised
\ben
 p_i=\frac{1-n_i}{2}\al_++\frac{1-m_i}{2}\al_-
\een
we write the corresponding primary fields as $\phi_{n_i,m_i}$. This notation 
allows us to state a hidden symmetry or duality in the minimal models 
(in fact, in any DF construction of degenerate models) as
\ben
 \langle\prod_{i=1}^4\phi_{n_i,m_i}\rangle\ \ \propto\ \ \langle\prod_{i=1}^4
  \phi_{\hat{n}_i,\hat{m}_i}\rangle\nn
\label{hiddual}
\een
where the fields are subject to the charge conservations
\ben
 p_1+p_2+p_3-p_4+N\al_++M\al_-=\hat{p}_1+\hat{p}_2+\hat{p}_3-\hat{p}_4
  +N\al_++M\al_-=0
\een
and where
\bea
 \hat{n}_1=\frac{n_1+n_2-n_3-n_4}{2}
  &\spa&\hat{m}_1=\frac{m_1+m_2-m_3-m_4}{2}\nn
 \hat{n}_2=\frac{n_1+n_2+n_3+n_4}{2}
  &\spa&\hat{m}_2=\frac{m_1+m_2+m_3+m_4}{2}\nn
 \hat{n}_3=\frac{-n_1+n_2+n_3-n_4}{2}
  &\spa&\hat{m}_3=\frac{-m_1+m_2+m_3-m_4}{2}\nn
 \hat{n}_4=\frac{-n_1+n_2-n_3+n_4}{2}
  &\spa&\hat{m}_4=\frac{-m_1+m_2-m_3+m_4}{2}
\eea
This relation is denoted 'mysterious' by Andreev \cite{An}. Nevertheless,
in the next section we will present a proof of it \cite{PRY2}. 

\subsection{An Integral Identity}
Here we want to establish the integral identity \cite{PRY2} underlying
(\ref{hiddual}) in the DF constructions.\\[.2 cm]
{\bf Proposition}
\bea
 &&\int_0^1\prod_{k=1}^Ndv_kv_k^{a'}(1-v_k)^{b'}(1-zv_k)^{c'}\prod_{k<k'}
  (v_k-v_{k'})^{2\rho'}\nn
 &\cdot&\prod_{i=1}^Mdw_iw_i^a(1-w_i)^b(1-zw_i)^c\prod_{i<i'}
  (w_i-w_{i'})^{2\rho}\prod_{k,i}^{N,M}(v_k-w_i)^{-2}\nn
 &=&K_{NM}\int_0^1\prod_{k=1}^Ndv_kv_k^{a'-\delta'}(1-v_k)^{b'+\delta'}
  (1-zv_k)^{c'-\delta'}\prod_{k<k'}(v_k-v_{k'})^{2\rho'}\nn
 &\cdot&\prod_{i=1}^Mdw_iw_i^{a-\delta}(1-w_i)^{b+\delta}
 (1-zw_i)^{c-\delta}
  \prod_{i<i'}(w_i-w_{i'})^{2\rho}\prod_{k,i}^{N,M}(v_k-w_i)^{-2}
\label{minimal}
\eea
where
\bea
 a'=-\rho'a\hspace{1 cm}b'&=&-\rho'b\hspace{1 cm}c'=-\rho'c\hspace{1 cm}
 \delta'=-\rho'\delta\hspace{1 cm}\rho'=1/\rho\nn
 \delta&=&a+c+1-N+(M-1)\rho\nn
 \delta'&=&a'+c'+1-M+(N-1)\rho'
\label{prime}
\eea
and
\bea
  K_{NM}&=&\prod_{i=0}^{N-1}\frac{\Gamma(a'+1+i\rho')\Gamma(b'+1+i\rho')}
    {\Gamma(-c'+M+(-N+1+i)\rho')\Gamma(a'+b'+c'+2-M+(N-1+i)\rho')}\nn
  &\cdot&\prod_{i=0}^{M-1}\frac{\Gamma(a+1-N+i\rho)\Gamma(b+1-N+i\rho)}
   {\Gamma(-c+(-M+1+i)\rho)\Gamma(a+b+c+2-2N+(M-1+i)\rho)}
\label{KNMmin}
\eea
The left hand side of (\ref{minimal}) has the structure of the standard
integral realization for minimal models \cite{DF}, leaving out some irrelevant 
pre-factors (\ref{WNminmod4}).
There are $N,M$ screening charges of the two kinds and they are
at positions $v_k$ and $w_i$, except they have been scaled by $z$, as to scale
their integration regions $\int_0^z\rightarrow\int_0^1$. The fact that all
integrations are like that corresponds to considering only
one kind of conformal block, where all the integrations are between $0$ 
and $z$ \cite{DF}, see also Chapter 5.
The integrations are taken to be time ordered in the sense
\ben
 T\prod_{i=1}^n\int_0^1dt_i\sim\int_0^1dt_1\int_{t_1}^1dt_2...\int_{t_{n-1}}^1
  dt_n\sim\frac{1}{n!}\prod_{i=1}^n\int_0^1dt_i
\label{timeordered}
\een
{\bf Proof}\\
The proof \cite{PRY2} 
we present of this identity is by brute force and takes several 
lengthy calculations. For the understanding of (\ref{hiddual}) it is 
interesting in it's own right. Though, we shall mostly be interested in a
similar but much more complicated integral identity (\ref{pryan}) between 
$SL(2)$ blocks. However, the techniques involved are to a large extend the 
same in the two cases, so let us illustrate them by going over some of the 
details of the present proof.

The idea is simply to consider both sides of (\ref{minimal}) as functions 
of $z$, and notice that the only singularities occur when
$z\rightarrow 0,1,\infty$.

The limit $z\rightarrow 0$ is simple. In that limit both sides of 
(\ref{minimal}) are holomorphic in $z$ ($1-zw\neq0$)
and we may simply put $z=0$. Then both sides may be computed 
in terms of the Dotsenko-Fateev integral (\ref{dfintegral}).
This gives immediately the normalisation $K_{NM}$.

The limits $z\rightarrow 1,\infty$ are much more complicated. Here there will
be several different power singularities of the form $(1-z)^A$ and $z^B$. 
We must
isolate those and compute their strengths and demonstrate that we get the same 
results for both sides of (\ref{minimal}).\\[.2cm]
$\underline{z\rightarrow1}$\\
Now $(1-zw)$ is close to 0 for $w$ close to 1, so
we split the integration region from $0$ to $1$ using a small positive 
$\epsilon$ as follows (the integrations are time-ordered throughout)
\bea
 \int_0^1&=&\int_0^{1-\epsilon}+\int_{1-\epsilon}^1\nn
  \int_0^1\prod_{k=1}^Ndv_k\prod_{i=1}^Mdw_i&=&\sum_{n,m}
  \int_0^{1-\epsilon}\prod_{k=N-n+1}^Ndv_k\prod_{i=M-m+1}^Mdw_i
\int_{1-\epsilon}^1
   \prod_{l=1}^{N-n}dv_l\prod_{j=1}^{M-m}dw_j
\eea
It is not difficult to check that a particular $n,m$ term will give rise to a
particular power of $(1-z)$. One performs the following scalings of the 
$\int_{1-\epsilon}^1$ integration variables, $w\sim v_l,w_j$,
\bea
  w&\rightarrow&1-(1-z)\frac{1-w}{w}\nn
  dw&\rightarrow&\frac{1-z}{w^2}dw\nn
  1-w&\rightarrow&(1-z)\frac{1-w}{w}\nn
  1-zw&\rightarrow&(1-z)\frac{(1-z)w+z}{w}\sim\frac{1-z}{w}\nn
  \int_{1-\epsilon}^1&\rightarrow&\int_{\frac{1-z}{\epsilon+1-z}}^1\sim
  \int_0^1
\eea
One rather easily finds that the power of $(1-z)$ occurring on both sides of 
(\ref{minimal}) is
\ben
 (1-z)^{(N-n)(b'+c'+1)+(N-n)(N-n-1)\rho'+(M-m)(b+c+1)+(M-m)(M-m-1)\rho
  -2(N-n)(M-m)}
\een
The coefficient of this singularity may also be evaluated 
on both sides in terms
of DF integrals. It is not, however, immediately obvious that
these coefficients are equal. Both sides involves many products of terms 
involving ratios of gamma-functions. One employs over and over again the 
simple identity
\ben
 \frac{\Gamma(X)}{\Gamma(X-L)}=\prod_{j=0}^{L-1}(X-1-j)
\label{gammaid}
\een
Thus, it turns out for example that there are factors on both sides  
involving gamma-functions with argument involving 
$a'$ only (no $b,'c'$). On the left hand side we have
\ben
 \prod_{i=0}^{n-1}\Gamma(a'+1+i\rho')
\label{lhsa}
\een
whereas on the right hand side there are similar factors of the form
\ben
 \frac{\prod_{i=0}^{N-1}\Gamma(a'+1
   +i\rho')}{\prod_{i=0}^{N-n-1}\Gamma(a'+1-M+(n+i)\rho')}
\label{rhsa}
\een
Using the identity (\ref{gammaid}) we find that the ratio between
(\ref{lhsa}) and (\ref{rhsa}) is
\ben
 \prod_{i=0}^{N-n-1}\prod_{j=0}^{M-1}\frac{1}{(a'-j+(n+i)\rho')}
\een
Similarly one finds that the ratio of terms only involving $a$ is
\ben
 \prod_{i=0}^{N-n-1}\prod_{j=0}^{M-1}(a-n-i+j\rho)
\een
Up to powers of $\rho$ these cancel according to (\ref{prime}).
By going over the several 
other different factors on both sides and working out the
ratios, one finally shows that the product of all ratios equals $1$.

This completes the proof that the singularities are identical in the limit 
$z\rightarrow 1$.\\[.2cm]
$\underline{z\rightarrow\infty}$\\
The strategy is entirely analogous. Now $(1-zw)$ is close to 0 (or indefinite)
for $w$ close to 0, so we make the split (time-ordered integrations)
\bea
  \int_0^1&=&\int_0^{\epsilon}+\int_{\epsilon}^1\nn
 \int_0^1\prod_{k=1}^Ndv_k\prod_{i=1}^Mdw_i&=&\sum_{n,m}\int_0^{\epsilon}
 \prod_{k=N-n+1}^Ndv_k\prod_{i=M-m+1}^Mdw_i\int_{\epsilon}^1\prod_{l=1}^{
  N-n}dv_l\prod_{j=1}^{M-m}dw_j
\label{zsplitinf}
\eea
The $\int_0^{\epsilon}$ integration variables are scaled according to
\bea
 w&\rightarrow&\frac{1-w}{-zw}\nn
 \int_0^{\epsilon}dw&\rightarrow&\int_1^{\frac{1}{1-z\epsilon}}\frac{dw}{zw^2}
  \sim(-z)^{-1}\int_0^1\frac{dw}{w^2}
\label{wwrtinf}
\eea
This time there is a subtlety in the identification of the two sides since it
turns out that we must identify the left hand side with $n,m$ with the right
hand side with $N-n,M-m$. It is then simple to verify that the power of $z$
on both sides are
\ben
 (-z)^{-na'+(N-n)c'-n-n(n-1)\rho'-ma+(M-m)c-m-m(m-1)\rho+2nm}
\een
To check that the coefficients also agree, as before one carries out explicitly
the integrations in terms of DF integrals resulting in many 
products of ratios of gamma-functions. Finally, one carries out the
cumbersome computation showing that the ratios indeed multiply up to $1$.\\
$\Box$\\
Let us remark that the trivial identity 
\ben
 F(\al,\beta;\g;z)=F(\beta,\al;\g;z)
\label{hypid}
\een
for hyper-geometric functions
\ben
 F(\al,\beta;\g;z)=\frac{\G(\g)}{\G(\al)\G(\beta)}\sum_{n\geq0}\frac{
  \G(\al+n)\G(\beta+n)}{\G(\g+n)n!}z^n
\een
is less transparent at the level of integral representations
\ben
 F(\al,\beta;\g;z)=\frac{\G(\g)}{\G(\beta)\G(\g-\beta)}\int_0^1t^{\beta-1}
  (1-t)^{\g-\beta-1}(1-zt)^{-\al}dt
\een
However, it is the simplest example of the integral identity (\ref{minimal})
with $(N,M)=(1,0)$ or $(N,M)=(0,1)$. Hence, (\ref{minimal}) may be viewed
as a highly non-trivial generalisation of (\ref{hypid}).

\section{Affine Current Algebras}
For every Cartan-Weyl generator $j_a$ of a simple Lie algebra {\bf g}
let there be
associated a spin 1 field $J_a(z)$ (of conformal dimension (1,0)).
Dimensional analysis leads us to define the
operator product expansion of these (2$D$) current algebra generators
to be
\ben
 J_a(z)J_b(w)=\frac{\kappa_{ab}k}{(z-w)^2}+\frac{\fabc J_c(w)}{z-w}
\label{JaJb}
\een
where regular terms have been omitted. $k$ is the central extension 
(see below), commuting with all currents and takes on a 
constant value in any representation, so we will treat it as a constant.
This is analogue to the treatment of the algebra element $c$ (the central
charge) in the Virasoro algebra as a constant.
$\kn=\frac{2k}{\theta^2}$ is the level. The Chevalley currents (generators)
have OPE's
\bea
 H_i(z)H_j(w)&=&\frac{kG_{ij}}{(z-w)^2}\nn
 H_i(z)E_j(w)&=&\frac{A_{ij}}{z-w}E_j(w)\nn
 H_i(z)F_j(w)&=&\frac{-A_{ij}}{z-w}F_j(w)\nn
 E_i(z)F_j(w)&=&\frac{\frac{2k}{\al_i^2}\delta_{ij}}{(z-w)^2}+
  \frac{\delta_{ij}}{z-w}H_j(w)
\eea
It is wrt the Sugawara construction (originally considered in \cite{Sug, Som})
\bea
T(z)&=&\frac{1}{\theta^2(\kn+\hn)}\kappa^{ab}:J_aJ_b:(z)\nn
    &=&\frac{1}{t}:\sum_{\al>0}\frac{1}{\al^2}(E_\al F_\al+F_\al E_\al)
       +\frac{1}{2}(H,H):
\eea
that the currents have dimension (1,0). Here we have introduced the parameter
\ben
 t=\frac{\theta^2}{2}\left(\kn+\hn\right)
\een
where $\hn$ is the dual Coxeter number. The central charge is easily found to
be
\ben
 c=\frac{\kn d}{\kn+\hn}
\label{csugawara}
\een
The mode expansion reads
\ben
 J_a(z)=\sum_{n\in\Z}J_a^nz^{-n-1}
\een
In terms of the modes the algebra is expressed as
\ben
 [J_a^n,J_b^m]=\fabc J_c^{n+m}+\kappa_{ab}kn\delta^{n+m,0}
\label{AKM}
\een
For $n=m=0$ one obtains a subalgebra isomorphic to {\bf g}. It is
often denoted the horizontal algebra.
One obtains an affine Lie algebra 
(special case of a Kac-Moody algebra, see \cite{Kac, GO, BdeK, Fuc, MP}) 
by including a derivation element $D$ satisfying $[D,J_a^n]=nJ_a^n$ and
$[D,k]=0$. In view of the fact that the currents have spin one 
\ben
 [L_n,J_a^m]=-mJ_a^{n+m}
\een
we may take $D=-L_0$ whenever we consider affine Lie algebras in connection
with the Virasoro algebra. $D$ plays an important role in the study of the 
root system of the affine Lie algebra.
Precisely as the Virasoro algebra is a central extension of the Witt
algebra, (\ref{AKM}) is a central extension \cite{PS, KR87} of the loop algebra
{\bf g}$_t=$ {\bf g} $\otimes\ \C[t,t^{-1}]$, whose basis elements are written
$j_a^n=j_a\otimes t^n=j_a(n)$ and satisfies
\ben
 [j_a(n),j_b(m)]=[j_a,j_b](n+m)
\een
The restriction from $t\in\C\setminus\{0\}$ to $t\in S^1$ reveals the 
terminology of loop algebra since then {\bf g}$_t$ may be viewed as the
space of smooth mappings, loops, from $S^1$ to the Lie algebra {\bf g}.

Generalisations of the Sugawara construction is in the 
basis of recent developments, see \cite{HKOC} and references therein. It has
given insight in irrational CFT.

\subsection{Fields and States}
Having introduced the affine Lie (current) algebra we are in a situation
with an enlarged chiral algebra, being the semi-direct sum of the Virasoro
algebra (based on the Sugawara construction) and the affine Lie algebra
(with corresponding simple Lie algebra {\bf g}). By the (anti-)chiral algebra
we mean the (anti-)holomorphic part of the symmetry algebra of the theory.
It is natural to try to classify the field content not only wrt the
Virasoro algebra but wrt the whole chiral algebra. This is part of 
the foundation of the study of $W$ algebras which generally 
are (non-linear) extensions of the Virasoro algebra.
Here we confine ourselves to the case of extending by an affine current
algebra and refer to the pioneering work on $W_3$ algebra \cite{Zam} by 
Zamolodchikov, some
work on Hamiltonian reduction \cite{DS, FRRTW, BO, FeFr}, the review paper
\cite{BS}, the reprint collection \cite{BS95} and \cite{Ras}. 
The latter includes several details and unpublished
material on Hamiltonian reduction of affine $SL(3)$ current algebra and
on free field realization of $W_3$ algebra. In \cite{FZ87, FL} $W_N$ minimal
models are considered. The references mentioned are
mainly concerning bosonic extensions. However, the literature includes
a huge amount of material on fermionic extensions such as the $N=1$ 
\cite{Ram, NS, Eic, BKT, FQS85}
and $N=2$ (see e.g. \cite{FS} and references therein) superconformal algebras. 

Now, let us define a primary field (multiplet) $\phi$
wrt the current algebra by 
\ben
 J_a(z)\phi(w)=\frac{t_a}{z-w}\phi(w)
\label{JprimOPE}
\een
Here $t_a\phi$ is a shorthand notation for the action of the matrix $t_a$
on the field multiplet, $t_a\phi^i={(t_a)^i}_j\phi^j$. Here we do not specify
any further the representation that $\phi$ forms
of the Lie algebra {\bf g} generated by the horizontal subalgebra.
Similarly, the highest weight state 
\ben
 \ket{\lambda}=\lim_{z\rightarrow0}\phi(z)\ket{0}
\een
(see (\ref{primdim}) for choice of notation) 
provides a representation of $\{J_a^0\}$ satisfying
\bea
 J_a^0\ket{\lambda}&=&t_a\ket{\lambda}\nn
 J_a^n\ket{\lambda}&=&0\spa n>0
\eea
where, in addition to (\ref{vacuum}), the vacuum satisfies
\ben
 J_a^n\ket{0}=0\spa n\geq0
\een
Note that just like $T$ is not a conformal primary field, $J_a$ is not
an affine primary field.
The (affine) descendants are obtained by acting on the highest weight
$\ket{\lambda}$ with the raising operators $J_a^{-n}$, $n>0$.
One may specify the highest weight state further using the triangular
decomposition of {\bf g}
\ben
 E_\al^0\ket{\lambda}=0=J_a^n\ket{\lambda}\ \ \mbox{for}
  \ \ n>0\spa H_i^0\ket{\lambda}=
  (\al_i^\nu,\lambda)\ket{\lambda} 
\label{hwvaffine}
\een
Integrable (unitary) highest weight representations satisfy
\ben
 0\leq\lambda\cdot\theta\leq k\spa k^\nu\in\Z
\label{integrable}
\een

Let us end this section by briefly considering 
the GKO or coset construction \cite{GKO}. We have a situation with two 
Lie algebras {\bf g, h} where 
{\bf h}$\subset${\bf g}. In the corresponding affine Lie algebras we choose
the generators of {\bf h} to be the first dim{\bf h} generators among the
generators $J_a$, $a=1,...,$dim{\bf g}, of {\bf g}. Wrt the difference
$T_{g/h}=T_g-T_h$ we then have
\ben
 T_{g/h}(z)J_a(w) =0\spa a=1,...,\mbox{dim {\bf h}}
\een
and 
\ben
 c_{g/h}=c_g-c_h
\een
The level $k^\nu_h$ of {\bf h} depends on the level $k^\nu_g$ of {\bf g}
through the Dynkin index $I_{h\subset g}$ of the embedding 
{\bf h}$\subset${\bf g}, $k^\nu_h/k^\nu_g=I_{h\subset g}$ \cite{SW, BB, Fuc}.
Of particular interest are the cases 
\ben
 \mbox{{\bf g}}=(A_1)_l\oplus(A_1)_1\spa \mbox{{\bf h}}=(A_1)_{l+1}
\een
(the out-most subscripts are the integer levels) 
which produces the discrete series of central charge values (\ref{cunitary})
in the unitary minimal models
\ben
 c_{g/h}=\frac{3l}{l+2}+\frac{3}{1+2}-\frac{3(l+1)}{l+1+2}=1-\frac{6}{(
  l+2)(l+3)}
\een
It is the GKO construction that proves the existence of unitary minimal
models ($p=l+2$). Similar constructions of the $N=1$ minimal models
are also considered in \cite{GKO} while the $N=2$ minimal models
are considered in \cite{DiVPYZ, ZF, BFK, ET}. 

\subsection{The Knizhnik-Zamolodchikov Equations}
The currents $J_a$ generate the affine transformations as $T$ generates the
conformal transformations ($\eps(z)=\eps^a(z)t_a$)
\ben
 \delta_\eps F(\phi(w))=\oint\dz\eps^a(z)J_a(z)F(\phi(w))
\label{Wardaffine}
\een  
This is sometimes referred to as the affine Ward identity.
Notably, one has
\ben
 \langle J_a(z)\phi_1(w_1)...\phi_N(w_N)\rangle=\sum_{j=1}^N
  \frac{t_a}{z-w_j}\langle\phi_1(w_1)...\phi_j(w_j)...\phi_N(w_N)\rangle  
\label{Wardaffine2}
\een
The index $j$ on $\phi$ in $\phi_j$ is just a labelling.
Let us now see what we learn from performing the OPE $T\phi$ using the
Sugawara construction and then compare the result with the definition
(\ref{OPEprim}) of $\phi$ being a conformal primary field. From the double
pole we find
\ben
 \Delta(\phi)=\frac{\kappa^{ab}t_at_b}{\theta^2(\kn+\hn)}=\frac{(\lambda,
  \lambda+2\rho)}{2t}
\label{primdim}
\een
where $\lambda$ is the highest weight of the representation $t_a$, thus
$\phi$ is characterised and sometimes labelled by this weight
($\phi=\phi_\lambda$). From the single pole we find
\bea
 \pa\phi(w)&=&\frac{\kappa^{ab}t_a}{t}:J_b(w)\phi(w):\nn
 &=&\frac{\kappa^{ab}t_a}{t}\oint_w\dz\frac{J_b(z)\phi(w)}{z-w}
\eea
Inserting this in an $N$-point function,
deforming the contour into a sum of contours encircling the other fields
in the correlator and using (\ref{Wardaffine2}), we obtain
\ben
 \left(t\pa_{w_j}+\sum_{i\neq j}\frac{\kappa^{ab}t_a(\lambda_j)t_b(\lambda_i)}{
  w_i-w_j}\right)\langle\phi_1(w_1)...\phi_N(w_N)\rangle=0
\label{KZ1}
\een
for all $w_j$. Here we have specified the representations by $t_a(\lambda_i)$.
This set of equations are called the Knizhnik-Zamolodchikov
equations due to their first appearance in the seminal work \cite{KZ}. 
In terms of modes (\ref{KZ1}) expresses the decoupling of the combined
(Virasoro/affine Lie algebra) singular state
\ben
 \left(L_{-1}-\frac{1}{t}\kappa^{ab}t_aJ_b^{-1}\right)\ket{\phi}
\een
Decoupling of pure affine null vectors are studied in \cite{GW}.
In a subsequent section we will follow \cite{AY} and
use affine singular vectors to determine
the fusion rules for $SL(2)$ current algebra.

The above considerations pertain to {\bf g} being simple. For a semi-simple
Lie algebra one simply has to add appropriate terms in various places,
in particular both the Casimir in (\ref{primdim}) and the second term in the
bracket in (\ref{KZ1}) become sums of similar terms.

\subsection{WZNW Models}
A Lagrangian realization of a CFT with an affine symmetry is the 
Wess-Zumino-Novikov-Witten (WZNW) theory \cite{WZ, Nov, Wit84}.
It is a non-linear $\sigma$-model with a compact semi-simple Lie group 
manifold $G$ as target space and is based on the action
\bea
 S=k^\nu S^\pm&=&\frac{k^\nu}{4\pi}\int_\Sigma d^2zTr(g^{-1}\pa g
  g^{-1}\bar{\pa}g)\nn
 &\pm&\frac{k^\nu}{12\pi}\int_Bd^3y\epsilon^{\al\beta\g}Tr(g^{-1}\pa_\al g
  g^{-1}\pa_\beta gg^{-1}\pa_\g g)
\eea
where $k^\nu$ is the level of the corresponding affine Lie algebra, while
$\pa B=\Sigma$. In the first term $g$ is a map from the two 
dimensional manifold $\Sigma$ to $G$, $g:\Sigma
\rightarrow G$, while in the second term we use the same symbol to represent 
the extension to the three manifold $B$, $g:B\rightarrow G$. The action
is conformally invariant for the ratio given of the normalisations of the 
two terms. Different extensions of $g$ from $\Sigma$ to $B$ may give
different values of the second term in the action, but following Witten
\cite{Wit84} these amount to additional constants $2\pi k^\nu n$, $n$
integer. Thus, for the path integral weight $e^{iS}$ to be unique,
$k^\nu$ is necessarily an integer \cite{Wit84, BCZ}.
This is related to the second and third integer homology groups of $G$ being 
$H_2(G;\Z)\simeq1$ and $H_3(G;\Z)\simeq\Z$ in the cases of interest \cite{Fuc}. 
The action respects the Polyakov-Wiegman identity \cite{PW, DiVDP}
\ben
 S^+(gh)=S^+(g)+S^+(h)+\frac{1}{2\pi}\int_\Sigma d^2zTr(g^{-1}\bar{\pa}g\pa h
  h^{-1})
\een
and similarly for $S^-$, found by $S^-(g)=S^+(g^{-1})$. Exploring the
identity, the action is seen to be invariant under
\ben
 g(z,\zb)\rightarrow \Omega(z)g(z,\zb)\bar{\Omega}(\zb)
\een
giving rise to the conserved currents
\ben
 J=k^\nu g^{-1}\pa g\spa \bar{J}=k^\nu \bar{\pa}gg^{-1}\spa \bar{\pa}J=\pa
  \bar{J}=0
\een
satisfying classical versions of (\ref{JaJb}).
The models possess unitary highest weight representations since the level
is integer. The Hilbert space of a WZNW model decomposes into a direct
sum of highest weight representations \cite{GW}, and due to
$0\leq\lambda\cdot\theta^\nu\leq k^\nu$, the summation over $\lambda$ truncates
and the WZNW model based on a compact Lie group $G$ is a RCFT.

For non-compact Lie groups such as $SL(N,\R)$, one is more 'sloppy' and 
does not restrict $k^\nu$ to be integer, since Witten's argument is
no longer applicable, and the theory may be unbounded from below.
However, the final quantum theory of interest still makes
sense when gauging and constraining the WZNW theory \cite{KPSY, KS90, FRRTW}
restore the boundedness of the spectrum \cite{Gaw92, deB}.
Gauged WZNW theories appear e.g. in path integral realizations of the GKO
construction \cite{GK89, KPSY} and in the $G/G$ approach to non-critical
string theory.

\chapter{Free Field Realizations}

Since the work by Wakimoto \cite{Wak} on free field realizations of affine
$SL(2)$ current algebra much effort has been made in obtaining similar 
constructions in more general cases \cite{FF, Ku, BMP, GMMOS, Ito}. 
Many partial results are found 
in the literature but a general construction is still lacking. However, here 
we present an explicit (generalised) Wakimoto free field realization
of affine Lie algebras based on simple Lie algebras \cite{PRY3}. 
The method employed is a generalisation of and a much more detailed version 
than the one by Awata {\em et al} \cite{ATY}. First one finds differential 
operator realizations of the simple Lie algebras using Gauss decompositions, 
which we work out in details. This was carried
out to second order in $x$ in \cite{ATY} and by similar methods to all
orders in \cite{Tay}, though some essential parts are only presented there as
recursion relations. 
The framework of isotopic $x$ variables in these works is a generalisation
of the situation in $SL(2)$ \cite{FZ, FGPP, PRY1, An} to any simple group.
Secondly one quantises the realization by translating
into free fields and then adding appropriate anomalous terms to the lowering
operators \cite{FF, Ku, ATY}. 
Such terms were previously only known explicitly in the case of simple roots
\cite{Ito} and for $SL(n)$ \cite{BMP, FF}, 
but here explicit expressions for all roots are presented. The anomalous terms
take care of multiple contractions, or in other words of the normal ordering.

As shown by Dotsenko and Fateev \cite{DF} in the case of minimal models and
briefly reviewed in Chapter 2, the 
use of screening currents in constructing conformal blocks and Greens functions
is essential. In chapters 4 and 5 we shall generalise their work to 
admissible representations \cite{KW} (to be introduced in Chapter 4) of affine 
$SL(2)$ current algebras. Thereby we find use of the screening current of the 
second kind originally introduced by Bershadsky and Ooguri \cite{BO}. Here
we recapitulate the known results \cite{BMP, ATY} 
on screening currents of the first kind in the general case.
To our knowledge, a completely satisfactory proof of their existence is
lacking. We shall present some sufficient conditions which are then checked
in some details. To be able to treat admissible representations of affine
$SL(2)$ current algebra we shall find it necessary in chapters 4 and 5 to 
include also the screening currents of the second kind. Hence, we shall 
undertake a discussion of those in the general case. In the case of $SL(n)$ we 
show that they indeed exist if screening currents of the first kind exist.
We also comment on a quantum group
structure in the braiding-commutation algebra of the screening currents
following ideas of Gomez and Sierra \cite{GS1, RRR, GS2}.

Finally, one needs explicit free field realizations of primary fields.
In the framework of the $x$ variables, 
such realizations are known for $SL(2)$ \cite{FGPP, PRY1}
but for more general groups little is known. Here we initiate
the discussion and present some partial results on $SL(n)$ and the complete
solution for $SL(3)$ and review the solution for $SL(2)$.

Throughout we choose to avoid normalising the root system of the Lie algebra 
by keeping the length of the highest root a free parameter.

\section{Differential Operator Realization}

Following the idea of \cite{ATY} we here discuss \cite{PRY3}
a differential operator realization of a simple Lie algebra
{\bf g} on the polynomial ring $\C[x^\al]$, given by the following right action
\bea
 E_\al(x,\pa)\langle\lambda|{\cal Z}&=&\langle\lambda|{\cal Z}e_\al\nn
 H_i(x,\pa,\lambda)\langle\lambda|{\cal Z}&=&\langle\lambda|{\cal Z}h_i\nn
 F_\al(x,\pa,\lambda)\langle\lambda|{\cal Z}&=&\langle\lambda|{\cal Z}f_\al
\eea
where ${\cal Z}=e^{x^\al e_\al}$ and where $\langle\lambda|$ is a lowest
weight vector
\ben
 \langle\lambda|f_\al=0\spa\langle\lambda|h_i=\langle\lambda|\lambda_i\spa
 \lambda_i=\langle\lambda,h_i\rangle=(\lambda,\al_i^\nu)
\een
There is an isotopic coordinate $x^\al$ for every positive root $\al>0$
and for brevity we sometimes write $x^i=x^{\al_i}$.
Before working out the explicit form of the $\lambda$ dependent first order
differential operators $E_\al,H_i,F_\al$, let us briefly
argue that they indeed represent a realization of {\bf g}. The verification
is based on the associativity
\ben
 \langle\lambda|\left({\cal Z}e^{sj_a}\right)e^{tj_b}=\langle\lambda|{\cal Z}
  \left(e^{sj_a}e^{tj_b}\right)
\een
where a comparison of terms linear in $st$ gives the desired commutator
of the differential operators $[J_a,J_b]=\fabc J_c$. As for the affine currents
we use the common notation $J_a$ to denote the differential operators.
We hope it does not lead to any confusion.

The Gauss decomposition of $\langle\lambda|{\cal Z}e^{tj_a}$ is 
(see also (\ref{Gauss2}))
\bea
 \langle\lambda|{\cal Z}\exp(te_\al)&=&\langle\lambda|\exp\left( 
  x^\g e_\g+t\vab(x)e_\beta+{\cal O}(t^2)\right)\nn
 &=&\langle\lambda|\exp\left(t\vab(x)\db+{\cal O}(t^2)\right){\cal Z}\nn
 \langle\lambda|{\cal Z}\exp(th_i)&=&\langle\lambda|\exp\left(th_i\right)
  \exp\left(x^\g e_\g+tV_i^\beta(x)e_\beta+{\cal O}(t^2)\right)\nn
 &=&\langle\lambda|\exp\left(t\left(V_i^\beta(x)\db+\lambda_i\right)
  +{\cal O}(t^2)\right){\cal Z}\nn 
 \langle\lambda|{\cal Z}\exp(tf_\al)&=&\langle\lambda|\exp\left(
  tQ_{-\al}^{-\beta}(x)f_\beta+{\cal O}(t^2)\right)\exp\left(
  t\paj(x)h_j+{\cal O}(t^2)\right)\nn
 &\cdot&\exp\left(x^\g e_\g+t\vmab(x)e_\beta+
  {\cal O}(t^2)\right)\nn
 &=&\langle\lambda|
  \exp\left(t\left(\paj(x)\lambda_j+\vmab(x)\db\right)
  +{\cal O}(t^2)\right){\cal Z}
\label{Gauss1}
\eea
hence
\bea
 E_\al(x,\pa)&=&\vab(x)\db\nn
 H_i(x,\pa,\lambda)&=&\vib(x)\db+\lambda_i\nn
 F_\al(x,\pa,\lambda)&=&\vmab(x)\db+\paj(x)\lambda_j
\label{defVP}
\eea
Here the notation $\db=\pa_{x^\beta}$ has been introduced.
Since $E_\al(x,\pa)$ is independent of $\lambda$ it may be defined through
a Gauss decomposition alone.
The remaining part of this section is devoted to determine the polynomials
$V$, $P$ and $Q$. 

The following versions \cite{Hel, Tay} of the Campbell-Baker-Hausdorff 
(CBH) formula will be used repeatedly\\[.2 cm]
{\bf Lemma}
\bea
 e^Ae^{tB}&=&\exp\left\{A+t\sum_{p\geq 0}M_p(\mbox{ad}_A)^pB
  +{\cal O}(t^2)\right\}\nn
 e^{tB}e^A&=&\exp\left\{A+t\sum_{p\geq0}M_p(-\mbox{ad}_A)^pB
  +{\cal O}(t^2)\right\}\nn
 e^{A+tB+tC}&=&e^{tB}\exp\left\{A+tC-t\sum_{p\geq1}M_p(-\mbox{ad}_A)^pB
  +{\cal O}(t^2)\right\}\nn
 e^Ae^{tB}&=&e^{tB}\exp\left\{A+t\left[A,B\right]+{\cal O}(t^2)\right\}\nn 
 M_p&=&(-1)^p\frac{B_p}{p!}
\label{CBH}
\eea
where the coefficients $B_p$ are the Bernoulli numbers
\bea
  B(u)&=&\frac{u}{e^u-1}=\sum_{n\geq 0}\frac{B_n}{n!}u^n\nn
  B_{2m+1}&=&0\ \ \ \ \ \ \ {\rm for}\ \ m\geq 1\nn
  B_0=1\ \ ,\ \ B_1&=&-\hf\ \ ,\ \ B_2=\frac{1}{6}\ \ ,\ \ B_4=-\frac{1}{30}\
   \ ,...
\label{Ber}
\eea
{\bf Proof}\\
In order to solve for $C$ the equation of the second version 
\ben
 e^{tB}e^A=e^{A+tC+{\cal O}(t^2)}
\een
one may introduce the two-parameter group element 
\ben
 g(t,s)=e^{s(A+tC)}
\een
and consider
\ben
 \pa_s\pa_t g(t,s)|_{t=0}=Cg(0,s)+A\pa_tg(t,s)|_{t=0}
\een
Furthermore, using
\ben
 e^{-s\mbox{\scriptsize{ad}}_A}B=e^{-sA}Be^{sA}
\een
following from
\ben
 \left(-s\mbox{ad}_A\right)^NB=\sum_n\binomial{N}{n}s^N(-1)^n
  A^nBA^{N-n}
\een
which is easily shown by induction, one has
\ben
 \pa_s\left(g(0,s)^{-1}\pa_tg(t,s)|_{t=0}\right)=
  e^{-s\mbox{\scriptsize{ad}}_A}C
\een
Now consider the integral
\ben
  e^{-\mbox{\scriptsize{ad}}_A}B= 
   \int^1_0\pa_s\left(g(0,s)^{-1}\pa_tg(t,s)|_{t=0}\right)ds=
  \frac{1-e^{-\mbox{\scriptsize{ad}}_A}}{\mbox{ad}_A}C
\een
{}From this it immediately follows that
\ben
 C=\sum_{n\geq0}M_n(-\mbox{ad}_A)^nB
\een
The remaining versions of the lemma are simple rewritings of the second
version.\\
$\Box$
 
In \cite{Tay} the $V$'s are determined by an approach very similar to the
one employed in the following. However, 
here and in \cite{PRY3}
the Gauss decomposition (\ref{Gauss1}) is carried out explicitly
whereby one is able to determine also the $P$'s and $Q$'s. 
In \cite{Tay} functions similar to
the $P$'s are given by recursion relations while functions similar to the 
$Q$'s are not discussed. 
In \cite{ATY} the polynomials $V_\al^\beta$ are worked out only to second 
order in $x$ while only the Chevalley generators amongst the lowering
operators are considered and similarly worked out to second order. 
As indicated in (\ref{Gauss1}) one
does not need the entire decomposition at this point but later on we will
make use of it.\\[.2 cm]
{\bf Proposition}
\bea
 \vab(x)&=&\delta_\al^\beta
  +\sum_{n\geq1}M_nC_{\al;\beta_1,...,\beta_n}^\beta x^{\beta_1}...
  x^{\beta_n}\nn
 \vib(x)&=&-(\al_i^\nu,\beta)x^\beta\nn
 \vmab(x)&=&\sum_{n\geq 1}\sum_{l=0}^{n-m(-\al,\beta_1,...,\beta_n)}
  \frac{B_l}{l!(n-l)!}
  C_{-\al;\beta_1,...,\beta_n}^\beta x^{\beta_1}...x^{\beta_n} \nn
 \paj(x)&=&\sum_{n\geq 1}\frac{1}{n!}C_{-\al;\beta_1,...,\beta_{n}}^j
  x^{\beta_1}...x^{\beta_{n}}\nn
  Q_{-\al}^{-\beta}(x)&=&\sum_{n\geq0}\frac{1}{n!}
  C_{-\al;\beta_1,...,\beta_{n}}^{-\beta}x^{\beta_1}...x^{\beta_{n}}
\eea
where the contracted structure constants are
\bea
 C_{a;\beta_1,...,\beta_n}^b&=&{f_{\beta_1a}}^{a_1}{f_{\beta_2a_1}}^{a_2}
  ...{f_{\beta_na_{n-1}}}^b\nn
 C_{a;\beta_1,...,\beta_n}^b&=&\delta_a^b\hspace{1cm}\mbox{for}\hspace{1cm}n=0
\label{defC}
\eea
and where $m=m(-\al,\beta_1,...,\beta_n)$ is 
defined for a given sequence of roots $(-\al,\beta_1,...,\beta_n)$
as the minimum integer for which $-\al+\beta_1+...+\beta_m>0$.\\[.2 cm]
{\bf Lemma}\\
In any (formal) expansion of the form
\ben
 \sum_{s\geq0}\left(-\sum_{n\geq1}b_nx^n\right)^s=\sum_{n\geq0}a_nx^n
\een
the following recursion relation is valid \cite{PRY3}
\bea
 a_0&=&1\nn
 a_n&=&-\sum_{l=0}^{n-1}b_{n-l}a_l\spa\mbox{for}\ n>0
\eea
{\bf Proof of lemma}\\
We have
\bea
 \sum_{n\geq0}a_nx^n&=&\sum_{s\geq0}\left(-\sum_{n\geq1}b_nx^n\right)^s\nn
 &=&1+\sum_{1\leq s}(-1)^s\sum_{1\leq i_1<i_2<...<i_s}b_{i_1}b_{i_2-i_1}...
  b_{i_s-i_{s-1}}x^{i_s}\nn
 &=&1+\sum_{0\leq s}(-1)^{s+1}\sum_{0=i_0<i_1<...<i_s<l}b_{i_1}b_{i_2-i_1}...
  b_{i_s-i_{s-1}}b_{l-i_s}x^l\nn
 &=&1-\sum_{1\leq n}b_nx^n-\sum_{n\geq 2}
  \sum_{0\leq s}(-1)^{s+1}\sum_{0=i_0<i_1<...<i_s<l<n}
  b_{i_1}b_{i_2-i_1}...b_{l-i_s}b_{n-l}x^n\nn
 &=&1-b_1x-\sum_{n\geq2}\left(b_n+\sum_{l=1}^{n-1}b_{n-l}a_l\right)x^n
\eea
and the lemma is easily read off.\\
$\Box$\\
{\bf Proof of proposition}\\
Let us notice the following almost trivial relations
\bea
 \left(\mbox{ad}_{x^\beta e_\beta}\right)^nj_a&=&C^b_{a;\beta_1,...,\beta_n}
  x^{\beta_1}...x^{\beta_n}j_b\nn
 C^b_{a;\beta_1,...,\beta_n}&=&C^c_{a;\beta_1,...,\beta_m}C^b_{c;\beta_{m+1},
  ...,\beta_n}
\eea
where $j_a\in\{ e_\al,h_i,f_\al\}$. Of course one could be more general
and let $j_a\in\mbox{span}\{ e_\al,h_i,f_\al\}$ but we may concentrate on
basis elements. It follows that
\ben
 {\cal Z}e^{tj_a}=\exp\left\{x^\beta e_\beta
  +t\sum_{n\geq0}M_nC_{a;\beta_1,...,\beta_n}^bx^{\beta_1}...x^{\beta_n}j_b+
  {\cal O}(t^2)\right\}
\een
and one may read off $\vab$ in the case of $a=\al$. For $a=i$ one may 
simply use (\ref{CBH}$d$) and thereby determine $\vib$. The last and by far
the most complicated case is $a=-\al$ corresponding to a lowering operator 
$j_a=f_\al$. Using the CBH formula one obtains
\bea
 {\cal Z}e^{tf_\al}&=&\exp\left\{t\sum_{l\geq1
  ,n_l>...>n_1\geq0}M_{n_1}M_{n_2-n_1}...M_{n_l-n_{l-1}}(-1)^{n_l-n_1+l-1}
  \right.\nn
 &\cdot&\left.
  C_{-\al;\beta_1,...,\beta_{n_1}}^{-\g_1}C_{-\g_1;\beta_{n_1+1},...,
  \beta_{n_2}}^{-\g_2}...C_{-\g_{l-1};\beta_{n_{l-1}+1},...,\beta_{n_l}}^{
  -\beta}
  x^{\beta_1}...x^{\beta_{n_l}}f_\beta+{\cal O}(t^2)\right\}\nn
 &\cdot&\exp\left\{t\sum_{l\geq1
  ,n_l>...>n_1\geq0}M_{n_1}M_{n_2-n_1}...M_{n_l-n_{l-1}}(-1)^{n_l-n_1+l-1}
  \right.\nn
 &\cdot&\left.
  C_{-\al;\beta_1,...,\beta_{n_1}}^{-\g_1}C_{-\g_1;\beta_{n_1+1},...,
  \beta_{n_2}}^{-\g_2}...C_{-\g_{l-1};\beta_{n_{l-1}+1},...,\beta_{n_l}}^i
  x^{\beta_1}...x^{\beta_{n_l}}h_i+{\cal O}(t^2)\right\}\nn
 &\cdot&\exp\left\{x^\beta e_\beta+t\sum_{l\geq1
  ,n_l>...>n_1\geq0}M_{n_1}M_{n_2-n_1}...M_{n_l-n_{l-1}}(-1)^{n_l-n_1+l-1}
  \right.\nn
 &\cdot&C_{-\al;\beta_1,...,\beta_{n_1}}^{-\g_1}C_{-\g_1;\beta_{n_1+1},...,
  \beta_{n_2}}^{-\g_2}...C_{-\g_{l-2};\beta_{n_{l-2}+1},...,\beta_{n_{l-1}}}^{
  a_-}C_{a_-;\beta_{n_{l-1}+1},...,\beta_{n_l}}^\beta\nn
 &\cdot&\left.x^{\beta_1}...x^{\beta_{n_l}}e_\beta+{\cal O}(t^2)\right\}\nn
 &=&\exp\left\{t\sum_{n\geq0}
  K_{n,n}C_{-\al;\beta_1,...,\beta_{n}}^{-\beta}
  x^{\beta_1}...x^{\beta_{n}}f_\beta+{\cal O}(t^2)\right\}\nn
 &\cdot&\exp\left\{t\sum_{n\geq1}
  K_{n,n}C_{-\al;\beta_1,...,\beta_{n}}^i
  x^{\beta_1}...x^{\beta_{n}}h_i+{\cal O}(t^2)\right\}\nn
 &\cdot&\exp\left\{x^\beta e_\beta
  +t\sum_{n\geq1}\sum_{\beta_1,...,\beta_n>0}
  K_{n,m}C_{-\al;\beta_1,...,\beta_n}^\beta
  x^{\beta_1}...x^{\beta_n}e_\beta+{\cal O}
  (t^2)\right\}
\label{Gauss2}
\eea
where $a_-\in\{-\al,i\}$ and where we have introduced the coefficients
\bea
 K_{n,m}&=&M_n+\sum_{1\leq l,0\leq n_1<...<n_l<m}(-1)^{n-n_1+l}
  M_{n_1}M_{n_2-n_1}...M_{n_l-n_{l-1}}M_{n-n_l}\nn
 &=&\left\{\begin{array}{ll} \delta_{n,1}&,\ m=1\\
    (-1)^nM_{n-1}+\sum_{l\geq1,1<n_1<...<n_l<m}(-1)^{n+l}M_{n_1-1}M_{n_2-n_1}
       M_{n-n_l}&,\ m>1
           \end{array}  \right.
\label{Knm}
\eea
where $m=m(-\al,\beta_1,...,\beta_n)$.
In the contraction of the contracted structure constants 
$C_{a;\beta_1,...,\beta_n}^b$ it has been used that the summation over $c$ in
(\ref{defC}) reduces to a summation over $-\g$ when $a=-\al$ and $b=a_-$.
Furthermore, we have introduced the parameter $m=m(-\al,\beta_1,...,\beta_n)$ 
in order to perform the 
contraction over $a_-$ in the last exponential in the Gauss decomposition.
The rewriting in (\ref{Knm}) is due to 
$\left(1+(-1)^{n+1}\right)M_n=\delta_{n,1}$.
For $m>1$ one now has the generating function
\ben
 \sum_{n\geq m\geq2}K_{n,m}y^{n-m}x^n=\sum_{l>k\geq0}(-1)^{l+1}M_l
  y^kx^{l+1}\sum_{s\geq0}\left(1-B(x)\right)^s
\een
In order to simplify the expression for $K_{nm}$ we use the lemma on 
\ben
 \sum_{s\geq0}\left(1-B(x)\right)^s=\sum_{s\geq0}\left(-\sum_{n\geq1}\frac{
  B_n}{n!}x^n\right)^s=\sum_{n\geq0}A_nx^n
\label{1minusB}
\een
and find the recursion relation
\ben
 A_n=-\sum_{l=0}^{n-1}\frac{B_{n-l}}{(n-l)!}A_l\spa A_0=1
\een
The solution is easily proven by induction to be
\ben
 A_n=\frac{1}{(n+1)!}
\een
The lemma also proves itself useful in considering
\ben
 B(x)=\frac{x}{e^x-1}=
  \sum_{s\geq0}\left(-\sum_{n\geq1}\frac{x^n}{(n+1)!}\right)^s=\sum_{n\geq0}
  a_nx^n
\een
which then yields
\ben
\sum_{l=0}^{n-1}B_l\binomial{n}{l}=\left\{ \begin{array}{l}0\spa n>1\\
                                     1\spa n=1\end{array}\right.
\label{sumbin1}
\een  
Finally, from the generating function we find using (\ref{sumbin1}) that
for $m>1$ 
\ben
 K_{n,m}=\sum_{l=0}^{n-m}\frac{B_l}{l!(n-l)!}
\een
while it follows immediately from (\ref{sumbin1}) that this relation
is also valid for $m=1$. \\
$\Box$\\
In the case of a simple root $\al=\al_j$, $V_{-\al}^\beta(x)$ reduces to
\ben
 V_{-\al_j}^\beta(x)={f_{\g,-\al_j}}^\beta x^\g-\sum_{n\geq 1}(-1)^nM_n
  C_{j;\beta_1,...,\beta_n}^\beta x^{\al_j}x^{\beta_1}...x^{\beta_n}
\label{Vmajb}
\een

\subsection{Differential Screening Operators}

By a left action we may define the differential operator $S_\al$
\bea
 \exp\{-te_\al\}{\cal Z}&=&\exp\{tS_\al(x,\pa)+{\cal O}(t^2)\}{\cal Z}\nn
 S_\al(x,\pa)&=&S_\al^\beta(x)\db
\eea
It is easily seen that
\ben
 S_\al(x,\pa)=E_\al(-x,-\pa)
\een
This indicates that
\ben
 S_\al^\beta(x)=-\vab(x)+{f_{\g\al}}^\beta x^\g
\een
The term differential screening operator is justified by the use of the 
polynomials $S_\al^\beta$ in constructing screening currents later in this
chapter. From the associativity property
\ben
 \langle\lambda|\left(e^{-se_\al}{\cal Z}\right)e^{tj_a}=\langle\lambda|
  e^{-se_\al}\left({\cal Z}e^{tj_a}\right)
\een
and the Gauss decomposition (\ref{Gauss1}),(\ref{Gauss2}) one deduces
\bea
 \left[E_\al(x,\pa),S_\beta(x,\pa)\right]&=&0\nn
 \left[H_i(x,\pa,\lambda),S_\beta(x,\pa)\right]&=&(\al_i^\nu,\beta)
  S_\beta(x,\pa)\nn
 \left[F_\al(x,\pa,\lambda),S_\beta(x,\pa)\right]&=&
 \paj(x)(\al_j^\nu,\beta)S_\beta(x,\pa)+Q_{-\al}^{-\g}(x)(\delta_{\beta\g}
  (\beta^\nu,\lambda)-{f_{\beta,-\g}}^\sigma S_\sigma(x,\pa))\nn  
 \left[S_\al(x,\pa),S_\beta(x,\pa)\right]&=&{f_{\al\beta}}^\g S_\g(x,\pa)
\label{Scomm}
\eea
The last commutator follows from the associativity of $e^{-se_\al}e^{-te_\beta}
{\cal Z}$.
For $\al=\al_i$ and $\beta=\al_j$ the commutator 
$\left[F_\al(x,\pa,\lambda),S_\beta(x,\pa)\right]$ becomes
\ben
 \left[F_{\al_i}(x,\pa,\lambda),S_{\al_j}(x,\pa)\right]=
  A_{ij}x^{\al_i}S_{\al_j}(x,\pa)+\lambda_i\delta_{ij}
\een

\subsection{Classical Polynomial Identities}

{}From the fact that $E_\al,H_i$ and $F_\al$ constitute a differential 
operator realization of {\bf g} one may easily deduce several interesting
polynomial identities amongst the $V$'s and $P$'s
\bea
 (\al_i^\nu,\beta-\al)V_\al^\beta(x)&=&(\al_i^\nu,\g)x^\g\pa_\g V_\al^\beta(x)
  \nn
 (\al_i^\nu,\g+\al)V_{-\al}^\g(x)&=&
  (\al_i^\nu,\beta)x^\beta\pa_\beta V_{-\al}^\g(x)\nn
 V_\al^\g(x)\pa_\g V_\beta^\sigma(x)-V_\beta^\g(x)\pa_\g V_\al^\sigma(x)&=&
  {f_{\al\beta}}^\g V_\g^\sigma(x)\nn
 V_\al^\g(x)\pa_\g V_{-\beta}^\sigma(x)-V_{-\beta}^\g(x)\pa_\g V_\al^\sigma(x) 
 &=&{f_{\al,-\beta}}^\g V_\g^\sigma(x)+{f_{\al,-\beta}}^{-\g}V_{-\g}^\sigma(x)
  -\delta_{\al\beta}(\al^\nu,\sigma)x^\sigma\nn
 V_{-\al}^\g(x)\pa_\g V_{-\beta}^\sigma(x)-V_{-\beta}^\g(x)\pa_\g 
  V_{-\al}^\sigma(x)&=&-{f_{\al\beta}}^\g V_{-\g}^\sigma(x)\nn
 V_{\al}^\beta(x)\db\paj(x)&=&G^{ij}(\al_i^\nu,\al^\nu)\nn
 V_\al^\g(x)\pa_\g P_{-\beta}^j(x)&=&{f_{\al,-\beta}}^{-\g}P_{-\g}^j(x)\nn
 (\al_i^\nu,\beta)x^\beta\pa_\beta\paj(x)&=&(\al_i^\nu,\al)\paj(x)\nn
 V_{-\al}^\g(x)\pa_\g P_{-\beta}^j(x)-V_{-\beta}^\g(x)\pa_\g\paj(x)&=&
  -{f_{\al\beta}}^\g P_{-\g}^j(x)
\label{classV}
\eea
These we will denote
classical identities since they are based on realizations of Lie algebras. 
Additional relations determined using the Wakimoto
free field realizations (of affine current algebras)
will be denoted quantum identities. This terminology is justified by
the point of view that the Wakimoto construction corresponds to a
quantisation of the differential operator realisation.
Similarly, (\ref{Scomm}) gives the classical identities
\bea
 V_\al^\g(x)\pa_\g S_\beta^\sigma(x)-S_\beta^\g(x)\pa_\g V_\al^\sigma(x)&=&0\nn
 (\al_i^\nu,\beta-\al)S_\al^\beta(x)&=&(\al_i^\nu,\g)x^\g\pa_\g S_\al^\beta(x)
  \nn
 V_{-\al}^\g(x)\pa_\g S_\beta^\sigma(x)-S_\beta^\g(x)\pa_\g
  V_{-\al}^\sigma(x)&=&\paj(x)(\al_j^\nu,\beta)S_\beta^\sigma(x)
  -Q_{-\al}^{-\g}(x){f_{\beta,-\g}}^\mu S_\mu^\sigma(x)\nn
 S_\al^\g(x)\pa_\g S_\beta^\sigma(x)-S_\beta^\g(x)\pa_\g S_\al^\sigma(x)&=&
  {f_{\al\beta}}^\g S_\g^\sigma(x)\nn
 S_\beta^\g(x)\pa_\g \paj(x)&=&-Q_{-\al}^{-\beta}(x)
  (\al_i^\nu,\beta^\nu)G^{ij}
\label{classS}
\eea

\section{Wakimoto Free Field Realization}

The free field realization \cite{FF, Ku, ATY}
is obtained from the differential operator realization by the substitution
\ben
 \pa_\al\rightarrow\beta_\al(z)\spa x^\al\rightarrow\g^\al(z)
  \spa\lambda_i\rightarrow\kvt\pa\varphi_i(z)
\een
and a subsequent renormalisation by adding an anomalous term, $F_\al^a(\g(z),
\pa\g(z))$, to the lowering part. This is summarized in\\
{\bf Proposition}
\bea
 E_\al(z)&=&:\vab(\g(z))\beta_\beta(z):\nn
 H_i(z)&=&:\vib(\g(z))\beta_\beta(z):+\kvt\pa\varphi_i(z)\nn
 F_\al(z)&=&:\vmab(\g(z))\beta_\beta(z):+\kvt\pa\varphi_j(z)\paj(\g(z))
  +F_\al^a(\g(z),\pa\g(z))\nn
 \Delta(J_a)&=&1
\label{Wakimoto}
\eea
where the anomalous part for a simple root is (this result was originally
found in \cite{Ito})
\bea
 F_{\al_i}^a(\g(z),\pa\g(z))&=&
  \left(\frac{2k}{\al_i^2}+\frac{\hn\theta^2/2-\al_i^2}{
  \al_i^2}\right)\pa\g^{\al_i}(z)\nn
 &=&\left(\frac{k+t}{\al_i^2}-1\right)
  \pa\g^{\al_i}(z)
\eea
while for a non-simple root, $\al=\al_{i_1}+...+\al_{i_n}\ ,\ n>1$, the 
anomalous part \cite{PRY3} is
\bea
 &&C_{\al_{i_1};\al_{i_2},...,\al_{i_n}}^{\al_{i_1}+...+\al_{i_n}}
  F^a_{\al_{i_1}+...+\al_{i_n}}(\g(z),\pa\g(z))\nn
 &=&(-1)^{n-1}D_nD_{n-1}...D_3\left(\left(\frac{k+t}{\al_{i_1}^2}-1\right)
  \pa\g^\sigma(z)\pa_\sigma
  V_{-\al_{i_2}}^{\al_{i_1}}(\g(z))\right)\nn
 &+&\sum_{m=1}^{n-1}(-1)^{n-m-1}     
  C_{\al_{i_1};\al_{i_2},...,\al_{i_m}}^{\al_{i_1}+...+\al_{i_m}}
  D_nD_{n-1}...D_{m+2}\nn
 &\cdot&\left(\pa\g^\sigma(z)
  \pa_\sigma\pa_\g V_{-\al_{i_{m+1}}}^\mu(\g(z))\pa_\mu 
  V_{-(\al_{i_1}+...+\al_{i_m})}^\g(\g(z))\right.\nn
 &-&\left. tG_{i_{m+1}l}\pa\g^{\al_{i_{m+1}}}(z)
  P^l_{-(\al_{i_1}+...+\al_{i_m})}
  (\g(z))\right)
\label{anomal}
\eea
Here we have introduced the differential operators
\ben
 D_m=V_{-\al_{i_m}}^\g(\g(z))\pa_\g+\pa\g^\sigma(z)\pa_\sigma V_{-\al_{i_m}}
  (\g(z))\pa_{\pa\g^\g}
\een
and defined the product $D_nD_{n-1}...D_{n+1}=1$, in the sense that 
$D_nD_{n-1}...D_3=1$ if $n=2$, and $D_nD_{n-1}...D_{m+2}=1$ for $m=n-1$. 
We recall (\ref{defC}) that for $m=1$ we have
$C_{\al_{i_1};\al_{i_2},...,\al_{i_m}}^{\al_{i_1}+...+\al_{i_m}}=
\delta_{\al_{i_1}}^{\al_{i_1}}=1$.
The anomalous part
$F_\al^a(\g(z),\pa\g(z))$
is seen to be linear in $\pa\g$. Indeed, this is necessary for $F_\al$ to be of
weight one.\\[.2 cm]
{\bf Lemma}
\bea
 &&:V_a^\mu(\g(z))\beta_\mu(z)::V_b^\nu(\g(w))\beta_\nu(w):\nn
 &=&\frac{1}{z-w}:
  (V_a^\nu\pa_\nu V_b^\mu-V_b^\nu\pa_\nu V_a^\mu)(\g(w))\beta_\mu(w):\nn
 &+&\frac{-1}{(z-w)^2}\left\{\pa_\mu V_a^\nu\pa_\nu V_b^\mu(\g(w))
  +(z-w)\pa\g^{\rho}(w)\pa_{\rho}\pa_\mu V_a^\nu\pa_\nu V_b^\mu(\g(w))\right\}
\eea
The proof of this is obvious.\\[.2 cm]
{\bf Proof of proposition}\\
The strategy is straightforward, namely one performs the OPE's
$J_a(z)J_b(w)$ and $T(z)J_a(w)$ using the lemma after which a comparison
with (\ref{JaJb}),
using the classical polynomial identities (\ref{classV}), reduces the problem
to the set of quantum polynomial identities of the next section. 
One of those is a recursion relation determining the anomalous part
(\ref{anomal}) from $F_{\al_i}^a$ 
while the remaining ones may be seen as consistency 
conditions on the solution (\ref{anomal}). 
The verification of the expression in the case of a simple root,
$F_{\al_i}^a$, is a fairly simple matter. Since it is known \cite{FF, Ku, ATY}
that the Wakimoto realization (\ref{Wakimoto}) exists and the anomalous part
is of the form found, $F_\al^a(\g,\pa\g)=G_{\al\beta}^a(\g)\pa\g^\beta$, we
may view the consistency conditions merely as interesting quantum 
polynomial identities. Referring to the next section for details on the
solution of the recursion relation, this completes the proof.\\
$\Box$\\
Let us emphasise that the new result in \cite{PRY3}
over \cite{FF, BMP, GMMOS, Ku, ATY} is the general and explicit
expression (\ref{anomal}) for the anomalous part of the lowering operator
in the Wakimoto realization (\ref{Wakimoto}), in addition to the explicit
expressions for the polynomials $V$ and $P$ (and $Q$). As earlier stated, 
these were already worked out to some extend in \cite{ATY, Tay}. 

\subsection{Quantum Polynomial Identities}

The only non-trivial quantum polynomial identities, besides the already
found classical identities,  arise when performing
OPE's involving a lowering operator $F_\beta$. We find
\bea
 \frac{2k}{\al^2}\delta_{\al,\beta}&=&
  -\pa_\sigma V_\al^\g\pa_\g V_{-\beta}^\sigma+V_\al^\g\pa_{\pa\g^\g}F_\beta^a
  \nn
 {f_{\al,-\beta}}^{-\g}F_{\g}^a&=&
  -\pa\g^\sigma\pa_\sigma\pa_\mu V_\al^\g\pa_\g V_{-\beta}^\mu+V_\al^\g\pa_\g
  F_\beta^a +\pa\g^\sigma\pa_\sigma V_\al^\g\pa_{\pa\g^\g}F_\beta^a\nn
 0&=&(\al_i^\nu,\sigma)\pa_\sigma V_{-\beta}^\sigma-(\al_i^\nu,\al)\g^\al
  \pa_{\pa\g^\al}F_\beta^a+tG_{ij}P_{-\beta}^j\nn
 (\al_i^\nu,\beta)F_\beta^a&=&
  (\al_i^\nu,\al)\g^\al\pa_\al F_\beta^a+(\al_i^\nu,\al)\pa\g^\al
  \pa_{\pa\g^\al}F_\beta^a\nn
 0&=&2(\rho,\al_j^\nu)P_{-\al}^j+\pa_\g V_{-\al}^\g\nn
 F_\al^a&=&\pa\g^\g\pa_{\pa\g^\g}F_\al^a\nn
 \pa_\g V_{-\al}^\sigma\pa_\sigma V_{-\beta}^\g&=&
  tG_{ij}P_{-\al}^iP_{-\beta}^j
  +V_{-\al}^\g\pa_{\pa\g^\g}F_\beta^a+V_{-\beta}^\g
  \pa_{\pa\g^\g}F_\al^a\nn
 {f_{\al\beta}}^\g F_\g^a&=&\pa\g^\sigma\pa_\sigma\pa_\g 
  V_{-\al}^\mu\pa_\mu V_{-\beta}^\g-V_{-\al}^\g\pa_\g 
  F_\beta^a+V_{-\beta}^\g\pa_\g F_\al^a\nn
 &-&tG_{ij}\pa\g^\sigma\pa_\sigma P_{-\al}^iP_{-\beta}^j
  -\pa\g^\sigma\pa_\sigma V_{-\al}^\g
  \pa_{\pa\g^\g}F_\beta^a-V_{-\beta}^\g\pa\g^\sigma\pa_\sigma
  \pa_{\pa\g^\g}F_\al^a
\label{quanpol}
\eea
from the OPE's $EF,HF,TF,FF$, two identities from each.
Not all the identities are independent, e.g. the second to last one follows
from the last. The last one is the aforementioned recursion relation
determining the anomalous part for a non-simple root. The proof of 
(\ref{anomal}) goes as follows. In the recursion relation we 
expand $\g=\al_{i_1}+...+\al_{i_n}$ in simple roots and let 
$\al=\al_{i_n}$, whereby the two terms including $F_\al^a$ vanish. We then
find
\bea
 {f_{\al_{i_n},\al_{i_1}+...+\al_{i_{n-1}}}}^{\al_{i_1}+...+\al_{i_n}}
  F_{\al_{i_1}+...+\al_{i_n}}^a
 &=&\pa\g^\sigma\pa_\sigma\pa_\g V_{-\al_{i_n}}^\mu
  \pa_\mu V_{-(\al_{i_1}+...+\al_{i_{n-1}})}^\g\nn
 &-&tG_{i_nj}\pa\g^{\al_{i_n}}P_{-(\al_{i_1}+...+\al_{i_{n-1}})}^j-D_n
  F_{\al_{i_1}+...+\al_{i_{n-1}}}^a
\eea
Now the procedure is to replace the anomalous part of RHS using the recursion
relation as above letting $\al$ be the 'last' simple root. In the final
step where $\g=\al_{i_1}+\al_{i_2}$ we encounter 
\ben
 D_2F_{\al_{i_1}}^a=\left(\frac{k+t}{\al_{i_1}^2}-1\right)  
  \pa\g^\sigma(z)\pa_\sigma V_{-\al_{i_2}}^{\al_{i_1}}(\g(z))
\een
explaining the origin of the first term in (\ref{anomal}).

\section{Screening Currents}

\subsection{Screening Currents of the First Kind}

The screening currents of the first kind are \cite{BMP, ATY}
\ben
 s_j(z)=:S_{\al_j}^\al
  (\g(z))\beta_\al(z)e^{-\frac{1}{\kvt}\al_j\cdot\varphi(z)}:
\label{sj}
\een
and they satisfy the OPE's\\[.2 cm]
{\bf Proposition}
\bea
  E_\al(z)s_j(w)&=&0\nn
  H_i(z)s_j(w)&=&0\nn
  F_{\al_i}(z)s_j(w)&=&-\frac{2t}{\al_j^2}\delta_{ij}\frac{\pa}{\pa w}\left(
    \frac{1}{z-w}:e^{-\frac{1}{\kvt}\al_j\cdot\var(w)}:\right)\nn
  T(z)s_j(w)&=&\frac{\pa}{\pa w}\left(\frac{1}{z-w}s_j(w)\right)
\eea
The last OPE expresses that the screening currents have conformal weights
$\Delta(s_j)=1$. In the formalism presented here the proof of the proposition 
is a matter of direct verification using the classical polynomial 
identities (\ref{classS}). In the case of $F_\al$ for $\al$ a non-simple
root it has not yet been proven that the singular part of 
$F_\al(z)s_j(w)$ is a total derivative. However, if we assume that there
exists a polynomial $A(\g(w))$ such that
\ben
 F_\al(z)s_j(w)=\frac{\pa}{\pa w}\left(\frac{1}{z-w}A(\g(w))
  :e^{-\frac{1}{\kvt}\al_j\cdot\var(w)}:\right)
\een
we find that a necessary condition is
\ben
 F_\al(z)s_j(w)=-\frac{2t}{\al_j^2}\frac{\pa}{\pa w}\left(
    \frac{1}{z-w}Q_{-\al}^{-\al_j}(\g(w))
  :e^{-\frac{1}{\kvt}\al_j\cdot\var(w)}:\right)
\label{Falsj}
\een
In the process we derive as sufficient conditions the following two relations
\bea
 \pa_\g V_{-\al}^\beta\pa_\beta S_{\al_j}^\g&=&
  2\left(1-\frac{t}{\al_j^2}\right)
  S_{\al_j}^\beta\pa_\beta P_{-\al}^j+S_{\al_j}^\g\pa_{\pa\g^\g}F_\al^a\nn
 S_{\al_j}^\beta\pa_\beta F_\al^a&=&\pa\g^\sigma\pa_\g V_{-\al}^\beta
  \pa_\beta\pa_\sigma S_{\al_j}^\g-A_{ij}\pa\g^\sigma\pa_\sigma S_{\al_j}^\beta
  \pa_\beta P_{-\al}^i
 -\pa\g^\sigma\pa_\sigma S_{\al_j}^\g\pa_{\pa\g^\g}F_\al^a
\label{sufffirst}
\eea
They have been checked explicitly for $\al=\al_{i_1}+\al_{i_2}$ \cite{PRY3}.
Furthermore \cite{PRY3}, it has been checked that for 
$\al=\beta_1+\beta_2$ ($\beta_i$ is
not necessarily a simple root) the first identity follows from the recursion
relation (\ref{quanpol}) and the assumption that (\ref{sufffirst}) is
satisfied for $\beta_1$ and $\beta_2$. However, the very final step in the
proof is still missing. The minimal
conclusion is that the first identity is satisfied for $\al$ being a sum
of 3 simple roots while the second identity is satisfied for $\al$ being
a sum of 2 simple roots.  
In the case of $SL(3)$ the highest root is $\theta=\al_1+\al_2$ so in that
case (\ref{sj}) and (\ref{Falsj}) are seen to be true. 

\subsection{Screening Currents of the Second Kind}

The only well-known screening current of the second kind
is the one by Bershadsky and Ooguri for $SL(2)$ \cite{BO, PRY1}.
In \cite{PRY3} a generalisation to $SL(n)$ and possibly to all simple groups
(see also \cite{Ito}) is discussed. The construction of 
screening currents of the second kind $\tilde{s}_j$ is summarised in\\[.2 cm]
{\bf Proposition}
\ben
   \tilde{s}_j(w)=:\left(S_{\al_j}^{\beta}(\g(w))\beta_{\beta}(w)
   e^{-\frac{1}{\kvt}\al_j\cdot\varphi(w)} \right)^{-\frac{2t}{\al_j^2}}:
\label{sjt1}
\een
\bea
  E_\al(z)\tilde{s}_j(w)&=&0\nn
  H_i(z)\tilde{s}_j(w)&=&0\nn
   F_{\al_i}(z)\tilde{s}_j(w)&=&-\frac{2t}{\al_j^2}
   \delta_{ij}\frac{\pa}{\pa w}\left(
  \frac{1}{z-w}:\left(S_{\al_j}^{\beta}(\g(w))\bb(w)\right)^{
   -\frac{2t}{\al_j^2}-1}
   e^{\kvt\al_j^\nu\cdot\var(w)}:\right)\nn
  T(z)\tilde{s}_j(w)&=&\frac{\pa}{\pa w}\left(\frac{1}{z-w}\tilde{s}_j(w)
  \right)
\label{sjt2}
\eea
Furthermore
\ben
 F_\al(z)\tilde{s}_j(w)=-\frac{2t}{\al_j^2}\frac{\pa}{\pa w}\left(\frac{1}{z-w}
  :Q_{-\al}^{-\al_j}(\g(w))
  \left(S_{\al_j}^{\beta}(\g(w))\bb(w)\right)^{-\frac{2t}{\al_j^2}-1}
   e^{\kvt\al_j^\nu\cdot\var(w)}: \right) 
\label{Fasjt}
\een
{\bf Proof}\\
It is straightforward to check that $H_i(z)\tilde{s}_j(w)=0$ and $\Delta(
\tilde{s}_j)=1$. Let us introduce the shorthand notation
\ben
 S_j^u(z)=:\left(S_{\al_j}^\beta(\g(z))\bb(z)\right)^u:
\een
Using the explicit expressions for $V_\al^\beta$ and $S_{\al_j}^\g$, we find
\bea
 E_\al(z)S_j^u(w)&=&\sum_{l\geq1}\frac{1}{(z-w)^l}(-1)^l\binomial{u}{l}
  \left(:\pa_{\g_1}...\pa_{\g_l}\vab\bb(z)S_{\al_j}^{\g_1}...S_{\al_j}^{\g_l}
  S_j^{u-l}:\right.\nn
  &-&l\left.:\pa_{\g_1}...\pa_{\g_{l-1}}\vab(z)S_{\al_j}^{\g_1}...S_{\al_j}^{
  \g_{l-1}}
  \db S_{\al_j}^{\g_l}\beta_{\g_l}S_j^{u-l}:\right)
\eea
and in the case of $SL(n)$ where a simple root (here $\al_j$) appears at most
once in the decomposition of a root, all terms for $l>1$ vanish.
Hence, in that case we find
\bea
 E_\al(z)S_j^u(w)&=&\frac{u}{z-w}:\left(V_\al^\g\pa_\g S_{\al_j}^{\beta}
  -S_{\al_j}^{\g}\pa_\g\vab\right)\bb S_j^{u-1}:\nn
 &=&0
\eea
Similarly for the lowering operator we find in the case of $SL(n)$
\bea
 :V_{-\al_i}^\beta(\g(z))\bb(z):S_j^u(w)&=&\frac{uA_{ij}}{z-w}:\g^{\al_i}
  S_j^u(w):-\frac{u\delta_{ij}}{(z-w)^2}\frac{\hn\theta^2/2-\al_i^2}{\al_i^2}
  S_j^{u-1}(w)\nn
 &+&\frac{u(u-1)}{(z-w)^2}\left(\hf
  :\pa_{\g_1}\pa_{\g_2}V_{-\al_i}^\beta(z)\bb(z)
  S_{\al_j}^{\g_1}(w)S_j^{\g_2}(w)S_j^{u-2}(w):\right.\nn
 &-&\left.\delta_{ij}
  :\pa_{\g_1}V_{-\al_i}^\beta(z)S_{\al_j}^{\g_1}(w)
  \pa_\beta S_{\al_j}^{\g_2}(w)\beta_{\g_2}(w)S_j^{u-2}(w):\right)
\eea
Including the more simple contributions from the $P_{-\al_i}^l$ and 
$F_{\al_i}^a$ parts we find
\bea
 F_{\al_i}(z)\tilde{s}_j(w)&=&\frac{u}{(z-w)^2}:\left( (2-u)\delta_{ij} 
  S_j^{u-1}+\frac{u-1}{2}S_{\al_j}^{\g_1}
  S_{\al_j}^{\g_2}\pa_{\g_1}\pa_{\g_2}V_{-\al_i}^\beta\bb S_j^{u-2}\right.\nn
 &-&\left.(u-1)\delta_{ij} S_{\al_j}^{\g_1}
   \pa_{\g_1}V_{-\al_i}^\beta\db S_{\al_j}^{\g_2}\beta_{\g_2}S_j^{u-2}\right)
   e^{\frac{1}{\kvt}\lambda\cdot\var(w)}:\nn
 &+&\frac{u}{z-w}:\left(\kvt\delta_{ij}\pa\var_i S_j^{u-1}
  +\frac{u-1}{2}\pa\g^{\al}
  \pa_\al
  \pa_{\g_1}\pa_{\g_2}V_{-\al_j}^\beta\bb S_{\al_j}^{\g_1}S_{\al_j}^{\g_2}
  S_j^{u-2}\right.\nn
 &+&\frac{u-1}{2}S_{\al_j}^{\g_1}S_{\al_j}^{\g_2}\pa_{\g_1}\pa_{\g_2}
  V_{-\al_i}^\beta\pa\bb 
   S_j^{u-2}\nn
 &-&\left.(u-1)\delta_{ij}\pa\g^{\al}\pa_\al\pa_{\g_1}V_{-\al_I}^\beta
  S_{\al_j}^{\g_1}\db S_{\al_j}^{\g_2}
   \beta_{\g_2}S_j^{u-2}\right)e^{\frac{1}{\kvt}\lambda\cdot\var(w)}:
\eea
where $u=-2t/\al_j^2$ and $\lambda=t\al_j^\nu$. A comparison
with (\ref{sjt2}) yields the consistency conditions
\bea
 \delta_{ij}S_{\al_j}^\beta&=&
 \hf S_{\al_j}^{\g_1}S_{\al_j}^{\g_2}\pa_{\g_1}\pa_{\g_2}V_{-\al_i}^\beta
  -\delta_{ij}S_{\al_j}^{\g_1}\pa_{\g_1}V_{-\al_i}^{\g_2}\pa_{\g_2}
  S_{\al_j}^\beta\nn
 \delta_{ij}\pa_\al S_{\al_j}^\beta&=&
 \hf\pa_\al\pa_{\g_1}\pa_{\g_2}V_{-\al_i}^\beta S_{\al_j}^{\g_1}
  S_{\al_j}^{\g_2}-\delta_{ij}\pa_\al\pa_{\g_1}V_{-\al_i}^{\g_2}S_{\al_j}^{
  \g_1}\pa_{\g_2}S_{\al_j}^\beta\nn
 \delta_{ij}S_{\al_j}^\beta&=& 
  \hf S_{\al_j}^{\g_1}S_{\al_j}^{\g_2}\pa_{\g_1}\pa_{\g_2}V_{-\al_i}^\beta
\eea
which are checked using the classical polynomial identities.
We may conclude that (\ref{sjt1}) and (\ref{sjt2}) work for $SL(n)$. 
For the more general statement in (\ref{Fasjt}) to be satisfied 
in the case of $SL(n)$, the following relations are sufficient conditions
\bea
 Q_{-\al}^{-\al_j}S_{\al_j}^\beta&=&\hf S_{\al_j}^{\g_1}S_{\al_j}^{\g_2}
  \pa_{\g_1}\pa_{\g_2}V_{-\al}^\beta
  -S_{\al_j}^\sigma\pa_\sigma V_{-\sigma}^\g
  \pa_\g S_{\al_j}^\beta\nn
 S_{\al_j}^\g\pa_\g F_\al^aS_{\al_j}^\beta&=&(u-1)\pa\g^\sigma\left(
  \hf S_{\al_j}^{\g_1}
  S_{\al_j}^{\g_2}\pa_\sigma\pa_{\g_1}\pa_{\g_2}V_{-\al}^\beta
  -S_{\al_j}^{\g_1}\pa_\sigma\pa_{\g_1}V_{-\al}^{\g_2}\pa_{\g_2}
  S_{\al_j}^\beta-Q_{-\al}^{-\al_j}\pa_\sigma S_{\al_j}^\beta\right) \nn
 &+&\pa\g^\sigma\left(S_{\al_j}^\g\pa_\sigma\pa_{\pa\g^\g}F_\al^a-
  \pa_\sigma Q_{-\al}^{-\al_j}
  -\pa_\sigma\pa_\g V_{-\al}^\mu\pa_\mu S_{\al_j}^\g-S_{\al_j}^\g
  \pa_\g\pa_\sigma P_{-\al}^i\lambda_i\right)S_{\al_j}^\beta\nn
 Q_{-\al}^{-\al_j}S_{\al_j}^\beta&=&\hf S_{\al_j}^{\g_1}S_{\al_j}^{\g_2}
  \pa_{\g_1}\pa_{\g_2}V_{-\al}^\beta
\label{suffsecond}
\eea
One can verify these conditions using the classical polynomial identities
together with the consistency conditions (\ref{sufffirst}). 
Hence, we conclude that in the case of $SL(n)$ the screening currents 
of the second kind (\ref{sjt2}) exist and are well-defined 
if the screening currents of the first kind (\ref{sj}) exist and are 
well-defined. Referring to the discussion of $\al=\al_{i_1}+\al_{i_2}$
in (\ref{sufffirst}) 
we find that both types of screening currents exist in the case of $SL(3)$. \\
$\Box$

\section{Primary Fields}

The primary field $\phi(w,x)$ is defined by
\bea
 J_a(z)\phi(w,x)&=&\frac{1}{z-w}\tilde{J}_a(x)\phi(w,x)\nn
 T(z)\phi(w,x)&=&\frac{\Delta(\phi)}{(z-w)^2}\phi(w,x)+\frac{1}{z-w}\pa\phi(
  w,x)
\label{primdef}
\eea
where 
\ben
 \tilde{J}_a=\sigma(J_a)
\een
and $\sigma$ is the involution
\bea
 \sigma(E_\al)&=&F_\al\nn
 \sigma(H_i)&=&H_i\nn
 \sigma(F_\al)&=&E_\al
\eea
$\Delta(\phi)$ is the conformal dimension of $\phi$.

In the simple case of $SL(2)$ a primary field of spin $j$, $\lambda=2j$, may
be written \cite{FGPP, PRY1}
\ben
 \phi(z,x)=(1+x\g(z))^\lambda V_\lambda(z)
\een
and has conformal dimension $\Delta(\phi)=\frac{1}{2t}(\lambda,\lambda+2\rho)$.
This is easily verified using
\bea
 \tilde{E}(x)&=&-xx\pa+\lambda x\nn
 \tilde{H}(x)&=&-2x\pa+\lambda\nn
 \tilde{F}(x)&=&\pa\nn
 E(z)&=&\beta(z)\nn
 H(z)&=&-2:\g\beta(z):+\kvt\pa\var(z)\nn
 F(z)&=&-:\g\g\beta(z):+\kvt\pa\var(z)\g(z)+\kn\pa\g(z)
\eea

In general the primary field is labelled by and depends on a highest weight
$\lambda$ (generalising the notion of spin) in such a way that for $x=0$
it becomes a highest weight field of weight $\lambda$, annihilated by the
raising operators $E_\al$. Highest weight fields are exactly the vertex
operators $V_\lambda$ and since the conformal dimension of the primary
field should remain unchanged when $x\neq0$ one should expect
\ben
 \phi(w,x)=\phi'(\g(w),x)V_\lambda(w)
\label{primans}
\een
Indeed such a field is conformally primary and has conformal dimension
$\Delta(\phi)=\frac{1}{2t}(\lambda,\lambda+2\rho)$. In order to comply with
(\ref{primdef}) for $J_a=H_i$, we find that $\phi'$ is symmetric in $\g(w)$
and $x$. Due to the fact that effectively one can forget the anomalous part
when contracting with $\phi'$, $F_\al^a\phi'\sim0$, it is then sufficient
to consider the cases $J_a=E_\al$, and we obtain the following sufficient 
conditions, one for each $\al>0$
\ben
 V_\al^\beta(\g)\pa_{\g^\beta}\phi'=V_{-\al}^\beta(x)\pa_{x^\beta}\phi'+P_{
  -\al}^j(x)\lambda_j\phi'
\label{primsuff}
\een
One can say even more. By induction in addition of positive roots one can
use the classical recursion relations (\ref{classV}) in $V_\al^\beta$, 
$V_{-\al}^\beta$ and $P_{-\al}^j$ to demonstrate that if $\phi$ is a
primary field wrt $E_\al$ and $E_\beta$, then it is a primary field wrt
${f_{\al\beta}}^\g E_\g$. This means that there are only $r$ sufficient
conditions a primary field (\ref{primans}) must satisfy:
\ben
 V_{\al_i}^\beta(\g)\pa_{\g^\beta}\phi'=V_{-\al_i}^\beta(x)\pa_{x^\beta}\phi'
  +x^{\al_i}\lambda_i\phi'
\label{primeq}
\een 
Only in a few cases, to be discussed in the following, solutions have been 
found \cite{PRY3}, though we believe that solutions exist
in the general case. 

\subsection{Case of $SL(3)$}

In the case of $SL(3)$ the primary field $\phi$ is given by\\[.2 cm]
{\bf Proposition}
\bea
 \phi(z,x)&=&\prod_{j=1}^2\left(1+\sum_{\al>0}P_{-\al}^j(x)P_{-\al}^j(\g(z))
  \right)^{\lambda_j}V_\lambda(z)\nn
 &=&\left(1+x^1\g^1(z)+(\hf x^1x^2+x^{12})(\hf\g^1(z)\g^2(z)+\g^{12}
   (z))
  \right)^{\lambda_1}\nn
 &\cdot&\left(1+x^2\g^2(z)+(\hf x^1x^2-x^{12})(\hf\g^1(z)\g^2(z)-\g^{12}(z))
   \right)^{\lambda_2}V_{\lambda}(z)
\label{prim3}
\eea
and has conformal dimension $\Delta(\phi)=\frac{1}{2t}(\lambda,\lambda+2\rho)$.
This can be checked using the relevant polynomials or the proposition
(\ref{propsln}) of the next section. Below we have listed
all the polynomials in the case of $SL(3)$ found by reducing the general
expressions of the preceding sections:
\bea
 V_{\al_1}^\beta(x)&=&\delta_{\al_1}^{\beta}-\hf x^2\dtb\nn
 V_{\al_2}^\beta(x)&=&\delta_{\al_2}^{\beta}+\hf x^1\dtb\nn
 V_1^{\beta}(x)&=&-2x^1\delta_{\al_1}^{\beta}+x^2\delta_{\al_2}^{\beta}-x^{12}
   \dtb\nn
 V_2^{\beta}(x)&=&x^1\delta_{\al_1}^{\beta}-2x^2\delta_{\al_2}^{\beta}
   -x^{12}\dtb\nn
 V_{-\al_1}^\beta(x)&=&-x^1x^1\delta_{\al_1}^{\beta}+\left(\hf x^1x^2-x^{12}
  \right)\delta_{\al_2}^{\beta}
   -\hf x^1\left(\hf x^1x^2+x^{12}\right)\dtb\nn
 V_{-\al_2}^\beta(x)&=&\left(\hf x^1x^2+x^{12}\right)
  \delta_{\al_1}^{\beta}-x^2x^2\delta_{\al_2}^{\beta}
  +\hf x^2\left(\hf x^1x^2-x^{12}\right)\dtb\nn
 V_{-\theta}^{\beta}(x)&=&-x^1\left(\hf x^1x^2+x^{12}\right)
  \delta_{\al_1}^{\beta}+x^2\left(\hf x^1x^2-x^{12}\right)
  \delta_{\al_2}^{\beta}-(\frac{1}{4}x^1x^1x^2x^2+x^{12}x^{12})
  \delta_{\theta}^{\beta}\nn
 P_{-\theta}^j(x)&=&\left(\hf x^1x^2+x^{12}\right)\delta_1^j
  +\left(-\hf x^1x^2+x^{12}\right)\delta_2^j\nn
 Q_{-\theta}^{-\beta}(x)&=&x^2\delta_{\al_1}^\beta-x^1\delta_{\al_2}^\beta
  +\delta_{\theta}^{\beta}\nn
 F_{\theta}^a(\g,\pa\g)&=&\hf\left(k^\nu+1\right)\left(\pa\g^1\g^2-\g^1\pa\g^2
  \right)+k^\nu\pa\g^{12}\nn
 S_{\al_1}^{\beta}(x)&=&-\delta_{\al_1}^{\beta}-\hf x^2\dtb\nn
 S_{\al_2}^{\beta}(x)&=&-\delta_{\al_2}^{\beta}+\hf x^1\dtb 
\eea
in addition to the trivial ones $V_\theta^\beta(x)=\dtb$, 
$S_\theta^\beta=-\dtb$, $P_{-\al_i}^j(x)=x^j\delta_i^j$, 
$Q_{-\al}^{-\al}(x)=1$ and $F_{\al_i}^a(\g,\pa\g)=(k^\nu+1/2)\pa\g^i$.

\subsection{Case of $SL(r+1\geq4)$}

Let us summarise some facts on the simply-laced Lie algebras $sl(r+1)=A_r$.
The positive roots are written
\ben
 \al_{ij}=\al_i+...+\al_j\spa1\leq i\leq j\leq r
\een
where $r$ is the rank. The dual Coxeter number is given by $\hn=r+1$ while
the Weyl vector can be written
\ben
 \rho=\hf\sum_{i=1}^ri(r+1-i)\al_i
\een
The structure coefficients are found using 
\ben
 [e_{\al_{ij}},e_{\al_{kl}}]=\delta_{k-j,1}e_{\al_{il}}-\delta_{i-l,1}e_{
  \al_{kj}}
\een
and the standard symmetries such as ${f_{\al,\beta}}^{\al+\beta}={f_{\beta,
-\al-\beta}}^{-\al}$, valid for any simply-laced algebra. 
The corresponding general relation
\ben
 \frac{{f_{\al,\beta}}^{\al+\beta}}{(\al+\beta)^2}=\frac{
  {f_{\beta,-\al-\beta}}^{-\al}}{\al^2}=\frac{{f_{-\al-\beta,\al}}^{-\beta}}{
  \beta^2}
\een
valid for all simple groups, follow from the Jacobi identities.
We find the following property of the Cartan matrix for $SL(n)$ very useful
\ben
 (\al_i^\nu,\al_{kl})=0\spa k<i<l
\label{CartanAr}
\een
In this section we use the shorthand notations
\ben
 D_\al=\pa_{\g^\al}\spa\pa_\al=\pa_{x^\al}\spa \pa_{ij}=\pa_{x^{ij}}
\een

A natural generalisation of the primary fields of $SL(2)$ and $SL(3)$ is
\ben
\phi(w,x)\sim\prod_{j=1}^r\left(R^j(\g(w),x)\right)^{\lambda_j}V_\lambda(w)
\label{nonprim}
\een
where $R^j(\g,x)$ is a polynomial in $\g$ and $x$.
We find that (\ref{primeq}) reduces to
\ben
  V_{\al_i}^\beta(\g)D_\beta R^j(\g,x)=V_{-\al_i}^\beta(x)\pa_\beta
   R^j(\g,x)+x^j
   \delta_i^jR^j(\g,x)
\label{nonprimeq}
\een
However, as it is stated in \cite{PRY3}, this is {\em not} a primary 
field. It is for the non-existence of solutions in the cases 
$2\leq j\leq r-1$ that we will now argue using that
$R^j(\g,x)$ can be expanded in powers of $x^{\theta}=x^{1r}$
\ben
  R^j(\g,x)=\sum_{l=0}^n\left(x^{1r}\right)^l R^j_{(l)}(\g,x)
\een
where $R^j_{(l)}(\g,x)$ is independent of $x^{1r}$. In the proof of this
'no-go theorem' we let $j$ be an arbitrary but fixed integer in the interval
given. Insertion of the 
expansion in (\ref{nonprimeq}) leads to the following set of equations
\bea
 V_{\al_i}^\al(\g)\sum_{l=0}^n\left(x^{1r}\right)^lD_\al R^j_{(l)}(\g,x)&=&
  V_{-\al_i}^{\al_{1r}}(x)
  \sum_{l=0}^nl\left(x^{1r}\right)^{l-1}R^j_{(l)}(\g,x)\nn
 &+&V_{-\al_i}^\beta(x)\sum_{l=0}^n\left(x^{1r}\right)^l\db R^j_{(l)}(\g,x)\nn
 &+&x^j\delta_i^j\sum_{l=0}^n\left(x^{1r}\right)^l R^j_{(l)}(\g,x)
\eea
Using (\ref{Vmajb}) to explore the dependence on $x^{1r}$ of $V_{-\al_i}^\beta$
we consider the powers $N$ of $x^{1r}$ higher or equal to $n$ and find\\
$i=1$; $N=n+1$:
\ben
  0=(-x^{1r})(x^{1r})^n\pa_{2r}\rjn
\een
$i=r$; $N=n+1$:
\ben
  0=x^{1r}(x^{1r})^n\pa_{1,r-1}\rjn
\een
$i=1$; $N=n$:
\ben
 V_{\al_1}^{\al}(\g)(x^{1r})^nD_\al\rjn=-\hf x^1x^{1r}n(x^{1r})^{n-1}\rjn
  +V_{-\al_1}^{\beta}(x)(x^{1r})^n\db\rjn 
\een
$i=j$; $N=n$:
\ben
  V_{\al_j}^{\al}(\g)(x^{1r})^nD_\al\rjn=V_{-\al_j}^\beta(x)(x^{1r})^n\db\rjn
  +x^j(x^{1r})^n\rjn
\een
$i=r$; $N=n$:
\ben
  V_{\al_r}^{\al}(\g)(x^{1r})^nD_\al\rjn=-\hf x^rx^{1r}n(x^{1r})^{n-1}\rjn
   +V_{-\al_r}^{\beta}(x)(x^{1r})^n\db\rjn
\een
$i\neq 1,j,r$; $N=n$:
\ben
  V_{\al_i}^\al(\g)(x^{1r})^nD_\al\rjn=V_{-\al_i}^\beta(x)(x^{1r})^n\db\rjn
\label{Ari}
\een
The last equation only exists for $r\geq4$. To indicate the strategy
in the most general case of $r\geq4$, let us go through the case of $r=3$ 
(where the only possibility is $j=2$) and begin by
summarising
\bea
 \pa_{13}\rton&=&\pa_{23}\rton=\pa_{12}\rton=0\nn
 V_{\al_1}^{\al}(\g)D_\al\rton&=&-\hf nx^1\rton+V_{-\al_1}^{\beta}(x)
   \db\rton\nn
 V_{\al_2}^{\al}(\g)D_\al\rton&=&
  V_{-\al_2}^{\beta}(x)\db\rton+x^2\rton\nn
 V_{\al_3}^{\al}(\g)D_\al\rton&=&-\hf nx^3\rton+V_{-\al_3}^{\beta}(x)
    \db\rton
\label{A3eqs}
\eea
Now we ask when $V_{-\al_1}^{\beta}$ for $\beta\neq\al_{13},\al_{12},
\al_{23}$ includes $x^{12}$ and find that it is only through the term
\ben
  x^{12}{f_{\al_{12},-\al_1}}^{\al_2}\delta_{\al_2}^{\beta}
\een
Using this fact in (\ref{A3eqs}) we find
\ben
  \pa_2\rton=0
\een
Similarly we ask when $V_{-\al_3}^{\beta}$ for $\beta\neq\al_{13},\al_{12},
\al_{23}$ includes $x^{23}$ and find the contribution
\ben
 x^{23}{f_{\al_{23},-\al_3}}^{\al_2}\delta_{\al_2}^{\beta}
\een
so we may once again conclude that $\pa_2\rton=0$. To see that $\pa_1\rton
=\pa_3\rton=0$ we consider the case $i=2(=j)$ which can be reduced to
\bea
  V_{\al_2}^{\al}(\g)D_\al\rton&=&V_{-\al_2}^{\al_1}(x)\pa_1\rton
  +V_{-\al_2}^{\al_3}(x)\pa_3\rton+x^2\rton\nn
  &=&\left(x^{12}{f_{\al_{12},-\al_2}}^{\al_1}
   -\hf x^2x^1(\al_2^{\nu},\al_1)
  \right)\pa_1\rton\nn
  &+&\left(x^{23}{f_{\al_{23},-\al_2}}^{\al_3}
   -\hf x^2x^3(\al_2^{\nu},\al_3)
  \right)\pa_3\rton\nn
  &+&x^2\rton
\label{A32}
\eea
and we immediately read off
\ben
  \pa_1\rton=\pa_3\rton=0
\een
We can therefore conclude that $\rton$ only depends on the $\g$'s but then
according to (\ref{A32})
\ben
 V_{\al_2}^{\al}(\g)D_\al\rton(\g)=x^2\rton(\g)
\een
which means that
\ben
 \rton(\g,x)=0
\een
Since this holds for the highest power, $n$, of $x^{1r}$ it must hold for 
all positive integers (or zero) less or equal to $n$ and we have
finally proven that
\ben
  R^2(\g,x)=0
\een
This means that for $\lambda_2\neq0$ (\ref{nonprim}) is not a primary field.
 
In the general case $r\geq4$ we have already seen that
\ben
 \pa_{1r}\rjn=\pa_{2r}\rjn=\pa_{1,r-1}\rjn=0
\een
For $2\leq i<j$ (if any; $j$ might be 2) we consider the appearance
of $x^{ir}$ in $V_{-\al_i}^\beta$ starting from 'the left' 
$i=2,...,j-1$ and find
in the first step
\ben
 x^{2r}{f_{\al_{2r},-\al_2}}^{\al_{3r}}\delta_{\al_{3r}}^{\beta}
\een
so
\ben
  \pa_{3r}\rjn=0
\een
We continue and conclude
\ben
  \pa_{ir}\rjn=0\spa1\leq i\leq j
\een
where we have included the facts $\pa_{1r}\rjn=\pa_{2r}\rjn=0$. From the
'other end' $i=r-1,...,j+1$ we consider the appearance of $x^{1i}$ and
conclude
\ben
  \pa_{1i}\rjn=0\spa j\leq i\leq r
\een
Next we consider the appearance of $x^{1,r-1}$ in $V_{-\al_1}^{\beta}$:
\ben
  x^{1,r-1}{f_{\al_{1,r-1},-\al_1}}^{\al_{2,r-1}}\delta_{\al_{2,r-1}}^{\beta}
\een
and we find
\ben
 \pa_{2,r-1}\rjn=0
\een
We continue by putting $i=2,...,j-1$ in $V_{-\al_i}^\beta$ 
and may then conclude that
\ben
  \pa_{i,r-1}\rjn=0\spa 1\leq i\leq j
\een
{}From the 'other end' $i=r,...,j+1$ we consider the appearance of $x^{2r}$
and conclude
\ben
  \pa_{2i}\rjn=0\spa j\leq i\leq r
\een
Then we consider $x^{1,r-2}$ in $V_{-\al_i}^\beta$ 
for $i=1,...,j-1$ and $x^{3r}$ in
$V_{-\al_i}^\beta$ for $i=r,...,j+1$, and conclude
\ben
  \pa_{i,r-2}\rjn=0\spa1\leq i\leq j
\een
and
\ben
  \pa_{3i}\rjn=0\spa j\leq i\leq r
\een
Following this procedure it allows one to eliminate the dependence of $\rjn$ on
all $x^{\al}$, $\al\geq\al_j$.

The procedure eliminating the remaining $x$'s may start by a consideration of
$i=j$ and the appearance of $x^{1j}$. Even though $V_{-\al_j}^\beta$ 
might include
$x^{1j}$ in many terms, it will only be the term
\ben
  x^{1j}{f_{\al_{1j},-\al_j}}^{\al_{1,j-1}}\delta_{\al_{1,j-1}}^{\beta}
\een
which will be of interest because the other terms will be for $\beta>\al_j$.
This consideration excludes  $x^{1,j-1}$ in $\rjn$. We continue by putting
$i=j-1$ and consider $x^{1,j-1}$, and so on. We conclude
\ben
  \pa_{1i}\rjn=0\spa1\leq i\leq j
\een
Then we look at $x^{2j}$ for $i=j$; then at $x^{2,j-1}$ for $i=j-1$ etc.
We may hereby eliminate dependence on all $x^{\al}$,
$\al$ being to 'the left' of $\al_j$. Similarly for the roots on 'the right'
of $\al_j$, starting by considering $x^{jr}$ for $i=j$ etc.
Finally we find that $\rjn$ only depends on the $\g$'s
\ben
  \rjn(\g,x)=\rjn(\g)
\een
But from 
\bea
 V_{\al_j}^\al(\g)D_\al\rjn(\g)&=&V_{-\al_j}^\beta(x)\db\rjn(\g)+x^j\rjn(\g)\nn
  &=&x^j\rjn(\g)
\eea
we conclude as for the case $SL(3)$ that
\ben
  R^j(\g,x)=0\spa2\leq j\leq r-1
\een
This concludes the proof of the 'no-go theorem'.

We {\em do} have a partial result in the general case of $SL(r+1)$, namely the
proposition\\[.2 cm]
{\bf Proposition}\\
For $\lambda_2=...=\lambda_{r-1}=0$ the primary field is given by
\ben
 \phi(w,x)=\left(1+\sum_{\al>0}P_{-\al}^1(\g(w))
  P_{-\al}^1(x)\right)^{\lambda_1}\left(1+\sum_{\al>0}P_{-\al}^r(\g(w))
  P_{-\al}^r(x)\right)^{\lambda_r}V_\lambda(w)
\label{propsln}
\een
{\bf Proof}\\
Firstly we note that the proposition comply with the negative result
above and with the cases of $SL(2)$ and $SL(3)$. Secondly we only need to
verify (\ref{nonprimeq}) for $j=1,r$ and we find that it reduces to
\bea
 \sum_{\al>0}V_{\al_i}^\beta(\g)D_\beta P_{-\al}^j(\g)P_{-\al}^j(x)
  &=&\sum_{\al>0}P_{-\al}^j(\g)V_{-\al_i}^\beta(x)\db \paj(x)\nn
 &+&x^j\delta_i^j\left(1+\sum_{\al>0}\paj(\g)\paj(x)\right)
\eea
Using the classical identities (\ref{classV}) we find the sufficient 
conditions
\ben
 \sum_{\al>0}\paj(\g)\left(V_{-\al}^{\al_j}(x)+x^j\paj(x)\right)\delta_i^j=0
   \spa j=1,r
\een
which obviously are respected for $i\neq j$. Due to the observation
\ben
 P_{-\al_{2,l}}^1(\g)=P_{-\al_{i,r-1}}^r(\g)=0
\een
the following are sufficient conditions
\bea
 V_{-\al_{1l}}^{\al_1}(x)&=&-x^1P_{-\al_{1l}}^1(x)\spa1\leq l\leq r\nn
 V_{-\al_{ir}}^{\al_r}(x)&=&-x^rP_{-\al_{ir}}^r(x)\spa1\leq i\leq r
\eea
These can be verified by induction using the recursion relations (\ref{classV})
in $V_{-\al}^\beta$ and $\paj$.\\
$\Box$

\section{A Quantum Group Structure}

In this section we shall discuss a quantum group
structure in the braiding-commutation algebra of the screening currents.
The original work of this kind is due to Gomez and Sierra \cite{GS1} for
the minimal models and was soon after carried out for affine algebras
\cite{RRR}, but that work was only based on screening currents of the 
{\em first} kind. As we will demonstrate, using the known results 
\cite{GS1, RRR, GS2} it is a simple matter to derive the quantum
group structure based on {\em both} kinds of screening currents.
For much more detailed accounts on quantum groups the reader is referred to
the books \cite{Fuc, CP, Kas, Maj, GS2} and references therein.

\subsection{Hopf Algebra}

A Hopf algebra $A$ is a vector space 
(over the field $F$) endowed with the following five operations
\bea
  M:&A\times A\rightarrow A\ \ &({\rm multiplication})\nn
  \eta:&F\rightarrow A &({\rm unit\ map})\nn
  \D:&A\rightarrow A\times A &({\rm co-multiplication})\nn
  \epsilon:&A\rightarrow F &({\rm co-unit\ map})\nn
  \g:& A\rightarrow A &({\rm antipode})
\eea
which possess the following five properties
\bea
 & M\circ(I\times M)=M\circ(M\times I) &({\rm associativity})\nn
 & M\circ(I\times\eta)=I=M\circ(\eta\times I) &({\rm existence\ of\ unit})\nn
 & (I\times\D)\circ\D=(\D\times I)\circ\D &({\rm co-associativity})\nn
 & (\epsilon\times I)\circ\D=I=(I\times\epsilon)\circ\D &({\rm existence
     \ of\ co-unit})\nn
 & M\circ(I\times\g)\circ\D=\eta\circ\epsilon=M\circ(\g\times I)\circ\D\ \ 
     &({\rm existence\ of\ antipode})
\label{Hopf}
\eea 
The permutation map $\pi$ from $A\times A$ to $A \times A$ is introduced as
\ben
   \pi(a\otimes b)=b\otimes a
\een
Using this we define the co-multiplication $\D'$ by
\ben
  \D'=\pi\circ\D
\een
A Hopf algebra is called quasi-triangular if the co-multiplications $\D$ and
$\D'$ are conjugate
\ben
  \D(a)=R\D'(a) R^{-1}\ \ \ \forall\ a\in A
\een
where $R\in A\times A$ is invertible and 
\bea
  (I\times\D)(R)&=&R_{13}R_{12}\nn
  (\D\times I)(R)&=&R_{13}R_{23}\nn
  (\g\times I)(R)&=&R^{-1}
\eea
Here we mean that when we write $R=\sum_iA_i\otimes B_i$ then $R_{12}
=\sum_iA_i\otimes B_i\otimes I$, $R_{13}R_{23}=\sum_{i,j}A_i\otimes A_j
\otimes B_iB_j$ etc.
The notion of quantum groups is usually meant
to cover the set of (non-co-commutative) quasi-triangular Hopf algebras. 
From the physicists point of view theses are interesting for example because
they provide solutions to the Yang-Baxter equation of integrable models.

\subsection{Screening Current Realization}

In \cite{GS2} Gomez and Sierra have outlined the derivation of the quantum
group structure in the general setting of
\bea
 S_i(z)S_j(w)&=&e^{i\pi \Omega_{ij}} S_j(w)S_i(z)\nn
 S_i(z)V_\lambda(w)&=&e^{i\pi\Omega_{i\lambda}} V_\lambda(w)J_i(z)\nn
 V_\lambda(z)V_{\lambda'}(w)&=&e^{i\pi\Omega_{\lambda\lambda'}}
   V_{\lambda'}(w)V_\lambda(z)
\label{braidcom}
\eea
where $\{S_i\}_{i=1,...,2r}$ denote the set of screening currents and
$\{V_\lambda\}$ denote the set of vertex operators. 
We use the notation $S_i=s_i$ and $S_{i+r}=\tilde{s}_i$ for $1\leq i\leq r$.
According to
\cite{GS2} there exists a quantum group structure when (\ref{braidcom}) 
is satisfied for $\Omega_{ij}$ and $\Omega_{\lambda\lambda'}$ being
symmetric. Since there is no relative braiding between $\beta$ and $\g$
we get $\beta(z)^a\g(w)^b=\g(w)^b\beta(z)^a$, hence the braiding-commutation
of the screening currents will be independent of the $\beta,\g$ part.
Therefore in our case, one may easily deduce that
\bea
 \Omega_{\lambda\lambda'}&=&\frac{1}{t}\lambda\cdot\lambda'\nn
 \Omega_{i\lambda}&=&\left\{\begin{array}{ll}-\al_i\cdot\lambda/t
           &\spa i\leq r\\
  \al_{i-r}^\nu\cdot\lambda&\spa i>r\end{array}\right.\nn
  \Omega_{ij}&=&\left\{\begin{array}{ll} \al_i\cdot\al_j/t&\spa i,j\leq r\\
    -\al_i\cdot\al_{j-r}^\nu&\spa i\leq r<j\\
    -\al_{i-r}^\nu\cdot\al_j&\spa j\leq r<i\\
      \al_{i-r}^\nu\cdot\al_{j-r}^\nu t&\spa r<i,j\end{array}\right.
\eea

The construction of the quantum group heavily relies on contour manipulations 
analogous to the
ones employed in chapters 4 and 5. Let us here confine ourselves to
outline the strategy \cite{GS2}. First one introduces (for every
$z$) a vector space ${\cal V}_{\lambda,z}$ spanned by screened vertex 
operators $U_{\lambda,I}(z)$, where $I=\left\{i_1,...,i_n\right\}$ is an
ordered set of labels of screening currents
\ben
 U_{\lambda,I}(z)=\int_{C_1}dw_1S_{i_1}(w_1)...\int_{C_n}dw_nS_{i_n}(w_n)
  V_\lambda(z)
\een
The contours all start at infinity, encircle counter-clockwise the branch
point $z$ once, and return to infinity. They do not intersect, nor do they 
cross the branch cut from $z$ to infinity. The ordering may be chosen such 
that $C_i$ lies inside $C_j$ for $i>j$. Let us denote such contours as
standard contours around $z$. The vertex operator itself belongs to
${\cal V}_\lambda$, namely $V_\lambda=U_{\lambda,\emptyset}$. It is
called the 'highest weight vector', distinguished by the absence of
contour structure. 

For every screening current $S_i$ we shall define a triplet of operators 
$F_i,K_i$ and $E_i$. The former is defined as the contour creating operator
\ben
 F_i\left(U_{\lambda,I}(z)\right)=\int_CdwS_i(w)U_{\lambda,I}(z)
\een
where $C$ is a standard contour around $z$ and encloses the ones in
$U_{\lambda,I}(z)$. Now choosing the branch cuts inherent in ${\cal V}_{
\lambda_1}(z_1)$ and ${\cal V}_{\lambda_2}(z_2)$ non-intersecting, we may
define the co-multiplication $\D(F_i)$ on ${\cal V}_{
\lambda_1}(z_1)\otimes{\cal V}_{\lambda_2}(z_2)$ by
\bea
 \D(F_i)\left(U_{\lambda_1,I_1}(z_1)\otimes U_{\lambda_2,I_2}(z_2)\right)
  &=&\int_CdwS_i(w)U_{\lambda_1,I_1}(z_1)U_{\lambda_2,I_2}(z_2)\nn
 &=&F_i\left(U_{\lambda_1,I_1}(z_1)\right)U_{\lambda_2,I_2}(z_2)\nn
 &+&K_i^{-1}\left(U_{\lambda_1,I_1}(z_1)\right)F_i\left(U_{\lambda_2,I_2}(z_2)
  \right)
\label{quaF}
\eea
where we have introduced the operator
\ben
 K_i\left(U_{\lambda,I}(z)\right)=e^{-i\pi\left(\sum_{i_l\in I}\Omega_{ii_l}
  +\Omega_{i\lambda}\right)}U_{\lambda,I}(z)
\een
The contour $C$ is of the standard type wrt the {\em pair} $z_1$ and
$z_2$. By an obvious contour deformation it may be written as a sum
$C=C_1+C_2$ where $C_1$ ($C_2$) is a standard contour around $z_1$ ($z_2$).
Thereby we performed the rewriting in (\ref{quaF}) which resulted in
\ben
 \D(F_i)=F_i\otimes1+K_i^{-1}\otimes F_i
\een
The co-multiplication of $K_i$ is trivial
\ben
 \D(K_i)=K_i\otimes K_i
\een

Next we want to introduce the raising operator $E_i$ which should be a contour
annihilating operator, in some sense dual to $F_i$. Recall the fundamental
property of the screening currents
\ben
 \left[ L_{-1},\int_C dwS(w)\right]=\int_Cdw\pa_wS(w)
\een
leaving only boundary terms. Thus, let us define implicitly the operator 
$\hat{E}_i$ by
\bea
 L_{-1}U_{\lambda,I}(z)&=&\left[ L_{-1},\prod_{i_l\in I}\int_{C_l}dw_l
  S_{i_l}(w_l)\right] V_\lambda(z)+\prod_{i_l\in I}\int_{C_l}dw_lS_{i_l}(w_l)
  L_{-1}V_\lambda(z) \nn
 &=&-\sum_{j\in J}\left(q_j-q_j^{-1}\right)S_j(\infty)\hat{E}_j
  U_{\lambda,I}(z)+\prod_{i_l\in I}F_{i_l}L_{-1}V_\lambda(z) 
\eea
The last term is a screened descendant of $V_\lambda$ and $J$ denotes
the total set of different screening currents. The factor $(q_i-q_i^{-1})$ is
for later convenience. Here we have introduced the deformation parameters
\ben
 q_i=e^{i\pi\Omega_{ii}/2}
\label{defpar}
\een
Let us here work out the explicit expression
for the action of $\hat{E}_i$ on $U_{\lambda,I}(z)$. This computation is
carried out in \cite{GS2} only for all screenings being identical. 
We find ($n=$dim $I$)
\bea
 \left[ L_{-1},\prod_{i_l\in I}\int_{C_l}dw_lS_{i_l}(w_l)\right] V_\lambda(z)
  &=&-\sum_{i_l\in I}(q_{i_l}-q_{i_l}^{-1})S_{i_l}(\infty)
  \frac{1-e^{2\pi i\left(\sum_{j>i_l}\Omega_{i_lj}+\Omega_{i_l\lambda}
  \right)}}{q_{i_l}-q_{i_l}^{-1}}\nn
 &\cdot&e^{i\pi\sum_{j<i_l}\Omega_{i_lj}}
  \prod_{i_{l'}\in I\setminus\left\{i_l\right\}}\int_{C_{l'}}dw_{l'}
  S_{i_{l'}}(w_{l'})V_\lambda(z)
\eea
from which we obtain
\ben
 \hat{E}_iU_{\lambda,I}=\sum_{i_l\in I,i_l\sim i}\frac{1-e^{2\pi i\left(
  \sum_{j>i_l}\Omega_{ij}+\Omega_{i\lambda}\right)}}{q_i-q_i^{-1}}
  e^{i\pi\sum_{j<i_l}\Omega_{ij}}U_{\lambda,I\setminus\left\{i_l\right\}}
\label{Ehatqua}
\een
Here we have used that $U_{\lambda,I\setminus\left\{i\right\}}=0$ for $i\not\in
I$. The restriction in the summation means $i_l\sim i$ if $S_{i_l}=S_i$.
In the special case of only one type of screening current, (\ref{Ehatqua})
reduces to ($U_{\lambda;n}=U_{\lambda,I}$)
\ben
 \hat{E}_iU_{\lambda;n}=\frac{1-q_i^{2(n-1)}
  e^{2\pi i\Omega_{i\lambda}}}{q_i-q_i^{-1}}[n]_{q_i}U_{\lambda;n-1}
\een
This is in agreement with \cite{GS2}. Here we have introduced
\ben
 [a]_q=\frac{1-q^a}{1-q}
\een

{}From the trivial co-multiplication of $L_{-1}$ it is not difficult to work 
out that
\ben
 \D(\hat{E}_i)=\hat{E}_i\otimes1+K_i^{-1}\otimes\hat{E}_i
\een
After redefining the raising operator by $E_i=K_i\hat{E}_i$ the 
co-multiplication reads
\ben
 \D(E_i)=E_i\otimes K_i+1\otimes E_i
\een
With the following definitions of an anti-pode $\g$ 
\ben
 \g(E_i)=-E_iK_i^{-1}\spa\g(K_i)=K_i^{-1}\spa \g(F_i)=-K_iF_i
\een
and a co-unit $\epsilon$ 
\ben
 \epsilon(E_i)=0\spa\epsilon(K_i)=1\spa \epsilon(F_i)=0
\een
it is straightforward to verify that the above is indeed a Hopf algebra.
The final task is to implement quasi-triangularity in the Hopf algebra $A$.
One defines $R=R_{\lambda_1,\lambda_2}$ to be the braiding matrix of
two screened vertex operators
\ben
 U_{\lambda_1,I_1}(z_1)U_{\lambda_2,I_2}(z_2)=\sum_{I_1',I_2'}\left[
  R_{\lambda_1,\lambda_2}\right]_{I_1,I_2}^{I_1',I_2'}
  U_{\lambda_2,I_2'}(z_2)U_{\lambda_1,I_1'}(z_1)
\een
We refer to \cite{GS2} for details.

Let us summarise the commutation relations, which are not difficult to
determine
\bea
 K_iK_j&=&K_jK_i\nn
 K_iE_j&=&e^{i\pi\Omega_{ij}}E_jK_i\nn
 K_iF_j&=&e^{-i\pi\Omega_{ij}}F_jK_i\nn
 \left[ E_i,F_j\right]&=&\delta_{ij}\frac{K_i-K_i^{-1}}{q_i-q_i^{-1}}
\eea
and review the defining (Chevalley) commutation relations for the 
$q$ deformed enveloping algebra $U_q$({\bf g}) of a simple Lie algebras {\bf g}
\bea
 k_ik_j&=&k_jk_i\nn
 k_ie_j&=&q_i^{A_{ij}}e_jk_i\nn
 k_if_j&=&q_i^{-A_{ij}}f_jk_i\nn
 \left[ e_i,f_j\right]&=&\delta_{ij}\frac{k_i-k_i^{-1}}{q_i-q_i^{-1}}
\eea
where $q_i=q^{D_i}$. $D=$ diag$(D_i)$ is a diagonal matrix symmetrising
the Cartan matrix, $D_iA_{ij}=D_jA_{ji}$.
In addition there are generalised (or quantum) Serre relations which will not
concern us here. 

In our case we find the deformation parameters (\ref{defpar}) to be
\bea
 q_j&=&e^{i\pi\Omega_{jj}/2}=\left\{ \begin{array}{lll}
   e^{i\pi\al_j^2/2t}&\spa &j\leq r\\
   e^{i\pi\left(\al_{j-r}^\nu\right)^2t/2}& \spa &j>r\end{array}\right.
\label{defpar2}
\eea
We immediately see that each screening current $S_i$, $i=1,...,2r$, gives rise
to a subalgebra ${\cal U}_{q_i}(sl(2))$ generated by $\left\{ E_i,K_i,F_i
\right\}$ with $q_i$ given by (\ref{defpar2}). Thus, the total quantum
group $A$ may be viewed as a semi-direct sum of these $2r$ subalgebras.
These are not the only subalgebras. Both {\em kinds} of screening currents
also give rise to a subalgebra each. One observes that with
\ben
 q=e^{i\pi /t}\spa D_i=\al_i^2/2
\een
the screening currents of the {\em first} kind ($S_i,\ i\leq r$) give
rise to the subalgebra ${\cal U}_{q}$({\bf g}), due to 
\ben
 \left(q^{D_i}\right)^{\pm A_{ij}}=e^{\pm i\pi\Omega_{ij}}
\een
Similarly, the screening currents of the {\em second} kind ($S_i,\ i>r$)
with
\ben
 \tilde{q}=e^{i\pi t}\spa \tilde{D}_i=\left(\al_{i-r}^\nu\right)^2/2
  =2/\al_{i-r}^2
\een
give rise to the subalgebra ${\cal U}_{\tilde{q}}$({\bf g}$^t$), due to
\ben
 \left(\tilde{q}^{\tilde{D}_i}\right)^{\pm A_{j-r,i-r}}
  =e^{\pm i\pi\Omega_{ij}}
\een
Here {\bf g}$^t$ is the dual Lie algebra to {\bf g}, obtained by
transposing the Cartan matrix $A_{ij}\rightarrow A_{ij}^t=A_{ji}$. Hence,
alternatively one may view $A$ as the semi-direct sum
\ben
 A={\cal U}_q(\mbox{{\bf g}})\oplus_{\mbox{\scriptsize{semi}}}{\cal U}_{
  \tilde{q}}(\mbox{{\bf g}}^t)
\een
That the sum is only semi-direct is due to the fact that 
$\Omega_{ij}$ is not a block matrix.
For admissible representations (see Chapter 4) $t$ is rational 
and the deformation parameters $q$ and $\tilde{q}$ are roots of unity. 

In the discussion on primary fields it was argued that
they are products of vertex operators and functions of $\g$. This means
that one may generalise the construction above from vertex operators of weight
$\lambda$ to primary fields labelled by $\lambda$.

\chapter{Chiral Blocks in $SL(2)$ Current Algebra}

$N$-point correlators of two dimensional conformal WZNW theories based on 
affine $SL(2)$ current algebra have been much studied
already. They are typically constructed either by applying the free field 
realization of Wakimoto \cite{Wak}, from which results have been given for 
example in \cite{BF,ATY,FGPP,D90}, or by solving the 
Knizhnik-Zamolodchikov (KZ)
equations \cite{KZ}, from which results have been given for example in 
\cite{KZ,FZ,A,CF,SV,FGPP,FIM}. Recently the structure of solutions of the 
KZ equations on higher genus Riemann surfaces has been
examined \cite{FV}. The results given in these various pieces of works are
quite complete as far as unitary, integrable representations \cite{GW}
are concerned, but appear surprisingly incomplete for the general case, 
including admissible representations \cite{KW}, see below. 
In general the WZNW theory (for $\widehat{sl}(2)_k$) is characterised by the 
level $k$ or equivalently by $t=k+2$. Then degenerate primary fields exist 
for representations characterised by spins $j_{r,s}$ given by \cite{KK,MFF}
\ben
 2j_{r,s}+1=r-st
\een
with $r,s$ integers. However, previous applications of the free field Wakimoto
realization are complete only for the case $s=0$ which is the full case only 
for integrable representations.
The reason for this restriction is fairly natural, since
the screening charge usually employed in the free field realization is capable
of screening just such primary fields. In fact, a possible second screening
operator capable of screening the general case was proposed by Bershadsky 
and Ooguri \cite{BO}, but since it involved fractional powers of the free
ghost fields, discussions on its interpretation have been only partly 
successful \cite{D90,FGPP}.

In Chapter 2 we outlined how the techniques
of fractional calculus \cite{Kir, McR, Ross, SKM} naturally provide a 
solution. As a result we are able to render 
the free field Wakimoto realization applicable in a straightforward way, and to
present how it leads to general integral formulas for the $N$-point chiral 
blocks on the sphere. In the process we introduce
auxiliary integration variables. We shall check that
the blocks provide exact solutions to the KZ equations. This is
merely a check of the procedure. The real merit, we believe, is that we  
manage to render the free field realization straightforward in cases where 
at first sight it appears very difficult to use.
It appears from comparing with known solutions 
in the mathematics literature \cite{SV,FIM}, that our formulas represent 
fairly powerful ways of dealing with such solutions.
We briefly explain why the solutions to the KZ equations
considered in \cite{FIM} constitute a class having only little 
overlap with ours and discuss a possible hybrid treatment.
Furthermore, we discuss the slightly non-trivial way in which projective 
invariances of our correlators are established.

Utilising a freedom in the choice of numbers of screenings needed, we 
introduce the notion of over-screening. In the case of 3-point functions
based on our free field realization, we are then able to re-derive
the fusion rules for admissible representations. 
In \cite{AY} the fusion rules are determined from considering decoupling of
singular vectors as they are described in \cite{MFF}. Shortly after the same 
two sets of fusion rules were obtained by cohomological methods \cite{FM}.
Only the first set corresponds to a direct generalisation of the situation
in integrable representations. Our derivation of the fusion rules brings
in the question of choosing contours for the 3-point functions, and we
find that a peculiar mix of known contours is needed. The discussion of 
choosing contours for 4-point blocks is postponed to Chapter 5.

The relation between the $SL(2)$ current algebra and the Virasoro algebra via
Hamiltonian reduction is well-known \cite{Bel, Pol}. In particular Bershadsky 
and Ooguri \cite{BO} have used the powerful BRST formalism for the reduction 
to establish equivalence between on the one hand $\widehat{sl}(2)_k$ WZNW 
theory after reduction, and on the
other hand conformal minimal theory labelled by $(p,q)$, provided $k+2=p/q$.
Here admissible representations are used in the WZNW case,
see also the work \cite{FF} by Feigin and Frenkel. This equivalence is
discussed in those references at the level of the algebra and of the BRST 
cohomology of physical states. A particularly simple and remarkable realization
of these ideas has been discussed by Furlan, Ganchev, Paunov and Petkova
\cite{FGPP} at the level of $N$-point blocks on the sphere.
Prior to our work \cite{PRY1, PRY1a}, the most detailed analysis of $N$-point
functions has not been in terms of the free field Wakimoto realization.
Rather it has been in terms of studying solutions to the KZ equations. Thus 
Furlan, Ganchev, Paunov and Petkova \cite{FGPP} presented a systematic 
approach whereby one makes use of representations of primary $SL(2)$ fields 
based on two variables $(z,x)$ as described in the previous chapter, and 
utilised in our approach too. The aforementioned
interesting proposal checked in many explicit examples \cite{FGPP}, is
that minimal model conformal blocks are obtained from the affine $SL(2)$ 
blocks by a simple substitution, one of identifying the $x_i$ variables 
related to the $i$'th $SL(2)$ representation, with the Koba-Nielsen 
variables $z_i$. Based on this, solutions of the KZ
equations are written as power series of $(x_i-z_i)$ \cite{FGPP}, and by
construction those solutions are selected for which the boundary conditions 
are, that $x_i-z_i=0$ reproduces the corresponding minimal model correlator. 
The expansion coefficients are given in terms of recurrence relations\footnote{
In fact, it appears that the sums may be explicitly performed, see the very
recent work \cite{FGP96}.}. To make sure that such a solution of the KZ 
equations really generates the WZNW correlator (up to normalisation), a study 
is performed in \cite{FGPP} of the null vector decoupling that follows from 
that solution, and whether that is as expected for a WZNW correlator. 
Although this has been checked in many examples \cite{FGPP} no explicit 
general proof has been provided. The relation between null vector decoupling 
in WZNW correlators and minimal model Virasoro correlators has been 
discussed for example in \cite{GP}.

Using our free field realization, the issue of whether the 
$x_i\rightarrow z_i$ limit gives rise to correlators in minimal models may 
be addressed. We shall provide a simple direct proof 
of this formulation of Hamiltonian reduction and evaluate a normalisation 
constant which occurs. We find that the result is obtained in general just by 
requiring that the $x_i$'s are {\em proportional} to the $z_i$'s, independent
of the common factor of proportionality. This is reasonable because such a 
proportionality constant would depend on normalisations of the fields.
Furthermore, we shall verify that our correlators based on free field
realizations indeed respect the proposed formulation of Hamiltonian
reduction. Finally we compare with more standard forms of Hamiltonian 
reduction. 

\section{Dual States}

Before considering correlators let us define our notations and 
choices as far as dual states are concerned \cite{FMS, D90, PRY1}. 
We are only going to consider the group $SL(2)$.
To comply with the notation in \cite{PRY1, PRY1a, PRY2} we will not
use the conventions of Chapter 3 on free fields but make the substitutions
\bea
  J^+&=&E\nn
  J^3&=&\hf H\nn
  J^-&=&F\nn
  k&=&\kn\nn
  \var&=&-\sqrt{\frac{\theta^2}{4}}\var_1\nn
  t&=&\frac{2}{\theta^2}t\nn
  j&=&\hf\lambda_1
\label{sl2conv1}
\eea
where the rhs refers to Chapter 3. This results in the following realizations
supplemented by the introduction of the following notation
\bea
 \sigma(J^+)(x)&=&D_x^+=-x^2\pa_x+2xj\nn
 \sigma(J^3)(x)&=&D_x^3=-x\pa_x+j\nn
 \sigma(J^-)(x)&=&D_x^-=\pa_x\nn
 J^+(z)&=&\beta(z)\nn
 J^3(z)&=&-:\gamma\beta:(z)-\sqrt{t/2}\pa\varphi(z)\nn
 J^-(z)&=&-:\gamma^2\beta:(z)+k\pa\gamma(z)-\sqrt{2t}\gamma\pa\varphi(z)\nn
 T(z)&=&:\beta\pa\gamma:(z)+\frac{1}{2}:\pa\varphi\pa\varphi:(z)+
  \frac{1}{\sqrt{2t}}\pa^2\varphi(z)\nn
 c&=&\frac{3k}{k+2}\nn
 s(z)&=&\beta(z)e^{\sqrt{2/t}\varphi(z)}\nn
 \tilde{s}(z)&=&\beta(z)^{-t}e^{-\sqrt{2t}\varphi(z)}\nn
 \phi_j(z)&=&(1+\gamma(z)x)^{2j}:e^{-j\sqrt{2/t}\varphi(z)}:
\label{sl2conv2}
\eea
where $\phi_j$ is a primary field with spin $j$ and $(\beta,\g)$ is a bosonic
ghost pair of
conformal weights (1,0). Also the phases of the screening currents have been
changed. Furthermore we have the OPE's
\bea
 \var(z)\var(w)&=&\ln(z-w)\nn
 \beta(z)\g(w)&=&\frac{1}{z-w}\nn
 J^+(z)J^-(w)&=&\frac{k}{(z-w)^2}+\frac{2}{z-w}J^3(w)\nn
 J^3(z)J^\pm(w)&=&\pm\frac{1}{z-w}J^\pm(w)\nn
 J^3(z)J^3(w)&=&\frac{k/2}{(z-w)^2}
\eea
The ghost number current $j=-:\beta\g:$ may be realized as $j(z)=\pa\phi(z)$
where $\phi(z)\phi(w)=-\ln(z-w)$. Notice that $\phi\neq\var$. 
We use the following mode expansions
\bea
 j(z)&=&\sum_{n\in\Z}j_nz^{-n-1}\nn
 \varphi(z)&=&q_\varphi +a_0\ln z +\sum_{n\neq 0}\frac{a_n}{-n}z^{-n}\nn
 \phi(z)&=&q_\phi+j_0\ln z+\sum_{n\neq 0}\frac{j_n}{-n}z^{-n}
\eea
with hermiticity properties (see (\ref{sl2bra}))
\ben
 a_n^\dagger=a_{-n}-\sqrt{\frac{2}{t}}\delta_{n.0}\spa
 j_n^\dagger=-j_{-n}+\delta_{n,0} 
\label{hermaj}
\een
The non-trivial commutation relations read
\bea
 {[}j_0,q_\phi{]}&=&-1\nn
 {[}j_n,j_m{]}&=&-n\delta_{n+m,0}\nn
 {[}a_0,q_\varphi{]}&=&1\nn
 {[}a_n,a_m{]}&=&n\delta_{n+m,0}
\eea
The ket-vacuum, invariant under both conformal and loop projective $SL(2)$, 
is $\ket{0}$ satisfying
\bea
 \beta_n\ket{0}=&0&=\gamma_n\ket{0}=a_n\ket{0}=j_n\ket{0}, \ n>0\nn
 \beta_0\ket{0}=&0&=a_0\ket{0}=j_0\ket{0}\nn
 \gamma_0\ket{0}\neq&0&\nn
 L_n\ket{0}=&0&, \ \ n\geq -1
\eea
with
\ben
 L_n=\sum_{m\in\Z}(-m:\beta_{n-m}\gamma_m:+ 
\frac{1}{2}:a_{n-m}a_{m}:)-(n+1)\sqrt{\frac{1}{2t}}a_n
\een
The normal ordering of modes places annihilation operators to the right
of creation operators.
Correspondingly the $SL(2)$ invariant bra-vacuum $\bra{sl_2}$ satisfies
\bea
\bra{sl_2}L_n&=&0 \spa n\leq 1\nn
\bra{sl_2}\beta_n&=&0 \spa n\leq 0\nn
\bra{sl_2}\gamma_n=\bra{sl_2}j_n=\bra{sl_2}a_n&=&0\spa n\leq -1\nn
\bra{sl_2}j_0&=&\bra{sl_2}\nn
\bra{sl_2}a_0&=&\sqrt{\frac{2}{t}}\bra{sl_2}\nn
\bra{sl_2}0\kt&=&0
\label{sl2bra}
\eea
The second to last equality follows from $\bra{sl_2}L_0=0$.
The last equality is due to the fact that the bra-vacuum defined above carries 
different charges comparing to the ket-vacuum. In what follows we shall 
define another bra-vacuum with all the charges at infinity screened. This
will be the dual vacuum we are mostly going to use in calculating the 
correlators.

We define the {\em dual vacuum state} $\bra{0}$ in the WZNW free field 
realization as
\ben
\bra{0}=\bra{sl_2}e^{-q_\phi}e^{\sqrt{2/t}q_\varphi}
\een
It satisfies
\bea
\bra{0}0\kt&=&1\nn
\bra{0}\gamma_0&=&0\nn
\bra{0}\beta_0&\neq&0\nn
\bra{0}a_0&=&0\nn
\bra{0}j_0&=&0\nn
\bra{0}\beta(z)\gamma(w)\ket{0}&=&\frac{1}{z-w}
\eea
{}From the dual vacuum we construct dual bra-states of lowest $SL(2)$ weight
\ben
\bra{j}=\bra{0}e^{j\sqrt{2/t}q_\varphi}
\label{dualbra}
\een
This state indeed satisfies the conditions for being a 
{\em lowest weight state} of the affine algebra
\bea
\bra{j}J^3_0&=&j\bra{j}\nn
\bra{j}J^-_n&=&0\spa n\leq 0\nn
\bra{j}J^3_n&=&0\spa n<0\nn
\bra{j}J^+_n&=&0\spa n<0
\eea
For the corresponding ket-states
\ben
\ket{j}=e^{-j\sqrt{2/t}q_\varphi}\ket{0}
\een
we have
\ben
\bra{j}j\kt=1
\een
and this ket-state is similarly a {\em highest weight state} 
of the affine algebra.
We notice that
\ben
\bra{0}J^+_0=\bra{0}\beta_0\neq 0
\een
Thus we are performing all calculations with an $SL(2)$ 
non-invariant bra-vacuum.
This gives rise to some complications when we wish to prove projective and 
global (loop projective) $SL(2)$ invariance of our correlators. 
In a subsequent section we shall explicitly demonstrate 
that the above state, $\bra{0}\beta_0$, while not being zero
is in fact a BRST \cite{BRST} exact state in the sense of Felder \cite{F}, 
and that therefore it must be expected to decouple from all correlators. 
This decoupling we then verify.

\subsection{Admissible Representations}
Here we introduce the notion of admissible representations \cite{KW}.
Most of the constructions in this chapter and in Chapter 5 pertain to 
such representations.

Degenerate highest weight representations \cite{KK} may be parameterised as
\bea
 2j^++1&=&r-st\hspace{1.3 cm},\hspace{1 cm}(r,s)\geq(1,0)\nn
 2j^-+1&=&-r+st\hspace{1 cm},\hspace{1 cm}(r,s)\geq(1,1)
\label{jpm}
\eea
Among these the admissible ones \cite{KW} are characterised by having
rational $t$ (or rational level $k=t-2$)
\ben
 t=\pm\frac{p}{q}\hspace{1 cm},\hspace{1 cm}(p,q)=1
\een
where $(p,q)=1$ means that $p$ and $q$ are co-prime. 
In the upper case ($t=+\frac{p}{q}$) it is possible to recover the minimal
models by Hamiltonian reduction (see later this chapter)
while the lower case corresponds to the Liouville series with
$c>25$ which is dual to the minimal models. 
Here we concentrate on the first case. Any $j_{r,s}^-$ may be written
\ben
 j_{r,s}^-=-j_{r,s}^+-1
\een
and in the case of admissible representations we have the translation 
symmetries
\bea
 j_{r,s}^-&=&\frac{p-r-1}{2}-\frac{q-s}{2}t=j_{p-r,q-s}^+\nn
 j_{r,s}^\pm&=&j_{r+np,s+nq}^\pm
\label{transl}
\eea
One notices the special case of $s=0$ (or the series $s=nq$ for admissible
representations) which is only present in the plus region.
Because of the translation symmetry we may choose to work
with $j^+$ parametrisations solely.

Among the admissible representations we find the better known integrable
representations where $2j$ is integer, so they simply correspond to the case
$s=0$. In the following sections we shall discuss how one constructs 
correlators using screening currents in the case of admissible 
representations, generalising the Dotsenko-Fateev 
construction \cite{DF} of minimal models. In order to be able to screen
both integer (the $r$ part) and fractional (the $s$ part) spins, we
need both kinds of screening currents. For integrable representations
the first kind suffices, and it is exactly the introduction of the second kind
of screening current that courses many of the complications due to the
fractional power of the $\beta$ ghost field (\ref{sl2conv2}).  

\section{3-point Functions}

Let us now consider the evaluation of the (chiral) 3-point function
\ben
 \bra{j_3}\phi_{j_2}(z,x)\ket{j_1}
\een
where the dual bra $\bra{j_3}$ and the ket $\ket{j_1}$ were defined 
in the preceding section. 
Using the free field realizations of $\phi_{j_2}(z,x)$, the 3-point 
function may be evaluated only provided the charges may be 
screened away in a way similar to the one employed in the case of minimal
models \cite{DF}, see above. Correspondingly 
$\phi_{j_2}(z,x)$ is replaced by the intertwining field  
$[\phi_{j_2}(z,x)]_{j_1}^{j_3}$ \cite{F, BF,BMP},
which maps a $j_1$ highest weight module into a $j_3$ highest weight module.
In this case \cite{PRY1} we are led to consider
\bea
 \left[\phi_{j_2}(z,x)\right]_{j_1}^{j_3}&=&\int\prod_{j=1}^s
  dv_j\prod_{i=1}^rdu_i\phi_{j_2}(z,x)P(u_1,...,u_r;v_1,...,v_s)\nn
 P(u_1,...,u_r;v_1,...,v_s)&=&\prod_{j=1}^s\tilde{s}(v_j)\prod_{i=1}^rs(u_i)\nn
 &=&\prod_{j=1}^s\beta^{-t}(v_j)
  :e^{-\sqrt{2t}\varphi(v_j)}:\prod_{i=1}^r\beta(u_i)
  :e^{\sqrt{2/t}\varphi(u_i)}:
\label{intertw}
\eea
The $a_0$ charge conservation reads
\ben
 j_1+j_2-j_3=r-st
\label{3rs}
\een
with $r$ and $s$ non-negative integers.
Let us notice that the definition (\ref{primdef}) of the primary field ensures
that it transforms covariantly 
under both conformal transformations and loop projective transformations, 
namely as an $\Delta(\phi_{j})$ 
tensor field for the former, and a $-j$ tensor field for 
the latter 
\bea
 z&\rightarrow& f(z)\nn
 x&\rightarrow& \frac{a(z)x+b(z)}{c(z)x+d(z)}\spa a(z)d(z)-b(z)c(z)=1
\eea
The last statement follows from the fact that the solution to 
\ben
 T_aJ^a(x)\phi_j(z,x)=\delta_\eps\phi_j(z,x)=
  T_a\eps^a(x)\pa_x\phi(z,x)-jT_a\pa_x\eps^a(x)\phi_j(z,x)
\een 
is a second order polynomial in $x$ which is indeed an infinitesimal (loop) 
projective transformation. Invariance under (loop) projective transformations,
notably the conservation of the ($J^3_0$) $L_0$ charge
\ben
 \bra{j_3}L_0\left[\phi_{j_2}(z,x)\right]_{j_1}^{j_3}
  \ket{j_1}=\bra{j_3}[L_0,\left[\phi_{j_2}(z,x)\right]_{j_1}^{j_3}]
   \ket{j_1}+\bra{j_3}\left[\phi_{j_2}(z,x)\right]_{j_1}^{j_3}L_0\ket{j_1}
\een
determines the singular $z$-($x$-)behaviour
\ben
 \bra{j_3}\left[\phi_{j_2}(z,x)\right]_{j_1}^{j_3}
  \ket{j_1}=C_{j_1j_2j_3}z^{-\D_1-\D_2+\D_3}
  x^{j_1+j_2-j_3}(1+{\cal O}(z,x))
\label{projinv3}
\een
It is obvious that this fact simply follows from putting $z_3,x_3=\infty$,
$z_2,x_2=z,x$ and $z_1,x_1=0$ (taking into account powers of $z,x$ as in
(\ref{infstate})) in the general expression for the 3-point function 
\ben
 \langle\phi_{j_3}(z_3,x_3)\phi_{j_2}(z_2,x_2)\phi_{j_1}(z_1,x_1)\rangle
  \ \propto\ \frac{(x_2-x_1)^{j_2+j_1-j_3}(x_3-x_2)^{j_3+j_2-j_1}
  (x_3-x_1)^{j_3+j+1-j_2}}{(z_2-z_1)^{\D_2+\D_1-\D_3}(z_3-z_2)^{\D_3+\D_2-\D_1}
  (z_3-z_1)^{\D_3+\D_1-\D_2}}
\een
which is indeed found using the projective invariances (see also 
(\ref{3pointgen})). In Chapter 5 there will be a discussion on the 
structure constants $C_{j_1j_2j_3}$ appearing as the proportionality
constants when including the anti-holomorphic sector..

The techniques to perform the $\var$-part of the Wick contractions were
described in the section on minimal models and we find
\bea
 W_3^\var&=&\int\prod_{i=1}^rdu_i\prod_{j=1}^sdv_j
  \prod_{i=1}^ru_i^{-2j_1/t}(1-u_i)^{-2j_2/t}\prod_{i_1<i_2}
  (u_{i_1}-u_{i_2})^{2/t}\nn
 &\cdot&\prod_{j=1}^sv_j^{2j_1}(1-v_j)^{2j_2}  
  \prod_{j_1<j_2}(v_{j_1}-v_{j_2})^{2t}\prod_{i,j}(u_i-v_j)^{-2}
\label{W3var}
\eea
Here we concentrate on explaining how to perform the $\beta,\g$-part. 
First we have to determine the asymptotic expansion in $\gamma$ within 
$\phi_{j_2}(z,x)$. By projective invariance (\ref{projinv3})
$x$ could be fractionally powered when $s$ (\ref{3rs}) is non-zero,
hence we should expand asymptotically 
\bea
 \phi_{j_2}(z,x)&=&(1+\gamma(z)x)_{(-st)}^{2j_2}
  :e^{-j_2\sqrt{2/t}\varphi(z)}:\nn
 &=&\sum_{n}\binomial{2j_2}{n-st}(\gamma(z)x)^{n-st}
  :e^{-j_2\sqrt{2/t}\varphi(z)}:
\eea
and as in (\ref{frbgcon})
\bea
 &&\beta(w)^{-t}(1+\gamma(z)x)_{(-st)}^{2j_2}\nn
 &=&:(\beta(w)+\frac{1}{w-z}\pa_{\gamma(z)})^{-t}
  \sum_{m}\binomial{2j_2}{m-st}(\gamma(z)x)^{m-st}:\nn
 &=&\sum_{n}
  \binomial{-t}{n}:\beta^n(w)(w-z)^{t+n}\pa_{\gamma(z)}^{-t-n}
  \sum_{m}\binomial{2j_2}{m-st}(\gamma(z)x)^{m-st}:\nn
 &=& \sum_{n}\binomial{-t}{n}:\beta^n(w)(w-z)^{t+n}
  \frac{\Gamma(2j_2+1)}{\Gamma(2j_2+t+n+1)}
  (1+\gamma(z)x)_{((1-s)t)}^{2j_2+t+n}:x^{-t-n} 
\label{preexp}
\eea
Similarly we deduce
\bea
 &&\prod_{i=1}^{s}\beta(w_i)^{-t}(1+\gamma(z)x)_{(-st)}^{2j_2}\nn
 &=&:\prod_{i=1}^{s}(\beta(w_i)+\frac{1}{w_i-z}\pa_{\gamma(z)})^{-t}
  \sum_{m}\binomial{2j_2}{m-st}(\gamma(z)x)^{m-st}:\nn
 &=&:\prod_{i=1}^{s}\sum_{n_i}
  \binomial{-t}{n_i}\beta^{n_i}(w_i)(w_i-z)^{t+n_i}\pa_{\gamma(z)}^{-t-n_i}
  \sum_{m}\binomial{2j_2}{m-st}(\gamma(z)x)^{m-st}:\nn
 &=&:\prod_{i=1}^{s}\sum_{n_i}
  \binomial{-t}{n_i}\beta^{n_i}(w_i)(w_i-z)^{t+n_i}
  \frac{\Gamma(2j_2+1)}{\Gamma(2j_2+st+\sum_i n_i+1)}\nn
 &\cdot&
  (1+\gamma(z)x)^{2j_2+st+\sum_i n_i}:x^{-st-\sum_i n_i} 
\label{preexp1}
\eea
Notice that in these equations $\beta$ and $\gamma$ appear within normal 
ordering signs with integral powers.
(\ref{preexp}) and (\ref{preexp1}) suggest that we consider some kind of 
generating function, which looks like the exponential function
\ben
 F(u)=\sum_{n}\frac{1}{\Gamma(n-a+1)}
  (1+\gamma(z)x)_{(\alpha)}^{n-a}u^{-n+a}
\een   
In the simple case of 3-point functions we do not gain much by this
introduction but for more general $N$-point function it turns out to
be of great value. We shall find it useful to use the following rather trivial
identity
\ben
 (1+\gamma(z)x)^{2j}=\Gamma(2j+1)\oint_0\dtp{u}\frac{1}{u}(u^{-1}D)^{-2j}
  \exp{[}(1+\gamma(z)x)/u{]}
\een
where $D$ converts the exponential function into the derivative of that 
function, in particular it acts on and only on 
the entire argument of that function. We now prove the following \\[.2 cm]
{\bf Lemma}
\ben
 \beta^a(w)\exp{[}(1+\gamma(z)x)/u{]}=:(\beta(w)+\frac{x/u}{w-z})^aD^a
  \exp{[}(1+\gamma(z)x)/u{]}:
\label{lemma}
\een
{\bf Proof}
\bea
 &&\beta^a(w)\exp{[}(1+\gamma(z)x)/u{]}\nn
 &=&\sum_{k,m,n}\binomial{a}{m}:\beta^m(w)(w-z)^{m-a}\pa_{\gamma(z)}^{a-m}
  \frac{1}{\Gamma(n+1)}\binomial{n}{k}(\gamma(z)x)^{n-k}:u^{-n}\nn
 &=&\sum_{k,m,n}
  \binomial{a}{m}:\beta^m(w)(w-z)^{m-a}\gamma^{n-k-a+m}(z):\nn
 &\cdot&\frac{1}{\Gamma(k+1)}\frac{1}{\Gamma(n-k-a+m+1)}x^{n-k-a+m}x^{a-m}
  u^{-n}\nn
 &=&\sum_{k,m,N}\binomial{a}{m}:\beta^m(w)(w-z)^{m-a}x^{a-m}u^{m-a}
  \gamma^{-a+N-k}(z):\nn
 &\cdot&\binomial{-a+N}{k}\frac{1}{\Gamma(-a+N+1)}x^{-a+N-k}u^{a-N}\nn
 &=&:(\beta(w)+\frac{x/u}{w-z})^aD^a\exp{[}(1+\gamma(z)x)/u{]}:
\eea
$\Box$\\
We may now calculate the $\beta\gamma$  contractions in the
3-point function
\bea
 &&\prod_{i=1}^r\beta(u_i)\prod_{j=1}^s\beta^{-t}(v_j)
  (1+\gamma(z)x)^{2j}\nn
 &=&\oint_0\dtp{u}\frac{u^{2j}}{u}
  :\prod_{i=1}^r{[}\beta(u_i)+\frac{x/u}{u_i-z}{]}
  \prod_{j=1}^s{[}\beta(v_j)+\frac{x/u}{v_j-z}{]}^{-t}\nn
 &\cdot&D^{-2j+r-st}\exp\{\frac{1+\gamma(z)x}{u}\}:\Gamma(2j+1)
\eea
When inserted between $\bra{j_3}$ and $\ket{j_1}$ to produce 
$W_3^{\beta\gamma}$ (the $\beta,\gamma$ part of the 3-point function) we
effectively put $\beta=0=\gamma$ (according to $j_0$ charge
conservation) whereupon the $u$ integration 
becomes trivial, and we find the result
\ben
 W_3^{\beta\gamma}=\frac{\Gamma(2j_2+1)}{\Gamma(2j_2-r+st+1)}x^{r-st}
  \int\prod_{i=1}^rdu_i(u_i-z)^{-1}\prod_{j=1}^sdv_j(v_j-z)^t
\een
Combining this with the $\var$-part (\ref{W3var}) and using projective 
invariances to put $z=1=x$ we obtain
\bea
 W_3&=&\frac{\Gamma(2j_2+1)}{\Gamma(2j_2-r+st+1)}
  \int\prod_{i=1}^rdu_i\prod_{j=1}^sdv_j
  \prod_{i=1}^ru_i^{-2j_1/t}(1-u_i)^{-2j_2/t-1}\prod_{i_1<i_2}
  (u_{i_1}-u_{i_2})^{2/t}\nn
 &\cdot&\prod_{j=1}^sv_j^{2j_1}(1-v_j)^{2j_2+t}  
  \prod_{j_1<j_2}(v_{j_1}-v_{j_2})^{2t}\prod_{i,j}(u_i-v_j)^{-2}
\label{W3}
\eea
In the derivation we did not pay any attention to the phases. Nevertheless,
in the sequel we will consider (\ref{W3}) as the 
defining expression for the 3-point functions. 

The final expression for the 3-point function is obviously depending
on the choice of contours. However, let us postpone a discussion of these 
to a subsequent section.

Due to projective invariance
\ben
 W(j_1,j_2,j_3)\ \propto\ W(P(j_1,j_2,j_3))
\label{perm}
\een
for any permutation $P$ where the proportionality constant is
non-vanishing and finite. There exists an involution  
\bea
 J_n^+&\rightarrow&J_n^{'+}=J_{1+n}^-\nn
 J_n^3&\rightarrow&J_n^{'3}=-J_n^3+\frac{t-2}{2}\delta_{n,0}\nn
 J_n^-&\rightarrow&J_n^{'-}=J_{-1+n}^+
\eea
Using this one must expect that
\ben
 \bra{j_3}\phi_{j_2}(z,x)\ket{j_1}\ \propto\ 
  \bra{-j_3-1+t/2}\phi'_{j_2}(z',x')\ket{-j_1-1+t/2}
\label{invol}
\een
again with a non-vanishing and finite proportionality constant and where
\ben
 \phi'_j(z',x')=\phi_j(z,x)z^jx^{-2j}\hspace{1 cm}z'=z,x'=z/x
\een
is seen to be a primary field in the primed notation in the same sense that
$\phi_j(z,x)$ is primary in the unprimed system

There is a considerable freedom in choosing the numbers of screenings
subject to the charge conservation (\ref{3rs}) since $p-qt=0$.
However, it turns out \cite{PRY2}
that we only need two different sets of screenings\\[.2 cm]
{\bf Standard Screening}
\bea
 2r&=&r_1+r_2-r_3-1\nn
 2s&=&s_1+s_2-s_3
\label{W3ss}
\eea
{\bf Over-screening}
\bea
 2r&=&r_1+r_2-r_3-1+p\nn
 2s&=&s_1+s_2-s_3+q
\label{W3os}
\eea
This is denoted over-screening due to the addition of $p,q$.
We will make use of both of these when determining the fusion rules below.

We conclude this section by making a comment on the possible $SL(2)$
representations carried by the intertwining field 
$[\phi_{j_2}(z,x)]_{j_1}^{j_3}$. On the ket-vacuum (see (\ref{phi0})),
\bea
 [\phi_{j_2}(z,x)]_{j_1}^{j_3}\ket{0} &=&
  [\phi_{j_2}(z,x)]_{0}^{j_2}\ket{0}\delta_{j_1,0}\delta_{j_3,j_2}\nn
 &=&e^{xJ_0^-}e^{zL_{-1}}\ket{j_2}\delta_{j_1,0}\delta_{j_3,j_2}
\eea
is in a highest weight representation of the $SL(2)$ current algebra with
the highest weight state $\ket{j_2}$. Similarly on the dual vacuum state,
$\bra{0}[\phi_{j_2}(z,x)]_{j_1}^{j_3}$ 
is in a lowest weight representation of the $SL(2)$ current algebra with
the lowest weight state $\bra{j_2}$. However, when sandwiched in the middle 
of the correlator, the intertwining field $[\phi_{j_2}(z,x)]_{j_1}^{j_3}$
might carry representations belonging to the continuous series 
of the $SL(2)$ algebra. A representation belong to this series when neither a
highest nor a lowest weight state exists. In our case the intertwining field
belong to such a representation when both $j_1+j_2-j_3$ and $j_1-j_2-j_3$ 
are non-integers. To understand this let us notice that the monodromy
of the field forces an expansion of it to be offset the integers by
$\al=r-st=j_1+j_2-j_3$
\ben
 [\phi_{j_2}(z,x)]_{j_1}^{j_3}=\sum_n\phi_{j_2,j_2-\al-n}(z,x)\spa
  \phi_{j_2,\lambda}(z,x)=x^{j_2-\lambda}\hat{\phi}_{j_2,\lambda}(z)
\een
This is a $J_0^3$ eigenstate decomposition
\bea
 \left[J_0^3,\phi_{j_2,\lambda}(z,x)\right]&=&\lambda\phi_{j_2,\lambda}(z,x)\nn
 \left[J_0^+,\phi_{j_2,\lambda}(z,x)\right]&=&(j_2+\lambda)x
  \phi_{j_2,\lambda}(z,x)\nn
 \left[J_0^-,\phi_{j_2,\lambda}(z,x)\right]&=&(j_2-\lambda)x^{-1}
  \phi_{j_2,\lambda}(z,x)
\eea
We see that the condition for $[\phi_{j_2}(z,x)]_{j_1}^{j_3}$ to belong to
the continuous series may equivalently be characterised by
$\lambda\not\in\Z\pm j_2$. Although in that case 
$[\phi_{j_2}(z,x)]_{j_1}^{j_3}$ does not correspond to a
highest weight representation, it maps a $j_1$ highest weight 
representation to a $j_3$ highest weight representation.  

\section{Fusion Rules}

\subsection{Decoupling of Singular States}

In \cite{AY} the fusion rules are determined from considering decoupling of
singular vectors as they are described in \cite{MFF}. Here we review this
derivation of the fusion rules.

Let $M_j$ denote the Verma module generated by the highest weight 
vector $\ket{j}$ (\ref{hwvaffine}), notably $J^3_0\ket{j}=j\ket{j}$.
An affine singular vector $\ket{\chi}\in M_j$ of grade (or at level) 
$N$ and charge $Q$ is defined by being a highest weight state itself, 
of weight $j+Q$ and by satisfying
$L_0\ket{\chi}=(\Delta_j+N)\ket{\chi}$, where $\Delta_j=j(j+1)/t$.
According to \cite{KK, MFF} there exists a unique
singular vector $\ket{\chi^\pm}$ of grade $N=rs$ and charge $Q^\pm=\mp r$
when one uses the parametrisation of (\ref{jpm}), and it is given by
\cite{MFF}
\bea
 \ket{\chi^+_{r,s}}&=&(F_0)^{r+st}(E_{-1})^{r+(s-1)t}...(E_{-1})^{r-(s-1)t}
  (F_0)^{r-st}\ket{j_{r,s}^+}\nn
 \ket{\chi^-_{r,s}}&=&(E_{-1})^{r+(s-1)t}(F_0)^{r+(s-2)t}...(F_0)^{r-(s-2)t}
  (E_{-1})^{r-(s-1)t}\ket{j_{r,s}^-}
\label{singaffine}
\eea
Note the consistency $N=(j^\pm\mp r)(j^\pm\mp+1)/t-j^\pm(j^\pm+1)/t=rs$.
For general $t\in\C\setminus\{0\}$ the exponents in
(\ref{singaffine}) are complex numbers,
nevertheless it is shown in \cite{MFF} that they make sense by analytic
continuation. This may be done by rearranging the terms and then show
that the product of generators lie in the enveloping algebra, thereby
rendering the above standard singular states.
Further discussions and relations to the Virasoro singular
vectors can be found in \cite{BaSo, KY}.
 
The analogue of (\ref{projinv3}) but involving a singular state is
\ben
 \bra{j_3}\left[\phi_{j_2}(z,x)\right]^{j_3}_{j_1^\pm}\ket{\chi_{r,s}^\pm}
  =C_{j_1j_2j_3}z^{-\D_1-\D_2+\D_3-rs}x^{j_1^\pm+j_2-j_3\mp r}
  f_{r,s}^\pm(j_1^\pm,j_2,j_3)(1+{\cal O}(z,x))
\een
Using (\ref{sl2conv2}) $(F_0=\pa_x,\ E_{-1}=z^{-1}(-x^2\pa_x+2xj))$ and the
rule (\ref{fracpol}) for fractional derivation one finds
\bea
 f_{r,s}^+&=&\prod_{n=0}^{r-1}\prod_{m=0}^s(j_1^++j_2-j_3-n+mt)\prod_{n=1}^r
  \prod_{m=1}^s(-j_1^++j_2+j_3+n-mt)\nn
 f_{r,s}^-&=&\prod_{n=0}^{r-1}\prod_{m=0}^{s-1}(-j_1^-+j_2+j_3-n+mt)
  \prod_{n=1}^r\prod_{m=1}^{s-1}(j_1^-+j_2-j_3+n-mt)
\eea
For a non-vanishing structure constant $C_{j_1j_2j_3}$, the decoupling of the
singular vector $\ket{\chi_{r,s}^\pm}$ is equivalent to $f_{r,s}^\pm=0$.
For admissible representations where $j_{r,s}^-=j_{p-r,q-s}^+$ (\ref{transl}),
we have the two independent singular vectors $\ket{\chi_{r,s}^+}$ 
and $\ket{\chi_{p-r,q-s}^+}$. Their decouplings result in (now $j_1=j_1^+$)
\bea
 0&=&\prod_{n=0}^{r-1}\prod_{m=0}^s(j_1+j_2-j_3-n+mt)\prod_{n=1}^r
  \prod_{m=1}^s(-j_1+j_2+j_3+n-mt)\nn
 &=&\prod_{n=0}^{p-r-1}\prod_{m=0}^{q-s-1}(-j_1+j_2+j_3-n+mt)
  \prod_{n=1}^{p-r}\prod_{m=1}^{q-s-1}(j_1+j_2-j_3+n-mt)
\eea
Now the fusion rules are readily obtained using that if
\bea
 x&=&n+m+c_x\spa a_n\leq n\leq b_n\nn
 y&=&n-m+c_y\spa a_m\leq m\leq b_m
\eea
then
\bea
 -x+c_x+c_y+2a_n\leq&y&\leq -x+c_x+c_y+2b_n\nn
 x-c_x+c_y-2b_m\leq&y&\leq x-c_x+c_y-2a_m
\eea
In conclusion, the fusion rules are\\[.2 cm]
{\bf Fusion Rule I}
\bea
 1+|r_1-r_2|\leq&r_3&\leq p-1-|r_1+r_2-p|\nn
 |s_1-s_2|\leq &s_3&\leq q-1-|s_1+s_2-q+1|
\label{FI}
\eea
{\bf Fusion Rule II}
\bea
 1+|p-r_1-r_2|\leq &r_3&\leq p-1-|r_1-r_2|\nn
 1+|q-s_1-s_2-1|\leq &s_3&\leq q-2-|s_1-s_2|
\label{FII}
\eea
In both cases, $r_3$ and $s_3$ jump in steps of 2. 
It is easily checked that both sets of fusion rules cannot be 
satisfied simultaneously.

In \cite{FM} the same set of fusion rules are obtained by cohomological
methods.

\subsection{Standard and Over-screening of 3-point Functions}

In this section we will discuss how the fusion rules
arise from the 3-point function (\ref{W3}) by appropriate choices of 
contours and numbers of screening currents \cite{PRY1, PRY2}.

Our considerations will rely on the famous Dotsenko-Fateev (DF) 
integral (last paper in Ref. \cite{DF}, appendix A, here 
with minor misprints corrected)
\bea
 {\cal J}_{nm}(a,b;\rho)&=&T\int_0^1\prod_{i=1}^ndu_i\prod_{j=1}^mdv_j
  u_i^{a'}(1-u_i)^{b'}
  \prod_{i_1<i_2}(u_{i_1}-u_{i_2})^{2\rho'}\nn
 &\cdot&v_j^a(1-v_j)^b\prod_{j_1<j_2}(v_{j_1}-v_{j_2})^{2\rho}
  \prod_{i,j}^{n,m}(u_i-v_j)^{-2}\nn
 &=&\rho^{2nm}\prod_{i=1}^n\frac{\Gamma(i\rho')}
  {\Gamma(\rho')}\prod_{i=1}^m\frac{\Gamma(i\rho-n)}{\Gamma(\rho)}\nn
 &\cdot&\prod_{i=0}^{n-1}\frac{\Gamma(1+a'+i\rho')\Gamma(1+b'+i\rho')}
  {\Gamma(2-2m+a'+b'+(n-1+i)\rho')}\nn
 &\cdot&\prod_{i=0}^{m-1}\frac{\Gamma(1-n+a+i\rho)\Gamma(1-n+b+i\rho)}
  {\Gamma(2-n+a+b+(m-1+i)\rho)}
\label{dfintegral}
\eea
where the parameters are subject to the relations
\ben
 a'=-\rho'a\spa b'=-\rho'b\spa\rho'=\rho^{-1}
\label{dfrelations}
\een
The integral is real and time ordered (\ref{timeordered}). 
In Chapter 5 we shall also need to consider another version of this
integral, namely
\bea
 \tilde{{\cal J}}_{nm}(a,b;\rho)&=&T\int_1^\infty
  \prod_{i=1}^ndu_i\prod_{j=1}^mdv_j
  u_i^{a'}(u_i-1)^{b'}
  \prod_{i_1>i_2}(u_{i_1}-u_{i_2})^{2\rho'}\nn
 &\cdot&v_j^a(v_j-1)^b\prod_{j_1>j_2}(v_{j_1}-v_{j_2})^{2\rho}
  \prod_{i,j}^{n,m}(u_i-v_j)^{-2}\nn
 &=&{\cal J}_{nm}(-a-b+2(n-1)-2(m-1)\rho,b;\rho)
\label{dfintegral2}
\eea
The last equality follows from changing variables $u_i\rightarrow1/u_i$,
$v_j\rightarrow1/v_j$ and using (\ref{dfrelations}).
It follows that the 3-point function (\ref{W3}) may be evaluated using the DF
integration. However, we need complex contours, both of the Felder \cite{F}
type (a set of closed circle-like contours passing through a common point
which is the only intersection point) and of the Dotsenko-Fateev 
\cite{DF} type (a set of open contours with common end points which 
are the only intersection points).  
The choice of contours give rise to certain pre-factors multiplying
the DF integral. Due to the relations (\ref{dfrelations}) one may interchange
or pull through each other a DF contour of a $u$ variable and a DF contour
of a $v$ variable \cite{DF}.   

Let us be more specific and consider (\ref{W3}) in
the case of admissible representations.
In the case that all contours are Felder contours (labelling 
by $1,...,r$ and $1,...,s$ the 'circles' from in-most to out-most) one finds
\cite{PRY2}
\ben
 W_{FF}^{r,s}=\chi_r^{(1)}(r_1;1/t)\chi_s^{(2)}(s_1;t)
  \frac{\Gamma(2j_2+1)}{\Gamma(2j_2-r+st+1)}{\cal J}_{rs}(2j_1,2j_2+t;t)
\label{W3FF}
\een
where the $\chi$-functions \cite{PRY2}
\bea
 \chi_r^{(1)}(r_1;1/t)&=&e^{i\pi r(r+1-2r_1)/t}
  \prod_{j=1}^r\frac{(1-e^{2\pi i(r_1-j)/t})(1-e^{2\pi i j/t})}{1-e^{2\pi i/t}}
  \nn
 &=&(2i)^re^{i\pi r(r-r_1)/t}\prod_{j=1}^r\frac{s((j-r_1)/t)s(j/t)}{s(1/t)}
\label{chir}
\eea 
\bea
 \chi_s^{(2)}(s_1;t)&=&e^{i\pi ts(s-1-2s_1)}
  \prod_{j=1}^s\frac{(1-e^{2\pi it(s_1+1-j)})(1-e^{2\pi itj})}
  {1-e^{2\pi it}}\nn           
 &=&(2i)^se^{i\pi ts(s-1-s_1)}\prod_{j=1}^s\frac{s((j-s_1-1)t)s(jt)}{s(t)}
\label{chis}
\eea
(with $s(x)=\sin(\pi x)$)
arise when performing contour deformations\footnote{It is not necessary to
introduce two $\chi$-functions since $\chi_n^{(1)}(a;t)=\chi_n^{(2)}(a-1;t)$.
However, it helps clarifying the origins of the expressions.}.
It turns out that the Felder contours alone cannot produce a well-defined
and non-vanishing 3-point function corresponding to fusion rule II
(see below). We
need the combination that the $r$ screening variables of the first kind are 
integrated along DF contours, while the $s$ screenings
of the second kind are taken along Felder contours (or vice versa). 
Let the convention for DF contours be that the labelling puts contours
with lower indices below contours with higher indices. 
Now contour manipulations give 
\ben
 W_{DFF}^{r,s}=\lambda_r(1/t) \chi_s^{(2)}(s_1;t)
  \frac{\Gamma(2j_2+1)}{\Gamma(2j_2-r+st+1)}{\cal J}_{rs}(2j_1,2j_2+t;t)
\label{W3DFF}
\een
where we have introduced the functions \cite{DF, PRY2}
\ben
 \lambda_r(1/t)=\prod_{j=1}^re^{-i\pi(j-1)/t}\frac{s(j/t)}{s(1/t)}
\label{lambda}
\een
Using the alternative combination where the $r$ screening variables
of the first kind are
integrated along Felder contours while the $s$ screening variables
are taken along DF contours, results in the pre-factor 
$\chi_r^{(1)}(r_1;1/t)\lambda_s(t)$.    
Similarly one may choose all contours of the DF type and then obtain the
pre-factor $\lambda_r(1/t)\lambda_s(t)$.

The fusion rules determine when a 3-point function is well-defined and
non-vanishing.
According to (\ref{perm}) and
(\ref{invol}) the fusion rules should be invariant under certain
transformations. Because of the translation symmetry (\ref{transl}) we may
choose to work with $j^+$ parameterisations whereby we choose the fusion rules
to be expressed in terms of such parameterisations.
Hence the transformations are
\ben
 (r_1,r_2,r_3;s_1,s_2,s_3)\rightarrow(P(r_1,r_2,r_3);P(s_1,s_2,s_3))
\label{perm2}
\een
and
\bea
 r_1&\rightarrow&p-r_1\nn
 r_2&\rightarrow&r_2\nn
 r_3&\rightarrow&p-r_3\nn
 s_1&\rightarrow&q-s_1-1\nn
 s_2&\rightarrow&s_2\nn
 s_3&\rightarrow&q-s_3-1
\label{invol2}
\eea
Of course these may be invoked in any order. The exact form of (\ref{invol2})
follows from (\ref{invol}) since $j_i\rightarrow -j_i-1+t/2$ corresponds
to $(r_i,s_i)\rightarrow(p-r_i,q-s_i-1)$.

Even before going into a detailed evaluation of the gamma- and sine-functions
in the 3-point function we can derive the fusion rules as {\em necessary}
conditions for a 3-point function to be well-defined.
Let us consider the case of over-screening (\ref{W3os}).
Assume that $r\geq p$ which means that $r_1+r_2-r_3-1+p\geq2p$ and therefore
$r_3\leq-p-1+r_1+r_2$. If this inequality is allowed by a set of fusion
rules then we may impose permutation invariance (\ref{perm2}) leading to
$p+1+|r_1-r_2|\leq r_3\leq-p-1+r_1+r_2$. Without loss of generality we let
$r_1\geq r_2$ and find $p+1+r_1-r_2\leq -p-1+r_1+r_2$ or equivalently
$p+1\leq r_2$ which takes us out of the standard region, so we
conclude that $r<p$. This leads to $2r\leq2p-2$ and $-p+1+r_1+r_2\leq r_3$. 
Imposing permutation invariance (\ref{perm2}) yields 
$-p+1+r_1+r_2\leq r_3\leq p-1-|r_1-r_2|$. Finally we impose the
involutionary invariance (\ref{invol2}) and obtain
\ben
 1+|p-r_1-r_2|\leq r_3\leq p-1-|r_1-r_2|
\een
Similarly the assumption $s\geq q$ will take us out of the standard region
by $q\leq s_2\leq s_1$. On the other hand $s<q$ leads to $-q+2+s_1+s_2
\leq s_3$ and as before we impose permutation invariance (\ref{perm2}) yielding
$-q+2+s_1+s_2\leq s_3\leq q-2-|s_1-s_2|$ and finally the involution  
(\ref{invol2}) gives
\ben
 1+|q-1-s_1-s_2|\leq s_3\leq q-2-|s_1-s_2|
\een
and we have obtained fusion rule II.
Fusion rule I follows as necessary conditions simply from imposing the
invariances on the conditions $r,s\geq0$.

Now we want to establish that the 3-point function (\ref{W3}) with Felder
contours (\ref{W3FF})
is well-defined and non-vanishing exactly when fusion rule I is
imposed. Furthermore, we want to establish 
that (\ref{W3}), with the combination of the $r$ screening
variables of the first kind being integrated along DF contours and the
$s$ screening variables taken along Felder contours (\ref{W3DFF}), 
is well-defined and non-vanishing exactly when fusion rule II is imposed.  
In the verification we allow cancellations of the form $\G(0)/\G(0)=1$.
Since we have already found that the fusion rules are necessary conditions
we only need to show that indeed they are sufficient conditions.
Thus, we simply count the numbers of $\G(0)$ and $1/\G(0)$ in 
${\cal J}_{rs}(2j_1,2j_2+t;t)$ and in the pre-factors when imposing
the fusion rules. The sufficient conditions are then equivalent to 
equality of these numbers. Notice that $s(0)\sim1/\G(0)$.
We find that for fusion rule I all terms are regular 
\ben
 \lambda_r(1/t)\sim\lambda_s(t)\sim\chi_r^{(1)}(r_1;1/t)\sim\chi_s^{(2)}(s_1;t)
  \sim \frac{\Gamma(2j_2+1)}{\Gamma(2j_2-r+st+1)}\sim
  {\cal J}_{rs}(2j_1,2j_2+t;t)\sim1
\een
while for fusion rule II we find
\bea
 \lambda_r(1/t)\sim\lambda_s(t)&\sim&1\nn
 \chi_r^{(1)}(r_1;1/t)\sim\chi_s^{(2)}(s_1;t)&\sim&1/\G(0)\nn
 \frac{\Gamma(2j_2+1)}{\Gamma(2j_2-r+st+1)}&\sim&1\nn
 {\cal J}_{rs}(2j_1,2j_2+t;t)&\sim&\left(\G(0)\right)^2/\G(0)
\eea
This concludes the derivation of the fusion rules from the 
free field realization of the 3-point function.

The above analysis shows that for fusion rule I, (\ref{W3}) is well-defined and
non-vanishing for all 4 combinations of contours $(FF,FDF,DFF,DFDF)$, while
for fusion rule II we need a combination of Felder and DF contours $(DFF,FDF)$.
Hence, for a given set of fusion rules 
a choice of contours among these allowed possibilities only
affects the normalisation of the 3-point function.

\subsection{Over-screening in Minimal Models}

The fusion rules for minimal models have been known in many years
\cite{BPZ, F}. However, the technique of over-screening might reveal the
existence of an unknown second set of fusion rules. This turns out not to
be the case as one should expect, since in minimal models the primary fields
are solely specified by their conformal weights, contrary to the primary
fields in $SL(2)$ current which are characterised by spins as well. 
Nevertheless, we will present an explicit analysis of this lack of a second
set of fusion rules.

Consider the minimal model 3-point function (in the limits $z_3,z_2,z_1
\rightarrow\infty,1,0$)
\bea
 W_3&\sim&\int\prod_{k=1}^Ndv_kv_k^{\al_{n_1,m_1}\al_+}(1-v_k)^{\al_{n_2,m_2}
  \al_+}\prod_{k<k'}(v_k-v_{k'})^{\al_+^2}\nn
 &\cdot&\prod_{i=1}^Mdw_iw_i^{\al_{n_1,m_1}\al_-}(1-w_i)^{\al_{n_2,m_2}\al_-}
  \prod_{i<i'}(w_i-w_{i'})^{\al_-^2}\prod_{k,i}^{N,M}(v_k-w_i)^{-2}
\label{w3minmod1}
\eea
where the charge conservation reads
\ben
 \al_{n_1,m_1}+\al_{n_2,m_2}-\al_{n_3,m_3}+N\al_++M\al_-=0
\een
Using the rationality condition $\al_-/\al_+=-p/q=-t$
(where $\al_-^2/2=t,\ \al_+^2/2=1/t$)
we consider the following screenings\\[.2 cm]
{\bf Standard Screening}
\ben
 N=\hf(n_1+n_2-n_3-1)\spa M=\hf(m_1+m_2-m_3-1)
\een
{\bf Over-screening}
\ben
 N=\hf(n_1+n_2-n_3-1+p)\spa M=\hf(m_1+m_2-m_3-1+q)
\een
Choosing all contours to be Felder contours and considering standard screening,
yields
\ben
 W_{FF}=\chi_N^{(1)}(n_1;1/t)\chi_M^{(2)}(m_1-1;t){\cal J}_{NM}
  (n_1-1-(m_1-1)t,n_2-1-(m_2-1)t;t)
\label{W3FFMM}
\een
This is seen to be non-vanishing and well-defined exactly when one employs the
fusion rules
\bea
 1+|n_1-n_2|\leq &n_3&\leq p-1-|p-n_1-n_2|\nn
 1+|m_1-m_2|\leq &m_3&\leq q-1-|q-m_1-m_2|
\label{fusionMM}
\eea
where $n_3$ and $m_3$ jump in steps of 2.
If we consider over-screening and choose the $N$ contours as DF contours
and the $M$ contours as Felder contours, we find
\ben
 W_{DFF}=\lambda_N(1/t)\chi_M^{(2)}(m_1-1;t){\cal J}_{NM}
  (n_1-1-(m_1-1)t,n_2-1-(m_2-1)t;t)
\een
An analysis of this along the lines indicated above, one finds the 
cancellation $(\G(0)/\G(0))^2$ exactly when
\bea
 1+|p-n_1-n_2|\leq&n_3&\leq p-1-|n_1-n_2|\nn
 1+|q-m_1-m_2|\leq&m_3&\leq q-1-|m_1-m_2|
\label{fusionMM2}
\eea
However, the point is that this set of fusion rules is obtained from 
(\ref{fusionMM}) simply by transforming $n_3,m_3\rightarrow p-n_3,q-m_3$,
while the conformal weights enjoy the symmetry 
\ben
 \D_{n,m}=\D_{p-n,q-m}
\label{Dsym}
\een
In conclusion, (\ref{fusionMM2}) is equivalent to (\ref{fusionMM}), and
the act of over-screening in minimal models merely reflects the freedom
(\ref{Dsym}) in parametrising the primary fields.

\section{$N$-point Functions}
We wish to evaluate the chiral block
\ben
 W_N=\bra{j_N}{[}\phi_{j_{N-1}}(z_{N-1},x_{N-1}){]}^{j_N}_{\iota_{N-2}}
  ...{[}\phi_{j_{n}}(z_{n},x_n){]}^{\iota_{n}}_{\iota_{n-1}}...
  {[}\phi_{j_{2}}(z_2,x_2){]}^{\iota_2}_{j_1}\ket{j_1}
\een
Thus we have primary fields at points
\ben
 z_1=0,z_2,...,z_{N-1},z_N=\infty
\een
having $x$ values
\ben
 x_1=0,x_2,...,x_{N-1},x_N=\infty
\een
From the pictorial version
\ben
\begin{picture}(300,70)
\put(0,0){$j_N$}  \put(10,10){\line(1,0){50}}  \put(55,0){$\iota_{N-2}$}
\put(35,10){\line(0,1){40}} \put(35,55){$j_{N-1}$} \put(110,10){\line(1,0){50}}
\put(110,0){$\iota_n$}   \put(155,0){$\iota_{n-1}$}  \put(135,55){$j_n$}
\put(135,10){\line(0,1){40}} \put(210,10){\line(1,0){75}} \put(280,0){$j_1$}
\put(235,10){\line(0,1){40}} \put(260,10){\line(0,1){40}}
\put(235,55){$j_3$}\put(260,55){$j_2$} \put(245,0){$\iota_2$}
\put(210,0){$\iota_3$}
\put(180,9){$\cdots$} \put(80,9){$\cdots$}
\end{picture}
\een
one reads off the following screening conditions
\bea
 j_1+j_2-\iota_2&=&\rho_2-\sigma_2 t\nn
 \iota_2+j_3-\iota_3&=&\rho_3-\sigma_3 t\nn
 &\vdots&\nn
 \iota_{n-1}+j_n-\iota_n&=&\rho_n-\sigma_n t\nn
 &\vdots&\nn
 \iota_{N-2}+j_{N-1}-j_N&=&\rho_{N-1}-\sigma_{N-1}t\nn
 2j_i+1&=&r_i-s_it
\eea
with $\sigma_n,\rho_n$ non-negative integers, while the last line is the usual
parametrisation of the weights.
We then get for the $\beta,\gamma$ part of the correlator, denoting by 
$w(n,i)$ and $v(n,k)$ the positions of the $i$'th and the $k$'th screening 
currents of the first and second kinds respectively around 
the $n$'th primary field
\bea
 W^{\beta\gamma}_N&=&\bra{j_N}\prod_{n=2}^{N-1}(1+x_n\gamma(z_n))^{2j_n}
  \prod_{k=1}^{\sigma_n}{[}\beta(v(n,k)){]}^{-t}
  \prod_{i=1}^{\rho_n}\beta(w(n,i))\ket{j_1}\nn
 &=&\oint_0 \prod_{n=2}^{N-1}\dtp{u_n}\prod_{i=1}^{\rho_n}
  \prod_{k=1}^{\sigma_n}\nn
 &&\bra{j_N}:\left ( \beta(w(n,i))+\frac{x_n/u_n}{w(n,i)-z_n}\right )
  \left(\beta(v(n,k))+\frac{x_n/u_n}{v(n,k)-z_n}\right )^{-t}\nn
 &\cdot& u_n^{2j_n}D_n^{-2j_n+\rho_n-\sigma_nt}
  \exp\left(\frac{1+x_n\gamma(z_n)}{u_n}\right)
  \Gamma(2j_n+1)\frac{1}{u_n}:\ket{j_1}\nn
 &=&\oint_0\prod_{n=2}^{N-1}\dtp{u_n}\prod_{i=1}^{\rho_n}
  \left(\sum_{\ell=2}^{N-1}
  \frac{x_\ell/u_\ell}{w(n,i)-z_\ell}\right )\prod_{k=1}^{\sigma_n}
  \left(\sum_{\ell=2}^{N-1}\frac{x_\ell/u_\ell}{v(n,k)-z_\ell}\right)^{-t}\nn
 &\cdot&u_n^{2j_n-1}D_n^{-2j_n+\rho_n-\sigma_nt}e^{(1/u_n)}
\Gamma(2j_n+1)
\eea
where we have used the techniques already developed for the  
3-point function.
In particular, in the second equality we applied the lemma of the previous
section, and in the last equality we kept doing that until normal ordering
signs surrounded all operators, at which point the calculation was completed by
putting $\beta$'s and $\gamma$'s under normal ordering signs equal to zero.
Conforming with the discussion in the previous section we may also throw
away all derivatives on the exponential (they are with respect to the full 
argument of the exponential), though their presence serve to remind us in some
cases, what representations would conveniently be used. In the sequel we drop
these derivatives.
It is straightforward to write down the contribution from the $\varphi$-part
of the free field realization. It is
\bea
 W^\varphi_N&=&\prod_{1\leq m<n\leq N-1}(z_m-z_n)^{2j_mj_n/t}
  \prod_{n=2}^{N-1}
  \prod_{i=1}^{\rho_n}\prod_{m=1}^{N-1}(w(n,i)-z_m)^{-2j_m/t}\nn
 &\cdot&\prod_{n=2}^{N-1}\prod_{k=1}^{\sigma_n}\prod_{m=1}^{N-1}
  (v(n,k)-z_m)^{2j_m}\prod_{(n,i)<(n',i')}(w(n,i)-w(n',i'))^{2/t}\nn
 &\cdot&\prod_{(n,k)<(n',k')}(v(n,k)-v(n',k'))^{2t}\prod_{(n,i),(n',k)}
  (w(n,i)-v(n',k))^{-2}
\eea
Here we have introduced a rather arbitrary ordering of indices, for example
as
\ben
 (n,i)<(n',i')
\een
if either $n<n'$ or $n=n', \ \ \ i<i'$.

Let us summarise our findings in a more compact notation: Let 
\bea
 M&=&\sum_{m=2}^{N-1}(\rho_m+\sigma_m)\nn
 w_i&&i=1,..., M
\eea
collectively denote the position of all screening charges
\ben
 \{w_i\}=\{w(n,i), v(n,k)\}
\een
Further let
\bea
 k_i&=&\left\{\begin{array}{rl}
   -1&i=1,...,\sum_m\rho_m\\
   t&i=\sum_m\rho_m+1,...,M
  \end{array}\right. \nn
 B(w_i)&=&\sum_{\ell=1}^{N-1}\frac{x_{\ell}/u_{\ell}}{w_i-z_\ell}
\label{B}
\eea
(here $x_1=0$).
Then the {\em integrand} of the $N$-point function is given by (we use the same
letters for the integrated expressions, we hope this will not cause confusion)
\ben
 W_N=W_BW^\varphi_NF
\label{npoint}
\een
with
\bea
 W_N^{\beta\gamma}&=&W_BF\nn
 W_B&=&\prod_{i=1}^M B(w_i)^{-k_i}\nn
 W^\varphi_N&=&\prod_{m<n}(z_m-z_n)^{2j_mj_n/t}\prod_{i=1}^M\prod_{m=1}^{N-1}
  (w_i-z_m)^{2k_ij_m/t}\prod_{i<j}(w_i-w_j)^{2k_ik_j/t}\nn
 F&=&\prod_{m=2}^{N-1}\Gamma(2j_m+1)u_m^{2j_m-1}e^{1/u_m}
\eea
We believe the above general closed expression for integral representation
of the $N$-point function to be useful for further development, in particular
integrations over the auxiliary variables, 
$u_\ell$, $\ell=2,...,N-1$ seem tractable as they stand.
If for some reason, one needs to get rid of these integrations, it is not
too difficult. As an example, we provide an explicit form \cite{PRY1}
for the result for integrable representations. 
First we define the following notation
\bea
 J_N&=&\{2,3,...,N-1\}\nn
 I_N&=&\{(n,i)|n=2,...,N-1, \ i=1,2,...,\rho_n\}\nn
 {\cal F}_N&=&\{\mbox{maps, $f$, from $I_N$ to $J_N$}\}
\eea
For $t=k+2$ integer, all $\sigma$'s are $=0$. In this case we may then write
\bea
 W^{\beta\gamma}_N&=&\oint_0\prod_{n=2}^{N-1}\dtp{u_n}u_n^{2j_n}
  \frac{e^{1/u_n}}{u_n}(2j_n)!\prod_{(n,i)\in I_N}\left(\sum_{\ell=2}^{N-1}
  \frac{x_\ell/u_\ell}{w(n,i)-z_\ell}\right)\nn
 &=&\oint_0\prod_{n=2}^{N-1}\dtp{u_n}u_n^{2j_n}\frac{e^{1/u_n}}{u_n}(2j_n)!
  \sum_{f\in {\cal F}_N}\prod_{(n,i)\in I_N}
  \left (\frac{x_{f(n,i)}/u_{f(n,i)}}{w(n,i)-z_{f(n,i)}}\right)\nn
 &=&\oint_0\prod_{n=2}^{N-1}\dtp{u_n}u_n^{2j_n}\frac{e^{1/u_n}}{u_n}(2j_n)!\nn
 &&\sum_{f\in {\cal F}_N}\prod_{\ell\in J_N}(x_\ell/u_\ell)^{|f^{-1}(\ell)|}
  \prod_{(n,i)\in f^{-1}(\ell)} (w(n,i)-z_{\ell})^{-1}\nn
 &=&\sum_{f\in {\cal F}_N}\prod_{\ell\in J_N}x_\ell^{|f^{-1}(\ell)|}
  \frac{(2j_\ell)!}{(2j_\ell-|f^{-1}(\ell)|)!}
  \prod_{(n,i)\in f^{-1}(\ell)} (w(n,i)-z_{\ell})^{-1}
\eea
A similar but even more complicated sum formula obtains in the general case.

\subsection{Different Realizations}
We now have two ways of calculating the chiral block corresponding to
$N$ primary fields. The way so far described is by using (part of) projective 
invariance (to be discussed further later) 
and global $SL(2)$ invariance to work it out as
\bea
 &W^{(I)}_N(z_N=\infty,x_N=\infty,z_{N-1},x_{N-1},...,z_2,x_2,z_1=0,x_1=0)&=\nn
 &\bra{j_N}{[}\phi_{j_{N-1}}(z_{N-1},x_{N-1}){]}^{j_N}_{\iota_{N-2}}...
  {[}\phi_{j_2}(z_2,x_2){]}^{\iota_2}_{j_1}\ket{j_1}&
\eea
However, obviously, we may also use our techniques to evaluate the same 
$N$-point chiral block as
\bea
 &W^{(II)}_N(z_N,x_N,z_{N-1},x_{N-1},...,z_2,x_2,z_1,x_1)&\nn
 =&\bra{0}{[}\phi_{j_N}(z_N,x_N){]}^0_{j_N}
  {[}\phi_{j_{N-1}}(z_{N-1},x_{N-1}){]}^{j_N}_{\iota_{N-2}}...
  {[}\phi_{j_2}(z_2,x_2){]}^{\iota_2}_{j_1}
  {[}\phi_{j_1}(z_1,x_1){]}^{j_1}_0 \ket{0}&
\label{WII}
\eea
We now want to demonstrate \cite{PRY1}
that up to normalisation these expressions are 
equivalent in the appropriate limits. 
Notice that the second form involves more screening charges around the last 
field than the first one. We shall see that these extra screenings give
rise to a constant contribution in the limit $z_N,x_N\rightarrow\infty$.
But for finite $z_N$ and $x_N$, unlike in the case of the conformal minimal 
models, there does not seem to be any 
simple way of getting a conjugate field, which would get
rid of the extra screening charges.

It is clear that in the limit $z_1,x_1\rightarrow 0$, the second formulation 
coincides with the first, regarding the ket-part. In 
particular, the second formulation involves no extra screening operators. 
Thus we shall concentrate on the limit $z_N,x_N\rightarrow\infty$.

As before, we let $w_i$ denote the position of screening operators {\em 
in the first
case,} $W_N^{(I)}$, and we let $i$ run over the same set as in the first case.
Further,
we let $z_n,x_n$ denote the arguments as in the first case and $n=1,2,...,N-1$
runs over the same set as in the first case. The new feature in the second 
case is:

(i) the appearance of 
$$j_N+j_N-0=r_N-1-s_Nt=\rho_N-\sigma_Nt$$
extra screening operators, the positions of which we denote by
$$w^N_{i_N}, \ i_N=1,...,\rho_N+\sigma_N$$
$$k^N_{i_N}=\left\{\begin{array}{rcl}
-1&,&i_N=1,...,\rho_N\\
t&,&i_N=\rho_N+1,...,\rho_N+\sigma_N
\end{array}\right.$$
and

(ii) an extra $u$ integration over a variable we call $u_N$.
Now we want to consider the limit as $z_N,x_N\rightarrow\infty$ (letting
$W_N^{(I)}$ stand for the integrand in an appropriate way) of
\bea
 &&z_N^{\frac{2j_N(j_N+1)}{t}}x_N^{-2j_N}W_N^{(II)}\nn  
 &=&z_N^{\frac{2j_N(j_N+1)}{t}}x_N^{-2j_N}\Gamma(2j_N+1)
  \prod_{i_N}B^{-k^N_{i_N}}(w^N_{i_N})
  \prod_{i_N<j_N}(w^N_{i_N}-w^N_{j_N})^{2k^N_{i_N}k^N_{j_N}/t}\nn
 &\cdot&\prod_{i,i_N}(w^N_{i_N}-w_i)^{2k^N_{i_N}k_i/t}
  \prod_{i_N}(z_N-w^N_{i_N})^{2k^N_{i_N}j_N/t}\prod_{i_N,n}
  (w^N_{i_N}-z_n)^{2k^N_{i_N}j_n/t}\nn
 &\cdot&\prod_n(z_N-z_n)^{2j_Nj_n/t}\prod_i(z_N-w_i)^{2j_Nk_i/t}
  u_N^{2j_N-1}e^{1/u_N}\dtp{u_N}W_N^{(I)}\prod_{i_N}dw^N_{i_N}
\label{WNII2}
\eea
where the function $B(w)$ is defined with one more term than for case (I), cf.
(\ref{B}).
We now use that
\bea
 -\sum_{i_N}k^N_{i_N}&=&\rho_N-\sigma_Nt= 2j_N\nn
  \sum_ik_i&=&-\sum_nj_n+j_N
\eea
In the limit $z_N\rightarrow\infty, x_N\rightarrow\infty$ we find
\bea
 w^N_{i_N}/z_N&=&\tilde{w}^N_{i_N} \nn
 w_i/z_N&\rightarrow& 0\nn
 z_n/z_N&\rightarrow& 0
\eea
with $\tilde{w}^N_{i_N}$ finite and
\ben
 B(w^N_{i_N})\sim \frac{x_N/u_N}{(1-\tilde{w}^N_{i_N})z_N}
\een
Hence, we have
\bea
 &&\lim_{z_N,x_N\rightarrow\infty}z_N^{\frac{2j_N(j_N+1)}{t}}
  x_N^{-2j_N}W_N^{(II)}\nn  
 &\sim&z_N^{2\frac{j_N(j_N+1)}{t}}x_N^{-2j_N}x_N^{-\sum_{i_N}k^N_{i_N}}
  u_N^{\sum_{i_N}k^N_{i_N}}z_N^{\sum_{i_N}k^N_{i_N}}\prod_{i_N}
  (1-\tilde{w}^N_{i_N})^{k^N_{i_N}}\nn
 &\cdot&\prod_{i_N < j_N}(\tilde{w}^N_{i_N}-\tilde{w}^N_{j_N})^
  {2k^N_{i_N}k^N_{j_N}/t}z_N^{\sum_{i_N < j_N}2k^N_{i_N}k^N_{j_N}/t}\nn
 &\cdot&\prod_{i_N,i}(\tilde{w}^N_{i_N})^{2k^N_{i_N}k_i/t}
  z_N^{\sum_{i_N,i}2k^N_{i_N}k_i/t}\prod_{i_N}(1-\tilde{w}^N_{i_N})^
  {2k^N_{i_N}j_N/t}z_N^{\sum_{i_N}2k^N_{i_N}j_N/t}\nn
 &\cdot&\prod_{i_N,n}(\tilde{w}^N_{i_N})^{2k^N_{i_N}j_n/t}
  z_N^{\sum_{i_N,n}2k^N_{i_N}j_n/t}
  z_N^{\sum_n2j_Nj_n/t}z_N^{\sum_i2j_Nk_i/t}\nn
 &\cdot&u_N^{2j_N-1}e^{1/u_N}\Gamma(2j_N+1)
  \dtp{u_N}W_N^{(I)}\prod_{i_N}d\tilde{w}^N_{i_N}
  z_N^{\rho_N+\sigma_N}
\eea
(with minor misprint corrected, here and in (\ref{WNII2}), 
compared to \cite{PRY1}). We may evaluate the total power of $z_N$ as
\bea
 &&\frac{2}{t}j_N(j_N+1)-2j_N+\frac{1}{t}{[}(\sum_{i_N}k^N_{i_N})^2-\sum_{i_N}
  (k^N_{i_N})^2{]}+\frac{2}{t}(-2j_N)(-\sum_nj_n+j_N)\nn
 &+&\frac{2}{t}j_N(-2j_N)+\frac{2}{t}(-2j_N)\sum_nj_n+\frac{2}{t}j_N\sum_nj_n
  +\frac{2}{t}j_N(-\sum_nj_n+j_N)+\rho_N+\sigma_N\nn
 &=&0
\eea
which merely shows that the intertwining field, 
${[}\phi_{j_N}(z_N,x_N){]}^0_{j_N}$ indeed has the right scaling dimension
in this formalism. Similarly the total power of zero for $x_N$ says that we 
treat the field with correct global $SL(2)$ properties.

Furthermore, the $u_N$-integrand becomes trivial, involving only
\ben
 u_N^{-1}e^{1/u_N}
\een
while the dependence on $\tilde{w}^N_{i_N}$ becomes
\ben
 \prod_{i_N}(\tilde{w}^N_{i_N})^{2k^N_{i_N}j_N/t}
  \prod_{i_N}(1-\tilde{w}^N_{i_N})^{k^N_{i_N}(2j_N+t)/t}\prod_{i_N<j_N}
  (\tilde{w}^N_{i_N}-\tilde{w}^N_{j_N})^{2k^N_{i_N}k^N_{j_N}/t}
\een
This yields an integral over the $\tilde{w}^N_{i_N}$'s which is 
independent of the remaining parameters of the correlator, except $j_N$, and
thus merely contributes to the normalisation of the state $\bra{j_N}$.
In conclusion we note that the alternative description (\ref{WII})
(in the appropriate limits) with the initial and final states being vacua,
is more symmetric and does not involve dual states (besides the bra-vacuum),
but in general it involves a larger number of screening integrations.

\subsection{The Knizhnik-Zamolodchikov Equations}
One may wonder whether the rules for contractions we have put forward, really
reproduce the structure of the conformal field theory. In order to settle this 
question in the affirmative we provide in this section an explicit proof 
\cite{PRY1} that
the $N$-point functions satisfy the Knizhnik-Zamolodchikov equations. In this
proof we should not, therefore, make any use of the rules of contractions.

The KZ equation (\ref{KZ1}) corresponding to the primary field at 
position $z_{m_0}$ may be written as 
\ben
 \left(t\pa_{z_{m_0}}+\sum_{m\neq m_0}
  \frac{D^+_{x_{m_0}}D^-_{x_m}+2D^3_{x_{m_0}}D^3_{x_m}
  +D^-_{x_{m_0}}D^+_{x_m}}{z_m-z_{m_0}}\right)W_N=0
\label{KZ}
\een
where $W_N$ is the $N$-point function after requisite integrals have been 
performed.

The structure of the proof is as follows.
For the selected position $z_{m_0}$ of the primary field at that position, we
shall define a function
\ben
 G(w)=\frac{1}{w-z_{m_0}}\left(D^+_{x_{m_0}}G^-(w)+2D^3_{x_{m_0}}G^3(w)+
  D^-_{x_{m_0}}G^+(w)\right)
\label{Gw}
\een
where the $G^a(w)$'s are functions to be defined and will turn out 
{\em a posteriori} to be
\ben
 G^a(w)=\br J^a(w){\cal O}\kt
\een
where ${\cal O}$ is the collection of free field realizations of all the
primary fields and screening charges. Indeed we shall evaluate
the $G^a(w)$'s using our contraction rules from that idea. However, the point 
about the proof is that the function $G(w)$ eventually written down 
(\ref{GKZ}) will only 
have {\em pole} singularities as a function of $w$, and will behave as 
${\cal O}(w^{-2})$
for $w\rightarrow\infty$, and thus the sum of residues will 
vanish. What we shall show explicitly is that the vanishing condition for 
this sum of residues is precisely the KZ equation on the 
$N$-point function. This should come as no surprise since this is merely the 
standard technique for proving the KZ equations. The point
is that in the standard proof one makes use of associativity properties of
the operators, and the purpose of the proof is exactly to establish that the
rules for contractions in fact conform to those.

Now it is clear how we build the functions $G^a(w)$. To use the contraction 
rules with the free field realizations (\ref{sl2conv2}), is very easy in our 
correlator, since one may establish the rules
\bea
 \beta(w)&\rightarrow&B(w)\nn
 \gamma(w)&\rightarrow&-\sum_{i=1}^M\frac{D_{B_i}}{w-w_i}\nn
  -\sqrt{t/2}\pa\varphi(w)&\rightarrow&\sum_{i=1}^M\frac{k_i}{w-w_i}+
  \sum_{m=1}^{N-1}\frac{j_m}{w-z_m}
\eea
where
$D_{B_i}=\frac{\pa}{\pa B(w_i)}$ and $k_i$ is defined as before.
Then we have (the notation here does not distinguish between integrands and 
integrated expressions)
\bea
 G^+(w)&=&B(w)W_BW^\varphi_NF\nn
 G^3(w)&=&\left(B(w)\sum_{i=1}^M\frac{D_{B_i}}{w-w_i}
  +\sum_{i=1}^M\frac{k_i}{w-w_i}
  +\sum_{m=1}^{N-1}\frac{j_m}{w-z_m}\right)W_BW^\varphi_NF\nn
 G^-(w)&=&\left(-B(w)\sum_{i,j=1}^M\frac{D_{B_i}D_{B_j}}{(w-w_i)(w-w_j)}
  +(t-2)\sum_{i=1}^M\frac{D_{B_i}}{(w-w_i)^2}\right.\nn
 &-&\left.2\sum_{i,j=1}^M\frac{k_iD_{B_j}}{(w-w_i)(w-w_j)} - 
  2\sum_{m,j=1}^{N-1,M}\frac{j_mD_{B_j}}{(w-z_m)(w-w_j)}\right)W_BW^\varphi_NF
\label{GKZ}
\eea
These expressions {\em define} the function $G(w)$ and from 
now on we may completely forget that they came from applying the contraction 
rules to certain correlators.

In the following calculations the structure of the auxiliary $u$ integrations
turns out to be very crucial. In fact, any dependence on an $x_m$ is via
the combination $x_m/u_m$ so that we may write
\ben
 x_m\pa_{x_m}=-u_m\pa_{u_m}
\een
and subsequently do a partial integration in  $u_m$, writing effectively
\ben
 x_m\pa_{x_m}W_BW^\varphi_NF\sim W_BW^\varphi_N\pa_{u_m}(u_mF)
\een
Let 
\ben
 D^-_{x_{m_0}}G^+_{z_{m_0}}=D^-_{x_{m_0}}\oint_{z_{m_0}}\dtp{w}
  \frac{1}{w-z_{m_0}}G^+(w)
\een
denote the contribution to the pole residue 
in $G(w)$ at $w=z_{m_0}$ coming from the term 
$D^-_{x_{m_0}}G^+(w)/(w-z_{m_0})$, etc. 
Then we find after some calculations for the pole at $z_{m_0}$
\bea
 D^-_{x_{m_0}}G^+_{z_{m_0}}&=&\sum_{\ell\neq m_0}
  \frac{x_\ell/u_\ell}{z_{m_0}-z_\ell}\pa_{x_{m_0}}W_B\cdot W^\varphi_NF\nn
 2D^3_{x_{m_0}}G^3_{z_{m_0}}&=&2\left(-\sum_{\ell\neq m_0}\frac{x_\ell/u_\ell}
  {z_{m_0}-z_\ell}u_{m_0}\pa_{x_{m_0}}W_B\cdot W^\varphi_N\right.\nn
 &-&\left.\pa_{z_{m_0}}W_B\cdot W^\varphi_N
  +\frac{t}{2j_{m_0}}W_B\cdot\pa_{z_{m_0}}W^\varphi_N\right)
  (-j_{m_0}+u^{-1}_{m_0})F\nn
 D^+_{x_{m_0}}G^-_{z_{m_0}}&=&\left(-\sum_{\ell\neq m_0}\frac{x_\ell/u_\ell}
  {z_{m_0}-z_\ell}\pa_{x_{m_0}}W_B\cdot 
  W^\varphi_N\left(
  2j_{m_0}u_{m_0}^{2j_{m_0}}-u_{m_0}^{2j_{m_0}-1}\right)\right.\nn
 &-&2\pa_{z_{m_0}}W_B\cdot W^\varphi_N\left((2j_{m_0}-1)u_{m_0}^{2j_{m_0}-1}-
  u_{m_0}^{2j_{m_0}-2}\right)\nn
 &+&(t-2+2j_{m_0})\pa_{z_{m_0}}W_B\cdot W^\varphi_N u_{m_0}^{2j_{m_0}-1}\nn
 &+&\left.\frac{t}{j_{m_0}}W_B\cdot \pa_{z_{m_0}}W_N^{\varphi}
  \left(2j_{m_0}u_{m_0}^{2j_{m_0}-1}-u_{m_0}^{2j_{m_0}-2}\right)\right)\nn
 &\cdot&e^{1/u_{m_0}}\Gamma(2j_{m_0}+1)\prod_{m\neq m_0}\Gamma(2j_m+1)
  u_m^{2j_m-1}e^{1/u_m}
\eea
This sums up to become
\ben
 t\pa_{z_{m_0}}W_N
\een
which is the first term in the KZ equation.

In a similar fashion the pole residue at $w=w_j$ may be calculated. However,
here we shall be more general and show that substituting $1/(w-z_{m_0})$ 
in the definition (\ref{Gw}) of $G(w)$ by any function $f(w)$
(free to depend on e.g. $z_m$ but independent of $w_j$), leads to
\bea
 G_f(w)&=&f(w)\left(D^+_{x_{m_0}}G^-(w)+2D^3_{x_{m_0}}G^3(w)+
  D^-_{x_{m_0}}G^+(w)\right)  \nn
 \oint_{w_j}\dtp{w}G_f(w)&=&D_{x_{m_0}}^+\oint_{w_j}\dtp{w}f(w)G^-(w)\nn
 &=&t \pa_{w_j}\left(D_{x_{m_0}}^+
  \left(f(w_j)\frac{W_BW^\varphi_NF}{B(w_j)}\right)\right)
\eea 
The first equality in the computation of the residue is easily verified,
while the second goes as follows
\bea 
 \oint_{w_j}\dtp{w}f(w)G^-(w)&=&
  \left(-2B(w_i)f(w_j)\sum_{i\neq j}\frac{D_{B_i}D_{B_j}}{(w_j-w_i)}
  -\pa_w{[}f(w)B(w){]}_{w_j}D^2_{B_j}\right.\nn
 &+&(t-2)f'(w_j)D_{B_j}-2\sum_{i\neq j}\frac{k_iD_{B_j}
  +k_jD_{B_i}}{w_j-w_i}f(w_j)-2f'(w_j)k_jD_{B_j}\nn
 &-&\left.2\sum_m\frac{j_mD_{B_j}}{w_j-z_m}f(w_j)\right)W_BW^\varphi_N F\nn
 &=&t \pa_{w_j}
  \left(f(w_j)\frac{W_BW^\varphi_NF}{B(w_j)}\right)
\label{G-f}
\eea
where we have used that 
\bea
 \pa_{w_j}W^\varphi_N&=&\left(\sum_m \frac{2j_mk_j/t}{w_j-z_m}+
  \sum_{i\neq j}\frac{2k_ik_j/t}{w_j-w_i}\right)W^\varphi_N\nn
  \pa_{w_j}W_B&=&-\frac{k_j}{B(w_j)}B(w_{j'})W_B
\eea
and that for both $k_j=-1$ and $k_j=t$
\ben
 k_j(k_j+1)-tk_j=t
\een
Putting $f(w)=1/(w-z_{m_0})$, we may now conclude that upon integration over 
the positions $w_j$, the contribution from the poles at $w_j$ vanishes.

Finally, the pole residues at the points $w=z_m\neq z_{m_0}$ 
give the remaining terms in the KZ equation (\ref{KZ}). Since there
are no other singularities in $w$, the KZ equation has been proven.

The above proof assumes that the integration contours of the screening 
charges are  closed. When parameters are such that this effectively is not 
the case, it is no longer true that the contribution from the total 
derivatives met at $w=w_j$ may be thrown away. However, in that case the 
derivative after $z_{m_0}$ contains an extra term coming from varying the 
end point of the integration contour, and it is possible to check that this 
extra contribution exactly cancels the contribution from the total derivative
term, so that the validity of the KZ equations remains true in those cases as
well.

\subsection{Projective Invariances}
In \cite{CF} it is shown, that solutions to the KZ 
equations are projectively invariant provided the primary fields can add up
to a singlet. In our case this is the requirement of global 
(loop projective) $SL(2)$ invariance.
Let us restrict to the case where the initial and final bra and ket
carry just the vacuum and the dual vacuum: $j_1=0=j_N$, i.e. we are really 
looking at an $(N-2)$-point function. 
Then global $SL(2)$ invariance is the statement that
\ben
\sum_{m=2}^{N-1}D^a_{x_m}W_{N-2}=0
\een
This is equivalent to the statement that in
\ben
G^a(w)=\br J^a(w){\cal O}\kt
\een
the leading behaviour as $w\rightarrow\infty$ is ${\cal O}(w^{-2})$ rather than
${\cal O}(w^{-1})$. From the expressions above, that is trivial for $G^-(w)$,
and for $G^3(w)$ it follows from the fact that
\ben
\sum_{i=1}^Mk_i=-\sum_{m=1}^{N-1}(\rho_m-t\sigma_m)=-\sum_{m=1}^{N-1}j_m
\een
for $j_N=0$.
For $G^+(w)$ it is more complicated to see. As previously discussed this is 
related to the fact that we are not using the projectively invariant vacuum
$\bra{sl_2}$ in our calculations, rather we are using the dual vacuum 
$\bra{0}$ for which $\bra{0}J^+_0=\bra{0}\beta_0\neq 0$.

What we are going to show \cite{PRY1} is that the state 
\ben
 \bra{0}\beta_0
\een
even though it is non-vanishing, is BRST exact in the sense of Felder \cite{F},
and that (hence) it decouples from correlators of BRST invariant operators.

First let us argue at the operator level, and subsequently at the level of our
correlators. We write
\ben
 \bra{0}\beta_0=\bra{0}e^{-\sqrt{2/t}q_\varphi}e^{\sqrt{2/t}q_\varphi}\beta_0
  =\bra{-1}\oint\dtp{z}:e^{\sqrt{2/t}\varphi(z)}:\beta(z)
\label{b0exact}
\een
where the bra state $\bra{-1}$ is the lowest weight state $\bra{j=-1}$
(\ref{dualbra}). Now this integral is in fact the appropriate BRST operator in 
Felder's formulation.
To see this, recall, that acting on the Fock space pertaining to $j_{r,s}$
and labelled $F_{r,s}$, the relevant BRST operator 
\cite{F} (for which the BRST current
is single valued) is 
\ben
 Q_r\sim \oint\dtp{v_0}...\dtp{v_{r-1}}S_1(v_0)...S_1(v_{r-1})
\een
with
\ben
 Q_r: \ F_{r,s}\mapsto F_{-r,s}
\een
For $j_N$ =0, this is the Fock space with $r=1,s=0$ ($2j_N+1=1-0\cdot t$). 
Therefore the relevant BRST operator on this space is $Q_1$ which is just the
one we obtained (\ref{b0exact}).

Next let us see how the argument works at the level of the correlator. We are
going to show that inserting the operator
\ben
 \beta_0=\oint_\infty\dtp{w}\beta(w)
\een
furthest to the left in a correlator with $j_N=0$, is equivalent to inserting 
the BRST charge operator
\ben
 \oint_\infty\dtp{w}\beta(w):e^{\sqrt{2/t}\varphi(w)}:
\een
furthest to the left of the operators and to the right of the bra 
$\bra{j_N=-1}$.
Indeed in the first case, using the rules for building correlators, we obtain
(up to normalisation)
\bea
 &&\oint_\infty
  \dtp{w}B(w)\prod_iB(w_i)^{-k_i}\prod_{i<j}(w_i-w_j)^{2k_ik_j/t} \nn
 &\cdot&\prod_{n<m}(z_n-z_m)^{2j_nj_m/t}\prod_{i,m}(w_i-z_m)^{2k_ij_m/t}
  \prod_m u_m^{2j_m-1}e^{1/u_m}
\eea
In the second case, we notice that the state labelled $\bra{-1}$ is exactly
the lowest weight state $\bra{j_N=-1}$ and we are formally looking
at an $(N-1)$-point function with $j_N=-1$. Since we have
\ben
 \iota_{N-2}+j_{N-1}-j_N=\rho_{N-1}-\sigma_{N-1}t
\een
we see that the value of $\rho_{N-1}$ will be one unit larger in the case
$j_N=-1$ than in the case $j_N=0$. This means we have one extra screening
charge of the first kind in that case, which we may 'lift off' the intertwining
field furthest to the left, in other words that calculation will just be the 
one we seek to carry out.
Using the rules developed we find in that case, letting $w$ denote the
position of the extra screening operator compared to the previous case (so in
the formulas below, the index $i$ runs over exactly the same set as before)
\bea
 &&\oint_\infty\dtp{w}B(w)\prod_i B(w_i)^{-k_i}\prod_i (w-w_i)^{-2k_i/t}
  \prod_n(w-z_n)^{-2j_n/t}\prod_{i<j}(w_i-w_j)^{2k_ik_j/t}\nn
 &\cdot&\prod_{n<m}(z_n-z_m)^{2j_nj_m/t}\prod_{i,n}(w_i-z_n)^{2k_ij_n/t}
  \prod_nu_n^{2j_n-1}e^{1/u_n}\nn
 &=&\oint_\infty\dtp{w}B(w)\prod_i B(w_i)^{-k_i}\prod_i (1-w_i/w)^{-2k_i/t}
  \prod_n(1-z_n/w)^{-2j_n/t}  w^{-2(\sum_ik_i+\sum_nj_n)/t}\nn
 &\cdot&\prod_{i<j}(w_i-w_j)^{2k_ik_j/t}\prod_{n<m}(z_n-z_m)^{2j_nj_m/t}
  \prod_{i,n}(w_i-z_n)^{2k_ij_n/t}\prod_n u_n^{2j_n-1}e^{1/u_n}
\eea
We now use that
\ben
 -\sum_ik_i =\sum_n\rho_n-t\sum_n\sigma_n=\sum_nj_n-j_N
\een
Since we use the notation that sums over $i$ and $n$ are pertaining to the
case with $j_N=0$, we see that the extra power of $w$ becomes zero, and for
a very large integration contour for $w$, all $w$ dependence drops out except
for the one in $B(w)$ so that we exactly prove the identity of the two cases
also at the level of our correlators.

Having come this far, we may move the Felder type BRST operator through
all intertwining fields in exactly the same way as for minimal models
\cite{F}, until in the end it hits the ket-vacuum. 
Since we are only using screening operators of
the first kind in the BRST operators, the procedure will work just as for the
minimal models.

In addition it is rather easy to verify that the functions $G^a(w)$ have
the expected pole residues for $\br J^a(w){\cal O}\kt$ at points
$w=z_m$. Further one verifies that there are no pole residues in $G^+(w)$
and $G^3(w)$ for $w=w_i$, the position of a screening charge. For $G^-(w)$
that residue is proportional to the total derivative
\ben
 \pa_{w_i}\left (D_{B_i}W_B\cdot W^\varphi_N\right)
\een
All of those remarks establish that the $N$-point chiral 
blocks have the correct projective and global $SL(2)$ invariance properties.

\subsection{Possible Hybrid}

This section is devoted to a very brief discussion of a generalisation
\cite{PRY1} of a work by Malikov {\em et al} \cite{FIM} in which a class
of solutions to the KZ equations are considered. 

Without going into the precise mathematical language employed in \cite{FIM}
it is easy enough to describe the approach by these authors in terms of ours.
Given some number of primary fields characterised by spins 
$j_1,j_2,...,j_{N-1}$
they begin by supplementing this set by an auxiliary primary field of spin
\ben
 j_N^{(0)}=\sum_{i=1}^{N-1}j_i
\een
carefully chosen so that no screening charges are required in the free field 
approach. In general, writing
\ben
 \rho-\sigma t=j_N^{(0)}-j_N
\een
our approach is applicable whenever $\rho$ and $\sigma$ are non-negative 
integers, in which case they represent the number of screening charges needed
of the two kinds. Obviously now, for $j_N=j_N^{(0)}$ no screening charges are 
needed, then the correlators are independent of the $x_i$'s and may be 
trivially written down. This expression is the starting point in \cite{FIM}. 
{}From this a larger and more interesting class of solutions to the KZ 
equations is obtained as follows. From the state $\ket{j_{r,s}}$ it is 
well-known (\ref{singaffine})
how to build the singular vectors in the Verma module generated by this
highest weight state, simply by multiplying $\ket{j_{r,s}}$ by certain
products of generators. Thus when this
operator is inserted into a correlator the state decouples and the correlator
vanishes. The trick in \cite{FIM} is to only make use of a subset of the 
operators, starting from the right and truncating the expansion at some point
to the left. Each factor performs a Weyl reflection and it may easily be shown
that any subset generates a formally singular state, although of course not
one that lies in the Verma module of $\ket{j_{r,s}}$. Hence, inserting
such a reduced set of operators will not lead to decoupling, but instead to
a new solution of the KZ equations \cite{FIM}.

What we shall argue now \cite{PRY1}, however, 
is that this procedure can never exhaust
the possibilities we wish to consider in conformal field theory. The point is
that when this procedure is applied to the state $\bra{j_N^{(0)}}$ the new
spin value for the state may only be modified relative to $j_N^{(0)}$ in a
way which is insufficient for producing all the spins needed by the 
fusion rules in a theory based on admissible representations. Indeed keeping
$m$ operators results in a new spin $j_N^{(m)}$ given by
\bea
 j_N^{(2n)}&=&j_N^{(0)}+nt\nn
 j_N^{(2n+1)}&=&j_N^{(0)}-r_N^{(0)}+(s_N^{(0)}-n)t
\eea
for $m$ even and odd respectively. 
We see that even though $j_N$ is not required
to be equal to $j_N^{(0)}$ the possibilities are too limited to allow 
considerations of all the relevant spin values in an admissible representation:
relative to $j_N^{(0)}$ only the $s$ value can be changed at will, 
the $r$ value
is fixed either at its value in $j_N^{(0)}$ or at minus that value. The above 
spins may be treated also in the formalism based on free fields and screening
operators. Indeed the number of screening operators are given by
\bea
 \rho_{2n}&=&0\nn
 \sigma_{2n}&=&n\nn
 \rho_{2n+1}&=&r_N^{(0)}\nn
 \sigma_{2n+1}&=&s_N^{(0)}-n
\eea
This is the number of integrations in the integral formulas (in addition to the
simpler ones having to do with $u$ integrations).

A possibly more interesting hybrid treatment 
based on the free field realization and \cite{FIM} might
be possible. In fact, if instead of starting with correlators needing no 
screening charges, one starts with correlators needing only the first kind of
screening charge for which extensive studies exist in the literature
(see e.g. \cite{BF, D90}), it would seem that the techniques of \cite{FIM} 
could provide a way alternative to the one discussed in the preceding sections 
to circumventing the use of the second screening operator. The result would
be a strange mixture of free field integral expressions and the ones considered
in \cite{FIM}, but it should agree with the ones given earlier up to 
normalisation. It is not clear 
that such  hybrid integral representations have advantages over what has 
otherwise been presented here, 
but at least they would also be complete, contrary to the ones in \cite{FIM}.

\section{Hamiltonian Reduction}
Here we want to discuss \cite{PRY1a} how conformal blocks in $SL(2)$ 
current algebra reduce to corresponding ones in minimal models. 

A general conformal block (on the sphere) in the affine theory is given by
\ben
 W_N=\bra{j_N}\phi_{j_{N-1}}(z_{N-1},x_{N-1})
  ...\phi_{j_{n}}(z_{n},x_n)...
  \phi_{j_{2}}(z_2,x_2)\prod_i \int dw_iS_{k_i}(w_i)\ket{j_1}
\label{wn}
\een
Different choices of integration contours for the screening charges define
different intertwining chiral vertex operators and different
conformal blocks. The two screening charges are
\bea
 S_{k_\pm}(w)&=&\beta^{-k_\pm}(w)S^\varphi_{k_\pm}(w)\nn
 S^\varphi_{k_\pm}(w)&=&:e^{-k_\pm\sqrt{2/t}\varphi(w)}:\nn
 k_+&=&-1\nn
 k_-&=&t
\label{screen}
\eea
The relation to minimal models is obtained by writing \cite{BO}
\bea
 2j_{r,s}+1&=&r-st\nn
 t&=&k+2=p/q \nn
 \alpha_+&=& \sqrt{\frac{2}{t}}=-2/\alpha_- \nn
 \alpha_{r,s+1}&=&-j_{r,s}\sqrt{\frac{2}{t}}
  =\frac{1}{2}((1-r)\alpha_+-s\alpha_-)\nn
 2\alpha_0&=&\alpha_++\alpha_-\nn
 \D_{r,s+1}&=& \frac{j_{r,s}(j_{r,s}+1)}{t}-j_{r,s}
  =\frac{1}{2}\alpha_{r,s+1}(\alpha_{r,s+1}-2\alpha_0)\nn
 \phi_{r,s+1}(z)&=& :e^{\alpha_{r,s+1}\varphi(z)}:
  =\phi_{j_{r,s}}(z)\nn
 V_{\alpha_{\pm}}(w)&=& :e^{\alpha_{\pm}\varphi(w)}:=S_{k_\pm}^{\varphi}(w)
\eea
It is now clear that if one truncates the $\beta$ dependence of the screening
currents and the $\gamma$ dependent factor in the primary fields, then the
minimal model correlators are obtained \cite{FGPP}. 
This is true despite the fact 
that the two theories, the WZNW model and the minimal model, have different
background charges for the $\varphi$ field:  namely 
$-\alpha_+=-\sqrt{\frac{2}{t}}$ (see (\ref{hermaj})) for the WZNW model and 
$-2\alpha_0=-\sqrt{\frac{2}{t}}+\sqrt{2t}$ for the minimal models
(see Chapter 2). However, this difference is of
no consequence in the practical evaluation of the free field correlators since
in both cases suitable dual bra-states are used to absorb those
background charges. In the following, we want to discuss how the Hamiltonian
reduction works at the level of correlators. Thereby we prove \cite{PRY1a}
the statement in \cite{FGPP} that the correlators in $SL(2)$ current algebra
reduce to corresponding ones in a particular minimal model in the limit 
where all $x$'s are put equal to the corresponding $z$'s.

\subsection{At the Level of Correlators}
The primary field (\ref{sl2conv2}) may be written as
\bea
 \phi_j(z,x)&=&e^{x\partial_y}\phi_j(z,y)|_{y=0}\nn
 &=&e^{xD_y^-}\phi_j(z,y)|_{y=0}\nn
 &=&e^{xJ^-_0}\phi_j(z,0)e^{-xJ^-_0}\nn
 &=&e^{xJ^-_0}:e^{-j\sqrt{2/t}\varphi(z)}:e^{-xJ^-_0}
\label{pfd2}
\eea
Here there is a subtlety in that 
the way the exponential function $e^{xJ^-_0}$ should be expanded,
must also respect the monodromy conditions in the $x$ variables. We shall come 
back to this subtlety presently.
We may further write
\bea
 \phi_j(z,x z)&=&e^{zx J^-_0}:e^{-j\sqrt{2/t}\varphi(z)}:e^{-zx J^-_0}\nn
 &=&e^{x zD^-_y}\phi_j(z,y)|_{y=0}\nn
 &=&e^{x J^-_1}:e^{-j\sqrt{2/t}\varphi(z)}:e^{-x J^-_1}
\label{pfd3}
\eea
and
\ben
 \phi_j(z,x)=e^{xJ_0^-}e^{zL_{-1}}:e^{-j\sqrt{2/t}\var(0)}:e^{-xJ_0^-}e^{-z
  L_{-1}}
\label{phi0}
\een
Consider the following conformal blocks
\ben
 W_N=\bra{j_N}\phi_{j_{N-1}}(z_{N-1},x z_{N-1})
  ...\phi_{j_{n}}(z_{n},x z_n)...
  \phi_{j_{2}}(z_2,x z_2)\prod_i \int dw_iS_{k_i}(w_i)\ket{j_1}
\een
Substituting (\ref{pfd3}) we may rewrite this as
\ben
 W_N=\bra{j_N}e^{x J^-_1}\phi_{j_{N-1}}(z_{N-1},0)
  ...\phi_{j_{n}}(z_{n},0)...\phi_{j_{2}}(z_2,0)
  \prod_i \int dw_iS_{k_i}(w_i)\ket{j_1}
\label{wn1}
\een
since $J_1^-\ket{j_1}=0$.
At this point however, there is a subtlety as to how adjacent exponentials
$e^{\pm x J^-_1}$ should be removed, and
we should examine how these exponentials are defined. Indeed as discussed
earlier the expansions of exponentials and other functions involving
the $\beta$ and $\gamma$ fields, depend on which monodromy the problem at hand
requires one to select. All these subtleties are dealt with using the following
two lemmas \cite{PRY1a}\\[.2 cm]
{\bf Lemma 1}\\
If the fractional part in powers of $x$ is $\alpha $, then we can expand
the last expression in (\ref{pfd2})
\ben
 e^{xJ^-_0}:e^{-j\sqrt{2/t}\varphi(z)}:e^{-xJ^-_0}=
  \sum_{n\in \Z}\frac{(xJ^-_0)^{\alpha-\beta+n}}{(\alpha-\beta +n)!}
  :e^{-j\sqrt{2/t}\varphi(z)}:
  \sum_{m\in \Z}\frac{(-xJ^-_0)^{\beta+m}}{(\beta +m )!}
\label{epd}
\een
for arbitrary complex number $\beta$.\\[.2 cm]
{\bf Lemma 2}
\ben
 1=\sum_{n\in \Z}\frac{(xJ^-_0)^{\alpha+n}}{(\alpha +n) !}
  \sum_{m\in\Z}\frac{(-xJ_0^-)^{-\alpha +m}}{(-\alpha +m)!}
  =e^{xJ^-_0}e^{-xJ^-_0}
\een

Before proving these lemmas we make the following remarks. Define
\bea
 \phi^{[n]}_j(z,0)&=&[J^-_0,\phi^{[n-1]}_j(z,0)]\nn
 \phi^{[0]}_j(z,0)&=& \phi_j(z,0)
\label{comm}
\eea
When $x$ is integrally powered, it is clear that we can expand $\phi_j(z,x)$ as
\bea
 e^{xJ^-_0}:e^{-j\sqrt{2/t}\varphi(z)}:e^{-xJ^-_0}
  &=&\sum_{n\geq 0}\frac{\phi^{[n]}_j(z,0)x^n}{n!}\nn
 &=&\sum_{n\geq 0}\frac{(D_y^-)^n\phi_j(z,y)x^n}{n!}|_{y=0}\nn
 &=&e^{xD_y^-}\phi_j(z,y)|_{y=0}\nn
 &=&\phi_j(z,x)
\label{scomm}
\eea
However, when $x$ is fractionally powered, we can no longer Taylor expand
$\phi_j(z,x)$, and the definition for both $\phi^{[n]}_j(z,0)$ in (\ref{comm})
and $(D_y^-)^n\phi_j(z,y)|_{y=0}$ in (\ref{scomm}) requires specification.
It is possible to generalise (\ref{comm}) and (\ref{scomm}) by defining
\ben
 \frac{\phi^{[N+\alpha+\beta]}_j(z,0)}{(N+\alpha+\beta)!}=
  \sum_{{n+m=N \atop n,m\in \Z}}\frac{(J^-_0)^{\alpha +n}}{(\alpha +n)!}
  :e^{-j\sqrt{2/t}\varphi(z)}:\frac{(-J_0^-)^{\beta+m}}{(\beta+m)!}
\label{comm1}
\een
Indeed, this is seen to satisfy the recursion relation in (\ref{comm}).
Although it looks like that the rhs of (\ref{comm1}) depends on both 
$\alpha$ and $\beta$, lemma 1 essentially means that it only depends on the 
combination $\alpha +\beta$.
The fractional derivatives at the origin may also be considered as analytical
continuations of their integral counterparts. Now
$\phi_j(z,x)=(1+\gamma(z)x)^{2j}:e^{-j\sqrt{2/t}\varphi(z)}:$, so for
non-negative integer $n$ we have
\ben
 \frac{(D^-_y)^n}{n!}\phi_j(z,y)|_{y=0}
  =\binomial{2j}{n}\gamma^n(z):e^{-j\sqrt{2/t}\varphi(z)}:
\een
We can analytically continue the variable $n$ in the above equation from
integers to complex numbers. Therefore $n$ could be any fractional
number and we have
\ben
 \frac{(D^-_y)^{n+\alpha}}{(n+\alpha)!}\phi_j(z,y)|_{y=0}=
  \binomial{2j}{n+\alpha}\gamma^{n+\alpha}(z):e^{-j\sqrt{2/t}\varphi(z)}:
\een
{\bf Proof of Lemma 1}
\bea
 && \sum_{n\in \Z}\frac{(xJ^-_0)^{\alpha-\beta+n}}{(\alpha-\beta +n)!}
  :e^{-j\sqrt{2/t}\varphi(z)}:
  \sum_{m\in \Z}\frac{(-xJ^-_0)^{\beta+m}}{(\beta +m )!}\nn 
 &=&\sum_{n\in \Z}\frac{(xJ^-_0)^{\alpha-\beta+n}}{(\alpha-\beta +n)!}
  \sum_{m\in \Z}\frac{(-xJ^-_0 + xD^-_y)^{\beta+m}}{(\beta +m) !}
  \phi(z,y)|_{y=0}\nn 
 &=&\sum_{N\in \Z}\sum_{m\in \Z}
  \frac{(xJ^-_0)^{\alpha-\beta+N-m}(-xJ^-_0 + xD^-_y)^{\beta+m}}
  {(\alpha-\beta +N-m)! (\beta +m )!}\phi(z,y)|_{y=0}\nn 
 &=&\sum_{N\in \Z}\frac{(xD^-_y)^{\alpha +N}}
  {(\alpha +N) !}\phi(z,y)|_{y=0}\nn
 &=&\sum_{N\in \Z}
  \binomial{2j}{N+\alpha}(\gamma(z)x)^{N+\alpha}:e^{-j\sqrt{2/t}\varphi(z)}:\nn
 &=&\phi_j(z,x)
\eea
$\Box$\\
{\bf Proof of Lemma 2}
\bea
 e^{xJ^-_0}e^{-xJ^-_0}&=&
  \sum_{n\in \Z}\frac{(xJ^-_0)^{\alpha+n}}{(\alpha +n) !}
  \sum_{m\in \Z}\frac{(-xJ^-_0)^{-\alpha+m}}{(-\alpha +m )!}\nn
 &=&\sum_{N\in \Z}\sum_{n+m=N}\frac{(xJ^-_0)^{N}}{(\alpha +n) ! (-\alpha +m) !}
  (-1)^{-\alpha+m}\nn
 &=&\sum_{N\in \Z}(xJ^-_0)^{N}\delta_{N,0}\nn
 &=& 1
\eea
$\Box$\\ 
Thus the manipulations leading to (\ref{wn1}) are justified. In the
proofs of the lemmas we have used versions of (\ref{1plusz2j}).

We may now go back to (\ref{wn1}). We see from that expression that the 
exponential 
\ben
 e^{x J^-_1}
\een
has to be expanded in powers offset from the integers by the amount
\ben
 -\sum k_i=r-st=\alpha
\een
with $r$ and $s$ the number of screening charges of the first and second
kind respectively, so this
is the combined power of the $\beta$ factors. Since the $\gamma$ factors
from the primary fields decouple for $x_i=0$, this particular expansion of the
exponential is required. One may then work out that \cite{PRY1a} 
(here corrected for minor misprint)
\ben
 \bra{j_N}e^{x J^-_1}
  =\bra{j_N}(1-x\gamma_1)^{k-2j_N} \frac{s(k-2j_N)}{s(k-2j_N-\alpha)}
  (-1)^{-\alpha}          
\label{jNexp}
\een
Notice that for $\alpha$ integer the ratio of sine-factors along with the
phase disappear. The result (\ref{jNexp}) is obtained  by writing 
(for $\alpha = r-st$)
\ben
 e^{xJ^-_1}=\sum_{n\in\Z}\frac{(xJ^-_1)^{n+\alpha}}{(n+\alpha)!}
\label{jNexp2}
\een
and observing that for any power
\ben
 \bra{j_N}(J^-_1)^{n+\alpha}=\bra{j_N}\gamma_1^{n+\alpha}
  \frac{\Gamma(2j_N-k+n+\alpha)}{\Gamma(2j_N-k)}
\label{jNexp3}
\een
The proportionality to a power of $\gamma_1$ follows from the free field 
realization and the properties of the vacuum. The 
value of the constant is obtained by consistency between $n+\al$ and $n+\al+1$ 
and by normalising with the result for $n+\alpha=1$.

We may now continue the calculation and obtain
\bea
 W_N&=&\bra{j_N}(1-x\gamma_1)^{k-2j_N}\frac{s(k-2j_N)}
  {s(k-2j_N-\alpha)}(-1)^{-\alpha}\nn
 &\cdot&\phi_{j_{N-1}}(z_{N-1},0)
  ...\phi_{j_{n}}(z_{n},0)...\phi_{j_{2}}(z_2,0)
  \prod_i \int dw_iS_{k_i}(w_i)\ket{j_1}\nn
 &=&C_N(\{j_m\},x)\bra{j_N}\phi_{j_{N-1}}(z_{N-1},0)
  ...\phi_{j_{n}}(z_{n},0)...\phi_{j_{2}}(z_2,0)
  \prod_i \int dw_iS_{k_i}^{\varphi}(w_i)\ket{j_1}\nn
 &=&C_N(\{j_m\},x)W_N^{\varphi}
\eea
where $C_N(\{j_m\},x)$ is the normalisation constant, and $W_N^{\varphi}$ is 
exactly the free field expression for the minimal model correlator
\ben
 W_N^{\varphi}=\bra{\D_{r_N,s_N+1}}\phi_{r_{N-1},s_{N-1}+1}(z_{N-1})
  ...\phi_{r_2,s_2}(z_2)
  \prod_i \int dw_iV_{\alpha_i}(w_i)\ket{\D_{r_1,s_1+1}}\nn
\een
The point is that since (\ref{jNexp})
now contains the only $\gamma$ dependence of the correlator, the 
$\beta$ dependence is effectively removed from the screening charges
since $\gamma_1$ interacts only with
$\beta_{-1}$ which is the {\em constant} ($w_i$ independent) mode. Thus 
\cite{PRY1a}
\bea
 C_N(\{j_m\},x)&=&\frac{\Gamma(k-2j_N+1)}{\Gamma(k-2j_N-r+st+1)}
  \frac{s(k-2j_N)}{s(k-2j_N-r+st)}(-1)^{-r+st}x^{r-st}\nn
 &=&\frac{\G(2j_N-k+r-st)}{\G(2j_N-k)}(-1)^{-r+st}x^{r-st}
\label{CN}
\eea
Of course, this may also be obtained without using (\ref{jNexp}) but inserting
(\ref{jNexp2}) and (\ref{jNexp3}) directly into the correlator.
In either way we use that
\bea
 \bra{0}\left[\g_1^\al,\beta_{-1}^\al\right]...\ket{0}&=&
  (-1)^{-\al}\G(\al+1)\bra{0}...\ket{0}\nn
 \bra{0}\left[\beta_{-1}^\al,\g_1^\al\right]...\ket{0}&=&
 (-1)^{-\al+1}\G(\al+1)\bra{0}...\ket{0}
\label{bgmodefrac}
\eea
where ... represents $\beta,\g$ independent insertions.
(\ref{bgmodefrac}) is justified by
\bea
 \left[\g_1,\beta_{-1}^\al\right]&=&-\al\beta_{-1}^{\al-1}\spa
  \left[\beta_{-1},\g_1^\al\right]=\al\g_1^{\al-1}\nn
 \bra{0}\left[\g_1^n,\beta_{-1}^m\right]...\ket{0}&=&
  (-1)^{-n}\frac{\G(m+1)}{\G(m-n+1)}\bra{0}\beta_{-1}^{m-n}...\ket{0}
  =(-1)^{-n}\G(n+1)\delta_{nm}
\eea
This concludes the simple proof of the statement that correlators in $SL(2)$
current algebra reduce to corresponding ones in a particular minimal model
in the limit where all $x$'s are put equal to the corresponding $z$'s
(\ref{wn}).

For $x=1$ one may check from the 3-point function (\ref{W3FF})
that (\ref{CN}) is indeed the relative constant to the 3-point function of 
minimal models (\ref{W3FFMM}), up to an irrelevant phase (which may be taken 
care of by reverting to the phase convention in Chapter 3 for the screening
currents, cf. the comment following (\ref{sl2conv2})). 
This is due to the following rewriting of (\ref{CN})
\ben
 C_3(j_1,j_2,j_3;x=1)=\frac{\G(2j_2+1)}{\G(2j_2-r+st+1)}\frac{{\cal J}_{rs}(
  2j_1,2j_2+t;t)}{{\cal J}_{rs}(2j_1,2j_2;t)}(-1)^{-r+st}
\een
and the fact that the pre-factors from the contour deformations are 
identical in the two 3-point functions. This is also true for the other  
choices of contours.

\subsection{Analysis of Explicit Correlators}
Here we want to verify explicitly that the conformal blocks for the 
WZNW model evaluated in the preceding sections satisfy the results above. 
To this end we consider the 'interpolating' correlator
\ben
 \bra{j_N}e^{-x_NJ^-_1}\phi_{j_{N-1}}(z_{N-1},x_{N-1})...
  \phi_{j_2}(z_2,x_2)\prod_{i=1}^M\int dw_iS_{k_i}(w_i)\ket{j_1}=
  \bra{j_N}{\cal O}\ket{j_1}
\label{interpol}
\een
with
\bea
 x_\ell &=&z_\ell x, \ \ \ell=1,...,N-1\nn
 x_N&=&x-1
\eea
Thus for $x=1$ we get the WZNW model with all $x_i$'s put equal to the $z_i$'s.
For $x=0$ we should get the minimal model correlator up to normalisation. 
We wish to show that this interpolating correlator is 
{\em independent} of $x$. Using notation of preceding sections, we find
\ben
 {\cal O}=W_BW_N^\varphi F
\label{correlator}
\een
where
\bea
 B(w)&=&\sum_{\ell =1}^{N-1}\frac{x_\ell/u_\ell}{w-z_\ell}-x_N/u_N\nn
 W_B&=&\prod_{i=1}^MB(w_i)^{-k_i}\nn
 F&=&\Gamma(k-2j_N+1)u_N^{k-2j_N-1}e^{1/u_N}\prod_{\ell=2}^{N-1}
  \Gamma(2j_\ell +1)u_\ell^{2j_\ell -1}e^{1/u_\ell} \nn
 W^\varphi_N&=&\prod_{m<n}(z_m-z_n)^{2j_mj_n/t}\prod_{i=1}^M\prod_{m=1}^{N-1}
  (w_i-z_m)^{2k_ij_m/t}\prod_{i<j}(w_i-w_j)^{2k_ik_j/t}
\eea
Here we used that
\bea
 \bra{j_N}e^{-x_NJ^-_1}&=&\bra{j_N}(1+x_N\gamma_1)^{k-2j_N}\nn
 &=&\lim_{z'\rightarrow\infty}\bra{j_N}(1+x_Nz'\gamma(z'))^{k-2j_N}
\label{jNexpalt}
\eea
so $x_Nz'$ plays the role of the $x$ in the additional term 
$(1+x_Nz'\gamma(z'))^{k-2j_N}$ of the type
of the $\g$ part in a primary field with spin $2j=k-2j_N$. This certainly
gives the contribution 
\ben
 \lim_{z'\rightarrow\infty}\frac{x_Nz'/u_N}{w-z'}=-x_N/u_N
\een
in $B(w)$.
The integrations over the auxiliary variables $u_\ell$ are understood in 
(\ref{correlator}).

The above expression (\ref{jNexpalt})
may seem in contradiction to the expansion (\ref{jNexp}),
since the ratio of sine-functions we had there is absent now. So let us 
explain the reason for this subtlety. The point is that for $x\neq 0$ the 
fractional powers of the $\beta$'s in the correlator is balanced by the 
fractional powers of the $\gamma$'s in the primary fields. Thus the exponential
in the present case has to be expanded in {\em integral} powers. 
This is in contrast
to the situation in (\ref{jNexp}) where the $\gamma$ dependence of the primary 
fields were suppressed since they involved the case $x_i=0$. One may then ask 
how it is that this ratio of sine-factors is recovered in the present context.
Indeed, naively putting $x=0$ in the expression for the $B$ factors renders
the $u$ integrals trivial and we recover erroneously the factor $C_N(\{j_m\})$
without the ratio of sine-factors. However, a more careful analysis of the cut 
structure and of the integration contours shows that in fact this ratio arises
again when care is exercised. To see this \cite{PRY1a}, 
it is convenient to change variable 
from $u_N$ to $U_N=1/u_N$. As a function of $U_N$ the integrand has 
branch point singularities at $U_N=0,\infty,U_N(x,w_i)$ with
\ben
 U_N(x,w_i)=\sum_{\ell =1}^{N-1}\frac{xz_\ell/u_\ell}{(w_i-z_\ell)(x-1)}
\een
Now we should remember that all the $u$ integrals are to be taken along 
small circles surrounding the origin. The $U_N$ contour 
is therefore some {\em large } circle. However, for $x$ close to $1$, 
we see that
$U_N(x,w_i)$ will lie {\em outside} this circle, which we therefore may deform 
along a contour running above and below the negative real axis. 
We should choose the cut structure from the branch points $U_N(x,w_i)$ in a
way not interfering with the integration contour for $U_N$ or the deformations
of it. As $x$ is 
decreased from $1$, the branch points at $U_N(x,w_i)$ move closer to the 
original integration circle for $U_N$ and eventually cross it. Therefore, 
by analytic continuation this circle has to be
deformed so as to keep the singularities always outside for $x>0$. 
A convenient way of 
doing  that is precisely by deforming it to immediately wrap around the 
negative axis. But then the contributions from above and below the axis 
will have different phases depending on the power of $U_N$ (which is
$2j_N-k-1$), and it is not difficult to see that in the limit $x=0$ these 
exactly reproduce the seemingly missing ratio of sine-factors, and we are
left with (\ref{CN}). Here one uses that the remaining $u$ integrations
become trivial
\ben
 \G(2j_\ell+1)\oint\dtp{u}u_\ell^{2j_\ell-1}e^{1/u_\ell}=1
\een 

Consider now \cite{PRY1a} the function $G(w)$
\bea
 G(w)&=&G^-(w)w\nn
 G^-(w)&=&\br J^-(w){\cal O}\kt\nn
 &=&\left \{-\sum_{i,j}B(w)\frac{D_{B_i}D_{B_j}}{(w-w_i)(w-w_j)}
  +(t-2)\sum_i\frac{D_{B_i}}{(w-w_i)^2}\right.\nn
 &-&\left.2\sum_{i,j}\frac{k_iD_{B_j}}{(w-w_i)(w-w_j)} -
  2\sum_{m,j}\frac{j_mD_{B_j}}{(w-z_m)(w-w_j)}\right\}W_BW^\varphi_NF
\eea
The function $G(w)$ has simple poles as a function of $w$. It is a rather 
simple matter to evaluate the pole residues along the lines described when 
considering the Knizhnik-Zamolodchikov equations. For $w=z_m,\infty$ the 
result is
\bea
 \oint_{z_m} \dtp{w}G^-(w)w&=&z_m\pa_{x_m}W_BFW^\varphi_N\nn
 \oint_\infty\dtp{w}G^-(w)w&=&\pa_{x_N}W_BFW^\varphi_N
\eea
These contributions add up to produce the total
derivative of the original correlator with respect to $x$
\ben
 \sum_mz_m\pa_{x_m}{\cal O}+\pa_{x_N}{\cal O}=\pa_x{\cal O}
\label{partx}
\een
The pole residue at $w=w_i$ is (see (\ref{G-f}))
\ben
 \oint_{w_i} \dtp{w}G^-(w)w=t\pa_{w_j}
  \left(w_j\frac{W_N^\varphi W_B F}{B(w_j)}\right )
\een
After integration over the $w_i$'s we see that the expression (\ref{partx}) 
will vanish since this merely is the condition that the total sum of pole 
residues vanishes (when the pole at infinity is included as it is here).
Hence, the interpolating correlator is indeed independent of $x$.

\subsection{Comparison with Standard Formulations of Hamiltonian Reduction}
Having proved the equivalence of the two apparently different kinds of 
correlators, we now want to understand this equivalence from the point of view
of quantum Hamiltonian reduction. We briefly review the background.
Setting the affine current
\ben
 J^+(z)=1
\label{cstr}
\een
in the the equation of motion derived from the $SL(2)$ WZNW theory, one
recovers the classical equation of motion for Liouville theory. In order 
to implement
the constraint (\ref{cstr}) at the quantum level, one introduces a Lagrangian
multiplier field $A(z)$ and follows the standard procedure for
Hamiltonian reduction \cite{BO}, where $A(z)$ is treated as a gauge field.
The final theory, after gauge fixing, involves Faddeev-Popov ghost
fields, which are supposed to cancel out unwanted degrees of freedom
in the original WZNW theory.
The BRST quantisation has now become a standard approach to constrained
Hamiltonian systems. As far as correlation functions on the sphere are
concerned, the BRST quantisation is equivalent to imposing the
constraint (\ref{cstr}) on the correlators. Suppose one writes the correlation
function on the sphere as an operator insertion
\ben
 \bra{0}\hat{O}\ket{0}
\label{crt}
\een
then for the constrained system satisfying (\ref{cstr}), we have
\ben
 \bra{0}\hat{O}(J^+(z)-1)\ket{0}=0
\label{crt2}
\een
(\ref{crt2}) is equivalent to the following conditions
\bea
 J^+_n\ket{0}&=&0\ \ \ \ \ n\geq 0\nn
 \bra{0}\hat{O}(J^+_{-n}-\delta_{n,1})&=&0\ \ \ \ \ n\geq 1
\label{cdt}
\eea
In order not to confuse the notations used here, $J^+(z)$ is always
considered to be a conformal spin 1 field to fit the WZNW theory, so that
it has the expansion
\ben
 J^+(z)=\sum_{n\in\Z}J^+_n z^{-n-1}
\een
As usual, to fix $J^+(z)$ to be a constant value would require $J^+(z)$
be a scalar field. In other words, the energy momentum tensor must be improved
from the Sugawara construction by adding a term $\partial_z J^3(z)$. In
that context, one should rename $J^+_n \rightarrow J^+_{n+1}$.

(\ref{cdt}) is called the physical state condition. In BRST quantisation
the physical state space is the same as the BRST cohomology space
$Ker(Q)/Im(Q)$,
where $Q$ is the BRST charge defined by
\ben
 Q=\oint \dtp{w} (J^+(w)-1)c(w)
\label{brs}
\een
Here $c(w)$ is a conformal spin 1 fermionic ghost field with respect to the 
improved energy momentum tensor. Its conjugate field $b(w)$ is the anti-ghost 
field of spin 0 satisfying
\ben
 b(w)c(z)=\frac{1}{w-z}
\een
(\ref{cdt}) is equivalent to the BRST condition, in which one requires that the
vacuum states
$\bra{0}$ and $\ket{0}$ be physical states, and $\hat{O}$ be a physical
operator which maps physical states into physical states (see e.g. 
\cite{KO, HT}). In other words
\ben
 \bra{0}Q=[Q,\hat{O}]=Q\ket{0}=0
\een

Now consider the most general form for a class of conformal blocks in
$SL(2)$ WZNW theory, which are proportional to those in the Virasoro minimal
models. They can be written in the following form
\bea
 &&\bra{j_{r_N,s_N}}F(J^-_1)\phi_{j_{r_{N-1},s_{N-1}}}(z_{N-1},0)...
  \phi_{j_{r_2,s_2}}(z_2,0)\prod_i \int dw_iS_{k_i}(w_i)
  \ket{j_{r_1,s_1}}\nn
 &=&C\bra{\D_{r_N,s_N+1}}\phi_{r_{N-1},s_{N-1}+1}(z_{N-1})
  ...\phi_{r_2,s_2}(z_2)
  \prod_i \int dw_iV_{\alpha_i}(w_i)\ket{\D_{r_1,s_1+1}}
\label{equiv2}
\eea
where the normalisation constant $C$ is found as before to be
\ben
 C =(-1)^{2j_N-\sum_{i=1}^Nj_i}\frac{\G(\sum_{i=1}^Nj_i-k)}{\G(2j_N-k)}
  \pa_y^{-2j_N+\sum_{i=1}^Nj_i}F(y)|_{y=0}
\een
where we used that
\ben
 -2j_N+\sum_ij_i=-\sum_ik_i
\een
In general $C$ depends on $t$ and the $j_i$'s. For some values of
$t$ and $j_i$'s, $C$ vanishes. Then the conformal blocks in the Virasoro 
minimal models can only be obtained by dividing out $C$.
In other cases $C$ becomes infinity.
Then the conformal blocks for the Virasoro minimal models can
be either finite or zero, and in the latter case the relation between the WZNW 
and the minimal conformal blocks is singular. Strictly speaking,
simply taking the limit $x_i \rightarrow z_i$ is not equivalent to quantum
Hamiltonian reduction (\ref{cstr}). Rather it is in accord with the constraint
\ben
 J^+(w)=J^+_{-1}
\label{cstr2}
\een
To go to the minimal model we must further impose the condition
\ben
 J^+_{-1}=1
\label{cstr3}
\een
To see this, let us consider the BRST charge for quantum Hamiltonian reduction
(\ref{cstr2})
\ben
 \tilde{Q}=\oint \dtp{w} (J^+(w)-J^+_{-1})c(w)
\label{brs2}
\een
The physical state space now becomes the BRST cohomology space
$Ker(\tilde{Q})/Im(\tilde{Q})$. It is clear that
$\phi_j(z,0)$ commutes with $\tilde{Q}$, hence maps a physical state into 
another physical state. Now consider the ket and the bra states.
Notice that the ket state $\ket{j_1}$ is a highest weight state and the
bra state $\bra{j_{r_N,s_N}}$ is a lowest weight state
\ben
 J^+_n\ket{j_1}=\bra{j_{r_N,s_N}}J^+_{-n-1}=0, \ \ \ \ n\geq 0
\label{jpktbr}
\een
For the $b$, $c$ ghost fields, we have the following condition
\ben
 c_n\ket{j_1}=b_{n+1}\ket{j_1}
  =\bra{j_{r_N,s_N}}c_{-n-1}=\bra{j_{r_N,s_N}}b_{-n}=0, \ \ \ \ n\geq 0
\een
It can be verified that
with respect to the BRST charge $\tilde{Q}$ in (\ref{brs2}),
$\ket{j_1}$ is a physical state, and
the bra state $\bra{j_{r_N,s_N}}F(J^-_1)$ is a physical state  for
any arbitrary function $F(J^-_1)$. However, this extra degree of
freedom is removed if we further impose the condition (\ref{cstr3}),
which would fix the function  $F(J^-_1)$ uniquely
\ben
 \bra{j_N}F(J_1^-)J_{-1}^+=\bra{j_N}F(J_1^-)\spa [e^{-\g_1},\beta_{-1}]=
  e^{-\g_1}
\een
and we recover exactly 
the conformal blocks in the Virasoro minimal models (here $\al=r-st$)
\bea
 \bra{j_{r_N,s_N}}F(J^-_1)&=&\bra{j_{r_N,s_N}}e^{-\gamma_1}\nn
 &=&\bra{j_{r_N,s_N}}\sum_{n\in \Z}
  \frac{\Gamma(2j_N-k)}{\Gamma(2j_N-k+n+\alpha)\Gamma(n+\alpha+1)}
  (-J^-_1)^{n+\alpha}
\eea
(here corrected for minor misprint, compared to \cite{PRY1a})
where $\gamma_n$ is conjugate to $J^+_{-n}$
\ben
 [J^+_{-n}, \gamma_m]=\delta_{n,m}
\een
If we were to use $\phi_j(z,z)$ to represent a primary field in the Hamiltonian
reduced system (strictly speaking, $\phi_j(z,z)$ does not transform as a
primary field for the Virasoro algebra in the reduced system), then we
should normalise the correlation function by
dividing out the normalisation constant $C$. Then, in the limit $C$ goes
to zero, the conformal block in the reduced system would remain finite.

In conclusion, the constraint $J^+(z)=1$ completely freezes the degrees
of freedom of the $J^+(z)$ field. However, we could proceed in two steps in
putting the constraint on the correlation functions. First set
$J^+(z)=J^+_{-1}$ and then let $J^+_{-1}=1$. The first step would result in
a class of correlation functions which are proportional to that of the
completely constrained system, like the ones considered in the previous 
sections. However, the remaining degrees of freedom
of the $J^+_{-1}$ mode is reflected by the arbitrariness of the
proportionality. If we normalise the correlation function by dividing
out the normalisation constant, which is equivalent to setting $J^+_{-1}=1$,
then we recover the corresponding
correlators in the completely reduced system.

\chapter{4-point Greens Functions in $SL(2)$ Current Algebra}

Once the chiral blocks are obtained the next step in the program is to 
determine the monodromy invariant Greens functions. These are the ones for
which physical applications can be made and they are necessary before for 
example an application to $2D$ quantum gravity can be made. It is the 
principal goal in this chapter to obtain these monodromy invariant 
combinations, and from the ensuing monodromy coefficients to determine the 
operator algebra coefficients. This is then a generalisation of the 
celebrated work \cite{DF} by Dotsenko and Fateev in conformal minimal models 
to CFT based on affine $SL(2)$ current algebra, and may be viewed as the 
completion of the solution of CFT (on the sphere) based on affine $SL(2)$ 
current algebra for admissible representations.  

The problem of determining the Greens functions is conveniently solved by 
means of the crossing matrix relating the conformal blocks in the
s- and t-channels (in fact, just a particular row and column of that matrix).
The conformal blocks for 4-point functions may be characterised in 
terms of couplings to intermediate states. These in turn are determined by the
fusion rules of the theory, discussed in the preceding chapter. We shall
find that both sets of fusion rules are operating.
Thus, the fusion rules provide a neat starting point for giving convenient 
bases for the conformal blocks in the s- and t-channels. One must then 
understand how the general integral
representations can reproduce these bases. Here we shall use either the ones 
provided in Chapter 4 (sometimes to be referred to as PRY), or the one for the
4-point function obtained by Andreev \cite{An}, and in both cases 
specify how integration contours must be chosen in order to generate specific 
members of s- or t-channel bases. In \cite{An} the contours are not specified.
In the case of the PRY integral representation, we show how to obtain chiral
blocks in the s-channel corresponding to fusion rule I, and how 
to obtain blocks in the t-channel corresponding to fusion rule II,
using contours where the integration of the auxiliary variable $u$
is carried out first. We also show how to obtain
blocks in the s-channel corresponding to fusion rule II, and how
to obtain blocks in the t-channel corresponding to fusion rule I, 
using contours where the $u$ integration is done last. In the integral 
representation of Andreev there is no $u$ variable to worry about
and the contours we find are more tractable.
That his 4-point blocks are equivalent to ours is a priori
rather clear since both he and we have checked that the blocks we write down
satisfy the Knizhnik-Zamolodchikov equations. Nevertheless, we find it
very instructive to attempt a direct analytic proof of how the equivalence may 
be obtained. Substantial evidence will be provided at the level of highly 
non-trivial integral manipulations.

Having written down the full s- and t-channel bases for conformal blocks and
understood the corresponding integration contours, we may go on to calculate 
the relevant parts of the crossing matrix. Certain ratios of matrix elements
determine the monodromy coefficients used in building the monodromy
invariant 4-point Greens function. 

In the first considerations it will be assumed that both vertices in the
4-point blocks pertain to the same fusion rule (I {\em or} II). Then the idea 
of over-screening used for the 3-point function may
be employed to obtain additional 4-point blocks in which there are different 
fusion rules (I {\em and} II) operating at the two vertices of the block. These
new 4-point blocks correspond to different sets of external spins from the
ones previously considered, and so they do not intertwine with these under 
crossing. Based on the calculations in the case with non-mixed fusion rules,
it is fairly easy to obtain the new monodromy coefficients.
We may finally write down the operator algebra coefficients and in
particular, utilising the idea of over-screening allows us to normalise
unambiguously the operator algebra coefficients for fusion rule II, contrary
to the results in \cite{An}. It turns out that in terms of appropriate 
parametrisations, the operator algebra coefficients are identical for
the two fusion rules up to a power of the parameter $t$, which is determined
as a consistency condition.

\section{4-point Functions}
In the 4-point function 
\ben
 \langle \phi_{j_4}(z_4,x_4)\phi_{j_3}(z_3,x_3)\phi_{j_2}(z_2,x_2)
  \phi_{j_1}(z_1,x_1)\rangle
\een
we consider as usual the limits
\bea
 &\ \ z_4\rightarrow \infty, \ \ & x_4\rightarrow \infty\nn
 &z_3\rightarrow  1, \ \ & x_3\rightarrow 1\nn
 &z_2\rightarrow  z, \ \ & x_2\rightarrow  x\nn
 &z_1\rightarrow  0, \ \ & x_1\rightarrow  0
\label{limit}
\eea
so that the 4-point conformal blocks will be 
(in general multi-valued) functions of $(z,x)$. We 
label s- and t-channel conformal blocks by tree graphs, the meaning of which is
that in the limit $z\rightarrow 0$ followed by $x\rightarrow 0$ the s-channel
block corresponding to Fig. 1 has the behaviour following from the OPE's
\ben
 S(z,x)\sim z^{\D-\D_1-\D_2}(-x)^{j_1+j_2-j}(1+{\cal O}(z,-x))
\label{schannel}
\een
whereas for the t-channel block we have in the limit $z\rightarrow 1$ 
followed by $x\rightarrow 1$
\ben
 T(z,x)\sim (1-z)^{\D-\D_2-\D_3}(x-1)^{j_2+j_3-j}(1+{\cal O}(1-z,x-1))
\label{tchannel}
\een
\begin{figure}
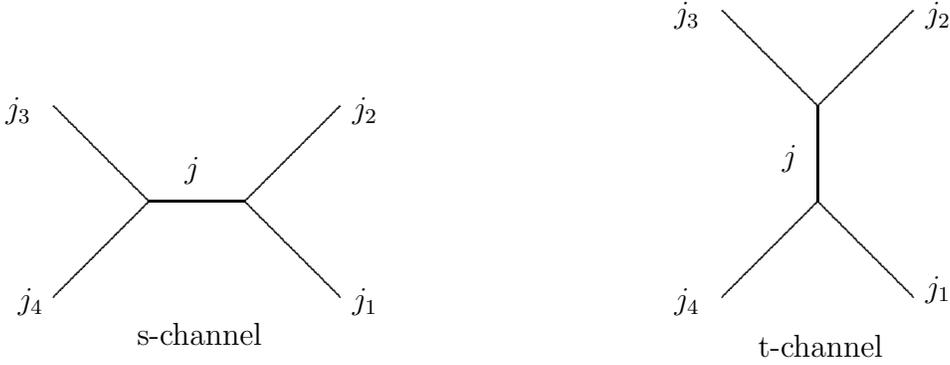

\font\thinlinefont=cmr5
\begingroup\makeatletter\ifx\SetFigFont\undefined
% extract first six characters in \fmtname
\def\x#1#2#3#4#5#6#7\relax{\def\x{#1#2#3#4#5#6}}%
\expandafter\x\fmtname xxxxxx\relax \def\y{splain}%
\ifx\x\y   % LaTeX or SliTeX?
\gdef\SetFigFont#1#2#3{%
  \ifnum #1<17\tiny\else \ifnum #1<20\small\else
  \ifnum #1<24\normalsize\else \ifnum #1<29\large\else
  \ifnum #1<34\Large\else \ifnum #1<41\LARGE\else
     \huge\fi\fi\fi\fi\fi\fi
  \csname #3\endcsname}%
\else
\gdef\SetFigFont#1#2#3{\begingroup
  \count@#1\relax \ifnum 25<\count@\count@25\fi
  \def\x{\endgroup\@setsize\SetFigFont{#2pt}}%
  \expandafter\x
    \csname \romannumeral\the\count@ pt\expandafter\endcsname
    \csname @\romannumeral\the\count@ pt\endcsname
  \csname #3\endcsname}%
\fi
\fi\endgroup
\mbox{\beginpicture
\setcoordinatesystem units <1.00000cm,1.00000cm>
\unitlength=1.00000cm
\linethickness=1pt
\setplotsymbol ({\makebox(0,0)[l]{\tencirc\symbol{'160}}})
\setshadesymbol ({\thinlinefont .})
\setlinear
%
% Fig POLYLINE object
%
\linethickness= 0.500pt
\setplotsymbol ({\thinlinefont .})
\putrule from  2.540 22.860 to  3.810 22.860
%
% Fig POLYLINE object
%
\linethickness= 0.500pt
\setplotsymbol ({\thinlinefont .})
\plot  3.810 22.860  5.080 24.130 /
%
% Fig POLYLINE object
%
\linethickness= 0.500pt
\setplotsymbol ({\thinlinefont .})
\plot  2.540 22.860  1.270 21.590 /
%
% Fig POLYLINE object
%
\linethickness= 0.500pt
\setplotsymbol ({\thinlinefont .})
\plot  3.810 22.860  5.080 21.590 /
%
% Fig POLYLINE object
%
\linethickness= 0.500pt
\setplotsymbol ({\thinlinefont .})
\plot 10.160 25.400 11.430 24.130 /
%
% Fig POLYLINE object
%
\linethickness= 0.500pt
\setplotsymbol ({\thinlinefont .})
\plot 11.430 24.130 12.700 25.400 /
%
% Fig POLYLINE object
%
\linethickness= 0.500pt
\setplotsymbol ({\thinlinefont .})
\putrule from 11.430 24.130 to 11.430 22.860
%
% Fig POLYLINE object
%
\linethickness= 0.500pt
\setplotsymbol ({\thinlinefont .})
\plot 11.430 22.860 10.160 21.590 /
%
% Fig POLYLINE object
%
\linethickness= 0.500pt
\setplotsymbol ({\thinlinefont .})
\plot 11.430 22.860 12.700 21.590 /
%
% Fig TEXT object
%
\put{\SetFigFont{12}{14.4}{rm}s-channel} [lB] at  2.381 20.955
%
% Fig POLYLINE object
%
\linethickness= 0.500pt
\setplotsymbol ({\thinlinefont .})
\plot  1.270 24.130  2.540 22.860 /
%
% Fig TEXT object
%
\put{$j_1$} [lB] at  5.239 21.431
%
% Fig TEXT object
%
\put{$j$} [lB] at 10.954 23.336
%
% Fig TEXT object
%
\put{$j_2$} [lB] at  5.239 23.971
%
% Fig TEXT object
%
\put{$j_3$} [lB] at  0.635 23.971
%
% Fig TEXT object
%
\put{$j_4$} [lB] at  0.794 21.431
%
% Fig TEXT object
%
\put{$j$} [lB] at  3.016 23.178
%
% Fig TEXT object
%
\put{\SetFigFont{12}{14.4}{rm}t-channel} [lB] at 10.636 20.796
%
% Fig TEXT object
%
\put{$j_1$} [lB] at 12.859 21.590
%
% Fig TEXT object
%
\put{$j_2$} [lB] at 12.859 25.241
%
% Fig TEXT object
%
\put{$j_3$} [lB] at  9.525 25.241
%
% Fig TEXT object
%
\put{$j_4$} [lB] at  9.525 21.431
\linethickness=0pt
\putrectangle corners at  0.635 25.546 and 12.859 20.739
\endpicture}

\caption{Graphs for s- and t-channel blocks}
\label{fig1}
\end{figure}
Here the conformal weights are given by the standard expression
$\D_i=j_i(j_i+1)/t$.

A very convenient way to think 
about the fusion rules consists in the following. Consider the 
s-channel coupling of $j_1,j_2$ to a $j$. When we parametrise 
$j=j_I$ for fusion rule I as $j_1+j_2-j_I=r-st$, the integers $r,s$ are 
related to the number of screenings of the first and second kinds around the
$j_1j_2j_I$ vertex. The singular behaviour of the s-channel block in the limit
$z\rightarrow 0, x\rightarrow 0$ is then
\ben
 z^{\D-\D_1-\D_2}(-x)^{j_1+j_2-j_I}=z^{\D-\D_1-\D_2}(-x)^{r-st}
\label{singFI}
\een 
with $\D=j_I(j_I+1)/t$. For fusion rule II we may then parametrise the internal
$j$ as
\ben
 j_{II}=-j_I-1
\een
Of course the conformal dimensions for $j_I$ and $j_{II}$ are the same, but we
find the singular behaviour of the s-channel block to be
\ben
 z^{\D-\D_1-\D_2}(-x)^{j_1+j_2-j_{II}}=z^{\D-\D_1-\D_2}(-x)^{2j_1+2j_2-r+st+1}
\label{singFII}
\een
All these statements follow by analysing the fusion rules (\ref{FI})
and (\ref{FII}).
By analysing the s-channel 4-point blocks in the limit 
$z\rightarrow 0, x\rightarrow 0$ we indeed find both of these singular 
behaviours (see below)
and hence verify that the blocks realize both fusion rules I and II.
In the t-channel the discussion is analogous, with $j_1\leftrightarrow j_3$,
$z\rightarrow 1-z$ and $x\rightarrow 1-x$, so that we consider the limits
$z\rightarrow 1$ followed by $x\rightarrow 1$.

Before specifying the contours needed to produce chiral blocks with
these singular behaviours, we reconsider the general result for
$N$-point functions in the case of $N=4$ in order to simplify the
integral expression. 
First of all we want to reduce the number of auxiliary integration 
variables $u$ from two to one. In due course, let us 
reconsider the $\beta,\gamma$ part of the 4-point function 
\ben
 \bra{j_4}[\phi_{j_3}(z_3,x_3)]_j^{j_4}[\phi_{j_2}(z_2,x_2)]_{j_1}^j 
  \ket{j_1}_{\beta\gamma}
\een
with the notation
\bea
 j_1+j_2-j&=&\rho_2-\sigma_2t\nn
 j+j_3-j_4&=&\rho_3-\sigma_3t
\eea
Thus we get
\bea
 W_4&\sim&\G(2j_3+1)\G(2j_2+1)\oint_0\dtp{u_2}\dtp{u_3}\frac{1}{u_2u_3}\nn
 &\cdot&\prod_{i_3=1}^{\rho_3}
  {[}\frac{x_2/u}{w(3,i_3)-z_2}+\frac{x_3}{w(3,i_3)-z_3}{]}
  \prod_{l_3=1}^{\sigma_3}
  {[}\frac{x_2/u}{v(3,l_3)-z_2}+\frac{x_3}{v(3,l_3)-z_3}{]}^{-t}\nn
 &\cdot&\prod_{i_2=1}^{\rho_2}
  {[}\frac{x_2/u}{w(2,i_2)-z_2}+\frac{x_3}{w(2,i_2)-z_3}{]}
  \prod_{l_2=1}^{\sigma_2}{[}\frac{x_2/u}{v(2,l_2)-z_2}
  +\frac{x_3}{v(2,l_2)-z_3}{]}^{-t}\nn
 &\cdot&u^{\rho_2-\sigma_2 t}(u_3^{-1}D_3)^{-2j_3+\rho_3-\sigma_3t}
  \exp(1/u_3)(u_2^{-1}D_2)^{-2j_2+\rho_2-\sigma_2t}
  \exp(1/u_2)
\eea
where we have introduced 
\ben
 u=u_2/u_3
\een
Again the (somewhat misleading) notation is that the $D$'s are derivatives with
respect to the entire argument of the relevant exponentials.

We want to consider a change of variables from $(u_2,u_3)$ to $(u_2,u)$. 
The integration measure is 
\ben
 \frac{du_2}{u_2}\frac{du_3}{u_3}=-\frac{du_2}{u_2}\frac{du}{u}
\een
and 
\ben
 2j_2+2j_3-\rho_3+\sigma_3t-\rho_2+\sigma_2t
  =2j_2+2j_3-(j_3+j-j_4)-(j_2+j_1-j)=j_2+j_3+j_4-j_1=J_1
\een
Here we have introduced the notation
\ben
 J_i=j_1+j_2+j_3+j_4-2j_i
\een
Using the generalised exponential identity (\ref{DDfrac}), 
we obtain the following $u_2$ dependence
\ben
 \frac{dudu_2}{uu_2}u^{-2j_3+\rho_2-\sigma_2t+\rho_3-\sigma_3t}
  u_2^{J_1}D^{-J_1}\exp\left(\frac{1+u}{u_2}\right)
\een
Now the integral over $u_2$ will produce the factor
\ben
 \frac{(1+u)^{J_1}}{\G(J_1+1)}
\een
We are left with the following integral
\bea
 &-&\int\dtp{u}
  \prod_{i_3=1}^{\rho_3}{[}\frac{x_2}{w(3,i_3)-z_2}
  +\frac{x_3u}{w(3,i_3)-z_3}{]}\prod_{l_3=1}^{\sigma_3}{[}
  \frac{x_2}{v(3,l_3)-z_2}+\frac{x_3u}{v(3,l_3)-z_3}{]}^{-t}\nn
 &\cdot&
  \prod_{i_2=1}^{\rho_2}{[}\frac{x_2}{w(2,i_2)-z_2}
  +\frac{x_3u}{w(2,i_2)-z_3}{]}\prod_{l_2=1}^{\sigma_2}{[}
  \frac{x_2}{v(2,l_2)-z_2}+\frac{x_3u}{v(2,l_2)-z_3}{]}^{-t}\nn
 &\cdot&u^{-2j_3-1}\frac{(1+u)^{J_1}}{\G(J_1+1)}\G(2j_3+1)\G(2j_2+1)
\eea

To write the final result for the 4-point function in a more compact form,
let us collectively denote the positions for both kinds of screening charges 
as
\ben
 w_i, \ i=1,...,\rho_2+\rho_3+\sigma_2+\sigma_3\equiv M 
\een
Then the complete expression for the 4-point function is 
\bea
 &&\bra{j_4}[\phi_{j_3}(z_3,x_3)]_j^{j_4}[\phi_{j_2}(z_2,x_2)]_{j_1}^j 
  \ket{j_1}\nn
 &=&- \prod_{m<n}^3(z_m-z_n)^{2j_mj_n/t}\G(2j_3+1)\G(2j_2+1)\nn
 &\cdot&\int\dtp{u}
  \prod_{i=1}^{M}\dtp{w_i}
  \left(\frac{x_2}{w_i-z_2}+\frac{x_3u}{w_i-z_3}\right)^{-k_i}
  \prod_{i<j}(w_i-w_j)^{\frac{2k_ik_j}{t}}\nn
 &\cdot&\prod_{i=1}^{M}\prod_{l=1}^{3}(w_i-z_l)^{\frac{2k_ij_l}{t}}
  u^{-2j_3-1}\frac{(1+u)^{J_1}}{\G(J_1+1)}   
\eea
where only {\em one} auxiliary $u$ integration is involved and we have 
included the $\var$ part. $k_i=-1$ for 
screenings of the first kind and $k_i=t$ for screenings of the second kind. 
In the limit (\ref{limit})
we obtain, letting $u\rightarrow -u$ and up to normalisation
\bea
 W_4&=&z^{2j_1j_2/t}(1-z)^{2j_2j_3/t}\int\dtp{u}\prod_{i=1}^{M}\dtp{w_i}
 w_i^{2k_ij_1/t}(w_i-z)^{2k_ij_2/t}(w_i-1)^{2k_ij_3/t}\nn
 &\cdot&\prod_{i<j}(w_i-w_j)^{2k_ik_j/t} 
 \left(-\frac{u}{w_i-1}+\frac{x}{w_i-z}\right)^{-k_i}u^{-2j_3-1}(1-u)^{J_1}
\label{4point1}
\eea

\subsection{Integration Contours}
In this section we shall specify the integration contours \cite{PRY2}
for the various variables in the integral representation (\ref{4point1})
of the 4-point function. 
The first contours we indicate will give rise to a set of 
s-channel conformal blocks corresponding to the intermediate state $j$
in Fig. 1 being given by fusion rule I, and to a set of conformal blocks in the
t-channel corresponding to the intermediate $j$ being given by fusion rule II.
In these cases the $u$ integration is carried out first.

\begin{figure}
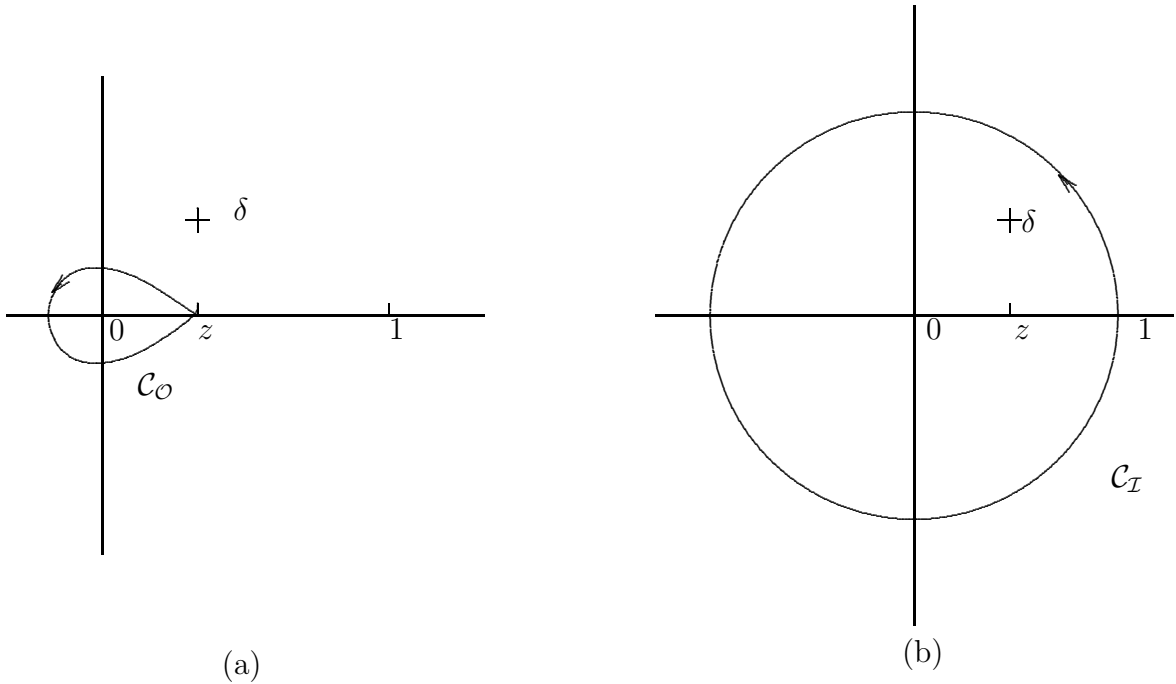

\font\thinlinefont=cmr5
\begingroup\makeatletter\ifx\SetFigFont\undefined
% extract first six characters in \fmtname
\def\x#1#2#3#4#5#6#7\relax{\def\x{#1#2#3#4#5#6}}%
\expandafter\x\fmtname xxxxxx\relax \def\y{splain}%
\ifx\x\y   % LaTeX or SliTeX?
\gdef\SetFigFont#1#2#3{%
  \ifnum #1<17\tiny\else \ifnum #1<20\small\else
  \ifnum #1<24\normalsize\else \ifnum #1<29\large\else
  \ifnum #1<34\Large\else \ifnum #1<41\LARGE\else
     \huge\fi\fi\fi\fi\fi\fi
  \csname #3\endcsname}%
\else
\gdef\SetFigFont#1#2#3{\begingroup
  \count@#1\relax \ifnum 25<\count@\count@25\fi
  \def\x{\endgroup\@setsize\SetFigFont{#2pt}}%
  \expandafter\x
    \csname \romannumeral\the\count@ pt\expandafter\endcsname
    \csname @\romannumeral\the\count@ pt\endcsname
  \csname #3\endcsname}%
\fi
\fi\endgroup
\mbox{\beginpicture
\setcoordinatesystem units <1.00000cm,1.00000cm>
\unitlength=1.00000cm
\linethickness=1pt
\setplotsymbol ({\makebox(0,0)[l]{\tencirc\symbol{'160}}})
\setshadesymbol ({\thinlinefont .})
\setlinear
%
% Fig POLYLINE object
%
\linethickness= 0.500pt
\setplotsymbol ({\thinlinefont .})
\putrule from  1.270 21.590 to  7.620 21.590
%
% Fig POLYLINE object
%
\linethickness= 0.500pt
\setplotsymbol ({\thinlinefont .})
\putrule from  2.540 24.765 to  2.540 18.415
%
% Fig POLYLINE object
%
\linethickness= 0.500pt
\setplotsymbol ({\thinlinefont .})
\putrule from  3.810 21.749 to  3.810 21.590
%
% Fig POLYLINE object
%
\linethickness= 0.500pt
\setplotsymbol ({\thinlinefont .})
\putrule from  6.350 21.749 to  6.350 21.590
%
% Fig POLYLINE object
%
\linethickness= 0.500pt
\setplotsymbol ({\thinlinefont .})
\plot  3.810 23.019  3.810 23.019 /
%
% Fig POLYLINE object
%
\linethickness= 0.500pt
\setplotsymbol ({\thinlinefont .})
\putrule from  3.810 23.019 to  3.810 22.701
%
% Fig POLYLINE object
%
\linethickness= 0.500pt
\setplotsymbol ({\thinlinefont .})
\putrule from  3.651 22.860 to  3.969 22.860
%
% Fig POLYLINE object
%
\linethickness= 0.500pt
\setplotsymbol ({\thinlinefont .})
\putrule from  9.890 21.590 to 16.780 21.590
%
% Fig POLYLINE object
%
\linethickness= 0.500pt
\setplotsymbol ({\thinlinefont .})
\putrule from 13.335 25.718 to 13.335 17.462
%
% Fig POLYLINE object
%
\linethickness= 0.500pt
\setplotsymbol ({\thinlinefont .})
\putrule from 14.605 21.749 to 14.605 21.590
%
% Fig POLYLINE object
%
\linethickness= 0.500pt
\setplotsymbol ({\thinlinefont .})
\putrule from 14.605 23.019 to 14.605 22.701
%
% Fig POLYLINE object
%
\linethickness= 0.500pt
\setplotsymbol ({\thinlinefont .})
\putrule from 14.446 22.860 to 14.764 22.860
%
% Fig POLYLINE object
%
\linethickness= 0.500pt
\setplotsymbol ({\thinlinefont .})
\plot 15.373 23.345 15.261 23.457 /
%
% arrow head
%
\plot 15.486 23.322 15.261 23.457 15.396 23.232 /
%
%
% Fig ELLIPSE
%
\linethickness= 0.500pt
\setplotsymbol ({\thinlinefont .})
\ellipticalarc axes ratio  2.710:2.710  360 degrees 
	from 16.045 21.590 center at 13.335 21.590
%
% Fig POLYLINE object
%
\linethickness= 0.500pt
\setplotsymbol ({\thinlinefont .})
\plot  1.990 22.013  1.877 21.901 /
%
% arrow head
%
\plot  2.012 22.126  1.877 21.901  2.102 22.036 /
%
%
% Fig TEXT object
%
\put{${\cal C}_{\cal I}$} [lB] at 15.954 19.304
\linethickness= 0.500pt
\setplotsymbol ({\thinlinefont .})
%
% Fig CONTROL PT SPLINE
%
% open spline
%
\plot	 3.810 21.590  3.334 21.907
 	 3.216 21.982
	 3.101 22.046
	 2.988 22.101
	 2.877 22.146
	 2.769 22.180
	 2.664 22.205
	 2.561 22.220
	 2.461 22.225
	 2.365 22.220
	 2.277 22.205
	 2.196 22.180
	 2.123 22.146
	 1.999 22.046
	 1.948 21.982
	 1.905 21.907
	 1.870 21.828
	 1.845 21.749
	 1.831 21.669
	 1.826 21.590
	 1.831 21.511
	 1.845 21.431
	 1.870 21.352
	 1.905 21.273
	 1.948 21.198
	 1.999 21.134
	 2.123 21.034
	 2.196 21.000
	 2.277 20.975
	 2.365 20.960
	 2.461 20.955
	 2.561 20.960
	 2.664 20.975
	 2.769 21.000
	 2.877 21.034
	 2.988 21.079
	 3.101 21.134
	 3.216 21.198
	 3.334 21.273
	 3.445 21.348
	 3.542 21.416
	 3.624 21.477
	 3.691 21.530
	 3.780 21.615
	 3.810 21.669
	 /
\plot  3.810 21.669  3.810 21.749 /
%
% Fig TEXT object
%
\put{$\delta$} [lB] at  4.286 22.860
%
% Fig TEXT object
%
\put{$z$} [lB] at  3.810 21.273
%
% Fig TEXT object
%
\put{$1$} [lB] at  6.350 21.273
%
% Fig TEXT object
%
\put{$1$} [lB] at 16.304 21.273
%
% Fig TEXT object
%
\put{\SetFigFont{12}{14.4}{rm}(a)} [lB] at  4.128 16.828
%
% Fig TEXT object
%
\put{$z$} [lB] at 14.664 21.273
%
% Fig TEXT object
%
\put{$\delta$} [lB] at 14.764 22.701
%
% Fig TEXT object
%
\put{$0$} [lB] at 13.494 21.273
%
% Fig TEXT object
%
\put{\SetFigFont{12}{14.4}{rm}(b)} [lB] at 13.176 16.986
%
% Fig TEXT object
%
\put{$0$} [lB] at  2.633 21.258
%
% Fig TEXT object
%
\put{${\cal C}_{\cal O}$} [lB] at  3.016 20.542
\linethickness=0pt
\putrectangle corners at  1.245 25.743 and 17.805 16.751
\endpicture}

\caption{The integration contours ${\cal C}_{\cal O}$ (a) and
${\cal C}_{\cal I}$ (b) for an s-channel block corresponding to fusion rule I.}
\label{figpryws}
\end{figure}

We first describe the situation in the s-channel corresponding to fusion rule
I. We write for the conformal
block 
\bea
 W^{(R,S)}_{(r,s)}(j_1,j_2,j_3,j_4;z,x)
 &=&z^{2j_1j_2/t}(1-z)^{2j_2j_3/t}
  \oint_{{\cal C}_{\cal I}} \prod_{i\in {\cal I}}\frac{dw_i}{2\pi i}
  \oint_{{\cal C}_{\cal O}}\prod_{j\in {\cal O}}\frac{dw_j}{2\pi i}
  \oint_{{\cal C}_u}\frac{du}{2\pi i}\nn
 &\cdot&w_i^{2k_ij_1/t}(w_i-z)^{2k_ij_2/t}(w_i-1)^{2k_ij_3/t}
  \prod_{\stackrel{i,j\in{\cal A}}{i<j}}(w_i-w_j)^{2k_ik_j/t}\nn
 &\cdot&\prod_{i\in{\cal A}}\left ( -\frac{u}{w_i-1}+\frac{x}{w_i-z}
  \right)^{-k_i}u^{-2j_3-1}(1-u)^{2j_2+2j_3-R +S t}
\label{pryblock}
\eea
Here we are considering an integral representation of the 4-point conformal 
block with a total of $R$ screening operators of the first kind 
and a total of $S$ screening operators of the second kind. 
The $w_i$'s are the positions of the screening operators. 
When $i\in{\cal O}$ the corresponding $w_i$ is integrated along the contour of 
Fig. \ref{figpryws}(a) (whether it is of the first or second kind) 
corresponding to a screening of the vertex $j_1j_2j$. Different 
$i,i'\in{\cal O}$ 
are taken along slightly different contours in order to avoid 
the singularity coming from $(w_i-w_{i'})^{2k_ik_{i'}/t}$. Similarly, the 
$w_j$'s for $j\in{\cal I}$ are integrated along the contour 
Fig. \ref{figpryws}(b)
corresponding to a screening of the vertex $jj_3j_4$. 
${\cal A}={\cal O}\cup{\cal I}$ is simply the combined index set. We denote the
numbers of screenings of the first kind at the $j_1j_2j$ and the $jj_3j_4$ 
vertices respectively as $r$ and $R-r$. Similarly the corresponding 
numbers of screenings of the second kinds at the two vertices are denoted 
$s$ and $S-s$. In the product of factors $(w_i-w_j)^{2k_ik_j/t}$
an arbitrary ordering of the indices is implied.

\begin{figure}
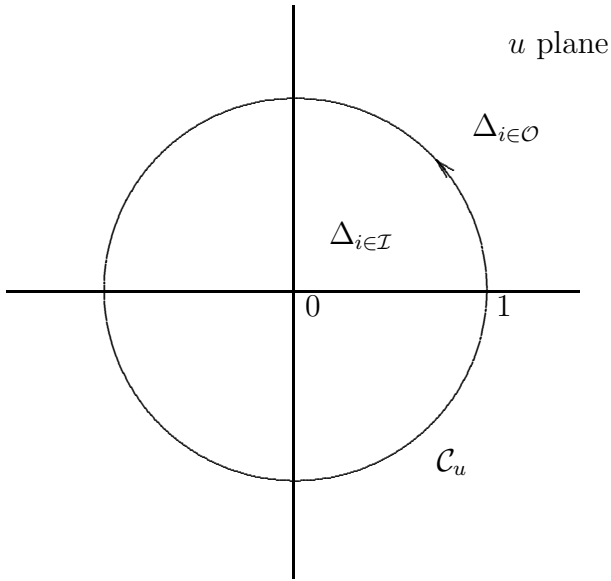


\font\thinlinefont=cmr5
\begingroup\makeatletter\ifx\SetFigFont\undefined
% extract first six characters in \fmtname
\def\x#1#2#3#4#5#6#7\relax{\def\x{#1#2#3#4#5#6}}%
\expandafter\x\fmtname xxxxxx\relax \def\y{splain}%
\ifx\x\y   % LaTeX or SliTeX?
\gdef\SetFigFont#1#2#3{%
  \ifnum #1<17\tiny\else \ifnum #1<20\small\else
  \ifnum #1<24\normalsize\else \ifnum #1<29\large\else
  \ifnum #1<34\Large\else \ifnum #1<41\LARGE\else
     \huge\fi\fi\fi\fi\fi\fi
  \csname #3\endcsname}%
\else
\gdef\SetFigFont#1#2#3{\begingroup
  \count@#1\relax \ifnum 25<\count@\count@25\fi
  \def\x{\endgroup\@setsize\SetFigFont{#2pt}}%
  \expandafter\x
    \csname \romannumeral\the\count@ pt\expandafter\endcsname
    \csname @\romannumeral\the\count@ pt\endcsname
  \csname #3\endcsname}%
\fi
\fi\endgroup
\mbox{\beginpicture
\setcoordinatesystem units <1.00000cm,1.00000cm>
\unitlength=1.00000cm
\linethickness=1pt
\setplotsymbol ({\makebox(0,0)[l]{\tencirc\symbol{'160}}})
\setshadesymbol ({\thinlinefont .})
\setlinear
%
% Fig POLYLINE object
%
\linethickness= 0.500pt
\setplotsymbol ({\thinlinefont .})
\putrule from  1.270 19.050 to  8.890 19.050
%
% Fig POLYLINE object
%
\linethickness= 0.500pt
\setplotsymbol ({\thinlinefont .})
\putrule from  5.080 22.860 to  5.080 15.240
%
% Fig POLYLINE object
%
\linethickness= 0.500pt
\setplotsymbol ({\thinlinefont .})
\plot  7.097 20.684  6.985 20.796 /
%
% arrow head
%
\plot  7.210 20.662  6.985 20.796  7.120 20.572 /
%
%
% Fig TEXT object
%
\put{$\Delta_{i\in{\cal O}}$} [lB] at  7.461 21.114
%
% Fig ELLIPSE
%
\linethickness= 0.500pt
\setplotsymbol ({\thinlinefont .})
\ellipticalarc axes ratio  2.544:2.544  360 degrees 
        from  7.654 19.078 center at  5.110 19.078
%
% Fig TEXT object
%
\put{$\Delta_{i\in{\cal I}}$} [lB] at  5.556 19.685
%
% Fig TEXT object
%
\put{$0$} [lB] at  5.239 18.733
%
% Fig TEXT object
%
\put{${\cal C}_u$} [lB] at  6.985 16.669
%
% Fig TEXT object
%
\put{\SetFigFont{12}{14.4}{rm}$u$ plane} [lB] at  7.938 22.225
%
% Fig TEXT object
%
\put{$1$} [lB] at  7.779 18.733
\linethickness=0pt
\putrectangle corners at  1.245 22.885 and  8.915 15.215
\endpicture}

\caption{The integration contour ${\cal C}_u$ for an s-channel block 
corresponding to fusion rule I.}
\label{figpryus}
\end{figure}

For fixed $w_i$'s the integrand has singularities in the $u$ plane at 
$u=0,1,\Delta_i$, where
\ben
\Delta_i=\frac{w_i-1}{w_i-z}x
\een
The integration contour for $u$ is to divide the singularities $\Delta_i$,
so that the ones for $i\in{\cal O}$ lie outside ${\cal C}_u$ and the ones for
$i\in{\cal I}$ lie inside ${\cal C}_u$, and ${\cal C}_u$ should pass through 
$u=1$. For $z$ and $x$ sufficiently small,
we may take ${\cal C}_u$ to be the unit circle, Fig. \ref{figpryus}. 
Remember that in order to 
identify the nature of the block and the value of the intermediate $j$, we are 
going to investigate the limit $z\rightarrow 0$ followed by $x\rightarrow 0$. 
The different positions of the singularities $\Delta_i$ in $u$ for 
$i\in{\cal O}$ and $i\in{\cal I}$ mean that they give rise to different 
singularities in the corresponding $w_i$ planes after the $u$ integration 
has been performed. 
In fact, for $i\in {\cal O}$ there occurs a pinching of singularities
when $\Delta_i$ collides with either $0$ of $1$ (the additional singularities 
in $u$). This happens for $w_i$ equal to $1$, and for $w_i$ equal to 
\ben
 \delta =\frac{x-z}{x-1}
\een
respectively. In particular no extra singularity is generated at $w_i=z$. 
This is why we may take the contour in $w_i$ to start from $z$ as indicated, 
since the
singularity for $w_i$ is what we term 'pure', meaning that it is of the form
\ben
 (w_i-z)^a(1+{\cal O}(w-z))
\een
One can check that this is enough to ensure that the corresponding block
will satisfy the 
KZ equations, going over the proof presented
in chapter 4 \cite{PRY1}. 
In contrast, the singularity at $w_i=0$ is 'non-pure': it 
is a mixture of different powers of $w_i$. Hence we cannot allow the contour to
end in $w_i=0$, it has to surround that point as indicated. 

Turning to the singularities in $w_i$ for $i\in{\cal I}$, we see that pinching
occurs only when $\Delta_i=1$, so that there is no extra singularity produced
at $w_i=1$: it remains pure, and we may take the integration contour to start 
in $w_i=1$ as indicated. If more convenient, one may take the contour to wrap
around the real axis form $1$ to $\infty$, which is a form closer to the one
used by Dotsenko and Fateev \cite{DF}.

Having established that the choice of contours 
indicated is allowed in the sense
that the conformal block will satisfy the KZ equations, it
is a relatively simple matter to find the leading singularity in the limit
$z\rightarrow 0$ followed by $x\rightarrow 0$. In fact, we may scale all the 
$w_i$'s with $i\in{\cal O}$ as
\ben
 w_i\rightarrow zw_i
\een
In the limit $z\rightarrow 0$ this is easily seen to result in a leading $z$ 
behaviour of the form
\ben
 W^{(R,S)}_{(r,s)}(z,x)\sim z^{-\D(j_1)-\D(j_2)+\D(j_I)}
\een
where $\D_j=\D(j)=j(j+1)/t$ and where 
\ben
 j_I=j_1+j_2-r+s t
\een
This is not enough to prove that indeed the intermediate state corresponds 
to a primary field with that value of $j$, since
\ben
 \D(j)=\D(-j-1)
\een
In fact, according to our earlier discussion, the difference between fusion 
rules I and II is exactly that for fusion rule I we should obtain $j=j_I$ 
whereas for fusion rule II we should obtain $j=j_{II}=-j_I-1$. In other words
the $z$ behaviour is precisely unable to distinguish between the two fusion 
rules. To distinguish we must investigate the leading $x$ behaviour 
in the limit
$x\rightarrow 0$ after we have taken $z\rightarrow 0$. However, it is an 
easy matter to do so and we find 
\ben
 W^{(R,S)}_{(r,s)}\sim z^{-\D(j_1)-\D(j_2)+\D(j_I)}(-x)^{r-s t}
\een
This is the proof that the conformal block we have constructed corresponds to
fusion rule I, since $r-s t=j_1+j_2-j_I$.

\begin{figure}
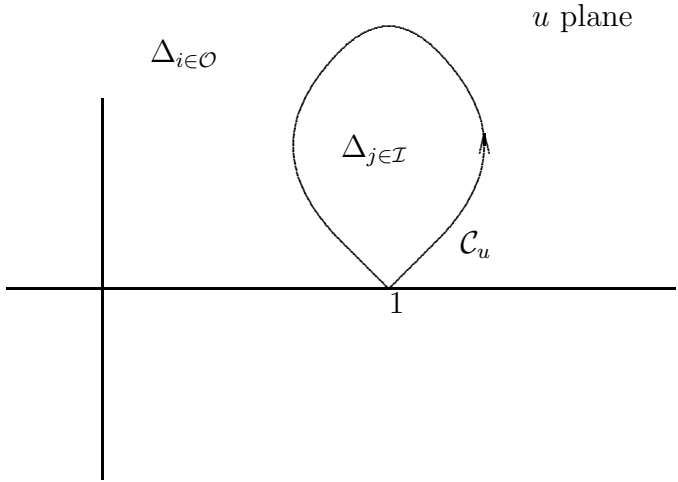
 

\font\thinlinefont=cmr5
\begingroup\makeatletter\ifx\SetFigFont\undefined
% extract first six characters in \fmtname
\def\x#1#2#3#4#5#6#7\relax{\def\x{#1#2#3#4#5#6}}%
\expandafter\x\fmtname xxxxxx\relax \def\y{splain}%
\ifx\x\y   % LaTeX or SliTeX?
\gdef\SetFigFont#1#2#3{%
  \ifnum #1<17\tiny\else \ifnum #1<20\small\else
  \ifnum #1<24\normalsize\else \ifnum #1<29\large\else
  \ifnum #1<34\Large\else \ifnum #1<41\LARGE\else
     \huge\fi\fi\fi\fi\fi\fi
  \csname #3\endcsname}%
\else
\gdef\SetFigFont#1#2#3{\begingroup
  \count@#1\relax \ifnum 25<\count@\count@25\fi
  \def\x{\endgroup\@setsize\SetFigFont{#2pt}}%
  \expandafter\x
    \csname \romannumeral\the\count@ pt\expandafter\endcsname
    \csname @\romannumeral\the\count@ pt\endcsname
  \csname #3\endcsname}%
\fi
\fi\endgroup
\mbox{\beginpicture
\setcoordinatesystem units <1.00000cm,1.00000cm>
\unitlength=1.00000cm
\linethickness=1pt
\setplotsymbol ({\makebox(0,0)[l]{\tencirc\symbol{'160}}})
\setshadesymbol ({\thinlinefont .})
\setlinear
%
% Fig POLYLINE object
%
\linethickness= 0.500pt
\setplotsymbol ({\thinlinefont .})
\putrule from  3.810 21.590 to 12.700 21.590
%
% Fig POLYLINE object
%
\linethickness= 0.500pt
\setplotsymbol ({\thinlinefont .})
\putrule from 10.160 23.495 to 10.160 23.654
%
% arrow head
%
\plot 10.224 23.400 10.160 23.654 10.097 23.400 /
\linethickness= 0.500pt
\setplotsymbol ({\thinlinefont .})
%
% Fig CONTROL PT SPLINE
%
% open spline
%
\plot    8.890 21.590  8.255 22.225
         8.178 22.304
         8.106 22.384
         8.039 22.463
         7.977 22.543
         7.920 22.622
         7.868 22.701
         7.821 22.781
         7.779 22.860
         7.742 22.939
         7.709 23.019
         7.682 23.098
         7.660 23.178
         7.642 23.257
         7.630 23.336
         7.622 23.416
         7.620 23.495
         7.622 23.574
         7.630 23.654
         7.642 23.733
         7.660 23.813
         7.682 23.892
         7.709 23.971
         7.742 24.051
         7.779 24.130
         7.821 24.209
         7.868 24.289
         7.920 24.368
         7.977 24.448
         8.039 24.527
         8.106 24.606
         8.178 24.686
         8.255 24.765
         8.334 24.839
         8.414 24.904
         8.493 24.958
         8.572 25.003
         8.652 25.038
         8.731 25.063
         8.811 25.078
         8.890 25.082
         8.969 25.078
         9.049 25.063
         9.128 25.038
         9.207 25.003
         9.287 24.958
         9.366 24.904
         9.446 24.839
         9.525 24.765
         9.602 24.686
         9.674 24.606
         9.741 24.527
         9.803 24.448
         9.860 24.368
         9.912 24.289
         9.959 24.209
        10.001 24.130
        10.038 24.051
        10.071 23.971
        10.098 23.892
        10.120 23.813
        10.138 23.733
        10.150 23.654
        10.158 23.574
        10.160 23.495
        10.158 23.416
        10.150 23.336
        10.138 23.257
        10.120 23.178
        10.098 23.098
        10.071 23.019
        10.038 22.939
        10.001 22.860
         9.959 22.781
         9.912 22.701
         9.860 22.622
         9.803 22.543
         9.741 22.463
         9.674 22.384
         9.602 22.304
         9.525 22.225
         /
\plot  9.525 22.225  8.890 21.590 /
%
% Fig POLYLINE object
%
\linethickness= 0.500pt
\setplotsymbol ({\thinlinefont .})
\putrule from  5.080 24.130 to  5.080 19.050
%
% Fig TEXT object
%
\put{\SetFigFont{12}{14.4}{rm}$u$ plane} [lB] at 10.795 25.082
%
% Fig TEXT object
%
\put{${\cal C}_u$} [lB] at  9.842 22.066
%
% Fig TEXT object
%
\put{$\Delta_{j\in{\cal I}}$} [lB] at  8.255 23.336
%
% Fig TEXT object
%
\put{$\Delta_{i\in{\cal O}}$} [lB] at  5.715 24.606
%
% Fig TEXT object
%
\put{$1$} [lB] at  8.890 21.273
\linethickness=0pt
\putrectangle corners at  3.785 25.387 and 12.725 19.025
\endpicture}

\caption{The integration contour ${\cal C}_u$ for a t-channel block 
corresponding to fusion rule II.}
\label{figpryut}
\end{figure}

\begin{figure}
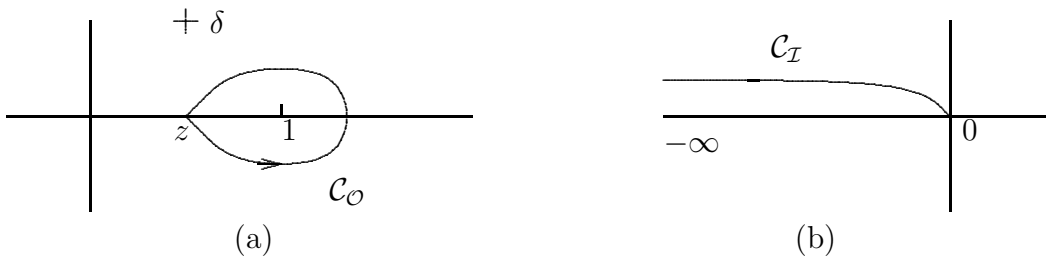

\font\thinlinefont=cmr5
\begingroup\makeatletter\ifx\SetFigFont\undefined
% extract first six characters in \fmtname
\def\x#1#2#3#4#5#6#7\relax{\def\x{#1#2#3#4#5#6}}%
\expandafter\x\fmtname xxxxxx\relax \def\y{splain}%
\ifx\x\y   % LaTeX or SliTeX?
\gdef\SetFigFont#1#2#3{%
  \ifnum #1<17\tiny\else \ifnum #1<20\small\else
  \ifnum #1<24\normalsize\else \ifnum #1<29\large\else
  \ifnum #1<34\Large\else \ifnum #1<41\LARGE\else
     \huge\fi\fi\fi\fi\fi\fi
  \csname #3\endcsname}%
\else
\gdef\SetFigFont#1#2#3{\begingroup
  \count@#1\relax \ifnum 25<\count@\count@25\fi
  \def\x{\endgroup\@setsize\SetFigFont{#2pt}}%
  \expandafter\x
    \csname \romannumeral\the\count@ pt\expandafter\endcsname
    \csname @\romannumeral\the\count@ pt\endcsname
  \csname #3\endcsname}%
\fi
\fi\endgroup
\mbox{\beginpicture
\setcoordinatesystem units <1.00000cm,1.00000cm>
\unitlength=1.00000cm
\linethickness=1pt
\setplotsymbol ({\makebox(0,0)[l]{\tencirc\symbol{'160}}})
\setshadesymbol ({\thinlinefont .})
\setlinear
%
% Fig POLYLINE object
%
\linethickness= 0.500pt
\setplotsymbol ({\thinlinefont .})
\putrule from  1.429 22.860 to  7.620 22.860
%
% Fig POLYLINE object
%
\linethickness= 0.500pt
\setplotsymbol ({\thinlinefont .})
\putrule from  5.080 23.019 to  5.080 22.860
%
% Fig POLYLINE object
%
\linethickness= 0.500pt
\setplotsymbol ({\thinlinefont .})
\putrule from 10.160 22.860 to 15.240 22.860
%
% Fig POLYLINE object
%
\linethickness= 0.500pt
\setplotsymbol ({\thinlinefont .})
\putrule from 13.970 24.130 to 13.970 21.590
%
% Fig POLYLINE object
%
\linethickness= 0.500pt
\setplotsymbol ({\thinlinefont .})
\putrule from 11.271 23.336 to 11.430 23.336
\putrule from 11.430 23.336 to 11.430 23.336
%
% Fig POLYLINE object
%
\linethickness= 0.500pt
\setplotsymbol ({\thinlinefont .})
\putrule from  4.763 22.225 to  5.080 22.225
%
% arrow head
%
\plot  4.826 22.162  5.080 22.225  4.826 22.289 /
\linethickness= 0.500pt
\setplotsymbol ({\thinlinefont .})
%
% Fig CONTROL PT SPLINE
%
% open spline
%
\plot	 3.810 22.860  4.128 23.178
 	 4.212 23.252
	 4.306 23.316
	 4.410 23.371
	 4.524 23.416
	 4.648 23.450
	 4.714 23.464
	 4.782 23.475
	 4.853 23.484
	 4.926 23.490
	 5.002 23.494
	 5.080 23.495
	 5.158 23.494
	 5.231 23.490
	 5.301 23.484
	 5.368 23.475
	 5.489 23.450
	 5.596 23.416
	 5.688 23.371
	 5.765 23.316
	 5.827 23.252
	 5.874 23.178
	 5.908 23.098
	 5.933 23.019
	 5.948 22.939
	 5.953 22.860
	 5.948 22.781
	 5.933 22.701
	 5.908 22.622
	 5.874 22.543
	 5.827 22.468
	 5.765 22.404
	 5.688 22.349
	 5.596 22.304
	 5.489 22.270
	 5.368 22.245
	 5.301 22.236
	 5.231 22.230
	 5.158 22.226
	 5.080 22.225
	 5.002 22.226
	 4.926 22.230
	 4.853 22.236
	 4.782 22.245
	 4.714 22.256
	 4.648 22.270
	 4.524 22.304
	 4.410 22.349
	 4.306 22.404
	 4.212 22.468
	 4.128 22.543
	 /
\plot  4.128 22.543  3.810 22.860 /
\linethickness= 0.500pt
\setplotsymbol ({\thinlinefont .})
%
% Fig CONTROL PT SPLINE
%
% open spline
%
\plot	 3.810 23.971  3.810 24.130
 	 /
\plot  3.810 24.130  3.810 24.289 /
\linethickness= 0.500pt
\setplotsymbol ({\thinlinefont .})
%
% Fig CONTROL PT SPLINE
%
% open spline
%
\plot	 3.651 24.130  3.810 24.130
 	 /
\plot  3.810 24.130  3.969 24.130 /
%
% Fig POLYLINE object
%
\linethickness= 0.500pt
\setplotsymbol ({\thinlinefont .})
\putrule from  2.540 24.130 to  2.540 21.590
\linethickness= 0.500pt
\setplotsymbol ({\thinlinefont .})
%
% Fig CONTROL PT SPLINE
%
% open spline
%
\plot	10.160 23.336 11.430 23.336
 	11.509 23.336
	11.586 23.336
	11.661 23.336
	11.735 23.335
	11.807 23.334
	11.878 23.333
	11.948 23.332
	12.015 23.331
	12.082 23.330
	12.146 23.328
	12.271 23.325
	12.389 23.321
	12.502 23.316
	12.608 23.311
	12.707 23.305
	12.801 23.299
	12.889 23.292
	12.970 23.284
	13.045 23.275
	13.114 23.266
	13.176 23.257
	13.290 23.236
	13.395 23.212
	13.489 23.186
	13.573 23.158
	13.648 23.127
	13.712 23.093
	13.811 23.019
	 /
\plot 13.811 23.019 13.970 22.860 /
%
% Fig TEXT object
%
\put{${\cal C}_{\cal I}$} [lB] at 11.589 23.654
%
% Fig TEXT object
%
\put{$z$} [lB] at  3.651 22.543
%
% Fig TEXT object
%
\put{$1$} [lB] at  5.080 22.543
%
% Fig TEXT object
%
\put{$\delta$} [lB] at  4.128 23.971
%
% Fig TEXT object
%
\put{$0$} [lB] at 14.129 22.543
%
% Fig TEXT object
%
\put{$-\infty$} [lB] at 10.160 22.384
%
% Fig TEXT object
%
\put{\SetFigFont{12}{14.4}{rm}(b)} [lB] at 11.906 21.114
%
% Fig TEXT object
%
\put{\SetFigFont{12}{14.4}{rm}(a)} [lB] at  4.445 21.114
%
% Fig TEXT object
%
\put{${\cal C}_{\cal O}$} [lB] at  5.715 21.749
\linethickness=0pt
\putrectangle corners at  1.403 24.306 and 15.265 21.038
\endpicture}

\caption{The integration contours for the $w_i$'s: ${\cal C}_{\cal O}$ (a) and
${\cal C}_{\cal I}$ (b) for a t-channel block corresponding to fusion rule II.}
\label{figprywt}
\end{figure}

We next describe how contours have to be chosen in order to produce a 
t-channel block corresponding to fusion rule II. This situation can occur
only provided there is at least one screening operator of the second kind. 
We use the same defining 
equation as in (\ref{pryblock}), but the sets of indices as well as $r$ 
and $s$ have different meanings. Again we have a total of $R$ 
and $S$ screenings of the first and second kinds. There are $r$ and
$s$ screenings associated with the upper vertex, and the corresponding
index set for the $w_i$'s is ${\cal O}$. There are $R-r$ and $S-s$ 
screening operators of the two kinds associated with the lower vertex and the 
corresponding index set for the $w_i$'s is ${\cal I}$.
The integration contour
${\cal C}_u$ is indicated in Fig. \ref{figpryut}. One checks that in the limit
$z\rightarrow 1$ followed by $x\rightarrow 1$ the two sets of singularities,
$\Delta_i$ for $i\in{\cal O}$ and $i\in{\cal I}$ respectively, are well 
separated, so that the contour may be taken to separate them as indicated.
The contours ${\cal C}_{\cal O}$ and ${\cal C}_{\cal I}$
for the two sets of $w_i$ variables, are shown 
in Fig. \ref{figprywt} (a) and (b).
In all cases one checks as for the s-channel block that the nature of 
singularities is such that the contours may be chosen as indicated, and that
the block thus defined will satisfy the KZ equations. 
Then we investigate the combined behaviour 
$z\rightarrow 1$ followed by $x\rightarrow 1$. To this end we perform the 
following scalings of the integration variables
\bea
 w_i&\rightarrow&\frac{w_i-1}{z-1}, \ \ i\in{\cal O}\nn
 w_i&\rightarrow&\frac{w_i}{w_i-1}, \ \ i\in{\cal I}\nn
 u&\rightarrow&\frac{u-\Delta_{j_0}}{1-\Delta_{j_0}}
\eea
where $j_0$ is an arbitrary index in the set ${\cal I}$, however with the 
restriction that $w_{j_0}$ is the position of a screening operator of the 
second kind. It is rather straightforward 
to check that this gives rise to the combined singular behaviour
\ben
 W^{(R,S)}_{(r,s)}(z,x)\sim (1-z)^{-\D(j_2)-\D(j_3)+\D(j_{II})}
 (x-1)^{2j_2+2j_3-r+s t+1}
\label{pryII}
\een
where 
\bea
j_{II}&=&-j_I-1\nn
j_I&=&j_2+j_3-r+s t
\eea
so that (\ref{pryII}) exactly demonstrates that we have fusion rule II, since
$j_2+j_3-j_{II}=2j_2+2j_3-r+s t+1$.

It follows that the conformal blocks defined on the basis of the free field 
realization elaborated in \cite{PRY1, PRY2}, indeed do give rise to both the
fusion rules.

\begin{figure}
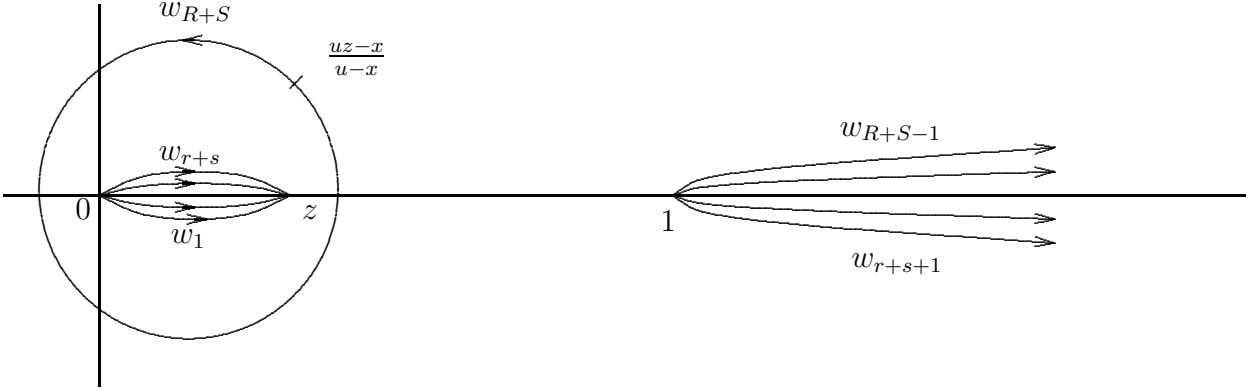

\font\thinlinefont=cmr5
\begingroup\makeatletter\ifx\SetFigFont\undefined
% extract first six characters in \fmtname
\def\x#1#2#3#4#5#6#7\relax{\def\x{#1#2#3#4#5#6}}%
\expandafter\x\fmtname xxxxxx\relax \def\y{splain}%
\ifx\x\y   % LaTeX or SliTeX?
\gdef\SetFigFont#1#2#3{%
  \ifnum #1<17\tiny\else \ifnum #1<20\small\else
  \ifnum #1<24\normalsize\else \ifnum #1<29\large\else
  \ifnum #1<34\Large\else \ifnum #1<41\LARGE\else
     \huge\fi\fi\fi\fi\fi\fi
  \csname #3\endcsname}%
\else
\gdef\SetFigFont#1#2#3{\begingroup
  \count@#1\relax \ifnum 25<\count@\count@25\fi
  \def\x{\endgroup\@setsize\SetFigFont{#2pt}}%
  \expandafter\x
    \csname \romannumeral\the\count@ pt\expandafter\endcsname
    \csname @\romannumeral\the\count@ pt\endcsname
  \csname #3\endcsname}%
\fi
\fi\endgroup
\mbox{\beginpicture
\setcoordinatesystem units <1.00000cm,1.00000cm>
\unitlength=1.00000cm
\linethickness=1pt
\setplotsymbol ({\makebox(0,0)[l]{\tencirc\symbol{'160}}})
\setshadesymbol ({\thinlinefont .})
\setlinear
%
% Fig POLYLINE object
%
\linethickness= 0.500pt
\setplotsymbol ({\thinlinefont .})
\putrule from  1.270 21.590 to 17.780 21.590
%
% Fig POLYLINE object
%
\linethickness= 0.500pt
\setplotsymbol ({\thinlinefont .})
\putrule from  2.540 24.130 to  2.540 19.050
%
% Fig POLYLINE object
%
\linethickness= 0.500pt
\setplotsymbol ({\thinlinefont .})
\putrule from  5.239 23.019 to  5.239 23.019
%
% Fig POLYLINE object
%
\linethickness= 0.500pt
\setplotsymbol ({\thinlinefont .})
\plot  5.080 23.019  5.239 23.178 /
%
% Fig POLYLINE object
%
\linethickness= 0.500pt
\setplotsymbol ({\thinlinefont .})
\putrule from  3.651 21.907 to  3.810 21.907
%
% arrow head
%
\plot  3.556 21.844  3.810 21.907  3.556 21.971 /
%
%
% Fig POLYLINE object
%
\linethickness= 0.500pt
\setplotsymbol ({\thinlinefont .})
\putrule from  3.651 21.749 to  3.810 21.749
%
% arrow head
%
\plot  3.556 21.685  3.810 21.749  3.556 21.812 /
%
%
% Fig POLYLINE object
%
\linethickness= 0.500pt
\setplotsymbol ({\thinlinefont .})
\putrule from  3.651 21.431 to  3.810 21.431
%
% arrow head
%
\plot  3.556 21.368  3.810 21.431  3.556 21.495 /
%
%
% Fig POLYLINE object
%
\linethickness= 0.500pt
\setplotsymbol ({\thinlinefont .})
\putrule from  3.810 21.273 to  3.969 21.273
%
% arrow head
%
\plot  3.715 21.209  3.969 21.273  3.715 21.336 /
%
%
% Fig POLYLINE object
%
\linethickness= 0.500pt
\setplotsymbol ({\thinlinefont .})
\putrule from  3.810 23.654 to  3.651 23.654
%
% arrow head
%
\plot  3.905 23.717  3.651 23.654  3.905 23.590 /
%
%
% Fig POLYLINE object
%
\linethickness= 0.500pt
\setplotsymbol ({\thinlinefont .})
\plot 11.589 26.035 11.589 26.035 /
\linethickness= 0.500pt
\setplotsymbol ({\thinlinefont .})
%
% Fig CONTROL PT SPLINE
%
% open spline
%
\plot	 2.540 21.590  2.857 21.669
 	 2.942 21.688
	 3.036 21.704
	 3.140 21.718
	 3.254 21.729
	 3.378 21.738
	 3.444 21.741
	 3.512 21.744
	 3.583 21.746
	 3.656 21.748
	 3.732 21.748
	 3.810 21.749
	 3.888 21.748
	 3.964 21.748
	 4.037 21.746
	 4.108 21.744
	 4.176 21.741
	 4.242 21.738
	 4.366 21.729
	 4.480 21.718
	 4.584 21.704
	 4.678 21.688
	 4.763 21.669
	 /
\plot  4.763 21.669  5.080 21.590 /
\linethickness= 0.500pt
\setplotsymbol ({\thinlinefont .})
%
% Fig CONTROL PT SPLINE
%
% open spline
%
\plot	 5.080 21.590  4.763 21.511
 	 4.678 21.492
	 4.584 21.476
	 4.480 21.462
	 4.366 21.451
	 4.242 21.442
	 4.176 21.439
	 4.108 21.436
	 4.037 21.434
	 3.964 21.432
	 3.888 21.432
	 3.810 21.431
	 3.732 21.432
	 3.656 21.432
	 3.583 21.434
	 3.512 21.436
	 3.444 21.439
	 3.378 21.442
	 3.254 21.451
	 3.140 21.462
	 3.036 21.476
	 2.942 21.492
	 2.857 21.511
	 /
\plot  2.857 21.511  2.540 21.590 /
\linethickness= 0.500pt
\setplotsymbol ({\thinlinefont .})
%
% Fig CONTROL PT SPLINE
%
% open spline
%
\plot	 2.540 21.590  2.857 21.749
 	 2.942 21.786
	 3.036 21.818
	 3.140 21.845
	 3.254 21.868
	 3.378 21.885
	 3.444 21.892
	 3.512 21.898
	 3.583 21.902
	 3.656 21.905
	 3.732 21.907
	 3.810 21.907
	 3.888 21.907
	 3.964 21.905
	 4.037 21.902
	 4.108 21.898
	 4.176 21.892
	 4.242 21.885
	 4.366 21.868
	 4.480 21.845
	 4.584 21.818
	 4.678 21.786
	 4.763 21.749
	 /
\plot  4.763 21.749  5.080 21.590 /
%
% Fig ELLIPSE
%
\linethickness= 0.500pt
\setplotsymbol ({\thinlinefont .})
\ellipticalarc axes ratio  1.988:1.988  360 degrees 
	from  5.717 21.670 center at  3.730 21.670
\linethickness= 0.500pt
\setplotsymbol ({\thinlinefont .})
%
% Fig CONTROL PT SPLINE
%
% open spline
%
\plot	 2.540 21.590  2.857 21.431
 	 2.942 21.394
	 3.036 21.362
	 3.140 21.335
	 3.254 21.312
	 3.378 21.295
	 3.444 21.288
	 3.512 21.282
	 3.583 21.278
	 3.656 21.275
	 3.732 21.273
	 3.810 21.273
	 3.888 21.273
	 3.964 21.275
	 4.037 21.278
	 4.108 21.282
	 4.176 21.288
	 4.242 21.295
	 4.366 21.312
	 4.480 21.335
	 4.584 21.362
	 4.678 21.394
	 4.763 21.431
	 /
\plot  4.763 21.431  5.080 21.590 /
%
% Fig TEXT object
%
\put{$w_{R+S}$} [lB] at  3.334 23.971
\linethickness= 0.500pt
\setplotsymbol ({\thinlinefont .})
%
% Fig CONTROL PT SPLINE
%
% open spline
%
\plot	10.160 21.590 10.398 21.669
 	10.490 21.689
	10.560 21.699
	10.646 21.709
	10.748 21.719
	10.867 21.729
	10.932 21.734
	11.001 21.739
	11.075 21.744
	11.152 21.749
	11.234 21.754
	11.319 21.759
	11.408 21.764
	11.502 21.769
	11.599 21.774
	11.701 21.779
	11.807 21.783
	11.916 21.788
	12.030 21.793
	12.147 21.798
	12.269 21.803
	12.395 21.808
	12.459 21.811
	12.525 21.813
	12.591 21.816
	12.658 21.818
	12.727 21.821
	12.796 21.823
	12.867 21.826
	12.938 21.828
	 /
\plot 12.938 21.828 15.240 21.907 /
%
% arrow head
%
\plot 14.988 21.835 15.240 21.907 14.984 21.962 /
\linethickness= 0.500pt
\setplotsymbol ({\thinlinefont .})
%
% Fig CONTROL PT SPLINE
%
% open spline
%
\plot	10.160 21.590 10.398 21.749
 	10.490 21.788
	10.560 21.808
	10.646 21.828
	10.748 21.848
	10.867 21.868
	10.932 21.878
	11.001 21.888
	11.075 21.898
	11.152 21.907
	11.234 21.917
	11.319 21.927
	11.408 21.937
	11.502 21.947
	11.599 21.957
	11.701 21.967
	11.807 21.977
	11.916 21.987
	12.030 21.997
	12.147 22.007
	12.269 22.017
	12.395 22.027
	12.459 22.032
	12.525 22.036
	12.591 22.041
	12.658 22.046
	12.727 22.051
	12.796 22.056
	12.867 22.061
	12.938 22.066
	 /
\plot 12.938 22.066 15.240 22.225 /
%
% arrow head
%
\plot 14.991 22.144 15.240 22.225 14.982 22.271 /
\linethickness= 0.500pt
\setplotsymbol ({\thinlinefont .})
%
% Fig CONTROL PT SPLINE
%
% open spline
%
\plot	10.160 21.590 10.398 21.511
 	10.490 21.491
	10.560 21.481
	10.646 21.471
	10.748 21.461
	10.867 21.451
	10.932 21.446
	11.001 21.441
	11.075 21.436
	11.152 21.431
	11.234 21.426
	11.319 21.421
	11.408 21.416
	11.502 21.411
	11.599 21.406
	11.701 21.401
	11.807 21.397
	11.916 21.392
	12.030 21.387
	12.147 21.382
	12.269 21.377
	12.395 21.372
	12.459 21.369
	12.525 21.367
	12.591 21.364
	12.658 21.362
	12.727 21.359
	12.796 21.357
	12.867 21.354
	12.938 21.352
	 /
\plot 12.938 21.352 15.240 21.273 /
%
% arrow head
%
\plot 14.984 21.218 15.240 21.273 14.988 21.345 /
\linethickness= 0.500pt
\setplotsymbol ({\thinlinefont .})
%
% Fig CONTROL PT SPLINE
%
% open spline
%
\plot	10.160 21.590 10.398 21.431
 	10.490 21.392
	10.560 21.372
	10.646 21.352
	10.748 21.332
	10.867 21.312
	10.932 21.302
	11.001 21.292
	11.075 21.282
	11.152 21.273
	11.234 21.263
	11.319 21.253
	11.408 21.243
	11.502 21.233
	11.599 21.223
	11.701 21.213
	11.807 21.203
	11.916 21.193
	12.030 21.183
	12.147 21.173
	12.269 21.163
	12.395 21.153
	12.459 21.148
	12.525 21.144
	12.591 21.139
	12.658 21.134
	12.727 21.129
	12.796 21.124
	12.867 21.119
	12.938 21.114
	 /
\plot 12.938 21.114 15.240 20.955 /
%
% arrow head
%
\plot 14.982 20.909 15.240 20.955 14.991 21.036 /
%
%
% Fig TEXT object
%
\put{$0$} [lB] at  2.223 21.273
%
% Fig TEXT object
%
\put{$z$} [lB] at  5.239 21.273
%
% Fig TEXT object
%
\put{$w_{r+s}$} [lB] at  3.334 22.066
%
% Fig TEXT object
%
\put{$w_1$} [lB] at  3.493 20.955
%
% Fig TEXT object
%
\put{$\frac{uz-x}{u-x}$} [lB] at  5.556 23.336
%
% Fig TEXT object
%
\put{$1$} [lB] at 10.001 21.114
%
% Fig TEXT object
%
\put{$w_{R+S-1}$} [lB] at 12.383 22.384
%
% Fig TEXT object
%
\put{$w_{r+s+1}$} [lB] at 12.541 20.637
\linethickness=0pt
\putrectangle corners at  1.245 26.060 and 17.805 19.025
\endpicture}

\caption{Integration contours for an s-channel block corresponding to fusion
rule II. The corresponding $u$ integration is simply along the
unit circle.}
\label{figprywsII}
\end{figure}

In \cite{PRY2} also realizations of conformal blocks 
corresponding to fusion rule
II in the s-channel and of ones corresponding to fusion rule I in the 
t-channel, have been found. However, these are not given by quite as 
simple contours as above. It is the appearance of non-pure
singularities which prevents one from finding such simple contours. 
The new idea is to carry out first the integrations of
the screening operators letting the contours depend on $u$. Then there are
only pure singularities in the $w_i$ planes and there will be no problems
caused by non-pure singularities. It turns out that it is possible to
find contours like that leaving, upon integration, a simple $u$ integral.
Let us first consider the conformal blocks in the s-channel corresponding
to fusion rule II, where the contours are depicted in Fig. \ref{figprywsII}
\bea
 &&W^{(R,S)}_{(r,s)}(j_1,j_2,j_3,j_4;z,x)\nn
 &=&z^{2j_1j_2/t}(1-z)^{2j_2j_3/t}
  \oint_{{\cal C}_u}\frac{du}{2\pi i}
  \oint_{{\cal C}_w}\frac{dw}{2\pi i}
  \int_1^\infty\prod_{i\in {\cal I}}dw_i
  \int_0^z\prod_{i\in {\cal O}} dw_i \nn
 &\cdot&w_i^{2k_ij_1/t}(w_i-z)^{2k_ij_2/t}(w_i-1)^{2k_ij_3/t}
  \prod_{\stackrel{i,j\in{\cal A}}{i<j}}(w_i-w_j)^{2k_ik_j/t}\nn
 &\cdot&\prod_{i\in{\cal A}}\left ( -\frac{u}{w_i-1}+
  \frac{x}{w_i-z}\right )^{-k_i}(1-u)^{2j_2+2j_3-R +S t}u^{-2j_3-1}
\label{pryb}
\eea
where 
\bea
 {\cal A}&=&{\cal I}\cup {\cal O}\cup \{ R+S\} \nn
 w &=& w_{R+S} \nn
 k_{R+S} &=& t
\eea
To see that the above formula produces the right singular behaviour 
in the limit
$z\rightarrow 0$ followed by $x\rightarrow 0$, we may scale all the 
$w_j$'s with $j\in{\cal O}$ as
\ben
 w_j\rightarrow zw_j, \ \ \ \ \ \ \mbox{for all } \ \ j \in {\cal O}
\een
and scale all the $w_i$'s with $i\in{\cal I}$ as
\ben
 w_i\rightarrow 1/w_i, \ \ \ \ \ \ \mbox{for all } \ \ i \in {\cal I}
\een
and also
\ben
 w\rightarrow \frac{uz-x}{u-x} w\een
One can then show that
\ben
 W^{(R,S)}_{(r,s)}(z,x)\sim z^{-\D(j_1)-\D(j_2)+\D(j_{II})}
  (-x)^{j_1+j_2-j_{II}}
\een
where 
\ben
 j_{II}=-j_1-j_2+r-s t-1
\een
This is precisely the expected singular behaviour.
\begin{figure}
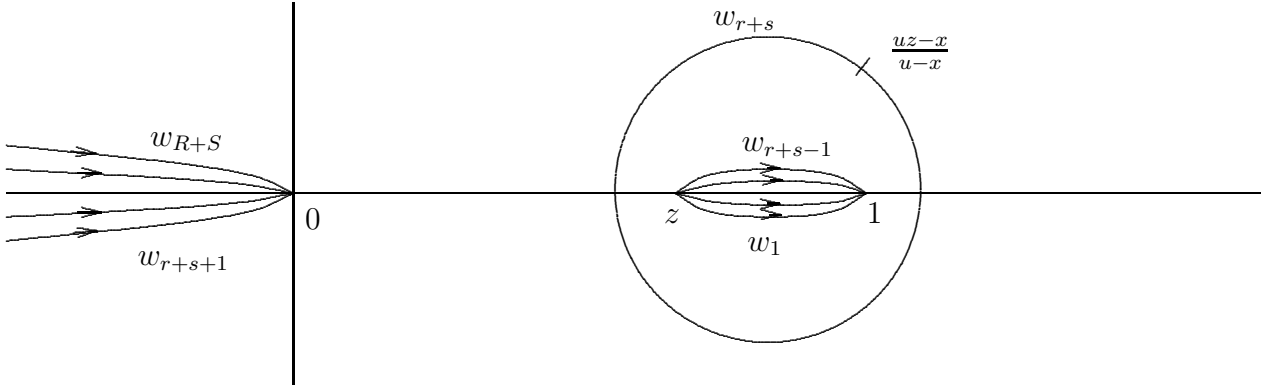

\font\thinlinefont=cmr5
\begingroup\makeatletter\ifx\SetFigFont\undefined
% extract first six characters in \fmtname
\def\x#1#2#3#4#5#6#7\relax{\def\x{#1#2#3#4#5#6}}%
\expandafter\x\fmtname xxxxxx\relax \def\y{splain}%
\ifx\x\y   % LaTeX or SliTeX?
\gdef\SetFigFont#1#2#3{%
  \ifnum #1<17\tiny\else \ifnum #1<20\small\else
  \ifnum #1<24\normalsize\else \ifnum #1<29\large\else
  \ifnum #1<34\Large\else \ifnum #1<41\LARGE\else
     \huge\fi\fi\fi\fi\fi\fi
  \csname #3\endcsname}%
\else
\gdef\SetFigFont#1#2#3{\begingroup
  \count@#1\relax \ifnum 25<\count@\count@25\fi
  \def\x{\endgroup\@setsize\SetFigFont{#2pt}}%
  \expandafter\x
    \csname \romannumeral\the\count@ pt\expandafter\endcsname
    \csname @\romannumeral\the\count@ pt\endcsname
  \csname #3\endcsname}%
\fi
\fi\endgroup
\mbox{\beginpicture
\setcoordinatesystem units <1.00000cm,1.00000cm>
\unitlength=1.00000cm
\linethickness=1pt
\setplotsymbol ({\makebox(0,0)[l]{\tencirc\symbol{'160}}})
\setshadesymbol ({\thinlinefont .})
\setlinear
%
% Fig POLYLINE object
%
\linethickness= 0.500pt
\setplotsymbol ({\thinlinefont .})
\putrule from  1.270 21.590 to 17.939 21.590
%
% Fig POLYLINE object
%
\linethickness= 0.500pt
\setplotsymbol ({\thinlinefont .})
\putrule from  5.080 24.130 to  5.080 19.050
%
% Fig POLYLINE object
%
\linethickness= 0.500pt
\setplotsymbol ({\thinlinefont .})
\plot  2.127 22.147  2.451 22.121 /
%
% arrow head
%
\plot  2.193 22.078  2.451 22.121  2.203 22.204 /
%
%
% Fig POLYLINE object
%
\linethickness= 0.500pt
\setplotsymbol ({\thinlinefont .})
\putrule from  2.206 21.861 to  2.523 21.861
%
% arrow head
%
\plot  2.269 21.797  2.523 21.861  2.269 21.924 /
%
%
% Fig POLYLINE object
%
\linethickness= 0.500pt
\setplotsymbol ({\thinlinefont .})
\putrule from  2.206 21.336 to  2.523 21.336
%
% arrow head
%
\plot  2.269 21.273  2.523 21.336  2.269 21.400 /
%
%
% Fig POLYLINE object
%
\linethickness= 0.500pt
\setplotsymbol ({\thinlinefont .})
\plot  2.155 21.063  2.470 21.090 /
%
% arrow head
%
\plot  2.223 21.005  2.470 21.090  2.212 21.132 /
%
%
% Fig POLYLINE object
%
\linethickness= 0.500pt
\setplotsymbol ({\thinlinefont .})
\putrule from 11.390 21.924 to 11.549 21.924
%
% arrow head
%
\plot 11.295 21.861 11.549 21.924 11.295 21.988 /
%
%
% Fig POLYLINE object
%
\linethickness= 0.500pt
\setplotsymbol ({\thinlinefont .})
\putrule from 11.422 21.766 to 11.580 21.766
%
% arrow head
%
\plot 11.326 21.702 11.580 21.766 11.326 21.829 /
%
%
% Fig POLYLINE object
%
\linethickness= 0.500pt
\setplotsymbol ({\thinlinefont .})
\putrule from 11.390 21.448 to 11.549 21.448
%
% arrow head
%
\plot 11.295 21.385 11.549 21.448 11.295 21.512 /
%
%
% Fig POLYLINE object
%
\linethickness= 0.500pt
\setplotsymbol ({\thinlinefont .})
\putrule from 11.405 21.304 to 11.563 21.304
%
% arrow head
%
\plot 11.309 21.241 11.563 21.304 11.309 21.368 /
%
%
% Fig POLYLINE object
%
\linethickness= 0.500pt
\setplotsymbol ({\thinlinefont .})
\plot 12.565 23.163 12.738 23.385 /
\linethickness= 0.500pt
\setplotsymbol ({\thinlinefont .})
%
% Fig CONTROL PT SPLINE
%
% open spline
%
\plot	 5.080 21.590  4.683 21.669
 	 4.567 21.689
	 4.495 21.699
	 4.415 21.709
	 4.327 21.719
	 4.229 21.729
	 4.123 21.739
	 4.008 21.749
	 3.885 21.759
	 3.820 21.764
	 3.753 21.769
	 3.684 21.774
	 3.612 21.779
	 3.539 21.783
	 3.463 21.788
	 3.385 21.793
	 3.305 21.798
	 3.222 21.803
	 3.138 21.808
	 3.051 21.813
	 2.962 21.818
	 2.871 21.823
	 2.778 21.828
	 /
\plot  2.778 21.828  1.270 21.907 /
\linethickness= 0.500pt
\setplotsymbol ({\thinlinefont .})
%
% Fig CONTROL PT SPLINE
%
% open spline
%
\plot	 5.080 21.590  4.763 21.749
 	 4.663 21.788
	 4.599 21.808
	 4.524 21.828
	 4.440 21.848
	 4.346 21.868
	 4.242 21.888
	 4.128 21.907
	 4.003 21.927
	 3.938 21.937
	 3.870 21.947
	 3.799 21.957
	 3.726 21.967
	 3.650 21.977
	 3.572 21.987
	 3.491 21.997
	 3.408 22.007
	 3.323 22.017
	 3.235 22.027
	 3.144 22.036
	 3.051 22.046
	 2.955 22.056
	 2.857 22.066
	 /
\plot  2.857 22.066  1.270 22.225 /
%
% Fig ELLIPSE
%
\linethickness= 0.500pt
\setplotsymbol ({\thinlinefont .})
\ellipticalarc axes ratio  2.032:2.032  360 degrees 
	from 13.422 21.639 center at 11.390 21.639
\linethickness= 0.500pt
\setplotsymbol ({\thinlinefont .})
%
% Fig CONTROL PT SPLINE
%
% open spline
%
\plot	 5.080 21.590  4.763 21.431
 	 4.663 21.392
	 4.599 21.372
	 4.524 21.352
	 4.440 21.332
	 4.346 21.312
	 4.242 21.292
	 4.128 21.273
	 4.003 21.253
	 3.938 21.243
	 3.870 21.233
	 3.799 21.223
	 3.726 21.213
	 3.650 21.203
	 3.572 21.193
	 3.491 21.183
	 3.408 21.173
	 3.323 21.163
	 3.235 21.153
	 3.144 21.144
	 3.051 21.134
	 2.955 21.124
	 2.857 21.114
	 /
\plot  2.857 21.114  1.270 20.955 /
%
% Fig TEXT object
%
\put{$w_{r+s}$} [lB] at 10.660 23.829
\linethickness= 0.500pt
\setplotsymbol ({\thinlinefont .})
%
% Fig CONTROL PT SPLINE
%
% open spline
%
\plot	 5.080 21.590  4.683 21.511
 	 4.567 21.491
	 4.495 21.481
	 4.415 21.471
	 4.327 21.461
	 4.229 21.451
	 4.123 21.441
	 4.008 21.431
	 3.885 21.421
	 3.820 21.416
	 3.753 21.411
	 3.684 21.406
	 3.612 21.401
	 3.539 21.397
	 3.463 21.392
	 3.385 21.387
	 3.305 21.382
	 3.222 21.377
	 3.138 21.372
	 3.051 21.367
	 2.962 21.362
	 2.871 21.357
	 2.778 21.352
	 /
\plot  2.778 21.352  1.270 21.273 /
\linethickness= 0.500pt
\setplotsymbol ({\thinlinefont .})
%
% Fig CONTROL PT SPLINE
%
% open spline
%
\plot	10.160 21.590 10.478 21.669
 	10.563 21.688
	10.661 21.704
	10.771 21.718
	10.894 21.729
	10.960 21.734
	11.029 21.738
	11.102 21.741
	11.177 21.744
	11.255 21.746
	11.337 21.748
	11.422 21.748
	11.509 21.749
	11.597 21.748
	11.681 21.748
	11.761 21.746
	11.837 21.744
	11.909 21.741
	11.978 21.738
	12.105 21.729
	12.216 21.718
	12.313 21.704
	12.395 21.688
	12.462 21.669
	 /
\plot 12.462 21.669 12.700 21.590 /
\linethickness= 0.500pt
\setplotsymbol ({\thinlinefont .})
%
% Fig CONTROL PT SPLINE
%
% open spline
%
\plot	10.160 21.590 10.398 21.749
 	10.466 21.786
	10.552 21.818
	10.655 21.845
	10.775 21.868
	10.842 21.877
	10.913 21.885
	10.988 21.892
	11.068 21.898
	11.152 21.902
	11.240 21.905
	11.333 21.907
	11.430 21.907
	11.527 21.907
	11.620 21.905
	11.708 21.902
	11.792 21.898
	11.872 21.892
	11.947 21.885
	12.018 21.877
	12.085 21.868
	12.205 21.845
	12.308 21.818
	12.394 21.786
	12.462 21.749
	 /
\plot 12.462 21.749 12.700 21.590 /
\linethickness= 0.500pt
\setplotsymbol ({\thinlinefont .})
%
% Fig CONTROL PT SPLINE
%
% open spline
%
\plot	10.160 21.590 10.398 21.431
 	10.466 21.394
	10.552 21.362
	10.655 21.335
	10.775 21.312
	10.842 21.303
	10.913 21.295
	10.988 21.288
	11.068 21.282
	11.152 21.278
	11.240 21.275
	11.333 21.273
	11.430 21.273
	11.527 21.273
	11.620 21.275
	11.708 21.278
	11.792 21.282
	11.872 21.288
	11.947 21.295
	12.018 21.303
	12.085 21.312
	12.205 21.335
	12.308 21.362
	12.394 21.394
	12.462 21.431
	 /
\plot 12.462 21.431 12.700 21.590 /
\linethickness= 0.500pt
\setplotsymbol ({\thinlinefont .})
%
% Fig CONTROL PT SPLINE
%
% open spline
%
\plot	10.160 21.590 10.478 21.511
 	10.563 21.492
	10.661 21.476
	10.771 21.462
	10.894 21.451
	10.960 21.446
	11.029 21.442
	11.102 21.439
	11.177 21.436
	11.255 21.434
	11.337 21.432
	11.422 21.432
	11.509 21.431
	11.597 21.432
	11.681 21.432
	11.761 21.434
	11.837 21.436
	11.909 21.439
	11.978 21.442
	12.105 21.451
	12.216 21.462
	12.313 21.476
	12.395 21.492
	12.462 21.511
	 /
\plot 12.462 21.511 12.700 21.590 /
%
% Fig TEXT object
%
\put{$w_{R+S}$} [lB] at  3.175 22.225
%
% Fig TEXT object
%
\put{$w_{r+s+1}$} [lB] at  3.016 20.637
%
% Fig TEXT object
%
\put{$0$} [lB] at  5.239 21.114
%
% Fig TEXT object
%
\put{$w_{r+s-1}$} [lB] at 11.041 22.147
%
% Fig TEXT object
%
\put{$w_1$} [lB] at 11.119 20.828
%
% Fig TEXT object
%
\put{$z$} [lB] at 10.008 21.194
%
% Fig TEXT object
%
\put{$1$} [lB] at 12.706 21.194
%
% Fig TEXT object
%
\put{$\frac{uz-x}{u-x}$} [lB] at 13.009 23.385
\linethickness=0pt
\putrectangle corners at  1.245 24.155 and 17.964 19.025
\endpicture}

\caption{Integration contours for the screening charges in the case of a 
t-channel block corresponding to fusion rule I. The $u$ integration is along 
a closed contour starting in 1 and surrounding $x$.}
\label{figprywtI}
\end{figure}

Finally we consider the conformal blocks in the t-channel corresponding to
fusion rule I, see Fig. \ref{figprywtI}
\bea
 &&W^{(R,S)}_{(r,s)}(j_1,j_2,j_3,j_4;z,x)\nn
 &=&z^{2j_1j_2/t}(1-z)^{2j_2j_3/t}
  \oint_{{\cal C}_u} \frac{du}{2\pi i}
  \oint_{{\cal C}_w}\frac{dw}{2\pi i}
  \int_0^{-\infty}\prod_{i\in {\cal I}}\frac{dw_i}{2\pi i}
  \int_z^1\prod_{i\in {\cal O}}\frac{dw_i}{2\pi i}\nn
 &\cdot&
   w_i^{2k_ij_1/t}(w_i-z)^{2k_ij_2/t}(1-w_i)^{2k_ij_3/t}
  \prod_{\stackrel{i,j\in{\cal A}}{i<j}}(w_i-w_j)^{2k_ik_j/t}\nn
 &\cdot&
  \prod_{i\in{\cal A}}\left ( -\frac{u}{w_i-1}+\frac{x}{w_i-z}\right )^{-k_i} 
  (1-u)^{2j_2+2j_3-R +S t}u^{-2j_3-1}
\label{pryt1a}
\eea
where 
\bea
 {\cal A}&=&{\cal I}\cup {\cal O}\cup \{ r+s\} \nn
 w &=& w_{r+s} \nn
 k_{r+s} &=& t
\eea
To see that the above formula produces the right singular behaviour 
in the limit
$z\rightarrow 1$ followed by $x\rightarrow 1$, we may scale all the 
$w_j$'s with $j\in{\cal O}$ as
\ben
 w_j\rightarrow 1-(1-z)w_j, \ \ \ \ \ \ \mbox{for all  } j \in {\cal O}
\een
and scale all $w_i$'s with $i\in{\cal I}$ as
\ben
 w_i\rightarrow w_i/(w_i-1), \ \ \ \ \ \ \mbox{for all } i \in {\cal I}
\een
and also
\ben
 w\rightarrow 1-\frac{(1-z)}{u-x}u w 
\een
We also scale $u$ as
\ben
 u \rightarrow x+(1-x)u 
\een
It should be noticed that the final $u$ contour starts at 1 and goes
along the unit circle such that it surrounds 0 and the other points
which are away from 0 by a distance of order $(1-z)/(1-x)$. This means that
we can not deform the $u$ contour to the form $\int_0^1du$, or in terms of the 
original $u$ variable, that cannot be deformed into $\int_x^1du$.
Using these scalings, we show that in the presence of at least one 
screening charge of the second kind in the scaling region (the region
close to $1$ and $z$) the singular behaviour is  
\ben
 W^{(R,S)}_{(r,s)}(z,x)\sim (1-z)^{-\D(j_2)-\D(j_3)+\D(j_I)}(x-1)^{j_2+j_3-j_I}
\een
where 
\ben
 j_I=j_2+j_3-r+s t
\een

What happens if there is no screening charge of the second kind in the scaling
region? Then the above method does not apply, but in that case
$j_2+j_3-j_I$ is an integer, and 
\ben
 W^{(R,S)}_{(r,0)}(z,x)\sim (1-z)^{-\D(j_2)-\D(j_3)+\D(j_I)}(x-1)^{j_2+j_3-j_I}
\label{nosec}
\een
is a polynomial in $x$.
There will be no extra singularities present in $w$'s,
such as at $\delta = \frac{x-z}{x-1}$, if we integrate over $u$
first. Thus we could choose the following contours ($w=w_{r+s}=w_r\in
{\cal O}$)
\bea
 \int_z^1dw_j&,&j\in {\cal O}\nn
 \int_{-\infty}^0dw_i&,& i \in {\cal I}\nn
 &\int_0^1du&
\eea
These contours are effectively closed in the sense that a total derivative
integrated along them vanishes, such as is required for the 
KZ equations to be satisfied.
Notice, however, that these contours are not closed (in the same sense) 
when there is a screening
charge of the second kind in the scaling region. However,  
it is difficult to determine explicitly the $(1-x)$ behaviour for these 
contours, but since we know that our formula is both projective and $SL(2)$ 
invariant, we could express the above formula in terms of $x_3=0$ and
$x_1=1$, where the $(1-x)$ behaviour is manifest. 

It may seem 
surprising that one could not make use of the $j_1\leftrightarrow j_3$ symmetry
to obtain t-channel contours from s-channel ones and vice versa. The reason is 
that the simple form of the 4-point function we have given with only one 
auxiliary $u$ integration, breaks this symmetry, since not all 4 primary fields
are treated on the same footing. For a more symmetric treatment, more $u$ 
integrations have to be introduced, which is also inconvenient, however.

\section{Andreev's Representation}
In this section we base our discussion on the integral realization of Andreev 
\cite{An}. In a subsequent section we discuss 
the equivalence between that realization and the one described in the 
preceding section \cite{PRY1, PRY2} and in the sequel denoted PRY. 
Here we show how to choose simple integration contours \cite{PRY2}
so that we produce both s-and 
t-channel blocks corresponding to both fusion rules I and II. It will turn out 
that the t-channel blocks are obtained in a very simple way from the s-channel 
blocks so we mostly concentrate on the latter. It is the specification of the 
integration contours which is the contribution here and in \cite{PRY3}
over \cite{An}. 
The advantage of the realization of \cite{An} is that contrary to the case
discussed so far, 
there is no auxiliary integration in addition to the integration 
over positions of screening charges. The disadvantage is that the 
Andreev representation (so far) has no underlying free field 
realization and is therefore only known for 4-point blocks.
For later purposes it is convenient to have different names for s- and 
t-channel blocks. We denote them by letters ${\cal S}$ or $S$ and 
${\cal T}$ or $T$. The differences will be explained.

\subsection{Case of Fusion Rule I}
\begin{figure}
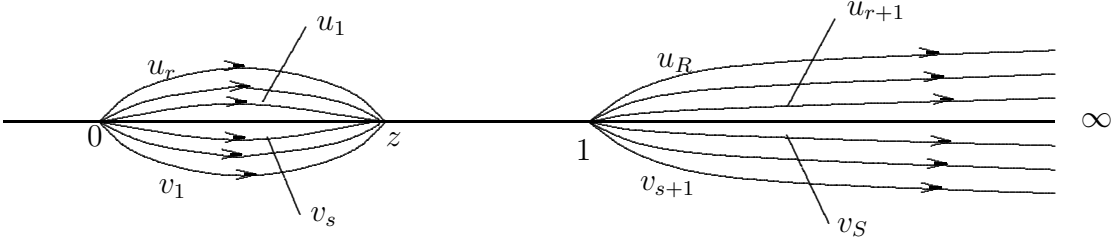

\font\thinlinefont=cmr5
\begingroup\makeatletter\ifx\SetFigFont\undefined
% extract first six characters in \fmtname
\def\x#1#2#3#4#5#6#7\relax{\def\x{#1#2#3#4#5#6}}%
\expandafter\x\fmtname xxxxxx\relax \def\y{splain}%
\ifx\x\y   % LaTeX or SliTeX?
\gdef\SetFigFont#1#2#3{%
  \ifnum #1<17\tiny\else \ifnum #1<20\small\else
  \ifnum #1<24\normalsize\else \ifnum #1<29\large\else
  \ifnum #1<34\Large\else \ifnum #1<41\LARGE\else
     \huge\fi\fi\fi\fi\fi\fi
  \csname #3\endcsname}%
\else
\gdef\SetFigFont#1#2#3{\begingroup
  \count@#1\relax \ifnum 25<\count@\count@25\fi
  \def\x{\endgroup\@setsize\SetFigFont{#2pt}}%
  \expandafter\x
    \csname \romannumeral\the\count@ pt\expandafter\endcsname
    \csname @\romannumeral\the\count@ pt\endcsname
  \csname #3\endcsname}%
\fi
\fi\endgroup
\mbox{\beginpicture
\setcoordinatesystem units <1.00000cm,1.00000cm>
\unitlength=1.00000cm
\linethickness=1pt
\setplotsymbol ({\makebox(0,0)[l]{\tencirc\symbol{'160}}})
\setshadesymbol ({\thinlinefont .})
\setlinear
%
% Fig POLYLINE object
%
\linethickness= 0.500pt
\setplotsymbol ({\thinlinefont .})
\plot  4.714 21.861  5.285 22.845 /
%
% Fig POLYLINE object
%
\linethickness= 0.500pt
\setplotsymbol ({\thinlinefont .})
\plot  4.777 21.385  5.207 20.369 /
%
% Fig POLYLINE object
%
\linethickness= 0.500pt
\setplotsymbol ({\thinlinefont .})
\putrule from  4.477 20.892 to  4.619 20.892
%
% arrow head
%
\plot  4.365 20.828  4.619 20.892  4.365 20.955 /
%
%
% Fig POLYLINE object
%
\linethickness= 0.500pt
\setplotsymbol ({\thinlinefont .})
\putrule from  4.286 21.162 to  4.428 21.162
%
% arrow head
%
\plot  4.174 21.099  4.428 21.162  4.174 21.226 /
%
%
% Fig POLYLINE object
%
\linethickness= 0.500pt
\setplotsymbol ({\thinlinefont .})
\putrule from  4.333 21.385 to  4.475 21.385
%
% arrow head
%
\plot  4.221 21.321  4.475 21.385  4.221 21.448 /
%
%
% Fig POLYLINE object
%
\linethickness= 0.500pt
\setplotsymbol ({\thinlinefont .})
\putrule from  4.350 21.861 to  4.492 21.861
%
% arrow head
%
\plot  4.238 21.797  4.492 21.861  4.238 21.924 /
%
%
% Fig POLYLINE object
%
\linethickness= 0.500pt
\setplotsymbol ({\thinlinefont .})
\putrule from  4.396 22.066 to  4.538 22.066
%
% arrow head
%
\plot  4.284 22.003  4.538 22.066  4.284 22.130 /
%
%
% Fig POLYLINE object
%
\linethickness= 0.500pt
\setplotsymbol ({\thinlinefont .})
\putrule from  4.320 22.337 to  4.462 22.337
%
% arrow head
%
\plot  4.208 22.274  4.462 22.337  4.208 22.401 /
%
%
% Fig POLYLINE object
%
\linethickness= 0.500pt
\setplotsymbol ({\thinlinefont .})
\plot  9.207 19.050  9.207 19.050 /
%
% Fig POLYLINE object
%
\linethickness= 0.500pt
\setplotsymbol ({\thinlinefont .})
\putrule from 13.540 22.511 to 13.540 22.511
\putrule from 13.540 22.511 to 13.699 22.511
%
% arrow head
%
\plot 13.445 22.447 13.699 22.511 13.445 22.574 /
%
%
% Fig POLYLINE object
%
\linethickness= 0.500pt
\setplotsymbol ({\thinlinefont .})
\putrule from 13.604 22.193 to 13.604 22.193
\putrule from 13.604 22.193 to 13.763 22.193
%
% arrow head
%
\plot 13.509 22.130 13.763 22.193 13.509 22.257 /
%
%
% Fig POLYLINE object
%
\linethickness= 0.500pt
\setplotsymbol ({\thinlinefont .})
\putrule from 13.731 21.893 to 13.731 21.893
\putrule from 13.731 21.893 to 13.890 21.893
%
% arrow head
%
\plot 13.636 21.829 13.890 21.893 13.636 21.956 /
%
%
% Fig POLYLINE object
%
\linethickness= 0.500pt
\setplotsymbol ({\thinlinefont .})
\putrule from 13.699 21.336 to 13.699 21.336
\putrule from 13.699 21.336 to 13.858 21.336
%
% arrow head
%
\plot 13.604 21.273 13.858 21.336 13.604 21.400 /
%
%
% Fig POLYLINE object
%
\linethickness= 0.500pt
\setplotsymbol ({\thinlinefont .})
\putrule from 13.650 21.035 to 13.650 21.035
\putrule from 13.650 21.035 to 13.809 21.035
%
% arrow head
%
\plot 13.555 20.972 13.809 21.035 13.555 21.099 /
%
%
% Fig POLYLINE object
%
\linethickness= 0.500pt
\setplotsymbol ({\thinlinefont .})
\putrule from 13.572 20.718 to 13.572 20.718
\putrule from 13.572 20.718 to 13.731 20.718
%
% arrow head
%
\plot 13.477 20.654 13.731 20.718 13.477 20.781 /
%
%
% Fig POLYLINE object
%
\linethickness= 0.500pt
\setplotsymbol ({\thinlinefont .})
\plot 11.699 21.812 12.317 22.955 /
%
% Fig POLYLINE object
%
\linethickness= 0.500pt
\setplotsymbol ({\thinlinefont .})
\plot 11.682 21.416 12.239 20.242 /
\linethickness= 0.500pt
\setplotsymbol ({\thinlinefont .})
%
% Fig CONTROL PT SPLINE
%
% open spline
%
\plot	 2.540 21.590  3.413 21.749
 	 3.523 21.767
	 3.634 21.783
	 3.746 21.797
	 3.860 21.808
	 3.974 21.817
	 4.090 21.823
	 4.207 21.827
	 4.326 21.828
	 4.446 21.827
	 4.567 21.823
	 4.689 21.817
	 4.812 21.808
	 4.937 21.797
	 5.063 21.783
	 5.126 21.776
	 5.190 21.767
	 5.254 21.758
	 5.318 21.749
	 /
\plot  5.318 21.749  6.350 21.590 /
\linethickness= 0.500pt
\setplotsymbol ({\thinlinefont .})
%
% Fig CONTROL PT SPLINE
%
% open spline
%
\plot	 2.540 21.590  3.493 21.431
 	 3.611 21.413
	 3.729 21.397
	 3.847 21.383
	 3.964 21.372
	 4.080 21.363
	 4.196 21.357
	 4.311 21.353
	 4.425 21.352
	 4.539 21.353
	 4.652 21.357
	 4.765 21.363
	 4.877 21.372
	 4.988 21.383
	 5.099 21.397
	 5.209 21.413
	 5.318 21.431
	 /
\plot  5.318 21.431  6.191 21.590 /
\linethickness= 0.500pt
\setplotsymbol ({\thinlinefont .})
%
% Fig CONTROL PT SPLINE
%
% open spline
%
\plot	 2.540 21.590  2.857 21.749
 	 2.942 21.787
	 3.036 21.823
	 3.140 21.857
	 3.254 21.888
	 3.378 21.916
	 3.444 21.930
	 3.512 21.942
	 3.583 21.954
	 3.656 21.966
	 3.732 21.977
	 3.810 21.987
	 3.887 21.996
	 3.958 22.005
	 4.083 22.022
	 4.186 22.035
	 4.266 22.046
	 4.366 22.066
	 4.485 22.046
	 4.576 22.035
	 4.693 22.022
	 4.761 22.014
	 4.834 22.005
	 4.914 21.996
	 5.001 21.987
	 5.088 21.977
	 5.173 21.966
	 5.255 21.954
	 5.333 21.942
	 5.408 21.930
	 5.481 21.916
	 5.550 21.902
	 5.616 21.888
	 5.739 21.857
	 5.849 21.823
	 5.947 21.787
	 6.032 21.749
	 6.106 21.712
	 6.166 21.679
	 6.251 21.630
	 6.271 21.590
	 /
\plot  6.271 21.590  6.191 21.590 /
%
% Fig POLYLINE object
%
\linethickness= 0.500pt
\setplotsymbol ({\thinlinefont .})
\putrule from  1.270 21.590 to 15.240 21.590
\linethickness= 0.500pt
\setplotsymbol ({\thinlinefont .})
%
% Fig CONTROL PT SPLINE
%
% open spline
%
\plot	 2.540 21.590  2.857 21.431
 	 2.942 21.393
	 3.036 21.357
	 3.140 21.323
	 3.254 21.292
	 3.378 21.264
	 3.444 21.250
	 3.512 21.238
	 3.583 21.226
	 3.656 21.214
	 3.732 21.203
	 3.810 21.193
	 3.887 21.184
	 3.958 21.175
	 4.083 21.158
	 4.186 21.145
	 4.266 21.134
	 4.366 21.114
	 4.485 21.134
	 4.576 21.145
	 4.693 21.158
	 4.761 21.166
	 4.834 21.175
	 4.914 21.184
	 5.001 21.193
	 5.088 21.203
	 5.173 21.214
	 5.255 21.226
	 5.333 21.238
	 5.408 21.250
	 5.481 21.264
	 5.550 21.278
	 5.616 21.292
	 5.739 21.323
	 5.849 21.357
	 5.947 21.393
	 6.032 21.431
	 6.106 21.468
	 6.166 21.501
	 6.251 21.550
	 6.271 21.590
	 /
\plot  6.271 21.590  6.191 21.590 /
%
% Fig TEXT object
%
\put{$v_S$} [lB] at 12.349 20.130
\linethickness= 0.500pt
\setplotsymbol ({\thinlinefont .})
%
% Fig CONTROL PT SPLINE
%
% open spline
%
\plot	 2.540 21.590  2.778 21.828
 	 2.845 21.886
	 2.927 21.942
	 3.024 21.996
	 3.135 22.046
	 3.262 22.095
	 3.331 22.118
	 3.403 22.141
	 3.479 22.163
	 3.559 22.184
	 3.643 22.205
	 3.731 22.225
	 3.820 22.244
	 3.909 22.260
	 3.999 22.273
	 4.088 22.285
	 4.177 22.293
	 4.266 22.299
	 4.356 22.303
	 4.445 22.304
	 4.534 22.303
	 4.624 22.299
	 4.713 22.293
	 4.802 22.285
	 4.891 22.273
	 4.981 22.260
	 5.070 22.244
	 5.159 22.225
	 5.247 22.205
	 5.331 22.184
	 5.411 22.163
	 5.487 22.141
	 5.559 22.118
	 5.628 22.095
	 5.755 22.046
	 5.866 21.996
	 5.963 21.942
	 6.045 21.886
	 6.112 21.828
	 /
\plot  6.112 21.828  6.350 21.590 /
\linethickness= 0.500pt
\setplotsymbol ({\thinlinefont .})
%
% Fig CONTROL PT SPLINE
%
% open spline
%
\plot	 2.540 21.590  2.778 21.352
 	 2.845 21.294
	 2.927 21.238
	 3.024 21.184
	 3.135 21.134
	 3.262 21.085
	 3.331 21.062
	 3.403 21.039
	 3.479 21.017
	 3.559 20.996
	 3.643 20.975
	 3.731 20.955
	 3.820 20.936
	 3.909 20.920
	 3.999 20.907
	 4.088 20.895
	 4.177 20.887
	 4.266 20.881
	 4.356 20.877
	 4.445 20.876
	 4.534 20.877
	 4.624 20.881
	 4.713 20.887
	 4.802 20.895
	 4.891 20.907
	 4.981 20.920
	 5.070 20.936
	 5.159 20.955
	 5.247 20.975
	 5.331 20.996
	 5.411 21.017
	 5.487 21.039
	 5.559 21.062
	 5.628 21.085
	 5.755 21.134
	 5.866 21.184
	 5.963 21.238
	 6.045 21.294
	 6.112 21.352
	 /
\plot  6.112 21.352  6.350 21.590 /
\linethickness= 0.500pt
\setplotsymbol ({\thinlinefont .})
%
% Fig CONTROL PT SPLINE
%
% open spline
%
\plot	 9.049 21.590  9.446 21.669
 	 9.504 21.679
	 9.581 21.689
	 9.675 21.699
	 9.788 21.709
	 9.851 21.714
	 9.918 21.719
	 9.990 21.724
	10.067 21.729
	10.148 21.734
	10.233 21.739
	10.323 21.744
	10.418 21.749
	10.517 21.754
	10.620 21.759
	10.728 21.764
	10.841 21.769
	10.958 21.774
	11.079 21.779
	11.205 21.783
	11.270 21.786
	11.336 21.788
	11.403 21.791
	11.471 21.793
	11.540 21.796
	11.610 21.798
	11.682 21.801
	11.754 21.803
	11.828 21.806
	11.903 21.808
	11.978 21.811
	12.055 21.813
	12.134 21.816
	12.213 21.818
	12.293 21.821
	12.375 21.823
	12.457 21.826
	12.541 21.828
	 /
\plot 12.541 21.828 15.240 21.907 /
\linethickness= 0.500pt
\setplotsymbol ({\thinlinefont .})
%
% Fig CONTROL PT SPLINE
%
% open spline
%
\plot	 9.049 21.590  9.446 21.511
 	 9.504 21.501
	 9.581 21.491
	 9.675 21.481
	 9.788 21.471
	 9.851 21.466
	 9.918 21.461
	 9.990 21.456
	10.067 21.451
	10.148 21.446
	10.233 21.441
	10.323 21.436
	10.418 21.431
	10.517 21.426
	10.620 21.421
	10.728 21.416
	10.841 21.411
	10.958 21.406
	11.079 21.401
	11.205 21.397
	11.270 21.394
	11.336 21.392
	11.403 21.389
	11.471 21.387
	11.540 21.384
	11.610 21.382
	11.682 21.379
	11.754 21.377
	11.828 21.374
	11.903 21.372
	11.978 21.369
	12.055 21.367
	12.134 21.364
	12.213 21.362
	12.293 21.359
	12.375 21.357
	12.457 21.354
	12.541 21.352
	 /
\plot 12.541 21.352 15.240 21.273 /
\linethickness= 0.500pt
\setplotsymbol ({\thinlinefont .})
%
% Fig CONTROL PT SPLINE
%
% open spline
%
\plot	 9.049 21.590  9.366 21.749
 	 9.451 21.787
	 9.545 21.823
	 9.649 21.857
	 9.763 21.888
	 9.887 21.916
	 9.953 21.930
	10.021 21.942
	10.092 21.954
	10.165 21.966
	10.241 21.977
	10.319 21.987
	10.404 21.997
	10.501 22.007
	10.610 22.017
	10.731 22.027
	10.795 22.032
	10.863 22.036
	10.934 22.041
	11.007 22.046
	11.084 22.051
	11.163 22.056
	11.245 22.061
	11.331 22.066
	11.419 22.071
	11.510 22.076
	11.604 22.081
	11.702 22.086
	11.802 22.091
	11.905 22.096
	12.011 22.101
	12.120 22.106
	12.231 22.111
	12.346 22.116
	12.464 22.121
	12.585 22.126
	12.708 22.131
	12.835 22.136
	12.899 22.138
	12.964 22.141
	13.030 22.143
	13.097 22.146
	 /
\plot 13.097 22.146 15.240 22.225 /
\linethickness= 0.500pt
\setplotsymbol ({\thinlinefont .})
%
% Fig CONTROL PT SPLINE
%
% open spline
%
\plot	 9.049 21.590  9.366 21.431
 	 9.451 21.393
	 9.545 21.357
	 9.649 21.323
	 9.763 21.292
	 9.887 21.264
	 9.953 21.250
	10.021 21.238
	10.092 21.226
	10.165 21.214
	10.241 21.203
	10.319 21.193
	10.404 21.183
	10.501 21.173
	10.610 21.163
	10.731 21.153
	10.795 21.148
	10.863 21.144
	10.934 21.139
	11.007 21.134
	11.084 21.129
	11.163 21.124
	11.245 21.119
	11.331 21.114
	11.419 21.109
	11.510 21.104
	11.604 21.099
	11.702 21.094
	11.802 21.089
	11.905 21.084
	12.011 21.079
	12.120 21.074
	12.231 21.069
	12.346 21.064
	12.464 21.059
	12.585 21.054
	12.708 21.049
	12.835 21.044
	12.899 21.042
	12.964 21.039
	13.030 21.037
	13.097 21.034
	 /
\plot 13.097 21.034 15.240 20.955 /
\linethickness= 0.500pt
\setplotsymbol ({\thinlinefont .})
%
% Fig CONTROL PT SPLINE
%
% open spline
%
\plot	 9.049 21.590  9.366 21.828
 	 9.451 21.886
	 9.545 21.942
	 9.649 21.996
	 9.763 22.046
	 9.887 22.095
	 9.953 22.118
	10.021 22.141
	10.092 22.163
	10.165 22.184
	10.241 22.205
	10.319 22.225
	10.404 22.245
	10.501 22.263
	10.610 22.282
	10.731 22.299
	10.795 22.308
	10.863 22.316
	10.934 22.325
	11.007 22.333
	11.084 22.341
	11.163 22.349
	11.245 22.356
	11.331 22.364
	11.419 22.371
	11.510 22.378
	11.604 22.386
	11.702 22.392
	11.802 22.399
	11.905 22.406
	12.011 22.412
	12.120 22.418
	12.231 22.425
	12.346 22.431
	12.464 22.436
	12.585 22.442
	12.708 22.448
	12.835 22.453
	12.899 22.456
	12.964 22.458
	13.030 22.461
	13.097 22.463
	 /
\plot 13.097 22.463 15.240 22.543 /
\linethickness= 0.500pt
\setplotsymbol ({\thinlinefont .})
%
% Fig CONTROL PT SPLINE
%
% open spline
%
\plot	 9.049 21.590  9.366 21.352
 	 9.451 21.294
	 9.545 21.238
	 9.649 21.184
	 9.763 21.134
	 9.887 21.085
	 9.953 21.062
	10.021 21.039
	10.092 21.017
	10.165 20.996
	10.241 20.975
	10.319 20.955
	10.404 20.935
	10.501 20.917
	10.610 20.898
	10.731 20.881
	10.795 20.872
	10.863 20.864
	10.934 20.855
	11.007 20.847
	11.084 20.839
	11.163 20.831
	11.245 20.824
	11.331 20.816
	11.419 20.809
	11.510 20.802
	11.604 20.794
	11.702 20.788
	11.802 20.781
	11.905 20.774
	12.011 20.768
	12.120 20.762
	12.231 20.755
	12.346 20.749
	12.464 20.744
	12.585 20.738
	12.708 20.732
	12.835 20.727
	12.899 20.724
	12.964 20.722
	13.030 20.719
	13.097 20.717
	 /
\plot 13.097 20.717 15.240 20.637 /
%
% Fig TEXT object
%
\put{$0$} [lB] at  2.381 21.273
%
% Fig TEXT object
%
\put{$z$} [lB] at  6.350 21.273
%
% Fig TEXT object
%
\put{$v_1$} [lB] at  3.334 20.637
%
% Fig TEXT object
%
\put{$u_r$} [lB] at  3.175 22.225
%
% Fig TEXT object
%
\put{$u_1$} [lB] at  5.429 22.813
%
% Fig TEXT object
%
\put{$v_s$} [lB] at  5.349 20.273
%
% Fig TEXT object
%
\put{$1$} [lB] at  8.873 21.082
%
% Fig TEXT object
%
\put{$\infty$} [lB] at 15.587 21.526
%
% Fig TEXT object
%
\put{$u_R$} [lB] at  9.953 22.320
%
% Fig TEXT object
%
\put{$u_{r+1}$} [lB] at 12.476 23.019
%
% Fig TEXT object
%
\put{$v_{s+1}$} [lB] at  9.730 20.686
\linethickness=0pt
\putrectangle corners at  1.245 23.247 and 15.587 19.025
\endpicture}

\caption{Integration contours for $u$ and $v$ variables for an s-channel block 
for fusion rule I.}
\label{figanI}
\end{figure}

We define the complex block in the s-channel for fusion rule I with $r$ 
screenings of the first kind and $s$ screenings of the second kind at the
right vertex as follows
\bea
 {\cal S}^{(R,S)}_{(r,s,0)}(z,x)&=&z^{2j_1j_2/t}(1-z)^{2j_2j_3/t}
  \int_0^z\prod_{i\in I_1,k\in J_1}du_idv_k
  \int_1^\infty\prod_{j\in I_2,l\in J_2}du_jdv_l\nn
 &\cdot&
  u_i^{a'}(1-u_i)^{b'}(z-u_i)^{c'}\prod_{i<i',\in I_1}(u_i-u_{i'})^{2\rho'}
  u_j^{a'}(u_j-1)^{b'}(u_j-z)^{c'}\nn
 &\cdot&\prod_{j<j',\in I_2}(u_j-u_{j'})^{2\rho'} 
  \prod_{i\in I_1,j\in I_2}(u_j-u_i)^{2\rho'}\nn
 &\cdot&v_k^a(1-v_k)^b(z-v_k)^c\prod_{k<k', \in J_1}(v_k-v_{k'})^{2\rho}
  v_l^a(v_l-1)^b(v_l-z)^c\nn
 &\cdot&\prod_{l<l',\in J_2}(v_l-v_{l'})^{2\rho}
  \prod_{k\in J_1,l\in J_2}(v_l-v_k)^{2\rho}\nn
 &\cdot&\prod_{i,k}(u_i-v_k)^{-2}
  \prod_{i,l}(u_i-v_l)^{-2}
  \prod_{j,k}(u_j-v_k)^{-2}\prod_{j,l}(u_j-v_l)^{-2}\nn
 &\cdot& \prod_{i,j,k,l}(u_i-x)(u_j-x)(v_k-x)^{-\rho}(v_l-x)^{-\rho}
\label{anblockI}
\eea
Here we have introduced the following index sets
\bea
 I_1&=&\{1,...,r\}\nn
 I_2&=&\{r+1,...,R\}\nn
 J_1&=&\{1,...,s\}\nn
 J_2&=&\{s+1,...,S\}
\eea
where $R$ and $S$ are the total numbers of screenings of the first and second 
kinds respectively. Variables $u$ and $v$ belong to screenings of the first and
second kind respectively, although this language is rather symbolic, since as 
yet there exists 
no known free field realization which directly gives this form.
Also the integrals are taken along complex Dotsenko-Fateev 
contours shown in Fig. \ref{figanI}. Notice that expressions of the form
$(u_i-u_{i'})^{2\rho'}$ have a phase defined by the fact that the first of the
two integration variables have a lower imaginary part than the last variable.
Finally
\bea
 a&=&-2j_3+t+R-St-1\nn
 b&=&-2j_1+t+R-St-1\nn
 c&=&2j_1+2j_2+2j_3-R+St+1\nn
 \rho&=&t, \ \ \rho'=1/t\nn
 a'&=&-a/t,\ \ \ b'=-b/t,\ \ \ c'=-c/t
\label{abc}
\eea
The integrand of this expression is provided in a slightly different form in 
\cite{An}. 
In fact, there the $j$'s are replaced by their parametrisations 
(\ref{jpm}) giving rise to 4 independent forms 
\cite{An} for the integrand depending on 
whether the $j_i^+$ or the $j_i^-$ form is used. The above form holds in 
general. By analysing the small $z$ and small $x$ behaviour of this form it
is easy to establish that this conformal block corresponds to the s-channel 
diagram Fig. \ref{fig1} 
with the intermediate $j$ given by fusion rule I. Indeed
by scaling $u_i\rightarrow zu_i,v_k\rightarrow zv_k, i\in I_1,k\in J_1$ we find
\ben
 {\cal S}^{(R,S)}_{(r,s,0)}(z,x)
  \sim z^{-\D(j_1)-\D(j_2)+\D(j_I)}(-x)^{j_1+j_2-j_I}
\een
with
\ben
 j_I=j_1+j_2-r+st
\een
The contours in Fig. \ref{figanI} are essentially equal to the contours in 
\cite{DF} for minimal models.

\subsection{Case of Fusion Rule II}
\begin{figure}
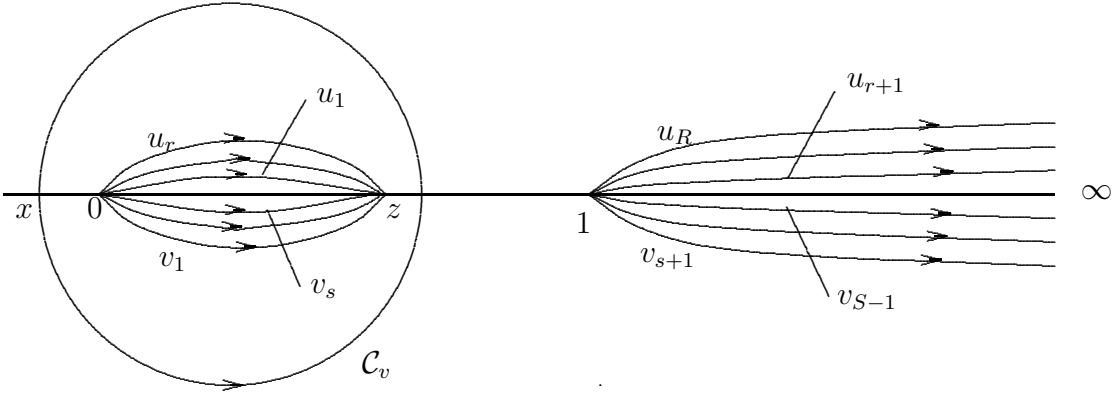

\font\thinlinefont=cmr5
\begingroup\makeatletter\ifx\SetFigFont\undefined
% extract first six characters in \fmtname
\def\x#1#2#3#4#5#6#7\relax{\def\x{#1#2#3#4#5#6}}%
\expandafter\x\fmtname xxxxxx\relax \def\y{splain}%
\ifx\x\y   % LaTeX or SliTeX?
\gdef\SetFigFont#1#2#3{%
  \ifnum #1<17\tiny\else \ifnum #1<20\small\else
  \ifnum #1<24\normalsize\else \ifnum #1<29\large\else
  \ifnum #1<34\Large\else \ifnum #1<41\LARGE\else
     \huge\fi\fi\fi\fi\fi\fi
  \csname #3\endcsname}%
\else
\gdef\SetFigFont#1#2#3{\begingroup
  \count@#1\relax \ifnum 25<\count@\count@25\fi
  \def\x{\endgroup\@setsize\SetFigFont{#2pt}}%
  \expandafter\x
    \csname \romannumeral\the\count@ pt\expandafter\endcsname
    \csname @\romannumeral\the\count@ pt\endcsname
  \csname #3\endcsname}%
\fi
\fi\endgroup
\mbox{\beginpicture
\setcoordinatesystem units <1.00000cm,1.00000cm>
\unitlength=1.00000cm
\linethickness=1pt
\setplotsymbol ({\makebox(0,0)[l]{\tencirc\symbol{'160}}})
\setshadesymbol ({\thinlinefont .})
\setlinear
%
% Fig POLYLINE object
%
\linethickness= 0.500pt
\setplotsymbol ({\thinlinefont .})
\putrule from  1.270 21.590 to 15.240 21.590
%
% Fig POLYLINE object
%
\linethickness= 0.500pt
\setplotsymbol ({\thinlinefont .})
\plot  4.714 21.861  5.285 22.845 /
%
% Fig POLYLINE object
%
\linethickness= 0.500pt
\setplotsymbol ({\thinlinefont .})
\plot  4.777 21.385  5.207 20.369 /
%
% Fig POLYLINE object
%
\linethickness= 0.500pt
\setplotsymbol ({\thinlinefont .})
\putrule from  4.477 20.892 to  4.619 20.892
%
% arrow head
%
\plot  4.365 20.828  4.619 20.892  4.365 20.955 /
%
%
% Fig POLYLINE object
%
\linethickness= 0.500pt
\setplotsymbol ({\thinlinefont .})
\putrule from  4.286 21.162 to  4.428 21.162
%
% arrow head
%
\plot  4.174 21.099  4.428 21.162  4.174 21.226 /
%
%
% Fig POLYLINE object
%
\linethickness= 0.500pt
\setplotsymbol ({\thinlinefont .})
\putrule from  4.333 21.385 to  4.475 21.385
%
% arrow head
%
\plot  4.221 21.321  4.475 21.385  4.221 21.448 /
%
%
% Fig POLYLINE object
%
\linethickness= 0.500pt
\setplotsymbol ({\thinlinefont .})
\putrule from  4.350 21.861 to  4.492 21.861
%
% arrow head
%
\plot  4.238 21.797  4.492 21.861  4.238 21.924 /
%
%
% Fig POLYLINE object
%
\linethickness= 0.500pt
\setplotsymbol ({\thinlinefont .})
\putrule from  4.396 22.066 to  4.538 22.066
%
% arrow head
%
\plot  4.284 22.003  4.538 22.066  4.284 22.130 /
%
%
% Fig POLYLINE object
%
\linethickness= 0.500pt
\setplotsymbol ({\thinlinefont .})
\putrule from  4.320 22.337 to  4.462 22.337
%
% arrow head
%
\plot  4.208 22.274  4.462 22.337  4.208 22.401 /
%
%
% Fig POLYLINE object
%
\linethickness= 0.500pt
\setplotsymbol ({\thinlinefont .})
\plot  9.207 19.050  9.207 19.050 /
%
% Fig POLYLINE object
%
\linethickness= 0.500pt
\setplotsymbol ({\thinlinefont .})
\putrule from 13.540 22.511 to 13.540 22.511
\putrule from 13.540 22.511 to 13.699 22.511
%
% arrow head
%
\plot 13.445 22.447 13.699 22.511 13.445 22.574 /
%
%
% Fig POLYLINE object
%
\linethickness= 0.500pt
\setplotsymbol ({\thinlinefont .})
\putrule from 13.604 22.193 to 13.604 22.193
\putrule from 13.604 22.193 to 13.763 22.193
%
% arrow head
%
\plot 13.509 22.130 13.763 22.193 13.509 22.257 /
%
%
% Fig POLYLINE object
%
\linethickness= 0.500pt
\setplotsymbol ({\thinlinefont .})
\putrule from 13.731 21.893 to 13.731 21.893
\putrule from 13.731 21.893 to 13.890 21.893
%
% arrow head
%
\plot 13.636 21.829 13.890 21.893 13.636 21.956 /
%
%
% Fig POLYLINE object
%
\linethickness= 0.500pt
\setplotsymbol ({\thinlinefont .})
\putrule from 13.699 21.336 to 13.699 21.336
\putrule from 13.699 21.336 to 13.858 21.336
%
% arrow head
%
\plot 13.604 21.273 13.858 21.336 13.604 21.400 /
%
%
% Fig POLYLINE object
%
\linethickness= 0.500pt
\setplotsymbol ({\thinlinefont .})
\putrule from 13.650 21.035 to 13.650 21.035
\putrule from 13.650 21.035 to 13.809 21.035
%
% arrow head
%
\plot 13.555 20.972 13.809 21.035 13.555 21.099 /
%
%
% Fig POLYLINE object
%
\linethickness= 0.500pt
\setplotsymbol ({\thinlinefont .})
\putrule from 13.572 20.718 to 13.572 20.718
\putrule from 13.572 20.718 to 13.731 20.718
%
% arrow head
%
\plot 13.477 20.654 13.731 20.718 13.477 20.781 /
%
%
% Fig POLYLINE object
%
\linethickness= 0.500pt
\setplotsymbol ({\thinlinefont .})
\plot 11.699 21.812 12.317 22.955 /
%
% Fig POLYLINE object
%
\linethickness= 0.500pt
\setplotsymbol ({\thinlinefont .})
\plot 11.682 21.416 12.239 20.242 /
%
% Fig POLYLINE object
%
\linethickness= 0.500pt
\setplotsymbol ({\thinlinefont .})
\putrule from  4.286 19.050 to  4.445 19.050
%
% arrow head
%
\plot  4.191 18.986  4.445 19.050  4.191 19.114 /
\linethickness= 0.500pt
\setplotsymbol ({\thinlinefont .})
%
% Fig CONTROL PT SPLINE
%
% open spline
%
\plot	 2.540 21.590  3.413 21.749
 	 3.523 21.767
	 3.634 21.783
	 3.746 21.797
	 3.860 21.808
	 3.974 21.817
	 4.090 21.823
	 4.207 21.827
	 4.326 21.828
	 4.446 21.827
	 4.567 21.823
	 4.689 21.817
	 4.812 21.808
	 4.937 21.797
	 5.063 21.783
	 5.126 21.776
	 5.190 21.767
	 5.254 21.758
	 5.318 21.749
	 /
\plot  5.318 21.749  6.350 21.590 /
\linethickness= 0.500pt
\setplotsymbol ({\thinlinefont .})
%
% Fig CONTROL PT SPLINE
%
% open spline
%
\plot	 2.540 21.590  3.493 21.431
 	 3.611 21.413
	 3.729 21.397
	 3.847 21.383
	 3.964 21.372
	 4.080 21.363
	 4.196 21.357
	 4.311 21.353
	 4.425 21.352
	 4.539 21.353
	 4.652 21.357
	 4.765 21.363
	 4.877 21.372
	 4.988 21.383
	 5.099 21.397
	 5.209 21.413
	 5.318 21.431
	 /
\plot  5.318 21.431  6.191 21.590 /
\linethickness= 0.500pt
\setplotsymbol ({\thinlinefont .})
%
% Fig CONTROL PT SPLINE
%
% open spline
%
\plot	 2.540 21.590  2.857 21.749
 	 2.942 21.787
	 3.036 21.823
	 3.140 21.857
	 3.254 21.888
	 3.378 21.916
	 3.444 21.930
	 3.512 21.942
	 3.583 21.954
	 3.656 21.966
	 3.732 21.977
	 3.810 21.987
	 3.887 21.996
	 3.958 22.005
	 4.083 22.022
	 4.186 22.035
	 4.266 22.046
	 4.366 22.066
	 4.485 22.046
	 4.576 22.035
	 4.693 22.022
	 4.761 22.014
	 4.834 22.005
	 4.914 21.996
	 5.001 21.987
	 5.088 21.977
	 5.173 21.966
	 5.255 21.954
	 5.333 21.942
	 5.408 21.930
	 5.481 21.916
	 5.550 21.902
	 5.616 21.888
	 5.739 21.857
	 5.849 21.823
	 5.947 21.787
	 6.032 21.749
	 6.106 21.712
	 6.166 21.679
	 6.251 21.630
	 6.271 21.590
	 /
\plot  6.271 21.590  6.191 21.590 /
%
% Fig ELLIPSE
%
\linethickness= 0.500pt
\setplotsymbol ({\thinlinefont .})
\ellipticalarc axes ratio  2.540:2.540  360 degrees 
	from  6.826 21.590 center at  4.286 21.590
\linethickness= 0.500pt
\setplotsymbol ({\thinlinefont .})
%
% Fig CONTROL PT SPLINE
%
% open spline
%
\plot	 2.540 21.590  2.857 21.431
 	 2.942 21.393
	 3.036 21.357
	 3.140 21.323
	 3.254 21.292
	 3.378 21.264
	 3.444 21.250
	 3.512 21.238
	 3.583 21.226
	 3.656 21.214
	 3.732 21.203
	 3.810 21.193
	 3.887 21.184
	 3.958 21.175
	 4.083 21.158
	 4.186 21.145
	 4.266 21.134
	 4.366 21.114
	 4.485 21.134
	 4.576 21.145
	 4.693 21.158
	 4.761 21.166
	 4.834 21.175
	 4.914 21.184
	 5.001 21.193
	 5.088 21.203
	 5.173 21.214
	 5.255 21.226
	 5.333 21.238
	 5.408 21.250
	 5.481 21.264
	 5.550 21.278
	 5.616 21.292
	 5.739 21.323
	 5.849 21.357
	 5.947 21.393
	 6.032 21.431
	 6.106 21.468
	 6.166 21.501
	 6.251 21.550
	 6.271 21.590
	 /
\plot  6.271 21.590  6.191 21.590 /
%
% Fig TEXT object
%
\put{${\cal C}_v$} [lB] at  6.032 19.209
\linethickness= 0.500pt
\setplotsymbol ({\thinlinefont .})
%
% Fig CONTROL PT SPLINE
%
% open spline
%
\plot	 2.540 21.590  2.778 21.828
 	 2.845 21.886
	 2.927 21.942
	 3.024 21.996
	 3.135 22.046
	 3.262 22.095
	 3.331 22.118
	 3.403 22.141
	 3.479 22.163
	 3.559 22.184
	 3.643 22.205
	 3.731 22.225
	 3.820 22.244
	 3.909 22.260
	 3.999 22.273
	 4.088 22.285
	 4.177 22.293
	 4.266 22.299
	 4.356 22.303
	 4.445 22.304
	 4.534 22.303
	 4.624 22.299
	 4.713 22.293
	 4.802 22.285
	 4.891 22.273
	 4.981 22.260
	 5.070 22.244
	 5.159 22.225
	 5.247 22.205
	 5.331 22.184
	 5.411 22.163
	 5.487 22.141
	 5.559 22.118
	 5.628 22.095
	 5.755 22.046
	 5.866 21.996
	 5.963 21.942
	 6.045 21.886
	 6.112 21.828
	 /
\plot  6.112 21.828  6.350 21.590 /
\linethickness= 0.500pt
\setplotsymbol ({\thinlinefont .})
%
% Fig CONTROL PT SPLINE
%
% open spline
%
\plot	 2.540 21.590  2.778 21.352
 	 2.845 21.294
	 2.927 21.238
	 3.024 21.184
	 3.135 21.134
	 3.262 21.085
	 3.331 21.062
	 3.403 21.039
	 3.479 21.017
	 3.559 20.996
	 3.643 20.975
	 3.731 20.955
	 3.820 20.936
	 3.909 20.920
	 3.999 20.907
	 4.088 20.895
	 4.177 20.887
	 4.266 20.881
	 4.356 20.877
	 4.445 20.876
	 4.534 20.877
	 4.624 20.881
	 4.713 20.887
	 4.802 20.895
	 4.891 20.907
	 4.981 20.920
	 5.070 20.936
	 5.159 20.955
	 5.247 20.975
	 5.331 20.996
	 5.411 21.017
	 5.487 21.039
	 5.559 21.062
	 5.628 21.085
	 5.755 21.134
	 5.866 21.184
	 5.963 21.238
	 6.045 21.294
	 6.112 21.352
	 /
\plot  6.112 21.352  6.350 21.590 /
\linethickness= 0.500pt
\setplotsymbol ({\thinlinefont .})
%
% Fig CONTROL PT SPLINE
%
% open spline
%
\plot	 9.049 21.590  9.446 21.669
 	 9.504 21.679
	 9.581 21.689
	 9.675 21.699
	 9.788 21.709
	 9.851 21.714
	 9.918 21.719
	 9.990 21.724
	10.067 21.729
	10.148 21.734
	10.233 21.739
	10.323 21.744
	10.418 21.749
	10.517 21.754
	10.620 21.759
	10.728 21.764
	10.841 21.769
	10.958 21.774
	11.079 21.779
	11.205 21.783
	11.270 21.786
	11.336 21.788
	11.403 21.791
	11.471 21.793
	11.540 21.796
	11.610 21.798
	11.682 21.801
	11.754 21.803
	11.828 21.806
	11.903 21.808
	11.978 21.811
	12.055 21.813
	12.134 21.816
	12.213 21.818
	12.293 21.821
	12.375 21.823
	12.457 21.826
	12.541 21.828
	 /
\plot 12.541 21.828 15.240 21.907 /
\linethickness= 0.500pt
\setplotsymbol ({\thinlinefont .})
%
% Fig CONTROL PT SPLINE
%
% open spline
%
\plot	 9.049 21.590  9.446 21.511
 	 9.504 21.501
	 9.581 21.491
	 9.675 21.481
	 9.788 21.471
	 9.851 21.466
	 9.918 21.461
	 9.990 21.456
	10.067 21.451
	10.148 21.446
	10.233 21.441
	10.323 21.436
	10.418 21.431
	10.517 21.426
	10.620 21.421
	10.728 21.416
	10.841 21.411
	10.958 21.406
	11.079 21.401
	11.205 21.397
	11.270 21.394
	11.336 21.392
	11.403 21.389
	11.471 21.387
	11.540 21.384
	11.610 21.382
	11.682 21.379
	11.754 21.377
	11.828 21.374
	11.903 21.372
	11.978 21.369
	12.055 21.367
	12.134 21.364
	12.213 21.362
	12.293 21.359
	12.375 21.357
	12.457 21.354
	12.541 21.352
	 /
\plot 12.541 21.352 15.240 21.273 /
\linethickness= 0.500pt
\setplotsymbol ({\thinlinefont .})
%
% Fig CONTROL PT SPLINE
%
% open spline
%
\plot	 9.049 21.590  9.366 21.749
 	 9.451 21.787
	 9.545 21.823
	 9.649 21.857
	 9.763 21.888
	 9.887 21.916
	 9.953 21.930
	10.021 21.942
	10.092 21.954
	10.165 21.966
	10.241 21.977
	10.319 21.987
	10.404 21.997
	10.501 22.007
	10.610 22.017
	10.731 22.027
	10.795 22.032
	10.863 22.036
	10.934 22.041
	11.007 22.046
	11.084 22.051
	11.163 22.056
	11.245 22.061
	11.331 22.066
	11.419 22.071
	11.510 22.076
	11.604 22.081
	11.702 22.086
	11.802 22.091
	11.905 22.096
	12.011 22.101
	12.120 22.106
	12.231 22.111
	12.346 22.116
	12.464 22.121
	12.585 22.126
	12.708 22.131
	12.835 22.136
	12.899 22.138
	12.964 22.141
	13.030 22.143
	13.097 22.146
	 /
\plot 13.097 22.146 15.240 22.225 /
\linethickness= 0.500pt
\setplotsymbol ({\thinlinefont .})
%
% Fig CONTROL PT SPLINE
%
% open spline
%
\plot	 9.049 21.590  9.366 21.431
 	 9.451 21.393
	 9.545 21.357
	 9.649 21.323
	 9.763 21.292
	 9.887 21.264
	 9.953 21.250
	10.021 21.238
	10.092 21.226
	10.165 21.214
	10.241 21.203
	10.319 21.193
	10.404 21.183
	10.501 21.173
	10.610 21.163
	10.731 21.153
	10.795 21.148
	10.863 21.144
	10.934 21.139
	11.007 21.134
	11.084 21.129
	11.163 21.124
	11.245 21.119
	11.331 21.114
	11.419 21.109
	11.510 21.104
	11.604 21.099
	11.702 21.094
	11.802 21.089
	11.905 21.084
	12.011 21.079
	12.120 21.074
	12.231 21.069
	12.346 21.064
	12.464 21.059
	12.585 21.054
	12.708 21.049
	12.835 21.044
	12.899 21.042
	12.964 21.039
	13.030 21.037
	13.097 21.034
	 /
\plot 13.097 21.034 15.240 20.955 /
\linethickness= 0.500pt
\setplotsymbol ({\thinlinefont .})
%
% Fig CONTROL PT SPLINE
%
% open spline
%
\plot	 9.049 21.590  9.366 21.828
 	 9.451 21.886
	 9.545 21.942
	 9.649 21.996
	 9.763 22.046
	 9.887 22.095
	 9.953 22.118
	10.021 22.141
	10.092 22.163
	10.165 22.184
	10.241 22.205
	10.319 22.225
	10.404 22.245
	10.501 22.263
	10.610 22.282
	10.731 22.299
	10.795 22.308
	10.863 22.316
	10.934 22.325
	11.007 22.333
	11.084 22.341
	11.163 22.349
	11.245 22.356
	11.331 22.364
	11.419 22.371
	11.510 22.378
	11.604 22.386
	11.702 22.392
	11.802 22.399
	11.905 22.406
	12.011 22.412
	12.120 22.418
	12.231 22.425
	12.346 22.431
	12.464 22.436
	12.585 22.442
	12.708 22.448
	12.835 22.453
	12.899 22.456
	12.964 22.458
	13.030 22.461
	13.097 22.463
	 /
\plot 13.097 22.463 15.240 22.543 /
\linethickness= 0.500pt
\setplotsymbol ({\thinlinefont .})
%
% Fig CONTROL PT SPLINE
%
% open spline
%
\plot	 9.049 21.590  9.366 21.352
 	 9.451 21.294
	 9.545 21.238
	 9.649 21.184
	 9.763 21.134
	 9.887 21.085
	 9.953 21.062
	10.021 21.039
	10.092 21.017
	10.165 20.996
	10.241 20.975
	10.319 20.955
	10.404 20.935
	10.501 20.917
	10.610 20.898
	10.731 20.881
	10.795 20.872
	10.863 20.864
	10.934 20.855
	11.007 20.847
	11.084 20.839
	11.163 20.831
	11.245 20.824
	11.331 20.816
	11.419 20.809
	11.510 20.802
	11.604 20.794
	11.702 20.788
	11.802 20.781
	11.905 20.774
	12.011 20.768
	12.120 20.762
	12.231 20.755
	12.346 20.749
	12.464 20.744
	12.585 20.738
	12.708 20.732
	12.835 20.727
	12.899 20.724
	12.964 20.722
	13.030 20.719
	13.097 20.717
	 /
\plot 13.097 20.717 15.240 20.637 /
%
% Fig TEXT object
%
\put{$0$} [lB] at  2.381 21.273
%
% Fig TEXT object
%
\put{$z$} [lB] at  6.350 21.273
%
% Fig TEXT object
%
\put{$v_1$} [lB] at  3.334 20.637
%
% Fig TEXT object
%
\put{$u_r$} [lB] at  3.175 22.225
%
% Fig TEXT object
%
\put{$u_1$} [lB] at  5.429 22.813
%
% Fig TEXT object
%
\put{$v_s$} [lB] at  5.349 20.273
%
% Fig TEXT object
%
\put{$1$} [lB] at  8.873 21.082
%
% Fig TEXT object
%
\put{$\infty$} [lB] at 15.587 21.526
%
% Fig TEXT object
%
\put{$u_R$} [lB] at  9.953 22.320
%
% Fig TEXT object
%
\put{$v_{s+1}$} [lB] at  9.730 20.686
%
% Fig TEXT object
%
\put{$v_{S-1}$} [lB] at 12.349 20.130
%
% Fig TEXT object
%
\put{$u_{r+1}$} [lB] at 12.476 23.019
%
% Fig TEXT object
%
\put{$x$} [lB] at  1.429 21.273
\linethickness=0pt
\putrectangle corners at  1.245 24.145 and 15.587 19.025
\endpicture}

\caption{Integration contours for $u$ and $v$ variables for an s-channel block
for fusion rule II.}
\label{figanII}
\end{figure}

Fig. \ref{figanII} shows 
the integration contours.
The s-channel block for fusion rule II is given by
\bea
 {\cal S}_{(r,s,1)}^{(R,S)}(z,x)&=&z^{2j_1j_2/t}(1-z)^{2j_2j_3/t}
  \int_0^z du_idv_k\int_1^\infty du_jdv_l
  \oint_{{\cal C}_v}\frac{dv}{2\pi i}\nn
 &\cdot&u_i^{a'}(1-u_i)^{b'}(z-u_i)^{c'}\prod_{i<i'}(u_i-u_{i'})^{2\rho'}
  u_j^{a'}(u_j-1)^{b'}(u_j-z)^{c'}\nn
 &\cdot&\prod_{j<j'}(u_j-u_{j'})^{2\rho'}
  \prod_{i,j}(u_j-u_i)^{2\rho'}\nn
 &\cdot&v_k^a(1-v_k)^b(z-v_k)^c\prod_{k<k'}(v_k-v_{k'})^{2\rho}
  v_l^a(v_l-1)^b(v_l-z)^c\prod_{l<l'}(v_l-v_{l'})^{2\rho}\nn
 &\cdot&\prod_{k,l}(v_l-v_k)^{2\rho}\prod_{i,k}(u_i-v_k)^{-2}\prod_{i,l}
  (u_i-v_l)^{-2}\prod_{j,k}(u_j-v_k)^{-2}\prod_{j,l}(u_j-v_l)^{-2}\nn
 &\cdot&(u_i-x)(u_j-x)(v_k-x)^{-\rho}(v_l-x)^{-\rho}\nn
 &\cdot&v^a(1-v)^b(v-z)^c(v-v_k)^{2\rho}(v_l-v)^{2\rho}(v-u_i)^{-2}(v-u_j)^{-2}
  (v-x)^{-\rho}\nn
\eea
Here the variables, $u_i,u_j,v_k,v_l$ are taken along approximately real 
(for $z$ real) contours as for fusion 
rule I. The indices indicate: 
$i=1,...,r; j=r+1,... ,R; k=1,...,s; l=s+1,...,S-1$, whereas $v$ runs
along the contour ${\cal C}_v$ which starts at $x$ and surrounds both $0$ and 
$z$, cf. Fig. \ref{figanII}.  In addition to the s-channel blocks we have 
defined above,
we define additional ones in analogy to the case for minimal models \cite{DF}.
Namely, instead of using complex contours close to 
the real axis (for real $z$), we
may use real time ordered integrations, with an ordering so that all terms
in the integral expression for ${\cal S}^{(R,S)}_{(r,s,0)}$ become real. 
This block is denoted $S^{(R,S)}_{(r,s,0)}$. Similar to the case of 3-point
functions one finds 
\bea
 {\cal S}^{(R,S)}_{(r,s,0)}(z,x)&=&
  \lambda_r(\rho')\lambda_{R-r}(\rho')\lambda_s(\rho)
  \lambda_{S-s}(\rho)S^{(R,S)}_{(r,s,0)}(z,x)\nn
 S^{(R,S)}_{(r,s,0)}(z,x)&=&s^{(R,S)}_{(r,s,0)}(z,x)N^{(R,S)}_{(r,s,0)}
\label{sblockI}
\eea
where $s^{(R,S)}_{(r,s,0)}(z,x)$ is normalised in such a way that the
behaviour as $z\rightarrow 0, x\rightarrow 0$ is
\ben
 s^{(R,S)}_{(r,s,0)}(z,x)=z^{-\D(j_1)-\D(j_2)+\D(j_I)}(-x)^{j_1+j_2-j_I}
  (1+{\cal O}(z,x))
\een
The $\lambda$-functions were defined in (\ref{lambda}).
Similarly for fusion rule II we write
\bea
 {\cal S}^{(R,S)}_{(r,s,1)}(z,x)&=&\lambda_r(\rho')\lambda_s(\rho)
  \lambda_{R-r}(\rho')\lambda_{S-s-1}(\rho)S^{(R,S)}_{(r,s,1)}(z,x)\nn
 S^{(R,S)}_{(r,s,1)}(z,x)&=&s^{(R,S)}_{(r,s,1)}(z,x)N^{(R,S)}_{(r,s,1)}\nn
 s^{(R,S)}_{(r,s,1)}(z,x)&=&z^{-\D(j_1)-\D(j_2)+\D(j_{II})}
  (-x)^{j_1+j_2-j_{II}}(1+{\cal O}(z,x))
\label{sblockII}
\eea
where
\ben
 j_{II}=-j_I-1
\een
The leading behaviour (for $z\rightarrow 0$ followed by $x\rightarrow 0$) 
in this case of fusion rule II is determined by the scalings
\bea
 u_i&\rightarrow &zu_i\nn
 v_k&\rightarrow &zv_k\nn
 v&\rightarrow &(-x)v
\eea
In the limit $z,x\rightarrow0$ the $v$ integral becomes trivial and is 
carried using (\ref{famous}).
The normalisation constants $N^{(R,S)}_{(r,s,0)}$ and  $N^{(R,S)}_{(r,s,1)}$ 
are found in terms of the Dotsenko-Fateev integral (\ref{dfintegral})
We shall need these normalisations in the calculation of crossing matrices. 
After some calculations we obtain
\bea
 N^{(R,S)}_{(r,s,0)}&=&(-)^{R-r+S-s}{\cal J}_{r,s}(a,c;\rho)
  {\cal J}_{R-r,S-s}(a+c-2(r-\rho s)-\rho,b;\rho)\nn
 &\cdot&\prod_{i=0}^{R-r-1}\frac{s(a'+c'-2(s-\rho'r)+1+i\rho')}
  {s(a'+b'+c'-2(s-\rho' r)+1+\rho'(R-r-1+i))}\nn
 &\cdot&\prod_{i=0}^{S-s-1}\frac{s(a+c-2(r-\rho s)-\rho+i\rho)}
  {s(a+b+c-2(r-\rho s)-\rho+\rho(S-s-1+i))}
\eea
and
\bea
 N^{(R,S)}_{(r,s,1)}&=&N^{(R,S)}_{(r,s+1,0)}
  \frac{\Gamma(\rho)\Gamma(1-\rho)\Gamma(2-2r+a+c+2s\rho)}
  {\Gamma(-a-c-2\rho s+2r)\Gamma(1-r+a+s\rho)
  \Gamma(1-r+c+s\rho)}\nn
 &\cdot&\frac{1}{\Gamma((s+1)\rho-r)\Gamma(2-r+a+c+(s-1)\rho)}
\eea
For the integral realization \cite{An} considered here it is trivial to obtain
the t-channel forms once the s-channel forms above are given. In fact we have
in an obvious notation ($\epsilon = 0,1$ for fusion rules I and II)
\ben
{\cal T}^{(R,S)}_{(r,s,\epsilon)}(z,x;j_1,j_2,j_3,j_4)=
{\cal S}^{(R,S)}_{(r,s,\epsilon)}(1-z,1-x;j_3,j_2,j_1,j_4)
\een
We notice the following. When in the integral realization, we also transform
all integration variables as $u\rightarrow 1-u, v\rightarrow 1-v$, the 
integrand for the t-channel block is identical to the one for the s-channel 
block, up to phases. In particular, whenever we have $(u-x)$ or $(v-x)^{-\rho}$ 
in the s-channel, we would have $(x-u)$ and $(x-v)^{-\rho}$ in the t-channel.
Also, after transformation of the variables, the integration contours in the 
t-channel are between $z$ and $1$ and between $0$ and $-\infty$, 
and the complex
contour for $v$ in the case of fusion rule II surrounds $z$ and $1$. The above
factors, $(u-x)$ etc. are real provided $x<0$ in the s-channel, or $x>1$ in the
t-channel. These two possibilities map to each other under $x\rightarrow 1-x$.
  
\section{Proof of Equivalence}  
As previously indicated there is no absolute need for proving the equivalence 
between the 4-point function by PRY and the one by Andreev since both 
satisfy the Knizhnik-Zamolodchikov equations. Nevertheless, it is of 
some interest to understand better how two such seemingly very different 
expressions can agree, and it is rather nice to be aware of the clarification
provided by the relation to the minimal model case treated in Chapter 2.
Here \cite{PRY2} we go over several of the steps needed for a direct 
analytic proof. In fact, we investigate the singularity structure of the two 
expressions in the double limits, 
$z,x\rightarrow 0, z,x\rightarrow 1, z,x\rightarrow \infty$ and in the single 
limit, $z\rightarrow x$. We restrict ourselves to just one of the s-channel 
conformal blocks.\\[.2 cm]
{\bf Proposition}
\bea
 &&\int_0^1du\prod_{k=1}^Ndv_kv_k^{a'}(1-v_k)^{b'}(1-zv_k)^{c'}\left(1-\frac{
  1-v_k}{1-zv_k}\frac{z}{x}u\right)\prod_{k<k'}(v_k-v_{k'})^{2\rho'}\nn  
 &\cdot&\prod_{i=1}^Mdw_iw_i^a(1-w_i)^b(1-zw_i)^c\left(1-\frac{1-w_i}{1-zw_i}
  \frac{z}{x}u\right)^{-\rho}\prod_{i<i'}(w_i-w_{i'})^{2\rho}\nn
 &\cdot&\prod_{k,i}^{N,M}(v_k-w_i)^{-2}u^{-c-1}(1-u)^{b+c-N+(M-1)\rho}\nn
 &=&K_{NM}^x\int_0^1\prod_{k=1}^Ndv_kv_k^{a'-\delta'}(1-v_k)^{b'+\delta'}
  (1-zv_k)^{c'-\delta'}\left(1-\frac{z}{x}v_k\right)\prod_{k<k'}(v_k-v_{k'})^{
  2\rho'}\nn
 &\cdot&\prod_{i=1}^Mdw_iw_i^{a-\delta}(1-w_i)^{b+\delta}(1-zw_i)^{c-\delta}
  \left(1-\frac{z}{x}w_i\right)^{-\rho}\prod_{i<i'}(w_i-w_{i'})^{2\rho}\nn
  &\cdot&\prod_{k,i}^{N,M}(v_k-w_i)^{-2}
\label{pryan}
\eea
where
\ben
 K_{NM}^x=\frac{\Gamma(-c)\Gamma(b+c+1-N+(M-1)\rho)}{\Gamma(b+1-N+(M-1)\rho)}
  K_{NM}
\een
and $K_{NM}$ is given by (\ref{KNMmin}).
Here, up to irrelevant common pre-factors, 
the left hand side is the PRY form of the
conformal block (\ref{pryblock}) for $r=R=N$ and $s=S=M$ in the s-channel. We
denote this by $S^{PRY}$.
Similarly up to the same pre-factors and the new normalisation constant, 
$K^x_{NM}$, the right hand side is essentially (\ref{anblockI}). 
We denote it by $S^A$.
Notice in particular that now we put 
\ben
 a=2j_1, \ b=2j_2+\rho, \ c=2j_3, \ \rho=t
\een
$\delta, \delta'$ are given by the same expressions as for the minimal models
\bea
 \delta&=&a+c+1-N+(M-1)\rho\nn
 \delta'&=&a'+c'+1-M+(N-1)\rho'
\eea
Then $a-\delta$ is what was called $a$ in previous sections, $b+\delta$ was 
previously called $c$ and $c-\delta$ was previously called $b$. In subsequent 
sections we shall revert to this notation, but in this section we stick to the
present notation in order to emphasise the similarity with minimal models. Also
notice, that because all integrations are between $0$ and $1$ (after scaling 
by $z$), it is possible to deform the $u$ integration in Fig. \ref{figpryus} to
being along the real axis from $0$ to $1$.\\[.2 cm]
{\bf Proof}\\
We demonstrate that both the left hand side and the right hand side of
the claimed identity have the same singularities in the limits 
$z,x\rightarrow0$,
$z,x\rightarrow 1$, $z,x\rightarrow \infty$ and $z\rightarrow x$. 
The $z$ limits are the {\em dominant} ones, the ones taken first. The proof 
turns out to be rather more laborious than for the minimal models. This
is due to the $x$ dependence and the $u$ integration in the case of $S^{PRY}$.
However, the general strategy is entirely analogous, so we will not go over
all the steps on the way.\\[.2cm]
$\underline{z,x\rightarrow0}$\\
The limit $z,x\rightarrow 0$ is simple to deal with and it gives rise to the
normalisation constant $K^x_{NM}$ differing from the one in the minimal 
models $K_{NM}$ because of the $u$ integration.\\[.2cm]
$\underline{z,x\rightarrow1}$\\
We first deal with $S^A$. Exactly as in the case of minimal models, we split
the ordered integration ranges for the $v$'s in $n$ $v_k$'s and $m$ $w_i$'s 
in $(0,1-\epsilon)$
and $N-n$ $v_l$'s and $M-m$ $w_j$'s in $(1-\epsilon,1)$. Omitting 
integration signs and products for brevity we find
\bea
S^A_{nm}
 &\sim&(1-z)^{(N-n)(b'+c'+1)+(N-n)(N-n-1)\rho'+(M-m)(b+c+1)
   +(M-m)(M-m-1)\rho-2(N-n)(M-m)}\nn
 &\cdot&(x-1)^{N-n-(M-m)\rho}K_{NM}^x\nn
 &\cdot&v_k^{a'-\delta'}(1-v_k)^{b'+c'+(N-n)2\rho'-2(M-m)}(x-v_k)(v_k-v_{k'}
  )^{2\rho'}\nn
 &\cdot&w_i^{a-\delta}(1-w_i)^{b+c+(M-m)2\rho-2(N-n)}(x-w_i)^{-\rho}(w_i-w_{
   i'})^{2\rho}(v_k-w_i)^{-2}\nn
 &\cdot&v_l^{-b'-c'-2-(N-n-1)2\rho'+2(M-m)}(1-v_l)^{b'+\delta'}(v_l-v_{l'})^{
  2\rho'}\nn
 &\cdot&w_j^{-b-c-2-(M-m-1)2\rho+2(N-n)}(1-w_j)^{b+\delta}(w_j-w_{j'})^{2\rho}
  (v_l-w_j)^{-2}
\label{sanm}
\eea
where we have performed the same scalings as for minimal models. 
The above has to
be summed over $n$ and $m$, but for a fixed value we pick up the pure $(1-z)$ 
and $(x-1)$ singularity indicated. The $l,j$ part of the integration gives 
immediately rise to a standard Dotsenko-Fateev integral. For the $k,i$ part we
perform the further split and scalings
\ben
 \int_0^1=\int_0^{1-\epsilon}+\int_{1-\epsilon}^1
\een
\bea
 v&\rightarrow&1-(1-1/x)\frac{1-v}{v}\nn
 \int_{1-\epsilon}^1dv&\rightarrow&\int_0^1(1-1/x)\frac{dv}{v^2}\nn
 \int_0^1\prod_{k=1}^ndv_k\prod_{i=1}^mdw_i&=&\sum_{n_0,m_0}
  \int_0^{1-\epsilon}\prod_{k_0=n-n_0+1}^{n}dv_{k_0}\prod_{i_0=m-m_0+1}^{m}
  dw_{i_0}\nn
 &\cdot&\int_{1-\epsilon}^1
  \prod_{k=1}^{n-n_0}dv_k\prod_{i=1}^{m-m_0}dw_i
\eea
In the limit $x\rightarrow 1$ we extract the $(1-x)$ power and find the 
coefficient again to be given by the product of two DF integrals.
Analysing the gamma-functions of these we see that we can only get a 
non-vanishing result if $(n_0,m_0)=(n,m)$ or if $(n_0,m_0)=(n,m-1)$. 
We denote these cases by 
$S^{AI}_{nm}$ and $S^{AII}_{nm}$. They will turn out to be related to fusion 
rules I and II. Combining everything we find the following singularities
in the limit $z,x\rightarrow 1$
\bea
 S^{AI}_{nm}&=&(1-z)^{(N-n)(b'+c'+1)+(N-n)(N-n-1)\rho'+(M-m)(b+c+1)
   +(M-m)(M-m-1)\rho-2(N-n)(M-m)}\nn
 &\cdot&(x-1)^{N-n-(M-m)\rho}N(S^{AI}_{nm})\nn
 S^{AII}_{nm}&=&(1-z)^{(N-n)(b'+c'+1)+(N-n)(N-n-1)\rho'+(M-m)(b+c+1)
   +(M-m)(M-m-1)\rho-2(N-n)(M-m)}\nn
 &\cdot&(x-1)^{b+c+1-N+n+(M-m-1)\rho}N(S^{AII}_{nm})
\eea
One checks that the singularities exactly correspond to fusion rules I and II.
The normalisations $N(S^{AI}_{nm})$ and $N(S^{AII}_{nm})$ are found explicitly
in terms of products of DF integrals to be lengthy expressions 
involving many products of ratios of gamma-functions.

We now turn to a similar analysis of $S^{PRY}$ in the same limit 
$z\rightarrow 1$ followed by $x\rightarrow 1$. We replace $u\rightarrow 1-u$ 
and perform the same split and the same scalings of the $v$ and $w$ variables
as in the case of minimal models. Omitting again integration signs and products
we find
\bea
 S^{PRY}_{nm}&\sim&(1-z)^{(N-n)(b'+c'+1)+(N-n)(N-n-1)\rho'+(M-m)(b+c+1)
   +(M-m)(M-m-1)\rho-2(N-n)(M-m)}\nn
 &\cdot&v_k^{a'}(1-v_k)^{b'+c'+(N-n)2\rho'-2(M-m)}(v_k-v_{k'}
  )^{2\rho'}\nn
 &\cdot&w_i^{a}(1-w_i)^{b+c+(M-m)2\rho-2(N-n)}(w_i-w_{
   i'})^{2\rho}(v_k-w_i)^{-2}\nn
 &\cdot&v_l^{-b'-c'-2-(N-n-1)2\rho'+2(M-m)}(1-v_l)^{b'}(x-(1-v_l)(1-u))
 (v_l-v_{l'})^{2\rho'}\nn
 &\cdot&w_j^{-b-c-2-(M-m-1)2\rho+2(N-n)}(1-w_j)^{b}(x-(1-w_j)(1-u))^{-\rho}
  (w_j-w_{j'})^{2\rho}\nn
 &\cdot&(v_l-w_j)^{-2}u^{b+c-N+(M-1)\rho}(1-u)^{-c-1}(x-1+u)^{n-m\rho}
\eea
Here the $k,i$ integrations are independent of $x$ and $u$ and are readily 
evaluated in terms of DF integrals, so we concentrate on the 
$l,j$ part. We perform the split and the scalings
\bea
 \int_0^1\prod_{l=1}^{N-n}dv_l\prod_{j=1}^{M-m}dw_j&=&\sum_{n_0,m_0}
  \int_0^{\epsilon}\prod_{l_0=N-n-n_0+1}^{N-n}dv_{l_0}
  \prod_{j_0=M-m-m_0+1}^{M-m}dw_{j_0}\nn
 &\cdot&\int_{\epsilon}^1
 \prod_{l=1}^{N-n-n_0}dv_l\prod_{j=1}^{M-m-m_0}dw_j\nn
 w&\rightarrow&(x-1)\frac{1-w}{w}\nn
 \int_0^{\epsilon}dw&\rightarrow&\int_1^{\frac{x-1}{\epsilon+x-1}}(1-x)
  \frac{dw}{w^2}\sim(x-1)\int_0^1\frac{dw}{w^2}
\eea
and similarly for $\int_0^1du$. To be able to distinguish we write
\bea
 \int_{\epsilon}^1du&\rightarrow&\int_0^1du\nn
 \int_0^{\epsilon}du&\rightarrow&(x-1)\int_0^1\frac{dy}{y^2}\nn
 u&\rightarrow&(x-1)\frac{1-y}{y}
\eea
and denote them the $u$ and $y$ cases respectively. 
In the $u$ case the arising DF integrals turn out to vanish unless 
$n_0=m_0=0$, and we find in the $u$ case
\bea
 S^{PRY,u}_{nm}&\sim&(1-z)^{(N-n)(b'+c'+1)+(N-n)(N-n-1)\rho'+(M-m)(b+c+1)
   +(M-m)(M-m-1)\rho}\nn
 &\cdot&(1-z)^{-2(N-n)(M-m)}\nn
 &\cdot&\prod_{k=1}^n\prod_{i=1}^m
  v_k^{a'}(1-v_k)^{b'+c'+(N-n)2\rho'-2(M-m)}(v_k-v_{k'}
  )^{2\rho'}\nn
 &\cdot&w_i^{a}(1-w_i)^{b+c+(M-m)2\rho-2(N-n)}(w_i-w_{
   i'})^{2\rho}(v_k-w_i)^{-2}\nn
 &\cdot&\prod_{l=1}^{N-n}\prod_{j=1}^{M-m}v_l^{-b'-c'-2-(N-n-1)2\rho'+2(M-m)}
  (1-v_l)^{b'}(v_l-v_{l'})^{2\rho'}\nn
 &\cdot&w_j^{-b-c-2-(M-m-1)2\rho+2(N-n)}(1-w_j)^b(w_j-w_{j'})^{2\rho}
  (v_l-w_j)^{-2}\nn
 &\cdot&(1-(1-v_l)(1-u))(1-(1-w_j)(1-u))^{-\rho}\nn
 &\cdot&u^{b+c-N+n+(M-m-1)\rho}
  (1-u)^{-c-1}
\label{PRYnmu}
\eea
Now we want to establish
\ben
  S^{PRY,u}_{NM}=S^{AI}_{NM}
\label{PRYNMANM1}
\een
and 
\ben
  S^{PRY,u}_{nm}=0\ \ \ ,\ \ \ (n,m)\neq (N,M)
\label{SPRY0}
\een
A straightforward analysis shows that the first one is satisfied.
The second identity follows from the first generalised DF integral in
(\ref{genDF}).

In the $y$ case we introduce a similar further splitting resulting in
objects $S^{PRY}_{nm;n_0m_0}$. It turns out to be possible
to demonstrate that
\bea
 S^{PRY}_{nm;00}&=&S^{AII}_{nm}\nn
 S^{PRY}_{nm;01}&=&S^{AI}_{nm}\spa m<M
\eea
The analysis contains no new ideas over the situation encountered for 
minimal models, but again the calculations involved are quite lengthy. 
In principle we should check that higher values of $n_0,m_0$ give zero. We
anticipate no interesting problems here \cite{PRY2}.
The seemingly missing PRY counterpart of $S_{nM}^{AI}$ for $n<N$ might be
'recovered' in the framework of the comments following (\ref{nosec}).\\[.2cm]
$\underline{z,z/x,x\rightarrow\infty}$\\
Again we first analyse the $S^A$ case. We introduce the same splitting
of integrations and the same variable transformations as for minimal models, 
and find
\bea
S^A_{nm} &\sim&K_{NM}^x(-z)^{n(-a'+\delta'-1)+(N-n)(c'-\delta'+1)-n(n-1)\rho'}
\nn
 &\cdot&(-z)^{m(-a+\delta-1)+(M-m)(c-\delta-\rho)-m(m-1)\rho+2nm}\nn
 &\cdot&x^{-(N-n)+(M-m)\rho}\nn
 &\cdot&v_k^{-a'-c'+2\delta'-2+2m-(n-1)2\rho'}(1-v_k)^{a'-\delta'}(v_k
  -v_{k'})^{2\rho'}\nn
 &\cdot&w_i^{-a-c+2\delta-2+2n-(m-1)2\rho}(1-w_i)^{a-\delta}(w_i-w_{i'}
  )^{2\rho}(v_k-w_i)^{-2}\nn
 &\cdot&\left(1+\frac{1-v_k}{xv_k}\right)\left(1+\frac{1-w_i}{xw_i}
  \right)^{-\rho}\nn
 &\cdot&v_l^{a'+c'-2\delta'+1-2m+n2\rho'}(1-v_l)^{b'+\delta'}(v_l-v_{l'}
  )^{2\rho'}\nn
 &\cdot&w_j^{a+c-2\delta-2n+(2m-1)\rho}(1-w_j)^{b+\delta}(w_j-w_{j'}
  )^{2\rho}(v_l-w_j)^{-2}
\eea
again omitting integration signs and products, which are just as for the case
of minimal models. The $l,j$ integration is seen to result in DF 
integrals. In the $k,i$ integrals we perform a split of integrals form $0$ to
$\epsilon$ and from $\epsilon$ to $1$. In the $\int_0^\epsilon$ we transform
variables like
\bea
 v&\rightarrow&\frac{1-v}{xv}\nn
 1+\frac{1-v}{xv}&\rightarrow&\frac{1}{1-v}-\frac{1}{x}\sim\frac{1}{1-v}\nn
 \int_0^{\epsilon}dv&\rightarrow&x^{-1}\int_0^1\frac{dv}{v^2}\nn
  \int_0^1\prod_{k=1}^ndv_k\prod_{i=1}^mdw_i&=&\sum_{n_0,m_0}
  \int_0^{\epsilon}\prod_{k_0=n-n_0+1}^{n}dv_{k_0}
  \prod_{i_0=m-m_0+1}^{m}dw_{i_0}\nn
 &\cdot&\int_{\epsilon}^1\prod_{k=1}^{n-n_0}dv_k\prod_{i=1}^{m-m_0}dw_i
\eea
An analysis of the coefficients of the singularities reveals that this is 
non-vanishing only if $(n_0,m_0)=(0,0)$ or if $(n_0,m_0)=(0,1)$.
These two cases we term again (we use the same notation as before, even though
now we consider a different limit), $S^{AI}_{nm}$ and $S^{AII}_{nm}$ for what
turns out to be fusion rules I and II. We find
\bea
 S^{AI}_{nm}&=&(-z)^{-(N-n)a'+nc'-(N-n)(N-n-1)\rho'-(M-m)a+mc
  -(M-m)(M-m)\rho+(M-m)(2N-2n-1)}\nn
 &\cdot&x^{-(N-n)+(M-m)\rho}N(S^{AI}_{nm})\nn
 S^{AII}_{nm}&=&(-z)^{-(N-n)a'+nc'-(N-n)(N-n-1)\rho'-(M-m)a+mc
  -(M-m)(M-m)\rho}\nn
 &\cdot&(-z)^{(M-m)(2N-2n-1)}\nn
 &\cdot&x^{-a-c-1+N-n+(-M+m)\rho}N(S^{AII}_{nm})
\eea
where the normalisations (different of course to the ones in the previous limit
$z,x\rightarrow 1$), $N(S^{AI}_{nm})$ and $N(S^{AII}_{nm})$ are given (in 
terms of DF integrals) by lengthy products of ratios of 
gamma-functions. The singularities shown indicate that indeed we are dealing
with fusion rules I and II.

We then treat the $S^{PRY}$ case. Again we first perform the same splittings
and variable transformations as for $S^A$ with the same meaning of 
$v_k,v_l,w_i,w_j$. The $i,k$ part is again simple, whereas the $l,j$ part is 
treated with a split of the ordered integrations as
\bea
 \int_0^1\prod_{l=1}^{N-n}dv_l\prod_{j=1}^{M-m}dw_j&=&\sum_{n_0,m_0}
  \int_0^{\epsilon}\prod_{l_0=N-n-n_0+1}^{N-n}dv_{l_0}
  \prod_{j_0=M-m-m_0+1}^{M-m}dw_{j_0}\nn
 &\cdot&\int_{\epsilon}^1
  \prod_{l=1}^{N-n-n_0}dv_l\prod_{j=1}^{M-m-m_0}dw_j\nn
\eea
followed by the scalings
\bea
 v_{l_0}&\rightarrow&\frac{1-v_{l_0}}{xv_{l_0}}\nn
 w_{j_0}&\rightarrow&\frac{1-w_{j_0}}{xw_{j_0}}
\eea
We then seek to demonstrate that
\bea
 S^{PRY}_{nm;00}&=&S^{AI}_{N-n,M-m}\nn
 S^{PRY}_{nm;01}&=&S^{AII}_{N-n,M-m}\nn
 S^{PRY}_{nm;n_0m_0}&=&0 ,\ \ \ (n_0,m_0)\neq (0,0),(0,1)
\eea
The proofs of the first two identities are lengthy, but with no new ideas 
introduced. An explicit proof of the last identity is lacking, but again
we anticipate no problems.\\[.2cm]
$\underline{z\rightarrow x}$\\
This case is the most complicated one. We omit nearly all the details, most of
which are similar to what have been described above. Let us outline 
the strategy. First one may 
check that the nature of the singularity is a linear combination of just two
different powers of $(z-x)$ namely either $(z-x)^0$ or $(z-x)^{(c+1-\rho)}$.
Second one must investigate whether the coefficients of these two powers are 
the same for the two sides of (\ref{pryan}). 
That coefficient is a function of $x$
in the limit $z\rightarrow x$, so we must investigate whether the coefficient 
functions defined by the two sides of (\ref{pryan}) are equal. As above the 
technique is to investigate the singularity structure in the singular limits
$x=0,1,\infty$. 

Let us return to the first issue. Here it is suggested to denote the channel
or basis corresponding to the limit $z\rightarrow x$ by v, inspired by
the alphabet s,t,u,v. Consider for fusion rule I the general real
s-channel block $S_{(r,s,0)}^{(R,S)}(z,x)$ and perform the transformations
$u_i,v_k\rightarrow zu_i,zv_k$ and $u_j,v_l\rightarrow 1/u_j,1/v_l$
whereby all integrations become $\int_0^1$ 
\bea
 S_{(r,s,0)}^{(R,S)}(z,x)&\sim&u_i^{a'}(1-zu_i)^{b'}(1-u_i)^{c'}
  (u_i-u_{i'})^{2\rho'}\nn
 &\cdot&u_j^{-a'-b'-c'}(1-u_j)^{b'}(1-zu_j)^{c'}(u_j-u_{j'})^{2\rho'}\nn
 &\cdot&u_j^{-2\rho'(R-r-1)}u_j^{-2\rho'}(1-zu_iu_j)^{2\rho'}u_j^{-2}\nn
 &\cdot&v_k^a(1-zv_k)^b(1-v_k)^c(v_k-v_{k'})^{2\rho}\nn
 &\cdot&v_l^{-a-b-c}(1-v_l)^b(1-zv_l)^c(v_l-v_{l'})^{2\rho}v_l^{-2\rho(S-s-1)}
  \nn
 &\cdot&v_l^{-2\rho}(1-zv_kv_l)^{2\rho}v_l^{-2}\nn
 &\cdot&(u_i-v_k)^{-2}(1-zu_iv_l)^{-2}v_l^2(1-zv_ku_j)^{-2}u_j^2
  (u_j-v_l)^{-2}u_j^{2(S-s-1)}v_l^{2(R-r-1)}\nn
 &\cdot&\left(1-\frac{z}{x}u_i\right)\left(1-\frac{z}{x}v_k\right)^{-\rho}
  u_j^{-1}(1-xu_j)v_l^{\rho}(1-xv_l)^{-\rho}
\eea
In these considerations we leave out integration signs and products.
We notice that only $u_i$ and $v_k$ may scale wrt $z/x$ and let us
denote the $r_0,s_0$ ones that do by $w_i$ and $w_k$
\bea
 u_i&\rightarrow& 1-\left(1-\frac{z}{x}\right)\frac{1-w_i}{w_i}\nn
 v_k&\rightarrow& 1-\left(1-\frac{z}{x}\right)\frac{1-w_k}{w_k}
\eea
In the limit $z\rightarrow x$ the $w$ integrations decouple 
\bea
 &&\prod_{i=1}^{r_0}dw_i\prod_{k=1}^{s_0}dw_kw_i^{-c'-2\rho'(r_0-1)-2+2s_0-1}
  (1-w_i)^{c'}(w_i-w_{i'})^{2\rho'}\nn
 &\cdot&w_k^{-c-2\rho(s_0-1)-2+2r_0+\rho}(1-w_k)^c(w_k-w_{k'})^{2\rho}
  (w_i-w_k)^{-2}
\eea
from which one deduces that $r_0=0$ and $s_0=0,1$. This gives that the only
possible powers of $z-x$ are
\ben
 (z-x)^{s_0(c+1-\rho)}\spa s_0=0,1
\een
A special feature arises for $s=0$ since then $s_0=0$ and 
in the corresponding rows there is only
one matrix element different from zero.
The two possible v-channel blocks are easily expressed in terms
of simple contours and will be denoted $V_1^{(R,S)}$ and $V_2^{(R,S)}$
\bea
 V_1^{(R,S)}&=&S_{(0,0,0)}^{(R,S)}\nn
 V_2^{(R,S)}&=&\int_1^\infty\prod_{j=1,l=1}^{R,S-1}du_jdv_l
  \int_z^xdw\nn
 &\cdot&u_j^{a'}(u_j-1)^{b'}(u_j-z)^{c'}(u_j-x)(u_j-u_{j'})^{2\rho'}\nn
 &\cdot&v_l^a(v_l-1)^b(v_l-z)^c(v_l-x)^{-\rho}
  (v_l-v_{l'})^{2\rho}(u_j-v_l)^{-2}\nn
 &\cdot&w^a(1-w)^b(w-z)^c(x-w)^{-\rho}(v_l-w)^{2\rho}(u_j-w)^{-2}
\eea
We have already seen that $V_1$ has the correct $(z-x)^0$ behaviour.
To see that $V_2$ has the correct $z-x$ singularity we make the transformations
$u_j,v_l\rightarrow 1/u_j,1/v_l$ and $v\rightarrow (x-z)v+z$ whereby all
integrations become $\int_0^1$ and one easily finds the singular behaviour
\ben
 (x-z)^{c-\rho+1}
\een

The situation for fusion rule II is very analogous. We make the transformations
$u_i,v_k,v\rightarrow zu_i,zv_k,xv$
and $u_j,v_l\rightarrow 1/u_j,1/v_l$ whereby we obtain (up to irrelevant
phases)
\bea
 S_{(r,s,1)}^{(R,S)}&\sim&\oint_{S^1}\prod\int_0^1
  u_i^{a'}(1-zu_i)^{b'}(1-u_i)^{c'}(u_i-u_{i'})^{2\rho'}\nn
 &\cdot&u_j^{-a'-b'-c'}(1-u_j)^{b'}(1-zu_j)^{c'}(u_j-u_{j'})^{2\rho'}
  u_j^{-2\rho'(R-r-1)}\nn
 &\cdot&u_j^{-2\rho'}(1-zu_iu_j)^{2\rho'}u_j^{-2}\nn
 &\cdot&v_k^a(1-zv_k)^b(1-v_k)^c(v_k-v_{k'})^{2\rho}\nn
 &\cdot&v_l^{-a-b-c}(1-v_l)^b(1-zv_l)^c(v_l-v_{l'})^{2\rho}v_l^{-2\rho
  (S-s-1)}\nn
 &\cdot&v_l^{-2\rho}(1-zv_kv_l)^{2\rho}v_l^{-2}\nn
 &\cdot&(u_i-v_k)^{-2}(1-zu_iv_l)^{-2}v_l^2(1-zv_ku_j)^{-2}u_j^2
  (u_j-v_l)^{-2}\nn
 &\cdot&u_j^{2(S-s-1)}v_l^{2(R-r-1)}\left(1-\frac{z}{x}u_i\right)\left(
  1-\frac{z}{x}v_k\right)^{-\rho}\nn
 &\cdot&u_j^{-1}(1-xu_j)v_l^{\rho}(1-xv_l)^{-\rho}\nn
 &\cdot&v^a(1-xv)^b\left(1-\frac{x}{z}v\right)^c
  \left(v-\frac{z}{x}v_k\right)^{2\rho}
  (1-xvv_l)^{2\rho}v_l^{-2\rho}\nn
 &\cdot&\left(v-\frac{z}{x}u_i\right)^{-2}(1-xvu_i)^{-2}u_j^2(1-v)^{-\rho}
\eea
As for fusion rule I it is relevant to scale $u_i,v_k$ but this time also $v$.
If $v$ is not scaled we find exactly the same decoupling of the transformed
variables (denoted $w_i,w_k$). Again this indicates that we only need
$V_1^{(R,S)}$ and $V_2^{(R,S)}$. However, if $v$ does scale the situation
is much more complicated to handle. Though, we anticipate that it does not
lead to any new v-blocks. 

Let us turn to the second issue. It turns out that the sought equality 
(\ref{pryan}) depends on the following identities
\bea
 &&\int_0^1 dwdy w^{-a-2+\rho}(1-w)^a(1-(1-w)(1-(1-x)y))^{-\rho}\nn
 &\cdot&y^{-b-c+N-2+(-M+2)\rho}(1-y)^{b+c-N+(M-1)\rho}\nn
 &=&(1-x)^{-a-1}\frac{\Gamma(a+1)\Gamma(b+c+1-N+(M-1)\rho)
  \Gamma(-a-b-c-2+N+(-M+2)\rho)}{\Gamma(\rho)}\nn
\eea
and
\ben
 \int_0^1 dw\int_0^w dy w^{-a}(1-w)^{a+c-2}(w-y)^{-c}y^{b+c-2}(1-y)^{-b}=0
\een
which are not too difficult to prove. Next define the Dotsenko-Fateev integrand
\bea
 &&DF(N,M;a,b,\rho;\{v_k\},\{w_i\})\nn
 &=&\prod_{k-1}^Nv_k^{a'}(1-v_k)^{b'}\prod_{k<k'}(v_k-v_{k'})^{2\rho'}
 \prod_{i=1}^Mw_i^a(1-w_i)^b\prod_{i<i'}(w_i-w_{i'})^{2\rho}
  \prod_{k,i}^{N,M}(v_k-w_i)^{-2}\nn
\eea
Then we find that the equality of the two sides of (\ref{pryan}) depends on 
the following three identities (generalised DF integrals)
\bea
 (I)&&\int_0^1 du\prod_{k=1}^N dv_k\prod_{i=1}^M dw_i 
  DF(N,M;a,b,\rho;\{v_k\},\{w_i\})\nn
 &\cdot&(1-(1-v_k)u)(1-(1-w_i)u)^{-\rho}u^{-c-1}(1-u)^{b+c-N+(M-1)\rho}\nn
 &=&\frac{\Gamma(-c)\Gamma(b+c+1-N+(M-1)\rho)\Gamma(a+b+c+2-N+(M-2)\rho)}
  {\Gamma(b+1-\rho)\Gamma(a+b+c+2-2N+(2M-2)\rho)}\nn
 &\cdot&\int_0^1\prod_{k=1}^N dv_k\prod_{i=1}^M dw_i 
  DF(N,M;a,b-\rho,\rho;\{v_k\},\{w_i\})\nn
 (II)&&\int_0^1 du\prod_{k=1}^n dv_k\prod_{i=1}^{m+1} dw_i\nn 
 &\cdot&DF(n,m+1;a,b+c-2N+2n+(2M-2m-1)\rho,\rho;\{v_k\},\{w_i\})\nn 
 &\cdot&(1-v_k-u)(1-w_i-u)^{-\rho}u^{-b-c+N-n-2+(-M+m+2)\rho}
  (1-u)^{b+c-N+(M-1)\rho}\nn
 &\sim&\frac{\Gamma(b+c+1-N+(M-1)\rho)\Gamma(a+b+c+2-N+(M-2)\rho)}
  {\Gamma(a+b+c+2-2N+(2M-2)\rho)}\nn
 &\cdot&\frac{\Gamma(-b-c+N-n-1+(-M+m+2)\rho)\Gamma(-N+n+(M-m)\rho)}
  {\Gamma(\rho)}\nn
 &\cdot&\int_0^1\prod_{k=1}^n dv_k\prod_{i=1}^m dw_i\nn 
 &\cdot&DF(n,m;a,b+c-2N+2n+(2M-2m-1)\rho,\rho;\{v_k\},\{w_i\})\nn
 (III)&&\int_0^1du\prod_{k=1}^n dv_k\prod_{i=1}^m dw_i\nn 
 &\cdot&DF(n,m;-a-c+2n-2+(-2m+2)\rho,a,\rho;\{v_k\},\{w_i\})\nn
 &\cdot&(1-v_ku)(1-w_iu)^{-\rho}u^{-c-1}(1-u)^{b+c-N+(M-1)\rho}\nn
 &=&\frac{\Gamma(-c)\Gamma(b+c+1-N+(M-1)\rho)\Gamma(a+b+c+2-N+(M-2)\rho)}
  {\Gamma(b+1-N+n+(M-m-1)\rho)\Gamma(a+b+c+2-N-n+(M+m-2)\rho)}\nn
 &\cdot&\int_0^1\prod_{k=1}^n dv_k\prod_{i=1}^m dw_i 
 DF(n,m;-a-c+2n-2+(-2m+2)\rho,a,\rho;\{v_k\},\{w_i\})\nn
\label{genDF}
\eea
All the final integrals are of course DF integrals.
In the second identity there is a phase depending on the precise choice of the
integration contour for $u$. These last three identities have not been 
proven directly, but are easily checked for low values of $N,M$. 
One may take the attitude that the
undoubted identity of the PRY realization  and that of Andreev, i.e. the 
unquestionable correctness of (\ref{pryan}), implies these 
somewhat remarkable integral identities in addition to the ones underlying
the cases discussed above.

Let us supply with some details on the limit $z\rightarrow x$.
We already know that the rhs of (\ref{pryan}) vanishes except in the two cases
$(r_0,s_0)=(0,0),(0,1)$. The notation used there is
different from the one used otherwise and to which
we will stick here.
The translation of $(r_0,s_0)=(0,0),(0,1)$ gives $(n,m)=(N,M),(N,M-1)$.
In conclusion we find for the Andreev representation
\bea
 S_{NM}^{A}&\sim&K_{NM}^x\prod_{k=1}^Nv_k^{a'-\delta'}(1-v_k)^{b'+\delta'+1}
  (1-xv_k)^{c'-\delta'}(v_k-v_{k'})^{2\rho'}\nn
 &\cdot&\prod_{i=1}^Mw_i^{a-\delta}(1-w_i)^{b+\delta-\rho}(1-xw_i)^{c-\delta}
  (w_i-w_{i'})^{2\rho}(v_k-w_i)^{-2}\nn
 &=&
  \frac{\Gamma(-c)\Gamma(b+c+1-N+(M-1)\rho)\Gamma(a+b+c+2-N+(M-2)\rho)}{
  \Gamma(b+1-\rho)\Gamma(a+b+c+2-2N+(2M-2)\rho)}\nn
 &\cdot&v_k^{a'}(1-v_k)^{b'+1}(1-xv_k)^{c'}(v_k-v_{k'})^{2\rho'}\nn
 &\cdot&w_i^a(1-w_i)^{b-\rho}(1-xw_i)^c(w_i-w_{i'})^{2\rho}(v_k-w_i)^{-2}
\label{ANMzx}
\eea
and
\bea
 S_{N,M-1}^{A}&\sim&(1-z/x)^{b+\delta-\rho+1}(1-x)^{c-\delta}\nn
 &\cdot&K_{NM}^x\prod_{k=1}^Nv_k^{a'-\delta'}(1-v_k)^{b'+\delta'-1}
  (1-xv_k)^{c'-\delta'}(v_k-v_{k'})^{2\rho'}\nn
 &\cdot&\prod_{i=1}^{M-1}
  w_i^{a-\delta}(1-w_i)^{b+\delta+\rho}(1-xw_i)^{c-\delta}
  (w_i-w_{i'})^{2\rho}(v_k-w_i)^{-2}\nn
 &\cdot&w^{-b-\delta+\rho-2}(1-w)^{b+\delta}\nn
 &=&(1-z/x)^{a+b+c+2-N+(M-2)\rho}(1-x)^{-a-1+N+(-M+1)\rho}\nn
 &\cdot&\frac{\Gamma(a+1)\Gamma(b+c+1-N+(M-1)\rho)\Gamma(-a-b-c-2+N+(-M+2)
  \rho)}{\Gamma(\rho)}\nn
 &\cdot&\prod_{k=1}^N
  v_k^{a'-1}(1-v_k)^{b'}(1-xv_k)^{c'-1}(v_k-v_{k'})^{2\rho'}\nn
 &\cdot&\prod_{i=1}^{M-1}
  w_i^{a+\rho}(1-w_i)^{b}(1-xw_i)^{c+\rho}(w_i-w_{i'})^{2\rho}(v_k-w_i)^{-2}
\label{ANM-1zx}
\eea 
The rewritings are due to the integral identity for minimal models 
(\ref{minimal}). Again we have used the same notation as in the other
limits, but as long we do not make any direct comparisons between results
obtained in the different limits, it should not lead to any confusion.

Now we turn to PRY. We will have to divide into two cases; 
$V^{PRY,u}$ where $u$ is not scaled and $V^{PRY,y}$ where $u$
is scaled
\ben
 u\rightarrow (1-z/x)\frac{1-y}{y}
\een
In both cases we let $n$ $v$'s and $m$ $w$'s transform and denote the
cases $S_{nm}^{PRY,u}$ and $S_{nm}^{PRY,y}$. Several lengthy and 
cumbersome manipulations using the various integral identities, lead to
the following identities
\ben
 S_{NM}^{A}\sim S_{00}^{PRY,u}\spa
 S_{N,M-1}^{A}\sim S_{01}^{PRY,y}
\een
Without going into these final and comprehensive computations we conclude that
the above proof presents substantial evidence for the claimed integral
identity (\ref{pryan}) at the level of direct verification. 
Taking the aforementioned attitude of accepting the equivalence of the
PRY and Andreev representations from the point of view of both 
representations producing
solutions to the KZ equations, the above assumptions are all justified.\\
$\Box$

\section{Over-screening}
The 4-point blocks considered so far are ones 
where we have either fusion rule I
operating at both vertices, or fusion rule II operating at both vertices. We
now describe how to obtain 4-point blocks for the case where we have either
fusion rule I for $j_1j_2j$ and fusion rule II for $jj_3j_4$, we denote
that case by (II,I), or fusion rule II for $j_1j_2j$ and fusion rule I for
$jj_3j_4$, we denote that case by (I,II). We emphasise that for a collection 
of spins considered so far, so that fusion rule I (or fusion rule II) 
is possible at both vertices, neither (I,II) nor (II,I) will be possible. Hence
there will be no mixing in the crossing matrix calculations.

Our technique is based on the discussion of fusion rules I and II for the 
3-point function in Chapter 4. Now the charge conservation reads 
\ben
 j_1+j_2+j_3-j_4=R-St
\een
and as for the charge conservation of the 3-point function there is a 
considerable freedom in choosing the numbers of screenings. Let us 
introduce the following three choices\\[.2 cm]
{\bf Standard Screening}
\bea
 2R&=&r_1+r_2+r_3-r_4-2\nn
 2S&=&s_1+s_2+s_3-s_4
\label{ss}
\eea
{\bf Over-screening}
\bea
 2R&=&r_1+r_2+r_3-r_4-2+p\nn
 2S&=&s_1+s_2+s_3-s_4+q
\label{os}
\eea
{\bf Reducible Over-screening}
\bea
 2R&=&r_1+r_2+r_3-r_4-2+2p\nn
 2S&=&s_1+s_2+s_3-s_4+2q
\label{ros}
\eea
Any 4-point function may be factorised into
a product of two 3-point functions, but in the previous work based on
(\ref{ss}) this fact has been somewhat unclear in the case of fusion
rule II. There is apparently a very simple explanation for that.
According to the discussion of 3-point functions,
the 'natural' way to obtain
a fusion rule II 4-point function is to use (\ref{ros}), an over-screening
by $p,q$ at both vertices. However, the total addition of $2p,2q$ might
be reducible since it involves only even numbers. Indeed this is 
what we find since we have been able to define 4-point functions for 
fusion rule II using the standard screening (\ref{ss}). 
The reducibility obscures the factorisation.

Let us first notice that (\ref{os}) indeed provides us with a new family
of 4-point functions (if we can find contours such that the blocks are
well-defined and non-vanishing). This follows from the fact that
there is no overlap between fusion rule I and II, such that
a block based on (II,I) will be different from blocks based on other
combinations. Here the notation (II,I) refers to the fusion rules involved
in the (left,right) vertices. There are two cases of interest\\[.2 cm]
{\bf (I,II)}
\bea
 (\rho_1,\sigma_1)&=&\left(\frac{r_1+r_2-r-1+p}{2},\frac{s_1+s_2-s+q}{2}\right)
  \nn
  (\rho_2,\sigma_2)&=&
  \left(\frac{r+r_3-r_4-1}{2},\frac{s+s_3-s_4}{2}\right)\nn
 (R,S)&=&(\rho_1+\rho_2,\sigma_1+\sigma_2)
\label{osI,II}
\eea
{\bf (II,I)}
\bea
  (\rho_1,\sigma_1)&=&\left(\frac{r_1+r_2-r-1}{2},\frac{s_1+s_2-s}{2}\right)
  \nn
   (\rho_2,\sigma_2)&=&
  \left(\frac{r+r_3-r_4-1+p}{2},\frac{s+s_3-s_4+q}{2}\right)\nn
 (R,S)&=&(\rho_1+\rho_2,\sigma_1+\sigma_2)
\label{osII,I}
\eea                
$(\rho_1,\sigma_1)$ are the numbers of screenings around the right vertex
$j_1j_2j$ while $(\rho_2,\sigma_2)$ are the numbers of screenings around 
the left vertex $jj_3j_4$. The internal state is parameterised as
\ben
 2j+1=r-st
\een
$R$ and $S$ are the total numbers of screenings.

At both vertices we choose contours similar to the 3-point functions.
Compared to the contours in Fig. \ref{figanI} this amounts
to substitute (in the case (I,II)) the $\sigma_1$ DF contours for the 
screenings of the second kind around the right vertex by Felder contours.
More precisely, we may take $v_1,...v_s$ to run
along circle like contours passing through the point $z$, and surrounding the 
contours for $u_1,...,u_r$, i.e. surrounding the point $0$, in such a way that
the $v_1$-contour lies inside the $v_2$-contour, etc. Similarly for the case
of (II,I), we modify the contours for $v_{s+1},...,v_S$ into circle like 
contours surrounding $0$ (actually surrounding $\infty$) and passing through 
the point $1$, in such a way that the contour for $v_{s+1}$ lies inside the 
contour for $v_{s+2}$ etc.

Thus, in the case (I,II) we want to establish the following\\[.2 cm]
{\bf (I,II)}
\ben
 \lambda_{\rho_1}(1/t)\chi_{\sigma_1}^{(2)}(S-s_3-1;t)\lambda_{\rho_2}(1/t)
  \lambda_{\sigma_2}(t)N_{(\rho_1,\sigma_1,0)}^{(R,S)}
  \neq0,\infty
\label{I,II}
\een
The argument of the $\chi$ function follows from the fact that the 
deformation of Felder contours only depends on the number of integration 
variables and (minus)
the fractional part of the common power of the $w_i$'s (not on the power
of $(1-w_i)$). In this case the power is $a=-2j_3+t+R-St-1$ so the
fractional part is $(s_3-S+1)t$. 
Using (\ref{dfintegral2}) one obtains the normalisation
\bea
 N_{(\rho_1,\sigma_1,0)}^{(R,S)}&=&{\cal J}_{\rho_1,\sigma_1}(a,c;t)
  \tilde{{\cal J}}_{R-\rho_1,S-\sigma_1}(a+c-2(\rho_1-\sigma_1 t)-t,b;t)\nn
 &=&{\cal J}_{\rho_1,\sigma_1}(a,c;t){\cal J}_{\rho_2,\sigma_2}(-a-b-c+2(R-1)
  -2(S-1)t+t,b;t)\nn
 &=&t^{2\rho_1\sigma_1}\prod_{j=1}^{\rho_1}\frac{\Gamma(j/t)}{\Gamma(1/t)}
  \prod_{j=1}^{\sigma_1}\frac{\Gamma(jt-\rho_1)}{\Gamma(t)}\nn      
 &\cdot&\prod_{j=0}^{\rho_1-1}\frac{\Gamma(S-s_3+(r_3-R+j)/t)}{
  \Gamma(1-2\sigma_1+s_1+s_2+(1+\rho_1-r_1-r_2+j)/t)}\nn
 &\cdot&\prod_{j=0}^{\rho_1-1}\Gamma(1+s_1+s_2
  +s_3-S+(2+R-r_1-r_2-r_3+j)/t)\nn
 &\cdot&\prod_{j=0}^{\sigma_1-1}\frac{\Gamma(1-\rho_1+R-r_3+(s_3-S+1+j)t)}
  {\Gamma(r_1+r_2-\rho_1+(\sigma_1-s_1-s_2+j)t)}\nn
 &\cdot&\prod_{j=0}^{\sigma_1-1}
  \Gamma(r_1+r_2+r_3-1-\rho_1-R+(S-s_1-s_2-s_3+j)t)\nn
 &\cdot&t^{2\rho_2\sigma_2}\prod_{j=1}^{\rho_2}\frac{\Gamma(j/t)}{\Gamma(1/t)}
  \prod_{j=1}^{\sigma_2}\frac{\Gamma(jt-\rho_2)}{\Gamma(t)}\nn      
 &\cdot&\prod_{j=0}^{\rho_2-1}\frac{\Gamma(S-s_2+(r_2-R+j)/t)\Gamma(S-s_1
  +(r_1-R+j)/t)}{\Gamma(2\sigma_1-s_1-s_2+(r_1+r_2-1-R-\rho_1+j)/t)}\nn
 &\cdot&\prod_{j=0}^{\sigma_2-1}\frac{\Gamma(1+\rho_1-r_2+(s_2+1-S+j)t)
  \Gamma(1+\rho_1-r_1+(s_1+1-S+j)t)}{\Gamma(2+R+\rho_1-r_1-r_2+(1-S-\sigma_1
   +s_1+s_2+j)t)}\nn
\eea
According to both fusion rule I and II
\ben
 \rho_i<p\ \ \ \ \ \ \ \ \ \ \sigma_i<q
\een
so
\ben
 \lambda_{\rho_i}\neq0\neq\lambda_{\sigma_i}
\een
and we may concentrate on $\chi$ and $N$.
The remaining analysis is straightforward (but cumbersome)
and equivalent to the one employed in the discussion of 3-point functions: 
using the relevant fusion rules, 
we consider the gamma- and sine-functions one by one in order to determine 
the netto number of $\Gamma(0)$'s ($\sim1/s(0)$'s).
As expected, one finds the cancellation $(\G(0)/\G(0))^2$ from the part
concerning the right vertex (fusion rule II). The expectation is due
to the factorisation into 3-point functions and
the similar experience from considering 3-point functions.

The analysis of the alternative possibility (\ref{osII,I}) is analogous.
Now we want to show \\[.2 cm]
{\bf (II,I)}
\ben
 (-1)^{\sigma_2}\lambda_{\rho_1}(1/t)\lambda_{\sigma_1}(t)\lambda_{\rho_2}
  (1/t)\chi_{\sigma_2}^{(2)}(S-s_2-1;t)N_{(\rho_1,\sigma_1,0)}^{(R,S)}
  \neq0,\infty
\een
This time we encounter the cancellation $(\G(0)/\G(0))^2$ at the left vertex,
again as expected.

Having performed the analysis of the 4-point function based on over-screening
(\ref{os}) it is a simple matter to verify that reducible over-screening 
(\ref{ros}) indeed leads to well-defined and non-vanishing 4-point functions.
We choose $\rho_1$ DF contours and $\sigma_1$ Felder contours around
the right vertex along with $\rho_2$ DF contours and $\sigma_2$ Felder
contours around the left vertex. The numbers of screenings are\\[.2 cm]
{\bf (II,II)}
\bea
 (\rho_1,\sigma_1)&=&\left(\frac{r_1+r_2-r-1+p}{2},\frac{s_1+s_2-s+q}{2}\right)
  \nn
 (\rho_2,\sigma_2)&=&\left(\frac{r+r_3-r_4-1+p}{2},\frac{s+s_3-s_4+q}{2}\right)
  \nn
 (R,S)&=&(\rho_1+\rho_2,\sigma_1+\sigma_2)
\label{osII,II}
\eea
The analysis of the corresponding 4-point function is most easily performed
using the techniques above
\ben
 {\cal S}_{(\rho_1,\sigma_1,(II,II))}^{(R,S)}
  =(-1)^{\sigma_2}\lambda_{\rho_1}(1/t)
   \chi_{\sigma_1}^{(2)}(S-s_3-1;t)\lambda_{\rho_2}(1/t)
   \chi_{\sigma_2}^{(2)}(S-s_2-1;t)N_{(\rho_1,\sigma_1,0)}^{(R,S)}
\label{SII,II}
\een
Remembering that for the over-screened 4-point function based on
(II,I) and (I,II) 
all $\Gamma(0)$'s appeared at the over-screened vertices, we may conclude
that in the present case there will be exactly 4 cancellations 
of the form $\Gamma(0)/\Gamma(0)$ leaving 
${\cal S}_{(\rho_1,\sigma_1,(II,II))}^{(R,S)}$ well-defined
and non-vanishing.

The representation (\ref{SII,II}) gives us an alternative form of that 
block from the one considered so far, but one with more screenings 
(more integrations in the case of Andreev's representation). 
Our previous treatment is the most economic as far
as the numbers of screenings are concerned, and indeed it is more
convenient when considering the crossing matrix (next section)
since our technique there is based on contour deformations and becomes 
'singular' when different numbers of contours are in play. We will have 
more to say about 4-point functions based on over-screening when
discussing operator algebra coefficients in a subsequent section.

\section{Crossing Matrix}
The crossing matrix $\alpha^{(R,S)}_{(r,s,\epsilon),(r',s',\epsilon')}$ 
is defined by the equation
\bea
S^{(R,S)}_{(r,s,\epsilon)}(z,x)&=&\sum_{r'=0}^R\sum_{s'=0}^S
\alpha^{(R,S)}_{(r,s,\epsilon),(r',s',0)}T^{(R,S)}_{(r',s',0)}(z,x)\nn
&+&\sum_{r'=0}^R\sum_{s'=0}^{S-1}
\alpha^{(R,S)}_{(r,s,\epsilon),(r',s',1)}T^{(R,S)}_{(r',s',1)}(z,x)
\eea
As explained 
in \cite{DF}, it is enough to calculate one column and one row
of this matrix in order to determine monodromy invariant Greens functions.
It turns out that a moderate modification of the techniques described 
there suffices for completing the corresponding calculations here following
\cite{PRY2}.
The main new feature is the fact that we have to observe also the $x$ 
dependence, and the presence of the complex contour in the case of 
conformal blocks for fusion rule II, cf. Fig. \ref{figanII}.

We will not discuss the case of over-screening in this context but refer
to the section below on Greens functions.
 
\subsection{The Column of the Transformation Matrix}
Following the idea of \cite{DF} we define the following object \cite{PRY2} 
(suppressing several variables)
\bea
 &&J(r_1,s_1,r_2,s_2,r_3,s_3)=z^{2j_1j_2/t}(1-z)^{2j_2j_3/t}\nn
 &\cdot&\int_0^z \prod_{i=1}^{r_1}du_i\prod_{k=1}^{s_1}dv_k\int_z^1
  \prod_{m=r_1+1}^{r_1+r_2}du_m\prod_{n=s_1+1}^{s_1+s_2}dv_n\int_1^\infty 
  \prod_{j=r_1+r_2+1}^{r_1+r_2+r_3}du_j\prod_{l=s_1+s_2+1}^{s_1+s_2+s_3}dv_l\nn
 &\cdot&\prod_i
  u_i^{a'}(1-u_i)^{b'}(z-u_i)^{c'}\prod_{i<i'}(u_i-u_{i'})^{2\rho'}(u_i-x)\nn
 &\cdot&\prod_m
  u_m^{a'}(1-u_m)^{b'}(u_m-z)^{c'}\prod_{m<m'}(u_m-u_{m'})^{2\rho'}(x-u_m)\nn
 &\cdot&\prod_j
  u_j^{a'}(u_j-1)^{b'}(u_j-z)^{c'}\prod_{j<j'}(u_j-u_{j'})^{2\rho'}(u_j-x)\nn
 &\cdot&\prod_{m,i}(u_m-u_i)^{2\rho'}\prod_{j,i}(u_j-u_i)^{2\rho'}\prod_{j,m}
  (u_j-u_m)^{2\rho'}\nn
 &\cdot&\prod_k 
  v_k^a(1-v_k)^b(z-v_k)^c\prod_{k<k'}(v_k-v_{k'})^{2\rho}(v_k-x)^{-\rho}\nn
 &\cdot&\prod_nv_n^a(1-v_n)^b(v_n-z)^c\prod_{n<n'}
  (v_n-v_{n'})^{2\rho}(x-v_n)^{-\rho}\nn
 &\cdot&\prod_l
  v_l^a(v_l-1)^b(v_l-z)^c\prod_{l<l'}(v_l-v_{l'})^{2\rho}(v_l-x)^{-\rho}\nn
 &\cdot&\prod_{n,k}(v_n-v_k)^{2\rho}\prod_{l,k}(v_l-v_k)^{2\rho}
  \prod_{l,n}(v_l-v_n)^{2\rho}\nn
 &\cdot&\prod_{\alpha,\beta}(v_\alpha-u_\beta)^{-2}
\eea
In other words, there are $r_1$ and $s_1$ $u$ and $v$ integrations between $0$ 
and $z$, $r_2$ and $s_2$ $u$ and $v$ integrations between $z$ and $1$ and
$r_3$ and $s_3$ $u$ and $v$ integrations between $1$ and $\infty$.
Also the variables, $u_i,v_k,u_j,v_l$ are taken along contours similar to the
ones in Fig. \ref{figanI}, whereas the variables, $u_m,v_n$ are taken along 
similar ones lying between $z$ and $1$.
We notice that 
\bea
 J(r,s,0,0,R-r,S-s)&=&{\cal S}^{(R,S)}_{(r,s,0)}\nn
 J(0,0,R,S,0,0)&=&{\cal T}^{(R,S)}_{(R,S,0)}
\eea
Therefore we may start from $J(r,s,0,0,R-r,S-s)$ and gradually move integration
contours by contour deformations on to the interval $(z,1)$. In the process
we pick up contributions from integrals between $-\infty$ and $0$, but these 
may be neglected in the calculation of the column, 
$\alpha^{(R,S)}_{(r,s,\epsilon),(R,S,0)}$. The calculational procedure 
consists in deforming upper and lower $u$ and $v$ contours in 
appropriate ways, and forming suitable linear combinations of the result. 
As explained in \cite{DF}, one may then derive identities 
for the functions, $J(r_1,s_1,r_2,s_2,r_3,s_3)$,
by carefully keeping track of the phases arising between the result of the 
deformations, and the definitions of the $J$'s. The useful identities turn out
to be after some calculations
\bea
 &&J(r_1,s_1,r_2,s_2,r_3,s_3)\nn
 &=&e^{i\pi\rho'(r_2-r_1+1)}\frac{s(b'+\rho'(r_2+r_3))}
  {s(b'+c'+\rho'(r_1-1+2r_2+r_3))}J(r_1-1,s_1,r_2+1,s_2,r_3,s_3)+...\nn
 &=&-e^{i\pi\rho(s_2-s_1+2)}\frac{s(b+\rho(s_2+s_3))}
  {s(b+c+\rho(s_1-1+2s_2+s_3))}J(r_1,s_1-1,r_2,s_2+1,r_3,s_3)+...\nn
 &\cdot&J(0,0,r_2,s_2,r_3,s_3)\nn
 &=&e^{i\pi\rho'(r_2-r_3+1)}\frac{s(c'+\rho'r_2)}
  {s(b'+c'+\rho'(2r_2+r_3-1))}J(0,0,r_2+1,s_2,r_3-1,s_3)+...\nn
 &=&-e^{i\pi\rho(s_2-s_3+2)}\frac{s(c+\rho s_2)}
  {s(b+c+\rho(2s_2+s_3-1))}J(0,0,r_2,s_2+1,r_3,s_3-1)+...
\label{usefulal}
\eea
Here the dots stand for terms that cannot contribute to the crossing matrix 
element. Let us illustrate the techniques involved by deriving the very first
identity in (\ref{usefulal}).
We deform the upper $u$ variable in the interval $(0,z)$ on to the intervals
$(-\infty,0),(1,z)$ and $(+\infty,1)$. 
This will result in a relation of the form
\ben
 J(r_1,s_1,r_2,s_2,r_3,s_3)=\phi_1 J(r_1-1,s_1,r_2+1,s_2,r_3,s_3)+
  \phi_2 J(r_1-1,s_1,r_2,s_2,r_3+1,s_3)+...
\een
where the dots stand for contributions from integration over the interval 
$(-\infty,0)$ which cannot have a projection onto the t-channel block we are 
interested in, which is $J(0,0,R,S,0,0)$. We call the upper $u$ variable 
simply $u$ and only consider the $u$ parts of the various integrands. 
Let us denote the sets of indices for the $u$ variables as $I_1,I_2,I_3$ when
the variables are in the intervals $(0,z),(z,1),(1,\infty)$ respectively. 
Likewise we denote the index sets for the $v$ variables as $J_1,J_2,J_3$ when
the $v$ variables are in the same intervals.
Then we have before we transform $J(r_1,s_1,r_2,s_2,r_3,s_3)$
\ben
 \int_0^zduu^{a'}(1-u)^{b'}(z-u)^{c'}\prod_{I_1}(u_i-u)^{2\rho'}\prod_{I_2}
  (u_m-u)^{2\rho'}\prod_{I_3}(u_j-u)^{2\rho'}(u-x)\prod_{J_1,J_2,J_3}
  (v_\alpha-u)^{-2}
\een
The $u$ part in $J(r_1-1,s_1,r_2+1,s_2,r_3,s_3)$ would be
\ben
 \int_z^1 du u^{a'}(1-u)^{b'}(u-z)^{c'}\prod_{I_1}(u-u_i)^{2\rho'}
  \prod_{I_2}(u_m-u)^{2\rho'}\prod_{I_3}(u_j-u)^{2\rho'}(x-u)
  \prod_{J_1,J_2,J_3}(v_\alpha-u)^{-2}
\een
Thus we find
\ben
 \phi_1=e^{-i\pi c'}e^{-i\pi 2\rho'(r_1-1)}(-1)^2
\een
The $u$ part in $J(r_1-1,s_1,r_2,s_2,r_3+1,s_3)$ would be
\ben
 \int_1^\infty du u^{a'}(u-1)^{b'}(u-z)^{c'}\prod_{I_1}(u-u_i)^{2\rho'}
  \prod_{I_2}(u-u_m)^{2\rho'}\prod_{I_3}(u_j-u)^{2\rho'}(u-x)
  \prod_{J_1,J_2,J_3}(v_\alpha-u)^{-2}
\een
so we find
\ben
 \phi_2=-e^{-i\pi (b'+c'+2\rho'(r_1-1)+2\rho'r_2)}
\een
We then transform the lowest $u$ contour similarly. Before transforming we 
have for the $u$ part of $J(r_1,s_1,r_2,s_2,r_3,s_3)$
\ben
 \int_0^z du u^{a'}(1-u)^{b'}(z-u)^{c'}\prod_{I_1}(u-u_i)^{2\rho'}
  \prod_{I_2}(u_m-u)^{2\rho'}\prod_{I_3}(u_j-u)^{2\rho'}\prod_{J_1,J_2,J_3}
  (v_\alpha-u)^{-2}(u-x)
\een
We want to find $\psi_1,\psi_2$ such that
\ben
 J(r_1,s_1,r_2,s_2,r_3,s_3)=\psi_1 J(r_1-1,s_1,r_2+1,s_2,r_3,s_3)
  +\psi_2J(r_1-1,s_1,r_2,s_2,r_3+1,s_3)+...
\een
The $u$ part of $J(r_1-1,s_1,r_2+1,s_2,r_3,s_3)$ would be
\ben
 \int_z^1 du u^{a'}(1-u)^{b'}(u-z)^{c'}\prod_{I_1}(u-u_i)^{2\rho'}
  \prod_{I_2}(u-u_m)^{2\rho'}\prod_{I_3}(u_j-u)^{2\rho'}\prod_{J_1,J_2,J_3}
  (v_\alpha-u)^{-2}(x-u)
\een
hence
\ben
 \psi_1=(-1)^2e^{i\pi(c'+2\rho' r_2)}
\een
The $u$ part of $J(r_1-1,s_1,r_2,s_2,r_3+1,s_3)$ would be
\ben
 \int_1^\infty du u^{a'}(u-1)^{b'}(u-z)^{c'}\prod_{I_1}(u-u_i)^{2\rho'}
  \prod_{I_2}(u-u_m)^{2\rho'}\prod_{I_3}(u-u_j)^{2\rho'}\prod_{J_1,J_2,J_3}
  (v_\alpha-u)^{-2}(u-x)
\een
so we find
\ben
\psi_2=-e^{i\pi(b'+c'+2\rho' r_2+2\rho' r_3)}
\een
Finally, we have
\ben
 (\phi_2^*-\psi_2^*)J(r_1,s_1,r_2,s_2,r_3,s_3)
  =(\phi_1\phi_2^*-\psi_1\psi_2^*)J(r_1-1,s_1,r_2+1,s_2,r_3,s_3)+...
\een
which gives the promised identity.

After several further but in principle straightforward calculations,
we obtain
\bea
 \alpha^{(R,S)}_{(r,s,0),(R,S,0)}&=& (-1)^Se^{i\pi S\rho}
  \alpha^{(R)}_{r,R}(a',b',c';\rho')\alpha^{(S)}_{s,S}(a,b,c;\rho)\nn
 \alpha^{(S)}_{s,S}(a,b,c;\rho)&=&
  \frac{\prod_{j=1}^Ss(j\rho)}{\prod_{k=1}^{S-s}s(k\rho)
  \prod_{m=1}^ss(m\rho)}\nn
 &\cdot&\prod_{j=0}^{s-1}\frac{s(b+\rho(S-s+j))}{s(b+c+\rho(S+j-1))}
  \prod_{l=0}^{S-s-1}\frac{s(c+\rho(s+l))}{s(b+c+\rho(s+S+l-1))}
\eea
and where $\alpha^{(R)}_{r,R}(a',b',c';\rho')$ is given by a completely similar
expression. The phase is the result of multiplying many phases together.
This completes the calculation of the matrix elements of the relevant column 
as far as fusion rule I is concerned. The result has a form identical to what
is found for minimal models \cite{DF}.

Concerning fusion rule II it turns out that a simple trick allows to obtain
the result rather easily. In fact, a suitable contour deformation of the
complex contour ${\cal C}_v$ allows one to obtain an equation of the form
\ben
 {\cal S}^{(R,S)}_{(r,s,1)}=-\frac{1}{\pi}\left(e^{i\pi\rho s}s(c+\rho s)
  {\cal S}^{(R,S)}_{(r,s+1,0)}+s(a+c+2\rho s)\int_x^0...\right)
\een
The integral from $x$ to $0$ (we imagine $x<0$ in the s-channel) cannot have a 
contribution with a $(1-z)$ singularity appropriate for 
${\cal T}^{(R,S)}_{(R,S,0)}$, and so we do not specify the integrand, and we 
drop the integral in the calculation. Now it is an easy matter to obtain the
missing matrix elements from the ones we have already given. One finds
\ben
 \alpha^{(R,S)}_{(r,s,1),(R,S,0)}=-\frac{1}{\pi}\frac{s(c+\rho s)s((s+1)\rho)}
  {s(\rho)}\alpha^{(R,S)}_{(r,s+1,0),(R,S,0)}
\een
where
\ben
 s=0,...,S-1
\een

\subsection{The Row of the Transformation Matrix}
The procedure is to consider the s-channel block $S^{(R,S)}_{(R,S,0)}$ and
then isolate the t-channel singularities in $(1-z)$ and $(x-1)$. The strengths
of these singularities will tell us which t-channel block is obtained. In this
way we determine modified crossing matrix elements
\bea
 S^{(R,S)}_{(r,s,\epsilon)}(z,x)&=&\sum_{r'=0}^R\sum_{s'=0}^S
  \alpha^{(R,S)'}_{(r,s,\epsilon),(r',s',0)}t^{(R,S)}_{(r',s',0)}(z,x)\nn
 &+&\sum_{r'=0}^R\sum_{s'=0}^{S-1}
  \alpha^{(R,S)'}_{(r,s,\epsilon),(r',s',1)}t^{(R,S)}_{(r',s',1)}(z,x)
\eea 
These matrix elements are related to the ones  we have previously considered
by the normalisation constants of the last section. Denoting the corresponding 
normalisations in the t-channel by $\tilde{N}^{(R,S)}_{(r,s,0)}(a,b,c;\rho)$,
we have
\bea
 \alpha^{(R,S)}_{(R,S,0),(r,s,0)}&=&\alpha^{(R,S)'}_{(R,S,0),(r,s,0)}/
  \tilde{N}^{(R,S)}_{(r,s,0)}(a,b,c;\rho)\nn
 &=&\alpha^{(R,S)'}_{(R,S,0),(r,s,0)}/
  N^{(R,S)}_{(r,s,0)}(b,a,c;\rho)
\eea
We consider 
the real time ordered form of the integral representation
\bea
 &&e^{i\pi(R-S\rho)}S^{(R,S)}_{(R,S,0)}\nn
 &=&z^{2j_1j_2/t}(1-z)^{2j_2j_3/t}\nn
 &\cdot&z^PT\int_0^1\prod_{I=1}^R du_I\prod_{K=1}^S dv_K u_I^{a'}
  (1-zu_I)^{b'}(1-u_I)^{c'}(x-zu_I)\prod_{I<I'}(u_I-u_{I'})^{2\rho'}\nn
 &\cdot&v_K^a(1-zv_K)^b(1-v_K)^c(x-zv_K)^{-\rho}
  \prod_{K<K'}(v_K-v_{K'})^{2\rho}\prod_{I,K}(u_I-v_K)^{-2}
\label{row}
\eea
The phase on the left hand side takes into account that s- and t-channel blocks
are defined with different phase conventions as far as the factors $(x-u)$ and
$(x-v)^{-\rho}$ are concerned.
The pre-factor $z^P$ is obtained from scaling the integration variables with 
$z$. When
$z\rightarrow 1$ it is regular and we shall ignore it in the following.

Next, we use analytic tricks similar to those
used when discussing the integral identity, in order to isolate the
singularities in $(1-z)$ and $(x-1)$. 
We consider the integration region, where the first $r$ $u_i$'s are 
integrated (ordered) from $1-\epsilon$ to $1$, ($\epsilon$ small $>0$) 
the first $s$ $v_k$'s 
similarly from $(1-\epsilon)$ to $1$, and the remaining variables from $0$ to
$1-\epsilon$. The first $u_i$'s and $v_k$'s are indexed by $i$ and $k$ and the 
remaining ones by $j,l$. The first $u_i$'s are transformed as 
$u_i\rightarrow 1-u_i$, followed by $u_i\rightarrow (1-z) u_i$, 
and likewise $v_k\rightarrow 1-v_k$ followed by 
$v_k\rightarrow (1-z) v_k$. Inserting that in (\ref{row}) we find the singular
behaviour as $z\rightarrow 1$
\ben
 (1-z)^{-h(j_2)-h(j_3)+h(j_I)}
\een
where
\ben
 j_I=j_2+j_3-r+st
\een
Furthermore one isolates the $x\rightarrow 1$ behaviour
\ben
 (x-1)^{r-st}=(x-1)^{j_2+j_3-j_I}
\een
Therefore we may calculate the coefficient of $t^{(R,S)}_{(r,s,0)}$ in the 
expansion of $S^{(R,S)}_{(R,S,0)}$. After some work one finds the result
\bea
 \alpha^{(R,S)'}_{(R,S,0),(r,s,0)}&=&e^{i\pi(-R+S\rho)}
  {\cal J}_{r,s}(-b-c+2(r-1-(s-1)\rho),c;\rho)\nn
 &\cdot&{\cal J}_{R-r,S-s}(b+c-\rho-2(r-\rho s),a;\rho)
\eea
Some further calculations give
\bea
 \alpha^{(R,S)}_{(R,S,0),(r,s,0)}&=&(-1)^Se^{i\pi S\rho}
  \alpha^{(R)}_{R,r}(a',b',c',d';\rho')\alpha^{(S)}_{S,s}(a,b,c,d;\rho)\nn
 \alpha^{(S)}_{S,s}(a,b,c,d;\rho)&=&
  \prod_{i=0}^{s-1}\frac{s(b+i\rho)}{s(b+c+(s-1+i)\rho)}\nn
 &\cdot&\prod_{i=0}^{S-s-1}\frac{s(a+b+c+d+2(S-1)\rho-i\rho)}
  {s(b+c+d+2(S-1)\rho-(S-s-1+i)\rho)}
\eea
For convenience of writing we have defined
\bea
 d&=&-\rho\nn
 d'&=&-d/\rho=1
\eea
The presence of the $d$ dependence 
is the only difference from the corresponding
expression in minimal models \cite{DF}. It originates directly from the factors
$(u-x)$ and $(v-x)^{-\rho}$ in the integral realization. Such factors are not
present in the case of minimal models.

In order to isolate the singularity which corresponds to the t-channel blocks
for fusion rule II, we supplement the above specification of the 
integration region by the requirement that the variable $v_{s+1}$ should be 
integrated between $1-\epsilon$ and $1$, and then transformed as 
$v_{s+1}\rightarrow 1-v_{s+1}$ followed by 
$v_{s+1}\rightarrow (x-1) v_{s+1}$. After some 
calculations we find
\bea
 \alpha^{(R,S)'}_{(R,S,0),(r,s,1)}&=&
  e^{i\pi(-R+S\rho)}{\cal J}_{r,s}(-b-c+2(r-s\rho)+2(\rho-1),c;\rho)\nn
 &\cdot&{\cal J}_{R-r,S-s-1}(a,b+c+\rho-2(r-\rho s);\rho)\nn
 &\cdot&\frac{\Gamma(\rho -1-b-c-2\rho s+2r)
  \Gamma(b+c+2\rho s-2r+1)}{\Gamma(\rho)}
\eea
and using the normalisations and various gamma-function identities
\bea
 \alpha^{(R,S)}_{(R,S,0),(r,s,1)}&=&\pi\alpha^{(R,S)}_{(R,S,0),(r,s+1,0)}
  \frac{s(\rho)}{s(b+\rho s)s(b+c+\rho (s-1))}
\eea

\section{Monodromy Invariant Greens Functions}
Following the discussion in \cite{DF}, monodromy invariant 4-point
Greens functions 
\ben
 G_{j_1,j_2,j_3,j_4}(z,\overline{z},x,\overline{x})
\een
can be obtained by writing
\ben
 G_{j_1,j_2,j_3,j_4}(z,\overline{z},x,\overline{x})=\sum_{r,s,\epsilon}
  |S^{(R,S)}_{(r,s,\epsilon)}(j_1,j_2,j_3,j_4;z,x)|^2X^{(R,S)}_{(r,s,\epsilon)}
\een
This form ensures single valuedness in the limits $z\rightarrow 0$ and 
$x\rightarrow 0$. Single valuedness in the limits $z\rightarrow 1$ and 
$x\rightarrow 1$ is ensured provided the $X$'s are chosen to satisfy \cite{DF}
\ben
 X^{(R,S)}_{(r,s,\epsilon)}\propto \frac{\alpha^{(R,S)}_{(R,S,0),
  (r,s,\epsilon)}(b,a,c,d;\rho)}{\alpha^{(R,S)}_{(r,s,\epsilon),(R,S,0)}
  (a,b,c;\rho)}
\label{aloveral}
\een
Using re-scaling tricks similar to \cite{DF} we may throw away terms
(products of sine-functions) independent of $r$ and $s$. In this case we
obtain \cite{PRY2}
\bea
 X^{(R,S)}_{(r,s,0)}&=&X^{(R)}_r(a',b',c',d';\rho')X^{(S)}_s(a,b,c,d;\rho)\nn
 X^{(S)}_s(a,b,c,d;\rho)&=&\prod_{i=1}^s s(i\rho)\prod_{i=1}^{S-s}s(i\rho)
  \prod_{i=0}^{s-1}
  \frac{s(a+i\rho)s(c+i\rho)}{s(a+c+(s-1+i)\rho)}\nn
 &\cdot&\prod_{i=0}^{S-s-1}\frac{s(b+i\rho)s(1-a-b-c-d-2(S-1)\rho+i\rho)}
  {s(1-a-c-d-2(S-1)\rho+(S-s-1+i)\rho)}
\label{Xmono}
\eea
where $X^{(R)}_r(a',b',c',d';\rho')$ is given by a quite similar expression.
Let us illustrate the re-scaling technique by demonstrating the appearance
of $\prod_{i=0}^{S-s-1}s(b+i\rho)$. From the ratio between the $\al$'s in
(\ref{aloveral}) we find that the only term involving only $b$ is
\bea
 \prod_0^{s-1}\frac{1}{s(b+\rho(S-s+j))}&=&\prod_{S-s}^{S-1}\frac{1}
  {s(b+\rho j)}\nn
 &=&\prod_0^{S-1}\frac{1}{s(b+\rho j)}\prod_0^{S-s-1}s(b+\rho j)
\eea
The factor with the product from 0 to $S-1$ may be scaled away and we
are left with the aforementioned term. Generally, it is allowed to 
renormalise the coefficients by arbitrary functions of $R,S$ and 
$a,b,c,d$.

The expression (\ref{Xmono}) is very similar to the result for minimal 
models except for the presence of the $d=-\rho$ and $d'=1$. Finally
\ben
 X^{(R,S)}_{(r,s,1)}=-\pi^2\frac{s^2(\rho)}
  {s(c+\rho s)s((s+1)\rho)s(a+\rho s)s(a+c+\rho(s-1))}X^{(R,S)}_{(r,s+1,0)}
\een

Consistency requires that the Greens functions thus defined automatically are 
single valued also around $z\rightarrow x$. It has been checked that indeed 
for one screening charge of the second kind this is the case \cite{PRY2}. 
It is expected to be true generally.

It is convenient for studies of the operator algebra coefficients to also 
introduce the following expansion
\ben
 G_{j_1,j_2,j_3,j_4}(z,\overline{z},x,\overline{x})=\sum_{r,s,\epsilon}
  |s^{(R,S)}_{(r,s,\epsilon)}(j_1,j_2,j_3,j_4;z,x)|^2f^{(R,S)}_{(r,s,\epsilon)}
\een
where $s^{(R,S)}_{(r,s,\epsilon)}$ 
are defined in (\ref{sblockI}) and (\ref{sblockII}).
The coefficients $f^{(R,S)}_{(r,s,\epsilon)}$ differ from the coefficients
$X^{(R,S)}_{(r,s,\epsilon)}$ by a factor $(N^{(R,S)}_{(r,s,\epsilon)})^2$. 
It is 
fairly straightforward to collect all the results and obtain the expressions
for $f^{(R,S)}_{(r,s,\epsilon)}$. On the way we use the following
symmetrised version of (\ref{dfintegral}), which is proven using the
identity (\ref{gammaid}),
\bea
 {\cal J}_{nm}(a,b;\rho)&=&\rho^{2nm}\prod_{i,j=1}^{n,m}\frac{1}{(-i+j\rho)}
  \prod_{i=1}^n\frac{\Gamma(i\rho')}{\Gamma(\rho')}\prod_{i=1}^m
  \frac{\Gamma(i\rho)}
  {\Gamma(\rho)}\nn
 &\cdot&\prod_{i,j=0}^{n-1,m-1}\frac{1}
  {(a+j\rho-i)(b+j\rho-i)(a+b+(m-1+j)\rho-(n-1+i))}\nn
 &\cdot&\prod_{i=0}^{n-1}\frac{\Gamma(1+a'+i\rho')\Gamma(1+b'+i\rho')}
  {\Gamma(2-2m+a'+b'+(n-1+i)\rho')}\nn
 &\cdot&\prod_{i=0}^{m-1}\frac{\Gamma(1+a+i\rho)\Gamma(1+b+i\rho)}
  {\Gamma(2-2n+a+b+(m-1+i)\rho)}
\eea
We then obtain
\bea
 &&f^{(R,S)}_{(r,s,0)}=\Lambda^{(R,S)}_{r,s}(\rho)\nn
 &\cdot&\prod_{i,j=0}^{r-1,s-1}\frac{1}
  {(a+j\rho-i)^2(c+j\rho-i)^2(a+c+\rho(s-1+j)-(r-1+i))^2}\nn
 &\cdot&\prod_{i,j=0}^{R-r-1,S-s-1}\frac{1}
  {(b-i+j\rho)^2(e-i+j\rho)^2(e+b-(R-r-1+i)+(S-s-1+j)\rho)^2}\nn
 &\cdot&\prod_{i=0}^{r-1}\frac{G(1+a'+i\rho')G(1+c'+i\rho')}
  {G(2-2s+a'+c'+(r-1+i)\rho')}
  \prod_{i=0}^{s-1}\frac{G(1+a+i\rho)G(1+c+i\rho)}{G(2-2r+a+c+(s-1+i)\rho)}\nn
 &\cdot&\prod_{i=0}^{R-r-1}\frac{G(1+b'+i\rho)G(1+e'+i\rho)}
  {G(2+e'+b'-2(S-s)+(R-r-1+i)\rho')}\nn
 &\cdot&\prod_{i=0}^{S-s-1}\frac{G(1+b+i\rho)G(1+e+i\rho)}
  {G(2+e+b-2(R-r)+(S-s-1+i)\rho)}
\label{fnormI}
\eea
where we have defined
\ben
 G(x)=\frac{\Gamma(x)}{\Gamma(1-x)}=\frac{1}{G(1-x)}
\label{Gdef}
\een
and where
\bea
 \Lambda^{(R,S)}_{r,s}(\rho)&=&
  \rho^{4rs+4(R-r)(S-s)}\prod_{i=1}^s G(i\rho)\prod_{i=1}^{S-s}G(i\rho)
  \prod_{i=1}^r G(i\rho')\prod_{i=1}^{R-r}G(i\rho')\nn
 &\cdot&\prod_{i,j=1}^{r,s}\frac{1}{(i-j\rho)^2}
  \prod_{i,j=1}^{R-r,S-s}\frac{1}{(i-j\rho)^2}
\eea
and where we have defined
\bea
 e&=&-a-b-c-d-2\rho(S-1)+2(R-1)\nn
 e'&=&-e/\rho
\eea
These expressions are like the ones for minimal models except for the 
appearance of the terms $d,d'$ in the definition of $e,e'$. Finally
we have
\bea
 f^{(R,S)}_{(r,s,1)}&=&f^{(R,S)}_{(r,s+1,0)}\nn
 &\cdot&\frac{G(2+a+c+2s\rho-2r)G(1+a+c+2s\rho-2r)G(1-(s+1)\rho+r)}
  {G(1-r+a+s\rho)G(1-r+c+s\rho)G(2-r+a+c+(s-1)\rho)}
\label{fnormII}
\eea

The following useful identity for the $G$-functions defined above
will be used repeatedly in the sequel
\ben
 G(x)=(-1)^nG(x-n)\prod_{i=1}^n(x-i)^2
\label{Gid}
\een

\subsection{Case of Mixed Fusion Rules}

The construction of crossing matrices and monodromy coefficients in the
case of mixed fusion rules (or over-screening) is made
essentially trivial by the following observations. We have previously seen that
the integral for the DF contours Fig. \ref{figanI} is related to
a corresponding integral for time ordered integrations by the factor
\ben
 \lambda_r(1/t)\lambda_s(t)\lambda_{R-r}(1/t)\lambda_{S-s}(t)
\een
Similarly for the present case of some
contours being of Felder type, we get instead a factor 
\ben
 \lambda_r(1/t)\chi^{(2)}_s(S-s_3-1;t)\lambda_{R-r}(1/t)\lambda_{S-s}(t)
\een 
for (I,II), and a factor
\ben
 \lambda_r(1/t)\lambda_s(t)\lambda_{R-r}(1/t)\chi^{(2)}_{S-s}(S-s_2-1;t)
\een
for (II,I). These rather 
trivial new normalisations allow us to follow completely the treatment for 
fusion rule I (i.e. the case (I,I)) described above and insert appropriate
$\frac{\chi}{\lambda}$ 
factors as normalisations. It is fairly easy to see, that for
the new monodromy coefficients $f^{(R,S)}_{(r,s,(I,II))}$ and
$f^{(R,S)}_{(r,s,(II,I))}$ (in a self explanatory notation), 
the only $\frac{\chi}{\lambda}$ factors multiplying $f^{(R,S)}_{(r,s,(I,I))}$
which survive, are
ones which do not depend on $(r,s)$ or $(R-r,S-s)$, but only on $(R,S)$ and 
hence may be absorbed into renormalisations.

\section{Operator Algebra Coefficients}

All properties of a CFT are encoded in the knowledge of the symmetry algebra
$W\oplus\bar{W}$ ($W$ is an extension of the Virasoro algebra $Vir$, here
a semi-direct sum of $sl(2)$ and $Vir$), the spectrum of primary fields
and the operator algebra (OA) coefficients or structure constants.
From the analysis in the preceding sections we are now in a position to
determine the OA coefficients and thereby 'solve' the theory.

In the $SL(2)$ theory the operator algebra coefficients 
$C_{\lambda_1\lambda_2}^{\lambda_3}$ are defined by
\ben
 \phi_{j_2}(z,\bar{z},x,\bar{x})\phi_{j_1}(0,0,0,0)=\sum_j
  \frac{|x|^{2(j_1+j_2-j)}}{|z|^{2(\D(j_1)+\D(j_2)-\D(j))}}
  C_{\lambda_1\lambda_2}^\lambda\phi_j(0,0,0,0)+...
\een
where the dots represent contributions from descendants and where 
$\lambda=2j+1$. The OA coefficients are only defined up to normalisation of 
the fields
\bea
 \phi_{j_i}(z,\bar{z},x,\bar{x})& \rightarrow & 
   \rho_{\lambda_i}\phi_{j_i}(z,\bar{z},x,\bar{x}),\ \ \ \ \rho_1=1\nn
 C_{\lambda_1\lambda_2}^{\lambda_3} &\rightarrow &
  \rho_{\lambda_1}\rho_{\lambda_2}\rho^{-1}_{\lambda_3}
  C_{\lambda_1\lambda_2}^{\lambda_3}
\eea
Define the conjugate field $\phi^j(z,\bar{z},x,\bar{x})$ by
\ben
 \bra{0}\phi^{j_i}(z,\bar{z},x,\bar{x}) \phi_{j_j}(0,0,0,0)\ket{0}
  =\delta^{\lambda_i}_{\lambda_j}|z|^{-4\D(j_i)}|x|^{4j_i}
\een
Then the OPE coefficients can be calculated by considering the 3-point 
function
\bea
 &&\bra{0} \phi^{j_3}(z_3,\bar{z}_3,x_3,\bar{x}_3)\phi_{j_2}(z_2,\bar{z}_2,x_2,
  \bar{x}_2)
  \phi_{j_1}(z_1,\bar{z}_1,x_1,\bar{x}_1)\ket{0} \nn
 &=&C_{\lambda_1\lambda_2}^{\lambda_3}
  |z_1-z_2|^{2(\D_3-\D_1-\D_2)}|z_2-z_3|^{2(\D_1-\D_2-\D_3)}
  |z_1-z_3|^{2(\D_2-\D_1-\D_3)}\nn 
 &\cdot&
  |x_1-x_2|^{2(j_1+j_2-j_3)}|x_2-z_3|^{2(j_2+j_3-j_1)}
  |x_1-x_3|^{2(j_1+j_3-j_2)}
\eea
Defining
\ben
 \overline{\phi^j}(z,\bar{z},x,\bar{x})=
  \phi^{j}(-1/z,-1/\bar{z},-1/x,-1/\bar{x})|z|^{4\D_j}|x|^{-4j}
\een
we see that
\ben
 C_{\lambda_1\lambda_2}^{\lambda_3}=
  \bra{0} \overline{\phi^{j_3}}(0,0,0,0)\phi_{j_2}(1,1,1,1)\phi_{j_1}(0,0,0,0)
  \ket{0}
\een
Consider now
\ben
 \bra{0}\phi_{j_i}(z,\bar{z},x,\bar{x}) \phi_{j_j}(0,0,0,0)\ket{0}
  =C_{\lambda_i\lambda_j}^1|z|^{-4\D(j_i)}|x|^{4j_i}
\een
In the $SL(2)$ theory the fields are self-conjugate so 
$C_{\lambda_i\lambda_j}^1$ is not only symmetric in $\lambda_i,\lambda_j$
but also diagonal, and it may be used as a metric, raising and lowering indices
\bea
 C_{\lambda_i\lambda_j}^1&=&c_{\lambda_i}\delta_{\lambda_i\lambda_j}=
  g_{\lambda_i\lambda_j}\nn
 C_{\lambda_i\lambda_j}^{\lambda_l}&=&g^{\lambda_l\lambda_k}
  C_{\lambda_i\lambda_j\lambda_k}=
  c_{\lambda_l}^{-1}C_{\lambda_i\lambda_j\lambda_l}
\eea
Here $C_{\lambda_i\lambda_j\lambda_k}$ is totally symmetric in the
$\lambda_i$'s, and $c_\lambda$ is a normalisation constant.
We also have 
\bea
 \phi^{j_i}(z,\bar{z},x,\bar{x})&=&g^{\lambda_i\lambda_j}
  \phi_{j_j}(z,\bar{z},x,\bar{x})\nn
 &=&c_{\lambda_i}^{-1}\phi_{j_i}(z,\bar{z},x,\bar{x})
\eea

One can fix the normalisation constant $\hat{c}_\lambda=1$ in a basis
where $\hat{\phi}_j=\rho_\lambda^{-1}\phi_j$, 
such that the fields are orthonormal
\ben
 \hat{C}_{\lambda_1\lambda_2}^1=\delta_{\lambda_1\lambda_2}
\een
Let the totally symmetric OA coefficients in the orthonormal basis be 
denoted as $\hat{C}_{\lambda_i\lambda_j\lambda_l}$, then in the other basis 
the OA coefficients
$C_{\lambda_i\lambda_j}^{\lambda_k}$ can be written as
\bea
 C_{\lambda_i\lambda_j}^{\lambda_k}&=&\rho_{\lambda_i}\rho_{\lambda_j}
  \rho_{\lambda_k}^{-1}
  \hat{C}_{\lambda_i\lambda_j\lambda_k} \nn
 \rho^2_{\lambda}&=&c_{\lambda}
\eea
Here the normalisation constants correspond to the original basis without
'hats'.

\subsection{Case of Fusion Rule I}

Since there is no overlap between the two sets of fusion rules, a certain
4-point Greens function 
\ben
 G_{j_1j_2j_3j_4}(z,\zb,x,\bar{x})=
  \bra{j_4}\phi_{j_3}(1,1,1,1)\phi_{j_2}(z,\zb,x,\bar{x})\ket{j_1}
\een
behaves, in the limit $z,x\rightarrow0$, either like
\bea
 G_{j_1j_2j_3j_4}(z,\zb,x,\bar{x})&=&\sum_{j}C_{\lambda_3\lambda}^{\lambda_4}
  C_{\lambda_1\lambda_2}^\lambda |z|^{2(-\D_{j_1}-\D_{j_2}+\D_j)}
  |x|^{2(j_1+j_2-j)}
  \left(1+{\cal O}(z,\zb,x,\bar{x})\right)\nn
 &=&\sum_{j}C_{\lambda_3\lambda}^{\lambda_4}(I)
  C_{\lambda_1\lambda_2}^\lambda(I) |z|^{2(-\D_{j_1}-\D_{j_2}+\D_j)}
  |x|^{2(j_1+j_2-j)}
  \left(1+{\cal O}(z,\zb,x,\bar{x})\right)\nn
 &+&\sum_{j}C_{\lambda_3\lambda}^{\lambda_4}(II)
  C_{\lambda_1\lambda_2}^\lambda(II) |z|^{2(-\D_{j_1}-\D_{j_2}+\D_j)}
  |x|^{2(j_1+j_2-j)}
  \left(1+{\cal O}(z,\zb,x,\bar{x})\right)\nn
\eea
or like
\bea
 G_{j_1j_2j_3j_4}(z,\zb,x,\bar{x})&=&
  \sum_{j}C_{\lambda_3\lambda}^{\lambda_4}(I)
  C_{\lambda_1\lambda_2}^\lambda(II) |z|^{2(-\D_{j_1}-\D_{j_2}+\D_j)}
  |x|^{2(j_1+j_2-j)}
  \left(1+{\cal O}(z,\zb,x,\bar{x})\right)\nn
 &+&\sum_{j}C_{\lambda_3\lambda}^{\lambda_4}(II)
  C_{\lambda_1\lambda_2}^\lambda(I) |z|^{2(-\D_{j_1}-\D_{j_2}+\D_j)}
  |x|^{2(j_1+j_2-j)}
  \left(1+{\cal O}(z,\zb,x,\bar{x})\right)\nn
\label{4Greensmixed}
\eea
depending on the 4-tuple $(j_1,j_2,j_3,j_4)$.
First we concentrate on the former case where
\ben
 f_{(r,s,\epsilon)}^{(R,S)}(j_1,j_2,j_3,j_4)=
  C_{\lambda_3\lambda_\epsilon}^{\lambda_4}
  C_{\lambda_1\lambda_2}^{\lambda_\epsilon}
\een
with $\lambda_\epsilon=\lambda_I,\lambda_{II}$ for $\epsilon =0,1$.
Such a split requires that the monodromy coefficients are properly normalised.
The normalisation adopted so far (\ref{fnormI}) (and thereby (\ref{fnormII}))
follows the prescription of Dotsenko and Fateev in the case of minimal models,
but turns out to be inadequate here
\cite{PRY2}. Indeed it is completely essential that the above 
factorisation takes place in such a way that the operator algebra 
coefficients only depend on the variables indicated and not on anything else.
In particular, $C_{\lambda_1\lambda_2}^{\lambda_\epsilon}$ is allowed to
depend on $r,s$ which are given in terms of the spins (the $\lambda$'s) 
indicated, but it is not allowed to depend on $R,S$ for example. Likewise
$C_{\lambda_3\lambda_\epsilon}^{\lambda_4}$ is allowed to
depend on $R-r,S-s$ but again, not on $R,S$. However, it is allowed 
(as utilised above) to renormalise the coefficients by arbitrary functions of
$R,S,\lambda_1,\lambda_2,\lambda_3$, ($j_4=j_1+j_2+j_3 -R+St$). It turns out 
\cite{PRY2} to be possible to devise such a normalisation with the above 
criterion satisfied. This we first do for fusion rule I. 
We have to use (cf. (\ref{abc}))
\bea
 a&=&-\lambda_3+R-St+t\nn
 b&=&-\lambda_1+R-St+t\nn
 e&=&-\lambda_2+R-St+t\nn
 c&=&\lambda_4+R-St\nn
 e+b&=&-\lambda_I+2(R-r)-1-2t(S-s-1)=+\lambda_{II}+2(R-r)-1-2t(S-s-1)\nn
 a+c&=&\lambda_I+2r-2st-1+t=-\lambda_{II}+2r-2st-1+t
\eea
Then we may express the monodromy coefficient as 
\bea
 f_{(r,s,0)}^{(R,S)}&=&C_{\lambda_1\lambda_2}^{\lambda_I}(I)
  C_{\lambda_3\lambda_I}^{\lambda_4}(I)\nn
 &=&t^{4rs}\prod_1^s G(it-r)\prod_1^r G(i/t)\nn
 &\cdot&\prod_0^{r-1}\frac{G(S+(\lambda_3-R+j)/t)G(S+1+(-\lambda_4-R+j)/t)}
  {G(1-(\lambda_I+1+j)/t)}\nn
 &\cdot&\prod_0^{s-1}\frac{G(1-\lambda_3+R-r-(S-s+j)t)
  G(1+\lambda_4+R-r-(S-s+1-j)t)}{G(1+\lambda_I+r-(s-j)t)}\nn
 &\cdot&t^{4(R-r)(S-s)}\prod_{1}^{S-s}G(it-R+r)\prod_1^{R-r}G(i/t)\nn
 &\cdot&\prod_0^{R-r-1}\frac{G(S+(\lambda_1-R+i)/t)G(S+(\lambda_2-R+i)/t)}
  {G((\lambda_I-1-i)/t)}\nn
 &\cdot&\prod_0^{S-s-1}\frac{G(1-\lambda_1+r-(s+i)t)G(1-\lambda_2+r-(s+i)t)}
  {G(1-\lambda_I+R-r-(S-s-1-i)t)}
\eea
This is certainly not in the form required, but we may renormalise it.
Thus we consider terms involving only $\lambda_1$
\bea
 &&\prod_{i=0}^{R-r-1}G(S+(\lambda_1-R+i)/t)\prod_{j=0}^{S-s-1}G(1-\lambda_1+r
  -(s+j)t)\nn
 &=&t^{2r(S-s)}\prod_{i=0}^{r-1}G(1-s+(1-\lambda_1+i)/t)
  \prod_{j=0}^{s-1}G(\lambda_1+jt)\nn
 &\cdot&\prod_{i=0}^{R-1}G(S+(\lambda_1-R+i)/t)
  \prod_{j=0}^{S-1}G(1-\lambda_1+(1-S+j)t)
\eea
The last line may now be scaled away. Symmetry arguments in the $\lambda_i$'s
render considerations of terms including $\lambda_2,\lambda_3,\lambda_4$
trivial, while the $\lambda_I$ part contributes as
\bea
 &&\prod_{i=0}^{R-r-1}G^{-1}((\lambda_I-1-i)/t)\prod_{j=0}^{S-s-1}G^{-1}
  (1-\lambda_I+R-r-(S-s-1-j)t)\nn
 &\cdot&\prod_{i=0}^{r-1}G^{-1}(1-(\lambda_I+1+i)/t)\prod_{j=0}^{s-1}
  G^{-1}(1+\lambda_I+r-(s-j)t)\nn
 &=&t^{-2(R-r)(S-s)-2rs}\prod_{i=0}^{R-r-1}G^{-1}(S-s-(1-\lambda_I+i)/t)
  \prod_{j=0}^{S-s-1}G^{-1}(1-\lambda_I-jt)\nn
 &\cdot&\prod_{i=0}^{r-1}G^{-1}(1+s-(1+\lambda_I+i)/t)
  \prod_{j=0}^{s-1}G^{-1}(1+\lambda_I-(j+1)t)
\eea
The total power of $t$ becomes
\ben
 t^{4RS-2(R-r)(S-s)-2rs}
\een
where we may absorb the $t^{4RS}$ in the normalisation. Finally we find
\bea
 C_{\lambda_1\lambda_2}^{\lambda}(r,s;I)&=&
  t^{-2rs}\prod_{i=1}^r G(i/t)\prod_{i=1}^s 
  G(it-r)\nn
 &\cdot&\prod_{i=0}^{r-1}\frac{G(1-s+(1-\lambda_1+i)/t)
  G(1-s+(1-\lambda_2+i)/t)}{G(1+s-(1+\lambda+i)/t)}\nn
 &\cdot&\prod_{i=0}^{s-1}
  \frac{G(\lambda_1+it)G(\lambda_2+it)}{G(1+\lambda-(1+i)t)}\nn
 C_{\lambda_3\lambda}^{\lambda_4}(R-r,S-s;I)&=&
  t^{-2(R-r)(S-s)}\prod_{i=1}^{R-r}G(i/t)\prod_{i=1}^{S-s}G(it-(R-r))\nn
 &\cdot&\prod_{i=0}^{R-r-1}\frac{G(1-(S-s)+(1-\lambda_3+i)/t)}{G(S-s-
  (1-\lambda+i)/t)}\nn
 &\cdot&\prod_{i=0}^{R-r-1}
  G(-(S-s)+(1+\lambda_4+i)/t)\nn
 &\cdot&\prod_{i=0}^{S-s-1}\frac{G(\lambda_3+it)G(-\lambda_4+(1+i)t)}
  {G(1-\lambda-it)}
\label{oac}
\eea
and notice the symmetry
\ben
 C_{\lambda_1\lambda_2}^{\lambda_3}(r,s;I)=
  C_{\lambda_1,-\lambda_3+t}^{-\lambda_2+t}(r,s;I)
\een
which reflects the involutionary invariance (\ref{invol}).
Here, for clarity we have indicated the dependencies on $r,s$ or $R-r,S-s$ 
and on the fusion rule (here I). In fact this is somewhat superfluous.
The point is, that from three
spins, it is always clear by which of the two fusion rules they couple,
and for each case there is a unique possible value of $r,s$ (or $R-r,S-s$).
Furthermore, a trivial consistency check reveals that indeed the two
expressions in (\ref{oac}) are equivalent. 

We may now compute the normalisation constants $c_\lambda$ 
and $\rho_\lambda$ for $\lambda=r-st$
\ben
 c_\lambda=t^{-2s}\frac{G(\lambda/t)}{G(1/t)}\spa \rho_\lambda=t^{-s}
  \frac{G^{1/2}(\lambda/t)}{G^{1/2}(1/t)}
\label{clambda}
\een
Using this we want to find the normalised and symmetric OA coefficient
$\hat{C}_{\lambda_1\lambda_2\lambda_3}(I)$. Let us introduce the notation
\bea
 \hat{R}&=&(r_1+r_2+r_3-1)/2\nn
 \hat{S}&=&(s_1+s_2+s_3)/2\nn
 R_i&=&\hat{R}-r_i\nn
 S_i&=&\hat{S}-s_i\nn
\label{RSconv}
\eea
where $\lambda_i=2j_i+1=r_i-s_it$. Employing (\ref{Gid}) one obtains
\bea
 \hat{C}_{\lambda_1\lambda_2\lambda_3}(I)&=&
  G^{1/2}(1/t)\prod_{i=1}^{\hat{R}} G(i/t)\prod_{j=1}^{\hat{S}} 
  G(j t-\hat{R})\nn
 &\cdot&\prod_{k=1}^3 \frac{t^{s_k}}{G^{1/2}(\lambda_k/t)}
  \frac{\prod_{i=1}^{R_k} G(i/t)\prod_{j=1}^{S_k} G(j t-R_k)}
  {\prod_{i=1}^{r_k-1}G(i/t)\prod_{j=1}^{s_k}G(j t-r_k+1)}
\label{ChatI}
\eea
In order to compare with Andreev's result we introduce the functions
$P(n,m)$ \cite{An}, here expressed in terms of the $G$-functions (\ref{Gdef}) 
\bea
 P(n,m)&=&\prod_{j=1}^{n-1}\prod_{i=1}^{m-1}(j t-i)^{-2}
  \prod_{j=1}^{n-1}G(j t)\prod_{i=1}^{m-1}G(i/t)\nn
 &=&(-1)^{(n-1)(m-1)}\prod_{i=1}^{m-1}G(i/t)\prod_{j=1}^{n-1}G(j t-m+1)
\eea
whereby we may rewrite (\ref{ChatI}) as
\ben
 \hat{C}_{\lambda_1\lambda_2\lambda_3}(I)=t^{2\hat{S}}(-1)^{\hat{S}}
  G^{1/2}(1/t)P(\hat{S}+1,\hat{R}+1)
  \prod_{k=1}^3 G^{1/2}(1-\lambda_k/t) 
  \frac{P(S_k+1,R_k+1)}{P(s_k+1,r_k)}
\een
This is in accordance with Andreev's result. However, so far we have
only considered the case of fusion rule I while the case of fusion II is 
somewhat more subtle. Indeed, in \cite{An} Andreev is able to determine
the OA coefficients in that case only up to normalisation. Nevertheless,
an inspection of 4-point functions with mixed fusion rules determines the
normalisation of the fields, after which  
(\ref{fnormII}) fixes unambiguously the
structure coefficients, as will become clear below. It should be stressed
that the fixing of the OA coefficients for fusion rule II is only 
{\em relative to the choice} of normalisation of the OA coefficients for 
fusion rule I (\ref{oac}). 

\subsection{Case of Fusion Rule II}
A priori it is not clear how to normalise the fields in a fusion rule
II coupling. This is due to the fact that the identity operator
($j=0;\ r=1,s=0$) is not present according to fusion rule II, hence
$C_{\lambda_i\lambda_j}^1(II)$ cannot
be used to determine the normalisations. However, when extracting the OA
coefficients from a 4-point Greens function with mixed fusion rules
(\ref{4Greensmixed}), the fields should have the same normalisations, 
whether they appear in the fusion rule I coefficient or in the fusion rule II 
coefficient. Therefore, the normalisation constants in the case of fusion 
rule II are inherited from the case of fusion rule I and are thus given by 
the same expressions (\ref{clambda}). 

In the notation (\ref{RSconv}) the numbers of screenings are $R_3+p/2,S_3+q/2$,
and we have, up to an overall renormalisation constant,
\bea
 C_{\lambda_1\lambda_2}^{\lambda_3}(II)&\propto&t^{-2(R_3+p/2)(S_3+q/2)}
  \prod_{i=1}^{R_3+p/2}G(i/t)\prod_{j=1}^{S_3+q/2}G(jt-(R_3+p/2))\nn
 &\cdot&\prod_{i=0}^{R_3+p/2-1}\frac{G(1-(S_3+q/2)+(1-\lambda_1+i)/t)
  G(1-(S_3+q/2)+(1-\lambda_2+i)/t)}{G(1+S_3+q/2-(1+\lambda_3+i)/t)}\nn
 &\cdot&\prod_{j=0}^{S_3+q/2-1}\frac{G(\lambda_1+jt)G(\lambda_2+jt)}{G(
  1+\lambda_3-(1+j)t)}\nn
 &=&t^{-4}\frac{G(1+\lambda_3)}{G(1+\hat{R}-\hat{S}t)}
  \prod_{i=1}^{R_3+p/2} G(i/t)\prod_{j=1}^{S_3+q/2} G(j t-R_3-p/2)\nn
 &\cdot&
  \frac{\prod_{i=1}^{p/2-1-R_2}G(i/t)\prod_{j=1}^{q/2-1-S_2}G(j t-p/2+R_2+1)}
  {\prod_{i=1}^{r_1-1}G(i/t)\prod_{j=1}^{s_1}G(j t-r_1+1)}\nn
 &\cdot&\frac{\prod_{i=1}^{p/2-1-R_1}G(i/t)
  \prod_{j=1}^{q/2-1-S_1}G(j t-p/2+R_1+1)}
  {\prod_{i=1}^{r_2-1}G(i/t)\prod_{j=1}^{s_2}G(j t-r_2+1)}\nn
 &\cdot&\frac{\prod_{i=1}^{\hat{R}-p/2}G(i/t)\prod_{j=1}^{\hat{S}-q/2-1}
  G(j t-\hat{R}+p/2)}
  {\prod_{i=1}^{p-r_3-1}G(i/t)\prod_{j=1}^{q-s_3}G(j t-p+r_3+1)}
\label{CIIpropto}
\eea
where we have used (\ref{Gid}). In order to determine the proportionality
constant, let us further rewrite (\ref{CIIpropto}). Thus, with the same
proportionality constant one finds
\bea
 C_{\lambda_1\lambda_2}^{\lambda_3}(II)&\propto&t^{-4-2(q-s_3)}\frac{
  G(1+\lambda_3)G(-\lambda_3/t)}{G(1+R_1-S_1t)G(1+R_2-S_2t)}
  C_{\lambda_1\lambda_2}^{-\lambda_3}(I)\nn
 &=&t^{-2(q-s_3)}\frac{G(2-\lambda_3/t)G(-R_3+S_3t)}{G(-\hat{R}+(\hat{S}+1)t)
  G(1+r_3-(1+s_3)t)}
  C_{\lambda_1\lambda_2}^{-\lambda_3+2t}(I)
\eea
Let us now make a comparison with (\ref{fnormII})
\bea
 f_{(r,s,1)}^{(R,S)}&=&C_{\lambda_3\lambda_{II}}^{\lambda_4}(II)
  C_{\lambda_1\lambda_2}^{\lambda_{II}}(II)\nn
 &=&C_{\lambda_3,-\lambda_4+t}^{-\lambda_{II}+t}(II)
  C_{\lambda_1\lambda_2}^{\lambda_{II}}(II)\nn
 &\propto&t^{-2q-6}C_{\lambda_3,-\lambda_4+t}^{\lambda_{II}-t}(I)
  C_{\lambda_1\lambda_2}^{-\lambda_{II}+2t}(I)\nn
 &\cdot&\frac{G(t-\lambda_{II})G(1-\lambda_{II}+t)G(-j_1-j_2+j_{II})}{G(
  -1+t-j_1-j_2-j_{II})G(1+j_3+j_4-j_{II})G(t-j_3-j_4-j_{II})}\nn
 &=&t^{-2q-6}f_{(r,s+1,0)}^{(R,S)}\nn
 &\cdot&\frac{G(1-\lambda_{II}+t)
  G(-\lambda_{II}+t)G(1-(s+1)t+r)}{G(1-\lambda_3+R-r-(S-s-1)t)
  G(1+\lambda_4+R-r-(S-s)t)G(1-\lambda_{II}+r-st)}\nn
\eea
{}From this we may read off the proportionality constant in (\ref{CIIpropto})
to be \cite{PRY2} 
\ben
 t^{q+3}
\een
In conclusion, with the relevant expressions for the numbers of screenings,
now denoted $r,s$
\ben
 r=(r_1+r_2-r_3-1+p)/2\spa s=(s_1+s_2-s_3+q)/2
\een 
the result (\ref{oac}) for $C_{\lambda_1\lambda_2}^{\lambda_3}$ is valid also 
for fusion rule II provided one multiplies the rhs by $t^{q+3}$.  

As a consistency check of the result $C_{\lambda_1\lambda_2}^{\lambda_3}$ in
(\ref{oac}) one may analyse the
products of $G$-functions in order to verify that up to cancellations of
the form $\G(0)/\G(0)$ raised to some power, (\ref{oac}) is non-vanishing
and well-defined for both sets of fusion rules. It turns out that for 
fusion rule I no cancellations are needed. For fusion rule II where
the extra factor of $t^{q+3}$ is of no importance in this analysis, one finds
\bea
 \G(0)&\sim&\prod_{i=0}^{r-1}G(1-s+(1-\lambda_1+i)/t)\sim
  \prod_{i=0}^{r-1}G(1-s+(1-\lambda_2+i)/t)\nn
 \G(0)&\sim&\prod_{i=0}^{r-1}\frac{1}{G(1+s-(1+\lambda_3+i)/t)}\nn
 1/\G(0)&\sim&\prod_{i=0}^{s-1}G(\lambda_1+it)\sim
 \prod_{i=0}^{s-1}G(\lambda_2+it)\sim
 \prod_{i=0}^{s-1}\frac{1}{G(1+\lambda_3-(1+i)t)}
\eea
leaving (\ref{oac}) non-vanishing and well-defined upon the cancellation
$\left(\G(0)/\G(0)\right)^3$.

In order to make a direct comparison with Andreev's result we use
(\ref{Gid}) to first obtain the symmetric OA coefficient
\bea
 \hat{C}_{\lambda_1\lambda_2\lambda_3}(II)&=&
  t^{(q-1)(4\hat{R}-2p-1)}G^{\frac{1}{2}}(1/t)
  \prod_{i=1}^{\hat{R}-p/2}G(i/t)\prod_{j=1}^{\hat{S}-q/2}G(jt-\hat{R}+p/2)\nn
 &\cdot&\prod_{k=1}^3 \frac{t^{s_k}}{G^{\frac{1}{2}}(\lambda_k/t)}
  \frac{\prod_{i=1}^{R_k+p/2} G(i/t)
  \prod_{j=1}^{S_k+q/2} G(j t-R_k-p/2)}
  {\prod_{i=1}^{p-r_k}G(i/t)\prod_{j=1}^{q-s_k-1}G(j t-p+r_k)}
\eea
where we have included the renormalisation constant $t^{q+3}$.
Furthermore, we should change our notation in (\ref{RSconv}) for $\lambda_3$
(the definitions for $\hat{R}$, $R_i$, $\hat{S}$, $S_i$ and 
$\lambda_{\{1,2\}}$ remain unchanged) 
\ben
 \lambda_3=p-r_3-(q-s_3)t=-r_3+s_3t
\een
We see that in terms of the present notation, the original one has been
transformed according to 
\bea
 r_1 \rightarrow r_1&\spa& s_1 \rightarrow s_1\nn
 r_2 \rightarrow r_2&\spa&  s_2 \rightarrow s_2\nn
 r_3 \rightarrow p-r_3&\spa& s_3 \rightarrow q-s_3\nn
 R_1 \rightarrow p/2-1-R_2&\spa& S_1 \rightarrow q/2-S_2\nn
 R_2 \rightarrow p/2-1-R_1&\spa& S_2 \rightarrow q/2-S_1\nn
 R_3 \rightarrow \hat{R}-p/2&\spa& S_3 \rightarrow \hat{S}-q/2\nn
 \hat{R}\rightarrow R_3+p/2&\spa& \hat{S} \rightarrow S_3+q/2
\eea
In this new notation one can establish the identity
\bea
 \frac{\prod_{i=1}^{R_k}G(i/t)\prod_{j=1}^{S_k-1}G(j t-R_k)}
  {\prod_{i=1}^{r_k-1}G(i/t)\prod_{j=1}^{s_k}G(j t-r_k+1)}
  &=&t^{2(q-1)(r_k-1-R_k)}\nn
 &\cdot&\frac{\prod_{i=1}^{p-1-R_k}G(i/t)\prod_{j=1}^{q-S_k}G(j t-p+1+R_k)}
  {\prod_{i=1}^{p-r_k}G(i/t)\prod_{j=1}^{q-s_k-1}G(j t-p+r_k)}
\eea
valid only for $k=1,2$. Finally one finds
\bea
 \hat{C}_{\lambda_1\lambda_2\lambda_3}(II)
 &=&t^{2S-1}
  \frac{G^{\frac{1}{2}}(\lambda_3/t)G^{\frac{1}{2}}(1/t)}
  {G^{\frac{1}{2}}(\lambda_1/t)G^{\frac{1}{2}}(\lambda_2/t)}
  \frac{G(1+\lambda_3)}{G(1+R_3-S_3t)}\nn
 &\cdot&\prod_{i=1}^{\hat{R}} G(i/t)\prod_{j=1}^{\hat{S}} G(j t-\hat{R})
  \prod_{k=1}^3\frac{\prod_{i=1}^{R_k}G(i/t)\prod_{j=1}^{S_k-1}G(j t-R_k)}
  {\prod_{i=1}^{r_k-1}G(i/t)\prod_{j=1}^{s_k}G(j t-r_k+1)}
\eea
while Andreev's formula may 
be written (in the notation employed here, and up to a power of $t$) as
\bea
 C_{\lambda_1\lambda_2\lambda_3}(II)&\propto & G(S_3t-R_3)
  \prod_{i=1}^{\hat{R}} G(i/t)\prod_{j=1}^{\hat{S}} G(j t-\hat{R})
  \prod_{k=1}^3\prod_{i=1}^{R_k}G(i/t)\prod_{j=1}^{S_k-1}G(j t-R_k)\nn
\eea

\chapter{Conclusions and Outlook}

In this thesis we have shown how to deal with fractional powers of free fields
by virtue of fractional calculus, and thereby we have elucidated
the meaning of the second screening charge
proposed by Bershadsky and Ooguri \cite{BO} for affine $SL(2)$ WZNW models.
The real merit of this is that it has enabled us to use free field 
realizations to build the most general chiral blocks for such 
theories on the sphere, even in the case of admissible representations with 
fractional levels. The ensuing integral formulas are tractable and have 
allowed us to verify many formal
properties, such as projective invariances and the fact
that the chiral blocks satisfy the Knizhnik-Zamolodchikov equations.
We have presented an explicit and simple proof of the proposal in
\cite{FGPP} for how Hamiltonian reduction works at the level of correlators,
reducing affine $SL(2)$ current algebra for admissible representations
to conformal minimal models. We have further verified that the blocks
found indeed reduce in that way. Also the relation to more standard
formulations of Hamiltonian reduction has been discussed. Employing the
notion of over-screening has allowed us to re-derive the fusion rules
on the basis of an analysis of 3-point functions in terms of our free field
realization.

The proposal in this thesis for how to treat free ghost fields
raised to fractional powers involves a variety of expansions of certain
functions, with different monodromies for the individual expansion terms.
The physical condition imposed is then that the final expression must respect
the known monodromy for the item in consideration. This idea has now
found applications in other respects, notably in the operator approach
to Liouville theory \cite{Schnitt}. 

In order to generalise a work \cite{DF} by
Dotsenko and Fateev on minimal models, we have investigated in
great detail the 4-point blocks. In particular we have devised integration
contours appropriate for suitable bases of blocks, both using our own
representation based on free fields and one by Andreev \cite{An} applicable
only to 4-point functions. The equivalence of these two representations
have been analysed thoroughly at the level of highly non-trivial
integral manipulations. As a by-product, we have presented a proof
of a remarkable duality in minimal models. In order to compute the monodromy
invariant Greens functions we have calculated the crossing matrix.
Based on the monodromy coefficients we have isolated 
the operator algebra coefficients of the theory for both
fusion rules. Again the notion of over-screening has been utilised, and in
such a way that we have been able to normalise the operator algebra
coefficients unambiguously, contrary to the results in \cite{An}. 

We have presented an explicit (generalised) 
Wakimoto free field realization of affine
Lie algebras by a Gauss decomposition and by determining anomalous 
(quantum) terms. We have discussed the screening currents of both kinds 
generalising the one of the second kind by Bershadsky and Ooguri,
and found partial proofs of their existence. Complete proofs were presented
in some cases. We have undertaken the study of primary fields in the 
framework of the isotopic $x$ variable representation and found 
solutions in special cases of $SL(n)$.
In the case of $SL(3)$ we have found the general solution for the primary 
fields and we have been able to prove the consistency conditions for
the existence of the screening currents. This means that we are now in a
position to generalise our program in chapters 4 and 5 for $SL(2)$ to $SL(3)$.
In principle we should then be able to determine the fusion rules based
on 3-point functions, find the 4-point Greens functions and eventually the 
operator algebra coefficients. We hope to come back to this elsewhere.

Very recently de Boer and  Feh\'{e}r \cite{deBF} have also obtained an
explicit construction of Wakimoto realizations of current algebras, based
on quantising a Poisson bracket realization obtained using Hamiltonian
reduction. However, a direct comparison is not obvious and their proof
is announced to be presented elsewhere. Furthermore, they write down the 
general result for screening currents of the {\em first} kind only and claim
to have and intend
to present a proof of that elsewhere. If this is the case we have then shown
in this thesis that screening currents of the {\em second} kind do exist
for $SL(n)$. We believe it should be possible to generalise our existence
proofs to the other simple groups. This belief is supported by our very
latest developments. Utilising a new and more compact notation to be 
presented in \cite{PRY3}, we hope to be able to carry out such
proofs and discuss further general results
on the issue of primary fields. Generalisations to supergroups are also
currently under investigation.

Some steps have already been taken in the direction of generalising to
supergroups. In \cite{BTBKT} the representation theory of the affine Lie 
superalgebra $\widehat{sl(2|1)}$ at fractional level and the free field 
realization of the corresponding affine current algebra have been discussed.

A future work lies in the computation of correlation functions in the
$SL(2)/SL(2)$ approach to non-critical strings where conformal minimal 
matter is coupled to gravity. Attempts in this direction are found in
\cite{An2}. Using the results established in this thesis, a possible
outline is the following. In the $SL(2)/SL(2)$ approach one deals with a
direct sum of 3 copies of affine $SL(2)$ current algebra $\widehat{sl(2)}$
with levels $k$, $-k-4$ and 4 respectively, $k=p/q-2$ is fractional and the 
co-prime integer pair $(p,q)$ parametrises the conformal minimal model. 
The former two copies correspond
to the original $SL(2)$ group and to the gauge group which is also an
$SL(2)$ but with shifted (dressed) level. The last copy with level 4
is expressed in terms of a fermionic ghost pair arising in a gauge 
fixing procedure. Utilising this splitting it should be possible to
calculate the correlation functions. However, subtleties from the BRST
conditions make the computation non-trivial. When the necessary techniques
eventually are developed, a generalisation to higher groups and
supergroups seems feasible.
\\[.6cm]
Though not final, yet indeed crucial steps in the direction of formulating 
non-critical strings in the $G/G$ framework, the work in this thesis presents 
some new and definite results, notably the completion of the solution of CFT 
based on affine $SL(2)$ current algebra for admissible representations in terms
of free field realizations, the 
completion of the Wakimoto construction for simple affine current algebras,
and discussions on free field realizations of screening currents and
primary fields for simple groups.

\end{document}